**CPP Magnetoresistance of Magnetic Multilayers: A critical review.**  (11/23/15)

Jack Bass

Department of Physics and Astronomy, Michigan State University, USA 48824

Abstract

We present a comprehensive, critical review of data and analysis of Giant (G) Magnetoresistance (MR) with Current-flow Perpendicular-to-the-layer-Planes (CPP-MR) of magnetic multilayers [F/N]$_n$ ($n$ = number of repeats) composed of alternating nanoscale layers of ferromagnetic (F) and non-magnetic (N) metal, or of spin-valves that allow control of anti-parallel (AP) and parallel (P) orientations of the magnetic moments of adjacent F-layers.  GMR, a large change in resistance when an applied magnetic field changes the moment ordering of adjacent F-layers from AP to P, was discovered in 1988 in the geometry with Current flow in the layer-Planes (CIP).  The CPP-MR has two advantages over the CIP-MR: (1) relatively simple two-current series-resistor (2CSR) and more general Valet-Fert (VF) models allow more direct access to the underlying physics; and (2) it is usually larger, which should be advantageous for devices.  When the first CPP-MR data were published in 1991, it was not clear whether electronic transport in GMR multilayers is completely diffusive or at least partly ballistic.  It was not known whether the properties of layers and interfaces would vary with layer thickness or number.  It was not known whether the CPP-MR would be dominated by scattering within the F-metals or at the F/N interfaces.  Nothing was known about: (1) spin-flipping within F-metals, characterized by a spin-diffusion length, $l_{sf}^F$; (2) interface specific resistances (AR = area A times resistance R) for N1/N2 interfaces; (3) interface specific resistances and interface spin-dependent scattering asymmetry at F/N and F1/F2 interfaces; and (4) spin-flipping at F/N, F1/F2 and N1/N2 interfaces.  Knowledge of spin-dependent scattering asymmetries in F-metals and F-alloys, and of spin-flipping in N-metals and N-alloys, was limited.  Since 1991, CPP-MR measurements have quantified the scattering and spin-flipping parameters that determine GMR for a wide range of F- and N-metals and alloys and of F/N pairs.  This review is designed to provide a history of how knowledge of CPP-MR parameters grew, to give credit for discoveries, to explain how combining theory and experiment has enabled extraction of quantitative information about these parameters, but also to make clear that progress was not always direct and to point out where disagreements still exist.  To limit its length, the review considers only collinear orientations of the moments of adjacent F-layers.  To aid readers looking for specific information, we have provided an extensive table of contents and a detailed summary.  Together, these should help locate over 100 figures plus 17 tables that collect values of individual parameters.  In 1997, CIP-MR replaced anisotropic MR (AMR) as the sensor in read heads of computer hard drives.  In principle, the usually larger CPP-MR was a contender for the next generation read head sensor.  But in 2003, CIP-MR was replaced by the even larger Tunneling MR (TMR), which has remained the read-head sensor ever since.  However, as memory bits shrink to where the relatively large specific resistance AR of TMR gives too much noise and too large an R to impedance match as a read-head sensor, the door is again opened for CPP-MR.  We will review progress in finding techniques and F-alloys and F/N pairs to enhance the CPP-MR, and will describe its present capabilities.

## 1. Introduction

### 1.1. General remarks and motivation for this review.

Experimental physics is a demanding discipline, especially when attacking a new subject where little is known. It is easy to be fooled by nature. To make this point, I tell new students two rules that are not unique to me (see, e.g., [1, 2]) but which I arrived at independently. (1) The first person to measure something often gets it wrong. (2) The second person to measure the same thing usually gets the same answer as the first person. The reason for rule (1) is that it is difficult to recognize and eliminate all systematic errors. To test the effects of changing many variables can require much extra work. And some variables are neither obviously important nor easy to control. The reason for rule (2) is that the second person trusts the first person. Initial results close to the first person's are accepted. Initial results not close are examined for reasons to make corrections to get close. Over time, investigators usually correct errors, eventually getting answers 'right'. I've used these rules to warn students to ask how their results could be wrong, and how they can be cross-checked for errors. I use them here to warn readers that not all published parameters are reliable. I have included a large number of figures to help the reader to gauge reliability.

To understand how a subject developed, one must know the assumptions initially made and how the incorrect ones were (often gradually) resolved. I will point out basic assumptions, and try to make clear which have been validated and how, and where full understanding is not yet available. I'll try to explain what we think we understand and why we think we understand it. Usually, I'll provide enough information so the reader can evaluate the conclusions that I reach. But to keep this review from being too long, I'll sometimes refer the reader to the original literature for more information. From here on, I'll shift from 'I' to the more colloquial 'we'.

As several reviews of the Current-Perpendicular-to-Plane (CPP) MR have already been published [3, 4] [5-13], one might ask 'Why this one?" The answer is, the prior reviews are either out of date or limited by space constraints, and none explicated the progress made toward device competitiveness of the CPP-MR. The present review is intended to be both more comprehensive on experiments and more critical in its discussion of limitations than any previous one. An extensive table of contents is designed to guide the reader to specific information about topics covered, including over 100 figures and 17 tables of measured values of basic parameters. The review is intended to be of use to readers as diverse as those who initially know nothing about CPP-MR (Sections 1-6 provide background), but want to know everything (sections 7-10 cover history, data, related analyses, and work toward devices), and those who want only to understand what can and has been learned, without too many details (each section starts with an overview and section 11 contains a summary and conclusions). Most of this review focuses upon the physics of CPP-MR at temperature T = 4.2K, which is simplest because electron-magnon and electron-phonon scattering are negligible. But section 10, which covers progress toward devices, focuses on room temperature (T = 293K). We also consider only collinear orientations of the moments of adjacent F-layers, except in passing in section 10. For the angular dependence of CPP-MR see, e.g., [14, 15]. For spin-transfer torque see, e.g., [16-19].

### 1.2. Electron Spin and Magnetic Moment.

An electron has not only charge, but also a quantum spin of ½, which gives a small magnetic moment that points opposite to the spin because the electron's charge is negative. Semiconductor electronics is based solely upon the electron's charge. The 1988 discovery [20, 21] of Giant

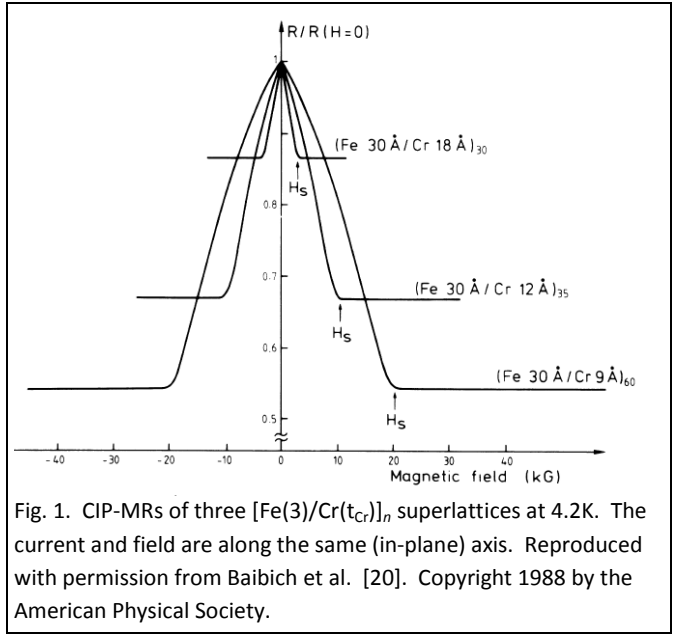

Fig. 1. CIP-MRs of three $[Fe(3)/Cr(t_{Cr})]_n$ superlattices at 4.2K. The current and field are along the same (in-plane) axis. Reproduced with permission from Baibich et al. [20]. Copyright 1988 by the American Physical Society.

Magnetoresistance (GMR)] in magnetic multilayers composed of alternating nanoscale layers of ferromagnetic (F) and non-magnetic (N) metals showed that the electron's spin (magnetic moment) could lead to large changes in electrical resistance with device applicability [22]. Outgrowths of this initial discovery have led to advances that spawned the term 'Spintronics'—electronics where the electron spin plays an important role [23]. We'll usually refer to the moment rather than to the spin.

### 1.3. CIP-MR and the Baibich model.

GMR involves a large change (usually a decrease) in the electrical resistance R of an $[F/N]_n$ or $[F/N]_N$ multilayer ($n$ or $N$ = number of repeats) when the magnetic moments of adjacent F-layers are reoriented by an applied in-plane magnetic field H from anti-parallel (AP) to each other (usually in low field), to parallel (P) above the saturation field, $H_s$, that aligns all of the F-layer moments. The original measurements that led to the name GMR are shown in Fig. 1 [20] as normalized R(H)/R(H=0) vs H at temperature T = 4.2K for $[Fe(3)/Cr(t_{Cr})]_n$ multilayers (layer thicknesses are in nm) with $t_{Cr}$ varying from 0.9 to 1.8 nm and $n$ varying from 30 to 60. These data were taken in the Current-In-Plane (CIP) geometry.

We define a general MR(H) as

$$MR(H) = [[R(H) - R(P)]/R(P)]\times100\% \tag{1a}$$

The largest MR is expected when the moments for R(H) are oriented AP to each other

$$MR = [[R(AP) - R(P)]/R(P)]\times100\% = [\Delta R/R(P)]\times100\%. \tag{1b}$$

Definition (1b) gives an MR $\cong$ 80% in Fig. 1 for $t_{Cr}$ = 0.9 nm, which earlier experiments by Grunberg et al. [24] had shown produces antiferromagnetic coupling between adjacent Fe layers, giving an AP state at H = 0. This leaves the question, what are the MRs defined by Eq. (1b) for $t_{Cr}$ = 1.2 nm and 1.8 nm? From only the data of Fig. 1 we cannot tell. The states of lowest R are P states, but the states of largest R are probably not fully AP states. Part of the reduction in MR is likely due to thicker Cr layers leading to a 'dilution' of the CIP-MR [25]. But, absent a reliable AP state, we can conclude only that the measured MRs for $t_{Cr}$ = 1.2 nm and 1.8 nm are lower bounds on the true MRs of Eq. 1b.

Baibich et al.'s qualitative explanation [20] of the GMR in Fig. 1 is illustrated in Fig. 2, except using the more transparent Current-Perpendicular-to-Plane (CPP) geometry. They made the following assumptions: (1) The GMR results from a change in the magnetic order of adjacent F layers from AP at H = 0 to P at large |H|; (2) The conduction electron current can be separated into two components, moment 'up' and moment 'down'; (3) These moments do not 'flip' (relax) as the conduction electrons traverse the multilayer; and (4) The scattering of conduction electrons as they pass through an F-layer is spin-dependent (asymmetric)—usually weaker when the electron's moment is along the moment of the F-layer and stronger when the electron's moment is opposite to that of the F-layer. Section 3 will show how all four assumptions were motivated by earlier research. In the top two pictures in Fig. 2, strong scattering is indicated by a change in direction of the electron's trajectory, and weak scattering by no change. In the AP state (left half of Fig. 2), electrons with each moment direction are scattered strongly in one F-layer and weakly in the other. The resulting scattering is 'intermediate'. In contrast, in the P state (right half of Fig. 2), electrons with up moments are scattered weakly in both F-layers, thereby 'shorting out' the sample. If the asymmetry of scattering is large, the difference in scattering for the AP and P states is also large, and the GMR is large.

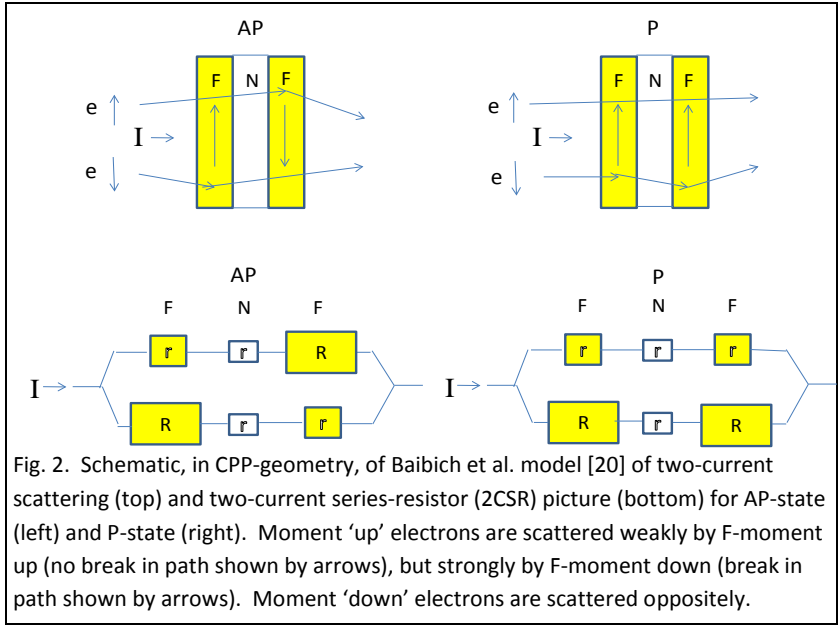

Fig. 2. Schematic, in CPP-geometry, of Baibich et al. model [20] of two-current scattering (top) and two-current series-resistor (2CSR) picture (bottom) for AP-state (left) and P-state (right). Moment 'up' electrons are scattered weakly by F-moment up (no break in path shown by arrows), but strongly by F-moment down (break in path shown by arrows). Moment 'down' electrons are scattered oppositely.

As shown in Fig. 1, GMR was discovered in the CIP geometry [20, 21]. For typical sample dimensions of length and width ~ mm and thickness ~ 0.1 μm, shape anisotropy guarantees that the moments will lie in the layer planes, and the multilayer CIP-R is ~ Ω, large enough to measure with standard instruments and to impedance match with typical device components. Indeed, within a decade of its discovery, CIP MR was used in the read heads in computer hard drives (section 10.1) and as sensors for various uses [22].

However, for quantifying the physics of GMR, the CIP-MR has two disadvantages relative to the CPP-MR. (a) CIP current flow through the multilayer is non-uniform. (b) When there is no spin-flipping, the CIP-MR requires three parameters not in the two-current series resistor (2CSR) model (section 1.4.1) for the CPP-MR, the mean-free-paths, $\lambda_N$, $\lambda_F^{\downarrow}$, and $\lambda_F^{\uparrow}$ in the N- and F-metals. These λs enter (and complicate) the CIP equations, because, while drifting on average along the layers, the electrons must traverse the distances across the F- and N-layers to transfer between adjacent F-layers information about relative moment orientations. Appendix A defines mean-free-paths and shows how to estimate their values. In sputtered samples at 4.2K they range from λ ~ 5-100 nm. Allowing spin-flipping adds two parameters each to the CIP- and CPP-MRs, the spin-diffusion lengths in F and N (Section 1.4.2)).

## 1.4. CPP-MR

This review focuses upon the alternative CPP geometry. Fig. 3 [26], from the first CPP-MR measurements, in 1991 by Pratt et al. on Co/Ag [26], shows that the CPP-MR is usually larger (often several times larger) than the CIP-MR. The more complex structure of MR(H) in Fig. 3 results from having Ag layers thick enough to make magnetic coupling between the Co layers weak. The use of weak coupling allows systematic studies of the CPP-MR over a wide range of layer thicknesses, $t_F$ and $t_N$, without complications of changing coupling. But reliable analysis requires determining (or at least estimating) MR(AP). We'll explain how to do so in sections 5 and 8.

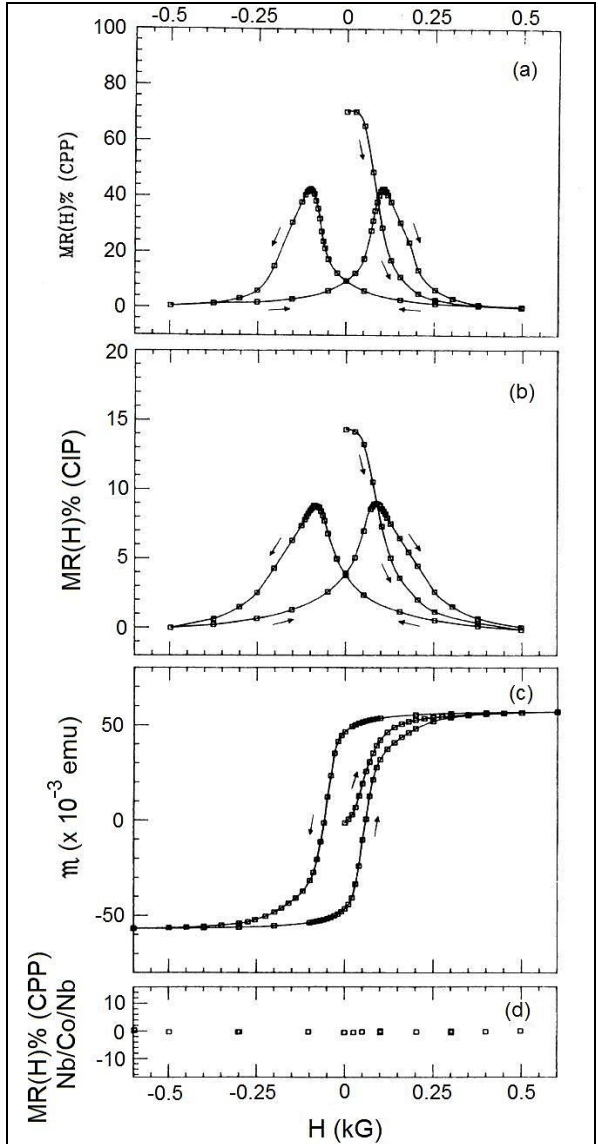

**Fig. 3. (a)** CPP-MR; (b) CIP-MR; and (c) Magnetization *M,* for [Co(6)/Ag(6)]$_{60}$ multilayers. (d) CPP-MR for single Co layer. MR(0) ≈ MR(AP), MR(Peak) = MR(Pk) ≈ MR(H$_c$) and MR(H$_s$) = MR(P), where AP = anti-parallel, P = parallel, (0) = virgin state, H$_c$ = coercive field, and H$_s$ = saturation field. Reproduced with permission from Pratt et al. [26]. Copyright 1991 by the American Physical Society.

In the CPP-MR geometry, the intrinsic quantity is the specific resistance, AR, the product of the resistance R and the area A through which the CPP current flows. Combining a typical AR ~ 10 fΩm$^2$ with the standard area A ~ 10$^{-6}$ m$^2$ listed above, gives a resistance R ~ 10$^{-8}$ Ω, much too small for devices and requiring special techniques to measure. Device applications require nanopillar areas, A ≤ 10$^{-14}$ m$^2$, which can now be made with advanced lithographies. CPP nanopillars with such areas are being tested for next generation read heads (see section 10).

In contrast, for quantifying physics the CPP-MR has advantages over the CIP-MR. First, with careful design, the CPP current flow can be made uniform. Second, the characteristic lengths are predicted (and usually seen) to be the spin-diffusion lengths, $l_{sf}$, (the average distance a conduction electron diffuses before its moment reverses direction). When CPP-MR measurements began in 1991, nothing was known about $l_{sf}^F$ in F-metals and little was known about $l_{sf}^N$ in N-metals (see Sections 3, 8, and Appendix 1). As explained in Appendix 1, $l_{sf}$ ~ 10λ was expected to be much longer than the layer thicknesses in most multilayers. If so, work by Zhang and Levy (ZL) [27], Lee et al. [28], and Valet and Fert [29] led to the hope that the CPP-MR might be described by simple two-current-series resistor (2CSR) equations where the only lengths are the layer thicknesses t$_F$ and t$_N$. In this model, AR(AP) and AR(P) for an [F/N]$_n$ multilayer are uniquely determined by t$_F$, t$_N$, *n*, and a set of fixed parameters for the F-metal, the N-metal, and the F/N interface. Because of their wide use in analyzing experiments, we first briefly describe the 2CSR [28, 29] and more general Valet-Fert (VF) [29] models, and define their parameters. Section 4 will cover details and limitations of the VF model, which adds moment flipping, and reduces to the 2CSR model when such flipping is negligible. An initial focus of CPP measurements was to see if such simple models could describe real data.

### 1.4.1. The Two-Current Series-Resistor (2CSR) Model and Its Parameters.

The 2CSR model of the CPP-MR is illustrated in Fig. 2, where scatterings in the top half are converted into resistances in the bottom half, including a small resistance (r) for the separating N-layer. If there is no moment-flipping, and transport is diffusive, the contribution to AR from within a metallic layer has the form ρt, where ρ is the layer resistivity and t is the layer thickness.   For the N-layer, only one resistivity is needed, $\rho_N$.  For the F-layer, two resistivities are needed, $\rho_F^\uparrow$ and $\rho_F^\downarrow$, where ↑, ↓ indicate conduction electron moment along (↑) or opposite to (↓) the F-layer moment.  For analysis, it is often more convenient to use two alternative parameters introduced in [28] and [29]: the dimensionless bulk moment-scattering asymmetry $\beta_F = (\rho_F^\downarrow - \rho_F^\uparrow)/(\rho_F^\downarrow + \rho_F^\uparrow)$, bounded by −1 and +1,  and the enhanced resistivity $\rho_F^* = (\rho_F^\uparrow + \rho_F^\downarrow)/4 = \rho_F/(1-\beta_F^2)$, where $\rho_F$ is the resistivity of the F-metal at 4.2K.  It has been found that $\rho_N$ and $\rho_F$ can often be estimated from van der Pauw measurements [30] on ~ 200 nm thick films of the N- or F-metal deposited in the same way as in the multilayer.  Since nothing was initially known about scattering at F/N interfaces, it made sense to allow such scattering to also be moment-dependent (asymmetric).  To do so requires two interface specific resistances, $AR_{F/N}^\uparrow$ and $AR_{F/N}^\downarrow$.  Here too, we use alternative parameters: the dimensionless interface scattering asymmetry $\gamma_{F/N} = (AR_{F/N}^\downarrow - AR_{F/N}^\uparrow)/(AR_{F/N}^\downarrow + AR_{F/N}^\uparrow)$ bounded by  −1 and +1, and twice the enhanced interface specific resistance $2AR^*_{F/N} = (AR_{F/N}^\downarrow + AR_{F/N}^\uparrow)/2$. [28, 29].    Except for the current leads to the multilayer, which depend upon the measurement geometry as will be discussed later, these 5 parameters characterize the 2CSR model.

To show the simplicity of the 2CSR model, we write its equations for an [F($t_F$)/N($t_N$)]$_n$ multilayer.  For convenience, we neglect lead resistances, which must be included to fit most experimental data.

We start with AR(AP).  Because of the symmetry of the AP state in the CPP geometry, the total AR for each electron moment direction is the same whether the direction is 'up' or 'down':

$$AR^{up,\ down}(AP) = n2\rho_N t_N + (n/2)\rho_F^\downarrow t_F + (n/2)\ \rho_F^\uparrow t_F + nAR_{F/N}^\downarrow + nAR_{F/N}^\uparrow. \tag{2}$$

Here we count two F/N interfaces for each F-layer, use $2\rho_N$ because each moment channel contains only half the total electrons, and neglect the unpaired F or N interface at each end of the sample. Rewriting Eq. 2 in terms of $\rho_F^*$ and $AR_{F/N}^*$, gives:

$$AR^{up,\ down}(AP) = n(2\rho_N t_N + 2\rho_F^* t_F + 4AR_{F/N}^*) \tag{3}$$

Adding the equal values of AR$^{up}$ and AR$^{down}$ in parallel gives a simple sum of bulk and interface terms:

$$AR(AP) = n(\rho_N t_N + \rho_F^* t_F + 2AR_{F/N}^*). \tag{4}$$

The simplicity of Eq. (4) shows the advantage of $\rho_F^*$ and $AR_{F/N}^*$ over $\rho_F^\downarrow$, $\rho_F^\uparrow$, $AR_{F/N}^\downarrow$, and $AR_{F/N}^\uparrow$.

With the same recipe, the 2CSR equations for AR(P) and A$\Delta$R are [28, 29]]:

$$AR(P) = AR(AP) - [n^2(\beta_F \rho_F^* t_F + 2\gamma_{F/N} AR_{F/N}^*)^2]/AR(AP), \text{ and} \tag{5}$$

$$A\Delta R = AR(AP) - AR(P) = [n^2(\beta_F \rho_F^* t_F + 2\gamma_{F/N} AR_{F/N}^*)^2]/AR(AP). \tag{6}$$

Lastly, multiplying both sides of Eq. 6 by AR(AP) and taking both square roots gives

$$\sqrt{(A\Delta R)AR(AP)} = n(\beta_F \rho_F^* t_F + 2\gamma_{F/N} AR_{F/N}^*). \tag{7}$$

The right hand side (rhs) of Eq. 7 is independent of the bulk properties of the N-metal. Also, for samples with fixed $t_F$, a plot of $\sqrt{(A\Delta R)AR(AP)}$ vs $n$ should yield a straight line passing through the origin. These two characteristics of Eq. 7 will be used below to extract important physics (see, e.g., sections 8.2.2, 8.5.1, 8.9, and 8.13).

To conclude this section, Eqs. 4-7 give direct access to the 5 parameters they contain, provided that one can measure AR(AP) and AR(P). As noted above, AR(P) is easy, just increase H to beyond $H_s$ for the F-layers. AR(AP), in contrast, presents a problem that early investigators had to solve. We'll see how they did so in section 5. In the meantime, we shall assume that AR(AP) can be determined.

**1.4.1.1. A$\Delta$R and the Relative Importance of Bulk vs Interfaces.**

The numerator of Eq. (6) lets us answer two important questions about the 2CSR CPP-MR: (a) What are its basic sources, and (b) What are the relative importances of Bulk vs Interface contributions?

(a) The fundamental sources of the 2CSR CPP-MR are the scattering asymmetries $\beta_F$ and $\gamma_{F/N}$, since A$\Delta$R vanishes if both asymmetries are zero. For more insight into the physics, note that the two terms in the numerator can be rewritten as: $\beta_F \rho_F^* = (\rho_F^\downarrow - \rho_F^\uparrow)/4$ and $2\gamma_{F/N} AR_{F/N}^* = (AR_{F/N}^\downarrow - AR_{F/N}^\uparrow)/2$.

(b) For thin F-layers, A$\Delta$R is usually dominated by the contribution from F/N interfaces. To estimate when Bulk F-metal contributions become important, we take rounded values for Co/Cu or Co/Ag of $\rho_{Co}^* \sim 100$ n$\Omega$m, $2AR_{Co/N}^* \sim 1$ f$\Omega$m$^2$, $\beta_{Co} \sim 0.5$, and $\gamma_{Co/N} \sim 0.8$ (see section 8.4). $\beta_F \rho_{Co}^* t_F$ will equal $2\gamma_{F/N} AR_{F/N}^*$ for $t_F \sim 16$ nm $\cong$ 80 monolayers (ML). Alternatively, since the numerator is squared, and $2\gamma_{F/N}AR_{F/N}^*$ is always present, increasing $t_{Co}$ from zero will double A$\Delta$R for $t_{Co} \sim 6$ nm, only about 30 ML. For the alternative Py/Cu, with larger $\rho_{Py}^* \sim 300$ n$\Omega$m, larger $\beta_{Py} \sim 0.8$, similar $2AR_{Py/Cu}^* \sim 1$ f$\Omega$m$^2$, and slightly smaller $\gamma_{Py/Cu} \sim 0.7$ (section 8.7), doubling A$\Delta$R requires only $t_{Py} \sim 3$ nm, or $\sim 15$ ML.

These parameters let us estimate the diameter d that will give the R $\sim$ 10 ohm resistances needed for devices. A minimum sample, with two F-layers separated by an N layer, has AR $\sim$ 10 f$\Omega$m$^2$ = 10 m$\Omega$($\mu$m)$^2$. A d = 1 $\mu$m pillar with AR = 10 f$\Omega$m$^2$ will give R $\sim$ 13 m$\Omega$, and R = 10 $\Omega$ needs d $\sim$ 36 nm.

**1.4.2. Additional parameters of the Valet-Fert (VF) model.**

To go beyond the 2CSR model, by letting moments flip (relax) in each layer, requires a more complex CPP-MR analysis, mostly requiring numerical solution. Such analysis is usually done using the model of Valet and Fert (VF), which provides recipes for calculating AR(AP), AR(P), and A$\Delta$R for general F/N multilayers [29]. We will describe the VF model in section 4.2. At this point, what is important is that it adds 3 more parameters to the 5 parameters above. Two are 'spin-diffusion lengths', $l_{sf}^N$ in the N-metal and $l_{sf}^F$ in the F-metal. The third, $\delta_{F/N}$, allows for moment-flipping at an F/N interface, giving the probability of moment flipping as $P = [(1 - \exp(-\delta)]$. $\delta_{F/N}$ has been investigated only recently, only for a few Co/N pairs, and has rarely been included in CPP analyses. Continuing to neglect current leads, adding these 3 parameters gives a total of 5 + 3 = 8 parameters for the VF model at 4.2K. When $l_{sf}^F$, $l_{sf}^N$ >> $t_F.t_N$, and $\delta_{F/N} = 0$, the VF model reduces to the 2CSR model.

When $t_F$ and $t_N$ in CPP multilayers are shorter than the layer mean-free-paths, one might expect transport across such layers to be ballistic. However, aside from a few cases of controversy, see expecially section 8.9, analyses have assumed diffusive transport. Diffusive transport has been justified by arguing that Intermixed F/N interfaces are too disordered and rough [31-35] to allow the coherent interactions between adjacent interfaces needed to see ballistic effects. We will see that early CPP-MR experiments exerted much effort to see if diffusive 2CSR and VF models could describe real data.

**1.5. Questions about the CPP-MR and some tentative answers.**

As a convenient framework for raising questions, we use the model of VF [29] that has been used to analyze most CPP-MR data. The model is based upon a Boltzmann Transport analysis that assumes diffusive transport, assumes the same spherical Fermi surfaces for both the F- and N-metals, and is strictly valid only in the limit $l_{sf} >> \lambda$. To characterize a given multilayer, VF analysis uses the set of 8 parameters given above, plus lead parameters as needed.

Crucial questions about the CPP-MR and these VF parameters include the following. (1) How does one obtain (or at least approximate) AP states? We'll see that the answer to this question for the earliest simple [F/N]$_n$ multilayer data of Fig. 3 is experimentally easy, but took substantial time and effort to justify. We'll describe alternatives, called spin-valves (SVs) that give reliable AP states. (2) Do CPP-MR data vary with layer thicknesses, $t_F$, $t_N$, and bilayer number, $n$, as predicted by VF theory? If so, when can data be described by the simpler 2CSR model? If the data vary as predicted, do they yield parameters, and thus properties of layers and interfaces, that are independent of $t_F$, $t_N$, and $n$? (3) How widely do the derived parameters vary for different F-metals and alloys and for different F/N pairs? (4) How sensitive are the parameters to structural details, to temperature, and to strictly satisfying the assumptions under which they were derived? (5) For samples with well-defined dominant scatterers, are CPP-MR parameters valid only for the CPP-MR? Or are they 'universal'—agreeing to within mutual uncertainties with the same parameters derived from very different kinds of measurements and/or from calculations with no adjustable parameters? (6) Do any experimental data clearly deviate from the 'predictions' of VF theory?

We'll argue that the VF model (including its 2CSR limit) can describe most or all CPP-MR data, and show that in some cases the VF parameters look to be closely 'universal'. These results suggest that CPP transport in present multilayers, where interfaces are disordered and somewhat intermixed, is (at least mostly) diffusive. They also show that, with care, the CPP-MR can provide fundamental information about electronic transport within magnetic layers and at metallic interfaces. When CPP-MR measurements began in 1991, little or nothing was known about moment flipping within F- or N-metals, or about the importance and any of the properties of F/N interfaces. In this review we try to explain what has been learned, the difficulties overcome in learning it, and what is still in dispute or not yet known. We'll also discuss progress toward CPP-MR devices.

**1.6. Organization of the Rest of the Review**.

The rest of the review is organized as follows. Section 2 answers the question: When should the VF parameters for layers and interfaces be intrinsic and when not. Section 3 gives background information needed to compare with CPP results. Section 4 gives an overview of theory of the CPP-MR. Prior reviews [3, 4, 8] have covered theory in detail. So we don't repeat complete coverage, but focus on the VF model that is used to analyze most CPP-MR data, and upon what other treatments tell us about its range of validity and expected limitations. Section 5 describes the different ways used to achieve AP states. Section 6 covers the three main techniques used to measure the CPP-MR, along with a more complex fourth one (Current at an Angle to the Plane--CAP) used for a few years. We first briefly describe each technique and outline its advantages and disadvantages. We then examine it in detail. Section 7 gives a historical timeline of what we deem to be especially significant CPP-MR results. This timeline is intended to show how understanding of the CPP-MR developed and to give credit for discoveries. Section 8 presents experimental results and analysis, organized roughly along the timeline, except that once we start a topic, we usually follow it to its end before turning to the next topic. Section 8 focuses upon the physics underlying the CPP-MR, leaving studies focused upon devices to section 10. Section 9 covers miscellaneous topics not easily categorized within section 8. Section 10 describes work toward CPP-MR devices. Section 11 contains a summary and conclusions. The review ends with two appendices, one on mean-free-paths and spin-diffusion lengths, and one on magnetic media and read-head sensors.

## 2. When Should Parameters be Intrinsic and When Not?

### 2.1. Bulk Parameters for well-defined alloys.

At 4.2K, if conduction electron scattering in an alloy is dominated by a known concentration of a chosen impurity, the bulk parameters within the alloy should be intrinsic. Such parameters should, thus, be reproducible over time in a given laboratory and also from laboratory to laboratory. They should also, in principle, agree with values for the same parameters determined with different experimental techniques. To check that a known impurity dominates the scattering, one must show that the residual resistivities, $\rho_o$, of deposited films of the alloy are significantly larger than those of deposited films of the host metal alone. To check stability over time, one must regularly recheck the $\rho_o$s of newly deposited thin films of the alloy made in the same way as in the multilayer.

At room temperature (293K), if scattering by phonons contributes a significant fraction of the total resistivity, then $\rho_N$, $\rho_F$, $l_{sf}^N$, and $l_{sf}^F$ should change with temperature roughly proportionally to the phonon contribution, $\rho$ increasing and $l_{sf}$ decreasing. $\beta_F$, $\gamma_{F/N}$, $2AR_{F/N}^*$, and $\delta_{F/N}$ may or may not change; their behaviors must be determined experimentally,

### 2.2. Bulk Parameters for nominally 'pure' metals.

At 4.2K, scattering in nominally pure metals is dominated by 'dirt' (unknown concentrations of unknown impurities and defects). The parameters are, thus, determined by a combination of the electronic structure of the metal and the 'dirt', and no parameter is intrinsic. Table 1 [36] shows that $\beta_f$ for alloys of Co, Ni, and Fe varies widely for different impurities, including in sign. The residual resistivities per at.% of different impurities in given F-metals also vary substantially [37, 38]. There is, thus, no fundamental reason for any of these parameters from different laboratories to agree. Even in a single laboratory, stability of $\rho_N$ and $\rho_F$ must be rechecked regularly over time on separately deposited thin films thick enough (e.g., 200 nm) to eliminate size effects. Such stability over years and different targets is a hopeful sign. But even then, other measurements must establish that the resistivity measured on a separately deposited film is close to that of the same metal as a thin layer in a multilayer. To show that it is, the resistivity in the multilayer must be treated as an unknown, and shown to agree to within uncertainties (probably 10-20% agreement is the best one can hope for) with the resistivity for separate films. Sections 3.5 and 8.4.1 contain examples of such tests. Lastly, stability of parameters over time in one laboratory says nothing about their use by another laboratory. The best one can hope is that parameters such as $l_{sf}^N$ and $l_{sf}^F$ might scale with the inverse resistivities, $1/\rho_N$ and $1/\rho_F$.

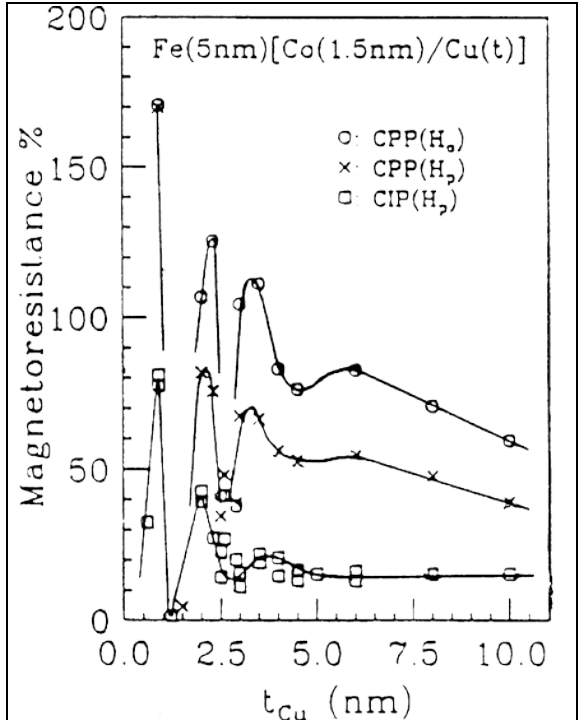

**Fig. 4.** CPP- and CIP-MRs vs $t_{Cu}$ for $[Co(1.5)/Cu(t_{Cu})]_n$ multilayers. Open circles are CPP-MR(0); crosses are CPP-MR(Pk); open squares are CIP-MR(Pk). From Schroeder et al. [39]. Reproduced with permission from Cambridge University Press.

At 293K, the parameters for a high purity metal in which phonon scattering is dominant should be intrinsic. But sputtered, MBE deposited, and electrodeposited films have typical Residual Resistance Ratios {RRR = R(293K)/R(4.2K)} ranging from ~ 4 to well below 2; so the contribution from 'dirt' stays important up to 293K, and the parameters of such 'pure' metals are not fully intrinsic even there.

### 2.3. Interface Parameters.

In general, the structure and properties of interfaces of metal A deposited on metal B need not be identical to those of interfaces of B deposited on A [35]. The interface properties derived from CPP-MR measurements are averages over the two interfaces, A/B and B/A.

As we'll see in sections 8.4.2 and 8.14, rough agreement of several derived parameters from different laboratories, and some surprisingly good agreements between measured interface specific resistances and no-free-parameter calculations for lattice-matched metal pairs (same lattice structure and nearly the same lattice parameter), suggest that at least some interface parameters might not be very sensitive to details of interface structure. But the importances of interface intermixing, of structural adjustments in lattice parameters across the interface, and of residual 'dirt' in each metal, have been only modestly examined. Effects of residual 'dirt' are likely minimal—i.e., effects of the electronic and physical structures probably dominate. Sections 4.4.1 and 8.14 will show evidence that for lattice matched pairs, the parameters $2AR_{N1/N2}$ and $2AR^*_{F/N}$ can be insensitive to interface intermixing. But the three studies where interface roughness was deliberately increased (in Co/Ag and Fe/Cr) gave conflicting results; in two the CPP-MR decreased with increasing interface roughness [40, 41], in the third it increased [42]. We'll examine these differences in section 8.13.

Before describing the theory underlying experimental analysis, we briefly cover some prehistory.

### 3. Prehistory—Some Important Results Prior to CIP-MR or to CPP-MR.

In this section we briefly examine four topics that were important for the discovery and interpretation of GMR, and which we will need to compare with CPP-MR results. (1) The discovery of antiferromagnetic coupling in F/N multilayers. (2) Early measurements of scattering asymmetry in F-alloys. (3) Conduction electron spin-resonance (CESR) measurements of spin-flipping cross-sections in Cu-and Ag-based alloys. (4) Early measurements of $l^N_{sf}$. For convenience, (5) we also collect here measurements of F/Nb interface specific resistances, $2AR_{F/Nb}$, that will be needed to analyze CPP-MR data taken with superconducting Nb cross-strips.

### 3.1. **Antiferromagnetic Coupling in F/N multilayers.**

Crucial to the discovery of GMR in Fe/Cr multilayers was the ability to produce AP ordering of the Fe/Cr multilayers at H = 0. In 1986, Grunberg et al. [24] found that Fe(10)/Cr($t_{Cr}$) trilayers with $t_{Cr} \approx$ 0.8 nm deposited by Molecular Beam Epitaxy (MBE) spontaneously adopted AP ordering (i.e., antiferromagnetic exchange coupling of adjacent Fe layers), and that it took a large field (~20 kG) to break that coupling and reverse the order to P. Fig. 1 shows the CIP-MR data for [Fe/Cr] multilayers at 4.2K that led to the designation Giant MR. The MRs of Fig. 1 decreased monotonically with increasing $t_{Cr}$ over the range of $t_{Cr}$ examined. In 1991, Parkin et al. [43] showed that similar AP coupling for Co(1)/Cu($t_{Cu}$) multilayers at $t_{Cu} \approx$ 0.9 nm was followed at larger $t_{Cu}$ by oscillatory coupling that weakened with increasing $t_{Cu}$, giving local maxima in MR for nearly AP coupling and local minima in MR for nearly P coupling. Fig. 4 illustrates such behavior, combining CIP and CPP data by Schroeder et al. in 1993 [39, 44] for multilayers with thin ($t_{Co}$ = 1.5 nm) Co layers. Note that some coupling seems to persist to at least $t_{Cu}$ = 5 nm.

### 3.2. **Earlier measurements of scattering asymmetry in F-alloys**.

The idea of scattering asymmetry in F-metals was introduced by Mott in 1936 [45]. In the 1970s, values of scattering asymmetries in Fe-, Ni-, and Co-based alloys were derived from measurements of Deviations from Matthiessen's Rule (DMR) in ternary F-based alloys, as collected by Campbell and Fert [36]. To simplify, if the scattering asymmetries for two impurities A, B in a given host F are similar, then the residual resistivities of ternary alloys of A and B in F should vary monotonically between the limits of just A in F and just B in F. If, instead, the asymmetries are very different (e.g., of opposite sign), the variation will be more complex. The asymmetries were characterized by a dimensionless ratio $\alpha = \rho_o^\downarrow / \rho_o^\uparrow$, where the subscript 'o' indicates a residual resistivity. This $\alpha$ is related to the asymmetry parameter defined in section 1.4.1 by $\beta_F = (\alpha - 1)/(\alpha + 1)$. Table 1 lists the DMR values of $\beta_F$ in the dilute alloy limit from [36] for selected F-alloys, some of which will be compared with CPP-MR values of $\beta_F$ for similar alloys. For each alloy, we list the minimum and maximum values given in [36] and the number (#) of different values given. Note the negative values for Cr, V, and Mn as impurities.

**Table 1. Selected values of the dilute alloy limit $\beta_F$ from DMR for Fe-, Co-, and Ni-based alloys.**

Table 1 lists the impurity, the minimum and maximum values of $\beta_F$, and the number of different values given [36] .

| Fe | | Co | | Ni | |
|---|---|---|---|---|---|
| (Ni ) | +0.5; +0.75   (2) | (Fe) | +0.85 | (Fe) | +0.76; +0.90   (3) |
| (Co) | 0; +0.57   (2) | (Mn) | -0.11- | (Co) | +0.86; +0.94   (6) |
| (Cr) | -0.46; -0.71   (2) | (Cr) | - 0.54 | (Cr) | -0.38;- 0.67   (6) |
| (V) | - 0.77, - 0.78   (2) | | | | |

### 3.3. **CESR measurements of spin-flipping cross-sections in Cu and Ag alloys.**

In section 8.5.1, we will test VF theory by comparing values of $l_{sf}^N$ derived for some Cu- and Ag-based alloys from CPP-MR with values derived independently from Conduction Electron Spin-Resonance (CESR) spin-flip cross-sections, $\sigma_{sf}$, for the same alloys [46].

This comparison requires equations given in Appendix A. A spin-flip mean-free-path is calculated from $\lambda_{sf} = 1/\zeta c \sigma_{sf}$, where $\zeta$ is the electron density in Cu or Ag and c is the impurity concentration. Then the spin-diffusion length is calculated from Eq. A.6.c, $l_{sf} = \sqrt{(\lambda \lambda_{sf})/6}$ , where $\lambda$ is the usual momentum mean-free-path, determined from Eq. A.3.

In addition to giving a list of spin-flip cross-sections, $\sigma_{sf}$, (which determine $\lambda_{sf}$) for various Cu- and Ag-based alloys, Monod and Schultz [46] also compared several of them with the momentum (i.e., resistivity) cross-sections that determine $\lambda$. Most of the spin-flip cross-sections were 100 to 1000 times smaller. The square root relation given just above then predicts $l_{sf} = (\sqrt{\frac{100}{6}})\lambda$ to $(\sqrt{\frac{1000}{6}})\lambda = 4\lambda$ to $13\lambda$. Such values led to the rough expectation when CPP-MR measurements began that $l_{sf} \sim 10\lambda$.

### 3.4. Spin-Diffusion Lengths prior to CPP-MR.

In section 3.3 and Appendix 1, simple arguments predict roughly $l_{sf} \sim 10\lambda$ for most non-magnetic (N) metals. Before CPP-MR began, only a little reliable information was available. From weak localization measurements at ~ 4.2K in 1982 [48] Bergmann inferred spin-diffusion lengths for Ag, Au, and Cu, and Santhanam et al. [[49, 50] in 1984 and 1987 inferred ones for Al. To automatically correct for different values of $\rho_o$, we calculate the ratios $\rho_o l_{sf}/\rho_t\lambda_b = l_{sf}^N/\lambda_N$ for their samples (the product $\rho_t\lambda_b$ is defined in Appendix 1). The ratios ranged from: ~ 20-40 for Al; ~ 30-40 for Ag; ~8 for Au; and ~ 40 for Cu, all consistent with the rough estimate of $l_{sf}/\lambda \sim 10$. In 1985 & 1988, Johnson-and Silsbee [[51, 52]] used a transverse geometry to directly measure $l_{sf}$ on a high purity Al foil. They found $\rho_o l_{sf}/\rho_t\lambda_b \sim 8\text{-}20$, also consistent with the rough estimate.

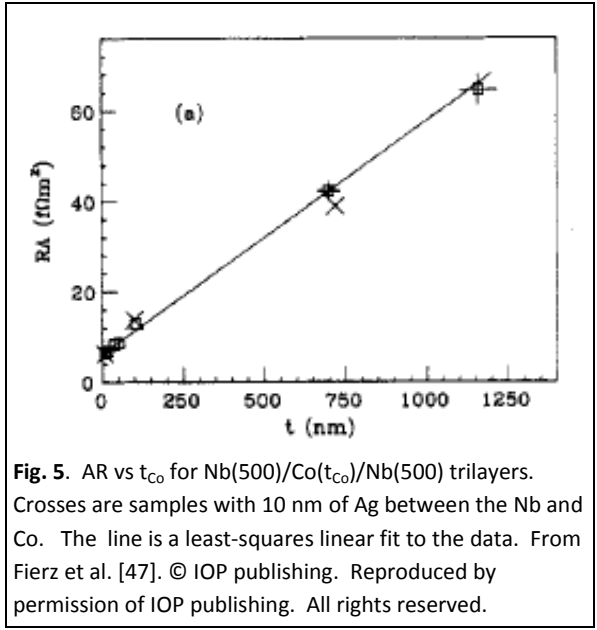

**Fig. 5.** AR vs $t_{co}$ for Nb(500)/Co($t_{co}$)/Nb(500) trilayers. Crosses are samples with 10 nm of Ag between the Nb and Co. The line is a least-squares linear fit to the data. From Fierz et al. [47]. © IOP publishing. Reproduced by permission of IOP publishing. All rights reserved.

Since the first CPP-MR measurements of $l_{sf}$ in 1994 [53], values for a wide range of N- and F-metals and alloys have been obtained from CPP-MR and other techniques such as lateral transport. These values are collected in [54]. In the present review, we limit ourselves to values from the CPP-MR, and emphasize that each must be coupled with its residual resistivity, $\rho_o$.

### 3.5. Contact Specific Resistances, $AR_{F/Nb}$, with Superconducting Nb.

Crucial to CPP-MR analysis is proper inclusion of the connecting leads (contacts) to the multilayer. When the leads are not superconducting, spin-accumulation extends into the leads from the multilayer, and one cannot simply add the total lead resistances in series. In contrast, at 4.2K, superconducting (S) Nb leads make the contact AR simple, because the Nb doesn't contribute. The only contribution is from the interface, $AR_{F/Nb}$. Values of $AR_{F/Nb}$ are determined by measuring $AR(t_F)$ vs $t_F$ for a simple Nb/F/Nb sandwich as shown in Fig. 5 [47] for F = Co from Fierz et al. in 1990. The slope of the resulting straight line gives $\rho_{Co} = 52 \pm 3$ nΩm and the intercept gives $2AR_{Co/Nb} = 6.1 \pm 0.3$ fΩm$^2$. A few additional points (crosses) show that inserting 10 nm of Ag between the Nb and the Co leaves the data unchanged, to within experimental uncertainties, as is also found for inserting 10 nm of Cu between Nb and Co or Py [55]. Similar measurements for Ni/Nb gave the results listed in Table 2, which also contains later F/Nb measurements for various F-metals [47] [41, 56-60]. An interesting feature of these results is that $2AR_{F/Nb} \cong 6.0 \pm 1.0$ fΩm$^2$ is similar for all of the F-metals and alloys in Table 2. The quantity $\rho_F$(film) was measured independently using the van der Pauw technique on separately sputtered thin (~ 200 nm thick) films. The overlaps of $\rho_F$(slope) and $\rho_F$(film) in Table 2 suggest that $\rho_F$(slope) and $\rho_F$(film) are both dominated by unknown impurities in the sputtered metals rather than by crystal defects, since the usual columnar growth would seem to give different defects for CIP films (the sides of columns), and for CPP layers (defects within a column). The overlaps also give hope that measurements on separately sputtered films can be used both to independently estimate layer resistivities and to check if the properties of sputtered Nb and F- and N-metals are stable over time. Similar agreements for other F-metals, and for N-metals, will be described in later sections.

The issue of diffusive vs ballistic transport will reappear in this review. To start the discussion, we ask if CPP transport for Co in Fig. 5 is diffusive or ballistic. From the information in Appendix A we estimate an effective Co transport mean-free-path of $\lambda_{Co} \sim 20\text{-}40$ nm for $\rho_{Co} \sim 50$ nΩm. AR in Fig. 5 grows linearly with $t_{Co}$ for $t_{Co}$ ranging from 10 nm to 1000 nm, i.e., from about $0.5\lambda_{Co}$ to $25\lambda_{Co}$. Combining this linear variation over the whole thickness range, with the agreements of $\rho_F(\text{slope})$ with $\rho_F(\text{film})$ (which surely involves diffusive transport), shows no obvious deviation in Fig. 5 from diffusive transport from $t_{Co}/\lambda_{Co} \leq 1$ to $t_{Co}/\lambda_{Co} \gg 1$.

Analyses of CPP-MR with superconducting cross-strips standardly take the F/S interfaces to have no spin-dependent scattering asymmetry, thus just adding a constant $2AR_{F/S}$ to both AR(AP) and AR(P) (more precisely, just adding $4AR_{F/S}$ to each spin-channel). Until 2004, this assumption was based only upon the ability to fit the published data. In 2004, Eid et al. [55] developed a way to look for spin-asymmetry, $\gamma_{F/XS}$. They made a Co/X/S interface part of a complex Py & Co based SV that was constructed to give AΔR near zero when X is absent. X is then inserted, and changes in AΔR are looked for as evidence of a non-zero

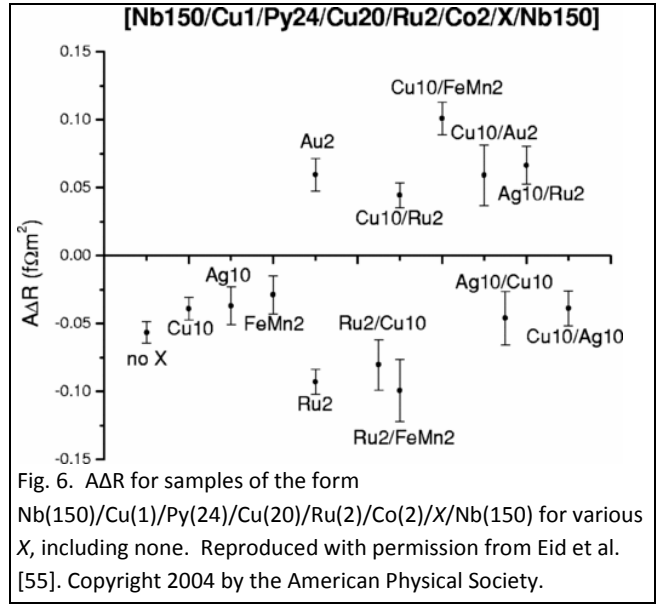

Fig. 6. AΔR for samples of the form Nb(150)/Cu(1)/Py(24)/Cu(20)/Ru(2)/Co(2)/X/Nb(150) for various X, including none. Reproduced with permission from Eid et al. [55]. Copyright 2004 by the American Physical Society.

$\gamma_{F/XNb}$. As shown in Fig. 6, X = Cu, Ag, FeMn, or combinations of Cu and Ag, gave no measurable change in AΔR, indicating γ = 0 to within uncertainty. In contrast, Ru, Au, or combinations involving Ru and Au, gave changes, Ru making AΔR more negative and Au making it more positive. The authors concluded that the F/S, F/Cu/S, F/Ag/S, or F/FeMn/S interfaces used in CPP-MR experiments are asymmetry free.

**Table 2. Values of $2AR_{F/Nb}$, $\rho_{Nb}$, $\rho_F(\text{slope})$, and $\rho_F(\text{film})$ for various F-metals.**

| Metal [ref.] | $2AR_{F/Nb}(f\Omega m^2)$ | $\rho_{Nb}(n\Omega m)$ | $\rho_F(\text{slope})$ (nΩm) | $\rho_F(\text{film})$ (nΩm) |
|---|---|---|---|---|
| Co [47] | 6.1 ± 0.3 | ~60 | 52 ± 3 | 58 ± 6 |
| Co [56] | 6.0 ± 1.0 | ~60 | | 68 ± 10 |
| Co [58] | 5.1 ± 0.2 | | | 25 ± 1 |
| Ni [47] | 4.8 ± 0.6 | ~60 | 35 ± 3 | 30 ± 3 |
| Fe [60] | 7.2 ± 0.5 | | 39 ± 2 | 37 ± 3 |
| Fe [41] | 6.0 ± 1.0 | ~60 | 40 ± 10 | |
| Py [57] | 6.0 ± 1.0* | | 123 ± 40 | 137 ± 30 |
| Py [59] | 6.1 ± 1.0 | ~60 | | 122 ± 20; 111±8 |

* See remarks in ref. [13] of [57].

#### 4. Theory Overview.

Several reviews have covered the theory of CPP-MR in detail [3, 4, 8]. So for this topic we discuss only those calculations that broke new ground or that raised issues that we discuss further. As our interest lies almost exclusively in AR(AP) and AR(P), we limit ourselves to collinear order of the F-layers.

#### 4.1. Theory Prior to Valet-Fert (VF).

The first CPP-MR calculation, by Zhang and Levy (ZL) in 1991 [27], presaged the use of a 2CSR model for CPP-MR data when moment-flipping is negligible. Their most important result was that the mean-free-paths in F and N dropped out of the final expression for AR. Paraphrasing, they found that, while the local resistivity $\rho(z)$ along the current axis is position dependent, with length scales set by the mean-free-path(s), $\lambda_N$, $\lambda_F^\downarrow$, and $\lambda_F^\uparrow$, these length scales disappear in the integral for $AR^{up,down} = \rho^{up,down}L$ (L is the total multilayer length), making the CPP resistance 'self-averaging'—i.e. AR depends only on the total scattering, and not upon its spatial distribution. This result suggests the 2CSR model shown schematically in the bottom half of Fig. 2, where the only lengths are $t_F$ and $t_N$.`

Later in 1991, Johnson [61] used a thermomagnetoelectric approach to correctly emphasize the importance of the spin-diffusion length for CPP analysis. For an isolated 'average' F/N interface, Johnson derived an accumulation interface specific resistance $AR_i$ that was determined by the products $\rho_F l_{sf}^F$ and $\rho_N l_{sf}^N$. He then assumed that $AR_T$ for a realistic $[F/N]_n$ multilayer would be just $n$ times $AR_i$. VF [29, 62] showed that this assumption is valid only in the limit $t_{F,t}$ $t_N >> l_{sf}^N$, $l_{sf}^F$, which is not the usual experimental case. In more realistic cases, where one or both of $t_F$, $t_N$ are comparable to $l_{sf}^N$, $l_{sf}^F$, effects of finite $l_{sf}^N$, $l_{sf}^F$ must be taken into account, but are more complex than simple addition. In the often used opposite limit, $t_F$, $t_N << l_{sf}^N$, $l_{sf}^F$, VF showed that cancellations from neighboring layers reduce the accumulation $AR_i$ to just the smaller products, $\rho_F^* t_F$ and $\rho_N t_N$, thereby giving precisely the 2CSR contributions from the F and N layers bracketing the F/N interfaces. In this limit, non-zero interface resistances result from interfacial scattering that is not included in the Johnson model.

In 1992, Bauer [63] presented a Landauer-Buttiker scattering formalism that presaged later quantitative calculations of interface ARs. In 1993, VF [29] published a Boltzmann Transport Equation (BTE)-based analysis that gave the formalism and parameters used to analyze most experiments.

#### 4.2. Valet-Fert (VF).

Starting with the BTE, VF took the F- and N-metal Fermi surfaces to be spherical and the same, thus neglecting band structure effects and initially also interfacial potential steps. Unlike ZL, VF allowed for moment-flipping in the F- and N-layers via spin-diffusion lengths, $l_{sf}^F$ and $l_{sf}^N$. They first derived a general BTE solution in powers of $(\lambda/l_{sf})$ and then showed that the equations reduced to macroscopic (or drift-diffusion) equations in the limit $(\lambda/l_{sf}) << 1$. From these latter equations (often called VF equations), they derived formulae for how the chemical potentials and currents in the F and N metals vary spatially with $l_{sf}^F$ and $l_{sf}^N$, and how the chemical potentials and currents are to be matched at the F/N interfaces. VF gave a closed form solution for only a multilayer with periodic boundary conditions. In the limits of $l_{sf}^F$, $l_{sf}^N >> t_F$, $t_N$, they noted that the VF model reduces to the 2CSR model. In the special case where randomly distributed up and down layer magnetizations give zero net magnetization, they noted that the 2CSR model should give the same AR as the AP state. We'll see in section 8.9 that this argument is invalidated if there is spin-flipping at F/N interfaces.

As VF did not give any closed form, general solutions for non-periodic samples, or including leads, solutions for specific samples have to be obtained using the VF formulae within individual layers and matching boundary conditions at interfaces. How this is done will be described in Section 4.2.2. As a warning, we briefly note a case of misunderstanding of the VF model. It starts with an incorrect use by Nakatani et al. in 2011 [64] of the VF periodic solution to derive VF parameters for finite $Co_2Fe(Al_{0.5}Si_{0.5})$/Ag multilayers. Not long afterward they realized their error, and in Taniguchi et al. [65] they reanalyzed Nakatani's data using the correct VF formulae and procedure for matching boundary conditions. Unfortunately, however, they then [65] mistakenly called the periodic solution 'VF theory', not realizing that VF theory was the correct detailed matching procedure that they had finally used. Tanaguchi also claimed, erroneously, that prior work by others had used the wrong (i.e. periodic) solution. They later corrected these errors in an erratum [66]. This work is discussed in section 10.5.2.

**4.2.1. Interfaces**. The assumption of identical, spherical Fermi surfaces for both the F and N metals meant that the VF equations could be solved without the presence of interface specific resistances—i.e., the chemical potentials could be taken continuous across the interfaces. Spin-dependent interface specific resistances were simply added in [28, 29], in [29] by adding scattering localized at infinitesimally-thin interfaces, producing steps in chemical potential at the interface proportional to the spin-dependent specific resistance, $AR_{F/N}^\perp$ or $AR_{F/N}^\uparrow$. Such resistances can have two physical sources.

(1) The first source is interface (I) intermixing, forming a concentrated alloy of high resistivity $\rho_I$, thickness $t_I$, and spin-diffusion length $l_{sf}^I$. For N1 and N2, these give $2AR_{N1/N2} = 2\rho_I t_I$ and $\delta_I = t_I/l_{sf}^I$. For F and N they give $AR_{F/N}^\perp = \rho_I^\perp t_I$, $AR_{F/N}^\uparrow = \rho_I^\uparrow t_I$, and $\delta_I = t_I/l_{sf}^I$. For fitting purposes, this interface alloy can be treated as an additional layer of thickness $t_I$ [67, 68].

(2) The second source is a potential step at the interface, which for an F/N interface is spin-dependent. Several authors, using different approaches (see, e.g. [69-71]) concluded that such a step should engender an exponential decay of the chemical potential out from the interface, on the scale of the mean-free-path, $\lambda$. The decay from a given interface should be affected if a second interface lies within $\lambda$ of the interface. Thus, when $t_F$ or $t_N$ are ~ $\lambda$, the apparent interface resistance could become layer thickness dependent, decaying exponentially with increasing $t$, and producing deviations from linear growth of AR(AP), AR(P), and A$\Delta$R with increasing $t_F$ or $t_N$. As such deviations should occur on the scale of $\lambda$, they are called mean-free-path (mfp) effects. We discuss the issues associated with mfp effects further in section 4.3.3, and consider them explicitly in sections 8.5.1, 8.8.1, 8.9, and 8.14.

To summarize, the VF model described in sections 1.4.1 and I.4.2 contains 8 parameters, 3 of which are interface parameters simply added to the 5 bulk parameters. Once added, the evidence so far is that the 8 parameters can adequately describe most or all published CPP-MR data.

**4.2.2. Solutions of the VF Equations**. The first solutions of the VF equations including both leads (superconducting cross-strips) and finite values of $l_{sf}^N$ were given in 1994 by Yang et al. [53]. We'll describe how they derived values of $l_{sf}$ for Cu- and Ag-based alloys in section 8.5.1. In 2000, Park et al. [67] generalized the VF formalism to include the parameter $\delta$ to describe spin-flipping at interfaces. Following the process mentioned above, they treated each interface (I) as an additional layer of assumed thickness $t_I$, resistivity $\rho_I$, $AR_I = \rho_I t_I$, and $\delta_I = t_I/l_{sf}^I$, and matched VF boundary conditions at both ends of this 'interface'. Fitting $AR_I$ and $\delta_I$ to data, caused $t_I$ to drop out of the analysis. Their procedure required developing computer programs to solve the VF equations for complex F/N multilayers and spin-valves (SVs)—see section 5. Their results for spin-flipping at the interfaces of N1/N2 multilayers are covered in section 8.11. Examples of VF solutions for several different multilayers were given in 2003 by Strelkov et al. [68], who developed a general code for solving VF for arbitrary multilayers. The examples included effects of scattering at the lateral edges of a thin nanowire or nanopillar when its diameter becomes comparable to the elastic mean-free-path, but not contributions from the normal metal leads at the ends of a nanopillar.

### 4.3. Limitations of Valet-Fert?

In this section we discuss studies of the limitations of VF analysis.

**4.3.1 Stringency of requirement of $l_{sf} \gg \lambda$?**  To investigate this stringency, in 2005, Penn and Stiles [72] solved the BTE numerically for two simplified trilayers with infinitely long outer layers separated by a spacer. The trilayers had the forms: (a) Cu/Py/Cu and (b) Py/Cu/Py, with Py = Permalloy = $Ni_{1-x}Fe_x$ with x = 0.2. Their goal was to explore when VF theory failed as the ratio $l_{sf}^{Py}/\lambda_{Py}^{\uparrow}$ decreased from $\gg 1$ to < 1 ($\lambda^{\uparrow}$ is the longer of the two mean-free-paths for Py). To do so, they calculated what they called the 'accumulation resistance' as a function of the spacer layer thickness, both from VF theory and by solving the BTE numerically for values of $l_{sf}^{Py}$ ranging from 100 nm down to 3.16 nm, more than 40% smaller than the best estimate of $l_{sf}^{Py}$ = 5.5 nm [73]. The accumulation resistance, A$\Delta$R/$\rho$ in nm, is the difference between the total AR, including effects of spin-accumulation, and the AR from just the sum of the resistivity contributions of the three layers, normalized to the resistivity of the infinite metal in the trilayer. They took as 'realistic parameters' for Cu and Py: $\lambda_{Cu}$ = 110 nm, $l_{sf}^{Cu}$ = 470 nm, $\lambda_{Py}^{\uparrow}$ = 5.5 nm, and $\lambda_{Py}^{\uparrow}$ = 1.8 nm. With these parameters, the best estimate of $l_{sf}^{Py}$ gives the ratio $l_{sf}^{Py}/\lambda_{Py}^{\uparrow}$ = 1, well below the VF 'requirement'. For a ratio as small as $l_{sf}^{Py}/\lambda_{Py}^{\uparrow}$ = 1.8 the difference between the VF and numerical BTE calculations of the accumulation resistance was only ~ 2%. Even for a ratio of $l_{sf}^{Py}/\lambda_{Py}^{\uparrow}$ = 0.6, the difference was only ~ 10%. Penn and Stiles concluded that VF theory seemed to remain valid even for $l_{sf}$ comparable to $\lambda^{\uparrow}$.

**4.3.2. VF Parameters.**  In 2011, Borlenghi et al. [74] applied a multiscale approach to spin transport in magnetic multilayers, embedding a tight-binding (TB) model (which describes the intrinsically quantum parts of the system at the quantum level), within a Continuous Random Matrix Theory (CRMT) (which treats the system semi-classically on scales larger than the elastic mean-free-path). They showed that CRMT is equivalent to circuit theory and, for collinear systems, also equivalent to VF theory, with a correction for Sharvin resistances at the leads. Since VF analyses of experimental data do not generally make this correction, it is made in the quantitative calculations (Eq. 8) described below. The Borlenghi calculations suggest that, so long as the CPP transport is "not perfectly ballistic (Fermi momentum mismatch at the interfaces, surface roughness, or impurity scattering)", the parameters of VF theory should be general, not limited by the simplified conditions under which they are derived.

**4.3.3. Ballistic vs diffusive scattering and mean-free-path (mfp) effects.**  In a single metal M, ballistic CPP transport is expected if the metal thickness $t_M$ is much shorter than the mean-free-path $\lambda_M$. In multilayers, in contrast, an effective $\lambda_{eff}$ depends upon scattering not only in both metals, but also at interfaces that are disordered [4, 75]. A ratio $t_M/\lambda_M \ll 1$ in one layer may be irrelevant if scattering in the other layer and/or at the interface is strong. To see ballistic effects probably requires at least partial quantum coherence in scattering between interfaces. We'll examine data relevant to this issue in section 8.9.

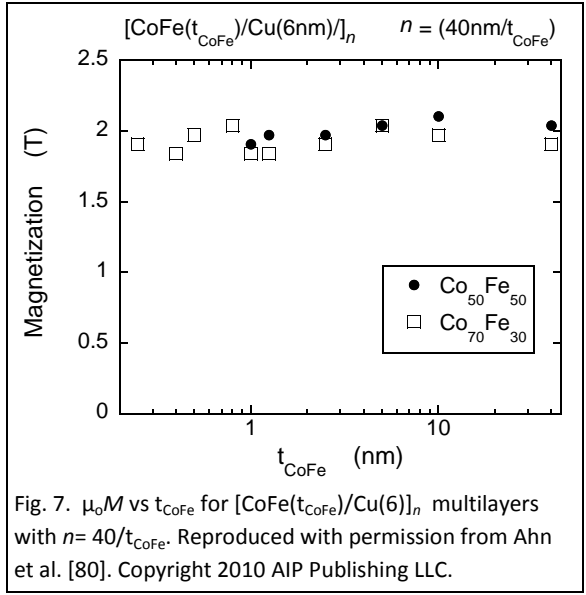

Fig. 7. $\mu_o M$ vs $t_{CoFe}$ for [CoFe($t_{CoFe}$)/Cu(6)]$_n$ multilayers with $n = 40/t_{CoFe}$. Reproduced with permission from Ahn et al. [80]. Copyright 2010 AIP Publishing LLC.

As noted, in section 4.2.1, several calculations suggest that the assertions by VF that the only characteristic lengths in the CPP-MR are the spin-diffusion lengths in the F- and N-metals are not always correct. These calculations suggest that the mean-free-path in a layer, λ, can also be a characteristic length in the CPP-MR, with the interface resistance predicted to decay on a scale of λ [69] [70] [76], thereby producing a change in interface AR with the distance between interfaces. Mfp effects have also been predicted when scattering is coherent enough to produce quantum well states [71, 77, 78].

Most experimenters have neglected mfp effects. However, some argue (e.g.,[8] [79]) that they are often dominant, and can be the true source of behaviors otherwise attributed to spin-relaxation. In sections 8.5.1, 8.8.1, 8.9, and 8.14, we consider both mfp effects and spin-relaxation, and ask if mfp effects have been isolated. To isolate mfp effects experimentally requires assumptions about the 'background'—i.e., is it just the 2CSR model, or the full VF theory with spin-relaxation and spin flipping at interfaces?

We will conclude that the evidence of mfp effects is, so far, inconclusive. All (or almost all) of the behaviors claimed as evidence for mfp effects can be explained in other ways. But not everyone agrees with us, and multilayers with more nearly perfect interfaces could enhance the visibility of mfp effects.

### 4.3.4 Uniformity of *M*?

Is the magnetic moment *M* uniform in magnitude and direction within an F-layer? If not, does any non-uniformity affect the VF interface parameters? The nature and structure of *M* in alloyed interfaces is not well established. Calculations [81, 82] and measurements [83] on some pairs favor dead layers. In contrast, a study of Co$_{70}$Fe$_{30}$/Cu interfaces showed no evidence of changes in *M* down to 0.2 nm = 1 ML ( Fig. 7) [80]. If dead layers are always present for a given F/N pair, they will presumably be reflected in the measured values of the interface parameters: $2AR^*_{F/N}$, $\gamma_{F/N}$, and $\delta_{F/N}$.

### 4.4. Quantitative calculations of VF parameters.

### 4.4.1. $2AR_{N1/N2}$, $AR^{\downarrow}_{F/N}$, and $AR^{\uparrow}_{F/N}$ for lattice matched pairs.

VF analysis applied to experimental data can give correct values for $AR_{N1/N2}$, $AR^{\downarrow}_{F/N}$ and $AR^{\uparrow}_{F/N}$ (or the alternatives $2AR^*_{F/N}$, and $\gamma_{F/N}$). But VF theory was never designed to calculate these quantities from 'first principles'. To do so, one must be able first to calculate the real electronic structures (correct Fermi surfaces and Fermi energies) of both the F- and N-metals, and then to use these Fermi surfaces and Fermi energies to calculate the transport coefficients $AR_{N1/N2}$, $AR^{\downarrow}_{F/N}$ and $AR^{\uparrow}_{F/N}$. We'll see in this section and section 8.14 that no-free-parameter calculations can give surprisingly good agreement with experiment for lattice matched pairs that have the same crystal structure and the same bulk lattice parameter to within ~ 1%. In contrast, so far, calculations for metals with different crystal structures, or larger differences in lattice parameters, give poor agreement. This lack of success is likely due to sensitivity of the interface properties to unknown details of the interfacial structure [84].

The Fermi surfaces and Fermi energies are calculated using density functional theory (DFT) with the coherent potential approximation (CPA), assuming crystal structures and lattice parameters for each metal. When bulk F and N (or N1 and N2) are lattice matched, the two metals and their interface can be described by a single crystal lattice having the bulk lattice parameter (or a slightly corrected one based upon x-ray measurements of the multilayer). Initial calculations used Linearized Muffin-Tin (LMT) potentials and s,p,d bases [85]. More recent calculations used full Muffin Tin (FMT) potentials and s,p,d,f bases [85]. Approximations such as those just noted, plus limitations on calculation of the Fermi energy, lead to uncertainties of at least 5-10% [86].

In 1997, Schep et al. [87] used Landauer formalism to derive for an isolated interface a general transport equation for $AR_{N1/N2}$ or $AR_{F/N}^{\downarrow}$ and $AR_{F/N}^{\uparrow}$ that contains a correction for the Sharvin resistance and is valid when scattering is diffuse in both metal layers, G & H, bounding the interface.

$$AR_{GH} = (Ah/e^2)[1/(\Sigma_{ij}(T_{ij})) - (1/2)[(1/N_G) + (1/N_H)]]. \qquad (8)$$

Here h is Planck's constant, e is the electron charge, $N_G$ and $N_H$ are the number of channels in G and H, $T_{ij}$ is the probability for eigenstate i in metal G to be transmitted through the interface into eigenstate j in metal H, and the two terms on the right are the Sharvin resistance corrections. Using Eq. 8 to calculate $AR_{Co/Cu}$ for (111) oriented Co/Cu gave semi-quantitative agreement with experiment assuming a perfect (flat and specular) interface as shown in Table 3. Similar agreement was found independently by Stiles and Penn in 2000 [88]. In contrast, modifying the Co/Cu calculation to couple specular interface scattering with ballistic transport in the two bounding layers gave strong disagreement (column 6 of Table 3]. For [111] oriented Co/Ni, or [011] oriented Fe/Cr, Table 3 shows that the perfect interface again gave only semi-quantitative agreement.

In 2001, Xia et al. [89, 90] extended this formalism to include more realistic disordered interfaces—approximated for Co/Cu and Fe/Cr as 2 monolayers (ML), and for Co/Ni as 2ML or 4ML, of a 50%-50% random alloy. The relatively modest changes in calculated values of $AR^{\uparrow,\downarrow}$ for perfect and disordered interfaces in Table 3 were attributed to approximate cancellation of two effects. Because a perfect interface is translationally invariant, $T_{ij}$ is limited by the constraint that state I of G can be transmitted into state j of H only if the component of the k-vector parallel to the interface, $k_{||}$, is conserved. Xia et al. called such transmission 'ballistic'. The loss of translational symmetry for a disordered interface removes the constraint of $k_{||}$ conservation. Scattering at the disordered interface then involves two competing effects, a reduction in 'ballistic' conductance, but an increase in 'diffuse' conductance in which $k_{||}$ is not conserved. If these two effects roughly cancel, the calculated result won't be sensitive to the amount of interfacial disorder.

To test for possible coherence, in 2001 Xia et al. [89] calculated $AR^{\downarrow}$ and $AR^{\uparrow}$ for Co/Cu from first principles assuming ballistic transport in the bounding metals, but now with the 2ML thick 50%-50% disordered interface. This calculation used only a 6x6 (area ~ 1 nm$^2$) supercell. For $AR^{\downarrow}$, interfacial scattering was strong enough to essentially remove effects of coherence. However, for $AR^{\uparrow}$, interfacial scattering was so weak that coherence persisted through several interfaces, giving values of $AR^{\uparrow}$ that decreased to a limit of $AR^{\uparrow} = 0.07$ fΩm$^2$ as the number of interfaces increased. Such coherence seems unlikely to persist over the much larger areas ($10^4$ nm$^2$ to mm$^2$) of real samples; but whether random disorder in separation between potential steps over such large areas will completely eliminate coherence is disputed (see, e.g., section 8.9).

**Table 3. Calculated $AR^\downarrow$ and $AR^\uparrow$ for lattice matched Co/Cu, Fe/Cr, and Co/Ni with different assumed interfaces.** The experimental values for Co/Cu are for sputtered (MSU)[6] samples. For details see section VIII.D.2. The experimental values for Fe/Cr [41] (for details see section 8.13) and for Co/Ni [91] (for details see section 8.14) are also both for sputtered samples. The two calculated values for Co/Cu and Fe/Cr perfect interfaces are due to slightly different procedures used in [89] and [88]. For Co/Cu, the sixth column shows results for ballistic bulk and a perfect interface [87]. For Co/Ni, the sixth column contains a calculation for 4 monolayers (ML) of 50%/50% disorder [91].

| AR (fΩm²) | MSU. Expt. | Perfect Interface | 2ML(50-50 | Ballistic bulk + perfect I. |
|---|---|---|---|---|
| $AR^\downarrow_{Co/Cu}$ | 1.84±0.14  [6]. | 1.46 [89];    1.95; [88] | 1.82  [89]] | 0.64    [87] |
| $AR^\uparrow_{Co/Cu}$ | 0.26±0.06  [6] | 0.39 [89]    0.43 [88]; | 0.41  [89] | 0.0001 [87] |
| | | | | |
| $AR^\downarrow_{Fe/Cr}$ | 0.5±0.2    [41] | 1.05 [90]   0.81 [88]] | 1.1    [90] | |
| $AR^\uparrow_{Fe/Cr}$ | 2.7±0.4    [41] | 2.74 [90]   2.11 [88] | 2.05   [90] | |
| | | | | |
| | | Perfect Interface. | 2ML(50-50.] | 4ML(50-50). |
| $AR^\downarrow_{Co/Ni}$ | 1.00±0.07  [91] | 0.73  [91] | 0.86   [91] | 1.19   [91] |
| $AR^\uparrow_{Co/Ni}$ | $0.03^{+0.02}_{-0.03}$ [91] | 0.015 [91] | 0.016  [91] | 0.018 [91] |

### 4.4.2. $l^{Py}_{sf}$.

A first principles calculation of $l^{Py}_{sf}$ was made in 2010 by Starikov et al. [92] Electronic structures were calculated using the local spin density approximation of density functional theory, with both spin-orbit coupling and chemical disorder, and relativistic effects were included via the Pauli Hamiltonian. The calculation assumed a disordered Py-layer sandwiched between two Cu layers, and examined how unpolarized states injected from one Cu-layer were transmitted into different spin-channels in the other Cu-layer as a function of the Py layer thickness. The calculated data points were fit with a VF-based equation containing 5 parameters: $\beta_{Py}$; $2AR^*_{Py/Cu}$; $\gamma_{Py/Cu}$; $l^{Py}_{sf}$; and $\delta_{Py/Cu}$, but the fit was most sensitive to $\beta_{Py}$ and $l^{Py}_{sf}$. The best values were $\beta_{Py} = 0.68$ and $l^{Py}_{sf} = 5.5 \pm 0.3$ nm, in agreement with the experimental values of $\beta_{Py} = 0.7 \pm 0.1$ and $l^{Py}_{sf} = 5.5 \pm 1$ nm given in column ( E ) of Table 8 in section 8.7.1 below.

## 5. Ways to Achieve AP states.

There are four different ways to achieve AP states. (A) Antiferromagnetic coupling. (B) Hybrid spin-valves (Hybrid SVs). (C) Exchange-Biased Spin-Valves (EBSVs). (D) Dipolar effect of fringing fields. We discuss each way and show an example of its CPP-AR(H).

### 5.1. Antiferromagnetic coupling.

As explained in section 3.1, and illustrated in Fig. 1 for $[Fe/Cr]_n$ and Fig. 4 for $[Co(1.5)/Cu(t_{Cu})]_n$, adjacent thin F-layers in F/N multilayers can couple antiferromagnetically (AF) for a particular small N-thickness ($t_N \sim 0.8 - 0.9$ nm) and the coupling can then oscillate with increasing $t_N$, giving local maxima in MR for AF-coupling and local minima for ferromagnetic (F) coupling. The strongest AF coupling should guarantee an AP state at H = 0. For weaker AF couplings the closeness to an AP state is less clear. Unfortunately, it takes a large H to overcome the strongest AF coupling. Also, since AF coupling occurs for only a few values of t$_N$. and weakens as t$_F$ grows, it is hard to use it to study systematically how CPP-MR varies with t$_F$ and t$_N$.

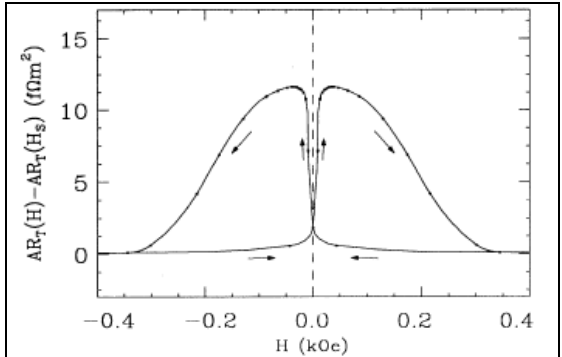

Fig. 8. AR(H) − AR$_↑$(H$_s$) vs H. Hysteresis curve for a [Co(3)/Cu(20)/Py(8)/Cu(20)]$_8$ hybrid SV. Reproduced with permission from Yang et al. [93]. Copyright 1995 by the American Phyiscal Society.

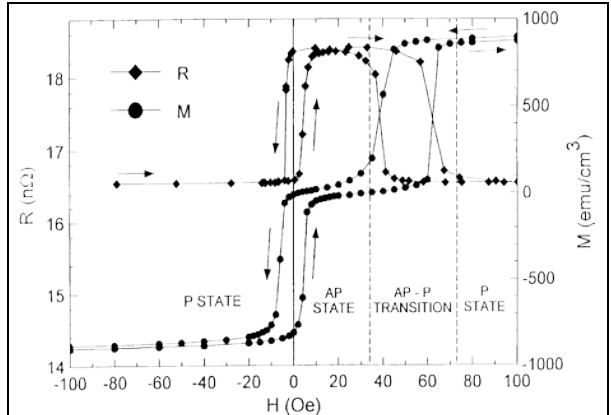

Fig. 9. Resistance (R) (diamonds and left scale) and magnetization (circles and right scale) vs H for an [FeMn(8)/Py(30)/Cu(20)/Py(30)] symmetric EBSV. The parallel (P) and antiparallel (AP) states are indicated. Agreement of R for oppositely directed P states shows that the FeMn does not contribute directly to the MR. Reproduced with permission from Steenwyk et al. [94]. Copyright 1997, AIP Publishing LLC.

### 5.2. Hybrid Spin-Valves.

A hybrid spin-valve (SV) unit has the form [F1/N/F2/N], where F1 and F2 are F-metals (or alloys) with coercive fields, H$_c$, different enough that the F1 and F2 moments reverse direction in very different fields, and thick N-layers magnetically decouple F1 and F2. Fig. 8 [93] shows AR vs H for a [Co/Cu/Py/Cu]$_8$ hybrid SV. F1 and F2 may also be different thicknesses of a single F-metal.

### 5.3. Exchange-Biased Spin-Valves.

An exchange biased spin-valve (EBSV) has the general form [AF/F1/N/F2], where F1 and F2 may be the same or different F-metals. Exchange-bias is produced by depositing a thick enough (typically ≥ 8 nm) antiferromagnetic (AF)-layer next to the F1-layer, heating the multilayer to above the blocking temperature of the AF-layer, applying a large enough magnetic field (typically a few hundred Oe), and cooling the multilayer to room temperature in the field. As illustrated in Fig. 9 [94], this process shifts the center of the hysteresis loop of the 'exchange-biased' or 'pinned' F1-layer to a high field, and usually widens the loop. At small fields sufficient to 'reverse' the moment of the free F2 layer, the moment of the biased F1 layer stays 'pinned', not reversing until much higher fields. For best results, the N-layer should be thick enough (typically 10 − 20 nm) to magnetically decouple F1 and F2, letting the 'free' F2-layer reverse at its usual H$_c$. Since the exchange-biasing interaction is basically an interface phenomenon, the shift of H$_c$ of the pinned F1-layer decreases with increasing t$_{F1}$. Two standard AFs are FeMn (Fe$_{50}$Mn$_{50}$) [67] and IrMn (Ir$_{20}$Mn$_{80}$) [95]. Both flip spins so strongly that the AF/F interface is standardly taken to randomize an incoming spin-polarization [67, 95, 96].

In a symmetric EBSV with F1 = F2 = F and $t_{F1}$ = $t_{F2}$ = $t_F$, when $t_F \gg l_{sf}^F$, A$\Delta$R 'saturates' to a constant limit that can be written in a simple closed form if one neglects all $\delta$s. F/N based EBSVs are so widely used that we write A$\Delta$R for crossed-superconductors, Nb/FeMn(8)/F($t_F$)/N($t_N$)/F($t_F$)/Nb, as:

$$A\Delta R \text{ (for } t_F \gg l_{sf}^F) = 4(\beta_F \rho_F^* l_{sf}^F + \gamma_{F/N} AR_{F/N}^*)^2/(2\rho_F^* l_{sf}^F + \rho_N t_N + 2AR_{F/N}^*). \qquad (9)$$

Eq. 9 is a variant of Eq. 6, but with only one F/N interface per F layer and the denominator simplified from the total AR(AP) to just the 'magnetically active center' = the middle N-layer plus the two adjoining F/N interfaces plus spin-diffusion lengths into the two F-layers. Importantly, the properties of the Nb/F and Nb/AF interfaces, and of the FeMn layer and the FeMn/F interface, have all dropped out of Eq. 9, so Eq. 9 is equally valid for an EBSV nanowire or nanopillar (so long as its current flow is uniform), or for an asymmetric EBSV, so long as both F-layer thicknesses are much longer than $l_{sf}^F$. In contrast, fitting A$\Delta$R from $t_F$ = 0 up to the constant limit requires input of all of the EBSV parameters.

### 5.4. Dipolar Effect of Fringing Fields, H$_f$.

### 5.4.1. Dipolar Effects in Nanowires and Nanopillars.

Small ellipsoidal magnets can be single domain. At applied H = 0 the moment orients along the long-axis. Spherical magnets can also be single domain, but the moment orientation at H = 0 depends upon the magnet's history. A circular, disc-shaped F-layer with thickness t small compared to its diameter d, and d not too large (e.g., t < 15 nm and d < 100 nm) [97] can be single domain at applied H = 0, most easily if the disc is elongated in one direction to give a convenient axis for the moment. If a disk-shaped F-layer is single domain, its moment should lie in the layer plane due to shape anisotropy, and its fringing magnetic field H$_f$ should be roughly dipolar. The H$_f$ of a dipole points out from the dipole's 'head', turns around as it moves back along the dipole, and points in at the dipole's 'tail'.

Given this information, one might hope to see dipolar effects in nanowires or nanopillars when the F-layers are much thinner than their diameters and their diameters are small enough. If two such thin, disc-shaped, magnetized F-layers, are separated by a not-too-thick N-layer, the dipolar interaction between them should tend to align their moments AP to each other. Even when the dipolar interaction is not strong enough to produce a full AP state in a multilayer, it will tend to anti-align the moments of adjacent layers, giving a non-zero MR. In Piraux et al. [98], Magnetic Force Microscopy (MFM) measurements on selected nanowires showed partial AP orderings ranging from 33% to 49% (see Fig. 45 in section 8.6.1). The ordering was parameterized by $p \sim 0.41 \pm 0.08$, an assumed AP fraction.

To try to improve $p$, nanowire multilayers were made of the form [Py($t_{Py}$)/Cu(10)/Py($t_{Py}$)/Cu(100)]$_n$. [99], with $t_{Py}$ < d, the wire diameter. Each pair of Py layers separated by only 10 nm of Cu should dipole couple, while the resulting widely separated 'tri layers' should be uncoupled. Indeed magnetization measurements showed greatly reduced remanent magnetizations, and hysteresis curves close to those in Fig. 1, consistent with $p \sim 0.85 \pm 0.15$ [99].

### 5.4.2. Dipolar Effects in Crossed-Superconductor Samples.

A more subtle dipolar effect turns out to be important in mm-square [Co/Ag]$_n$ and [Co/Cu]$_n$ multilayers studied with the crossed-superconductor geometry. From their hysteresis curves, illustrated by the one for CPP-MR in Fig. 3, it is not clear how to determine MR(AP). Since MR(AP) should be the largest MR, the obvious choice seems to be the initial 'virgin' value, MR(0), before the multilayer has been subjected to a magnetic field H. However, this initial value is not thereafter reproducible. Attempts to return to it by demagnetizing (i.e., starting from either $H_s$ or $H_{Pk}$ and reducing H in steps while alternating its sign) were rarely if ever successful. The usual result was an MR between MR(Pk) and MR(0). For comfort in using MR(0) ≅ MR(AP), one would like a mechanism to produce it.

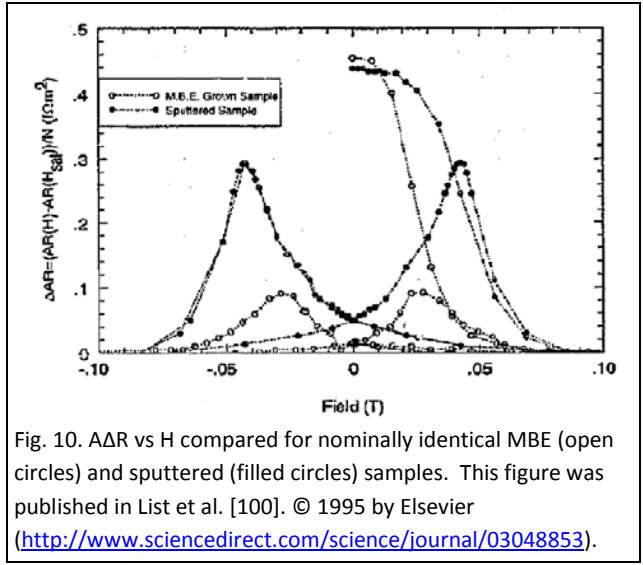

Fig. 10. A∆R vs H compared for nominally identical MBE (open circles) and sputtered (filled circles) samples. This figure was published in List et al. [100]. © 1995 by Elsevier (http://www.sciencedirect.com/science/journal/03048853).

In section 8.2.3.3, we'll see that a combination of Polarized Neutron Scattering (PNS) [101, 102] and Scanning Electron Microscopy with Polarization Analysis (SEMPA) studies [101, 103] of a [Co(6)/Ag(6)]$_{60}$ multilayer provides such a mechanism. Together they show that dipolar effects can explain why MR(0) in Fig. 3 is usually so much larger than MR($H_{pk}$) and often likely to approach an AP state. PNS showed strong antiferromagnetic (AP) ordering in the 'virgin' multilayer. SEMPA found the layer magnetizations to divide into micron-sized domains, with a strong tendency (~ 60%) for the domains in F-layers just above each other to be oriented AP (with the domain edges also similarly ordered). Figures showing these behaviors and further discussion are given in section 8.2.3.3. The following model is consistent with both sets of data. As the first thin F-layer of a multilayer grows in zero applied field, it develops micron-sized domains that have dipolar fringing fields. When that F-layer is completed, and a not-too-thick N-layer is deposited on top of it, a fringing field from a given domain within the F-layer extends to the top of the N-layer, where it points closely opposite to the moment of the originating domain in the bottom F-layer. If the magnetization of the newly growing next F-layer is sensitive to that local fringing field, then the moment of a domain in that layer will tend to grow AP aligned with the domain just below it. Continuing this process from F-layer to F-layer can lead to closely AP ordering within each domain column in the 'virgin' multilayer. Since the CPP current flows along the perpendicular-to-plane axis of each domain column, the CPP-MR can approximate that for full AP ordering, even when the ordering directions of different magnetic domains are oriented randomly in the layer plane. When most effective, such dipolar fringing fields can produce a 'virgin' state, ∆R(0), close to an AP state. However, if the situation was always simple, the ratio of R($H_o$)/R($H_c$) would be fixed. And, indeed, averages over data sets can give close to fixed ratios, as we'll see for [Co/Ag]$_n$ in section 8.4.1. However, as shown in Fig. 10 [100] and in ref. [104], individual samples can give ratios ranging from ≤ 1.5 to ≥ 2. This range of variation suggests that the mechanism can be fragile. Sections 8.2.3 and 8.4 show that this mechanism gives ∆R(0) ≅ ∆R(AP) for certain sample sets of Co/Ag and Co/Cu. But sections 8.3 and 8.7 show that it works less well for Py/Cu and doesn't work for Ni/Ag.

## 6. Experimental Techniques.

The three main techniques used to measure the CPP-MR are illustrated in Fig. 11. We discuss them in the order they were published. In each case we first describe the geometry and explain its advantages and disadvantages. We then go into more detail, including discussing potential problems.

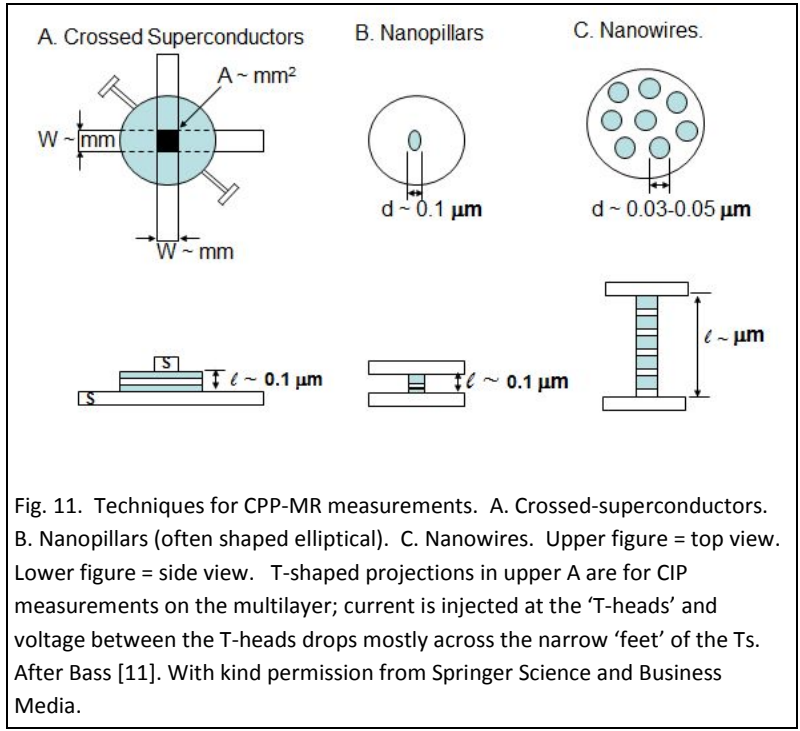

Fig. 11. Techniques for CPP-MR measurements. A. Crossed-superconductors. B. Nanopillars (often shaped elliptical). C. Nanowires. Upper figure = top view. Lower figure = side view. T-shaped projections in upper A are for CIP measurements on the multilayer; current is injected at the 'T-heads' and voltage between the T-heads drops mostly across the narrow 'feet' of the Ts. After Bass [11]. With kind permission from Springer Science and Business Media.

We conclude this section with a description of a fourth technique used for a few years to estimate the CPP-MR, Current-at-an-angle-to the Plane (CAP-MR). The CPP-MR is then estimated by extrapolating from CIP-MR and CAP-MR measurements.

### 6.1. Crossed Superconducting Leads.

### 6.1.1. The technique and its advantages and disadvantages.

Ever since it was used to first measure the CPP-MR [26], the crossed-superconducting lead technique has been the gold standard of CPP measurements, because it: (1) provides a uniform current; (2) has simple contact resistances, (3) allows measurements of AR(AP), AR(P) and A$\Delta$R, as well as the CPP-MR; (4) can be used with multilayers containing arbitrary combinations of metals; and (5) can control AR(AP) via hydrid-SVs and EBSVs.

As illustrated in Fig. 11A, the superconducting lead technique involves sandwiching an ~ 4 mm diam. multilayer of interest between crossed ~ 1.1 mm wide superconducting strips, typically 100-250 nm thick. Because they have zero resistance, the superconducting strips provide equipotential surfaces across the multilayer that ensure uniform CPP current flow through its mm$^2$ area A. This technique was first tried by Schuller and Schroeder at Argonne National Laboratory [105], where it failed because the need to open the sputtering system to air between depositing the Nb leads and the multilayer led to uncontrolled contamination of the Nb/multilayer interfaces. Subsequent users of simple cross-strips prepare their samples in situ with two different systems. The Michigan State University (MSU) group uses a sputtering system with 6 sputtering guns, a 10$^{-8}$ Torr base pressure, purified Argon sputtering gas, and an internal mask-changing system to allow sequential deposition of the lower Nb lead, the multilayer, and the upper Nb lead, without breaking vacuum [56, 106]. The Leeds University (Leeds) group uses a Molecular-Beam-Epitaxy (MBE) system with internal masking and the ability to deposit multiple metals without breaking ultra-high vacuum [60].

The advantages of the superconducting lead technique were already listed above.

The technique's disadvantages are: (1) its limitation to 4.2K; (2) The need to cool the samples to 4.2K for measurements; and (3) The need for a high sensitivity, high precision resistance measuring system, as the typical sample resistance is only ~ 10$^{-8}$ $\Omega$.

Four variants have combined Nb leads with lithography or a focused ion-beam (FIB) to get resistances at 4.2K large enough (m$\Omega$ to $\Omega$) to measure with standard meters. Highmore et al. [107] and Cyrille et al. [108] connected in series via Nb contacts 100 to 500 pillars with areas ranging from A = 4x4 ($\mu$m)$^2$ to A = 30x30 ($\mu$m)$^2$. We'll discuss the Cyrille results in section 8.13. Slater et al. [109] fabricated on the surface of a single multilayer, Nb pads with areas ranging down to 4 $\mu m^2$. Most recently, Bell et al. [110] used a Focused Ion-Beam (FIB) microscope to pattern an antiferromagnetically coupled [Cu(0.9)/Co(2)]$_{10}$ multilayer with area ~ 0.8 $\mu m^2$ sandwiched between 150 nm thick Nb contacts. At 0.35K, they found R ~ 0.1$\Omega$ and CPP-MR = 14%. Use of a simple cross-strip geometry with mm-wide Cu leads [111] instead of Nb, failed to give reliable results due to a combination of non-uniform current flow and large contact resistance.

### 6.1.2. Experimental Details.

To show what is needed to achieve reliable and reproducible results, we describe in detail the MSU system and techniques used for CPP-MR measurements.

The sample preparation system [56, 106] consists of an ~ 18" diam. sputtering chamber with ultra-high-vacuum (UHV) conflat flanges everywhere (including in 4 Simard 2.25" diam. sputtering guns), except for the main o-ring which is Viton, and smaller Viton o-rings on two 1" diam. sputtering guns. To absorb water vapor, the chamber contains a Meissner trap containing liquid nitrogen. Eight sample substrates are held in a sample positioning plate that can be cooled by means of a coil of copper tubing through which is circulated high pressure $N_2$ gas that has passed through the liquid $N_2$ in a dewar. Early tests showed that the data are most reproducible when the temperature of the positioning plate (measured by an attached thermocouple) is held between T = 243K and T = 303K. Sputtering is begun with the positioning plate at T ~ 243 K and stopped temporarily if T increases to 303K before the planned sputtering is done. Of 8 sample spaces, 6 or 7 are usually used for CPP samples, with the other 1 or 2 used for 200 nm thick single films to measure the resistivities of sample components using the van der Pauw technique. Each CPP sample requires a four site mask system consisting of a blank site to protect the substrate from the sputtering targets, a strip-shaped hole site of width ~ 1.1 mm for the first Nb strip, a larger circular hole site (with extensions for CIP-MR measurements—see Fig. 11A) for the multilayer, and a strip-shaped hole site of width ~ 1.1 mm for the second Nb strip oriented perpendicular to the first one. The four different sites are moved sequentially over a chosen substrate by pulling upon small metal posts using a vacuum sealed, externally moveable metallic finger. For the single films, a space contains two substrates, each of which can be moved behind either a blank site or a square hole site. With this system, 6 or 7 CPP samples and 2-4 single layer films can be made in a single day and the system can be opened and the substrates replaced that evening. With a light baking for several hours at ~ 373K at night, it initially takes 2 days to reach the desired base pressure ~ 1-2 x 10$^{-8}$ Torr after liquid $N_2$ is added and before the ~ 2.5 mTorr of Ar sputtering gas is admitted. After the first day's run, it usually takes only the overnight bakeout plus liquid $N_2$ to reach the desired base pressure for another run. But after the second run, the sputtering system is opened and the guns recleaned with acid. Then 2 days with bakeout and liquid $N_2$ are again needed to reach the desired base pressure. With these procedures, 24 to 28 CPP samples and 4 to 8 single layer films can be sputtered in five days.

To check for stability of the sputtered Nb over time, every few sample runs the resistivity of a separately sputtered 200 nm thick Nb film is measured at 12K (to eliminate superconductivity). Over two decades, the value stayed stable in the range $\rho_{Nb}$ = 60 ± 10 n$\Omega$m. For different experiments, the Nb layer thicknesses in the CPP-MR samples ranged from 100 nm to 500 nm. In 'dirty' Nb (i.e., $\rho_{Nb}$ ~ 60 n$\Omega$m), the current should flow only through a penetration depth ~ 100 nm. The standard measuring current of 100 mA gives a current density through the Nb leads ~ 10$^9$ A/m$^2$, about two orders of magnitude less than the expected zero field critical current density. In-plane H normally produces no finite lead resistance in fields to well above 1kG; typically not until H ~ 5 kG [41].

In the first years of research, each CPP sample had to be cooled to 4.2K in a separate liquid helium dewar that was precooled with liquid nitrogen before transferring liquid He from a storage dewar. This process limited CPP-MR measurements to one sample per day. Then Prof. W. P. Pratt Jr. designed a 'quick dipper' system, where the sample and its surrounding magnet are mounted at the end of a cylindrical stick in which are also located a Superconducting Quantum Interference Device (SQUID) null-detector and a reference resistor, as described in the next paragraph. The stick diameter was chosen to fit the o-ring at the top of a 100 liter liquid He storage dewar. This stick allowed the sample and magnet to be pre-cooled for ~ 20 minutes in the cold He gas at the neck of the dewar before being lowered into the liquid helium to cool to 4.2K. A typical CPP-MR measurement sweep takes about an hour. Then the stick is raised so that the sample and magnet are again in the He gas at the neck of the dewar, and this time the neck is warmed with a 'heat gun' for ~ 20 minutes, after which the sample and magnet are warm enough to be removed from the dewar free of any residual frost that might damage the sample or break a contact. The whole process takes less than 2 hours, thus allowing up to 5 CPP samples to be measured in a day. This quick-dipper system lets a week's worth (24-28) of sputtered samples be measured in the same week.

The typical small CPP resistance (~ $10^{-7}$ - $10^{-8}$ $\Omega$) is measured with a SQUID-based bridge circuit [112] containing the CPP sample of interest, an ~ 100 $\mu\Omega$ reference resistor, and a transformer to increase current sensitivity. An alternative ac system that gives adequate sensitivity is described in [113]. The Nb cross-strips protect the sandwiched CPP samples from degradation; remeasurements of the CPP-R on the same sample years apart are usually the same to within 1%. Connections to the sample are made with superconducting leads and indium (In) solder contacts. A magnetic field up to several kG, in the plane of the multilayer, is applied by means of a hand-wound cylindrical coil locked into place around the sample. The planar, thin film geometry of the Nb strips lets the field penetrate the sample uniformly (i.e., there is no significant 'field expulsion').

Determining the intrinsic quantity in the CPP geometry, AR, requires measuring A and R separately. As noted above, R and $\Delta$R can be measured to within ~ 1%, except when $\Delta$R approaches the measuring uncertainty. The largest uncertainty in AR is usually the uncertainty in A = $W_1$x$W_2$, where $W_1$ and $W_2$ are the widths of the two Nb cross-strips. $W_1$ and $W_2$ are measured with a profilometer, where a diamond stilus moves up and over a strip, reading out its profile [56]. Each width is typically measured four times, twice on each side of the sample, and the values averaged to give $W_1$ or $W_2$ and its uncertainty. Tests of repeated measurements of A for a given sample by a given student, and by different students for the same sample, usually give areas within a few percent of each other, but sometimes larger differences [114]. Since some portion ($\gtrsim$ 10 nm) of the Nb turns normal by 'anti-proximity effect' with the adjacent F-layer, there is also uncertainty as to what height above the substrate to use to measure A. Given both random and systematic uncertainties, an uncertainty of ± 5% is standardly assigned to each area, except when the calculated uncertainty is > 5%.

Fig. 11A shows the layers of the multilayers and the Nb contacts as being perfectly flat. In fact, sputtering or MBE growth of crystalline (as distinct from amorphous) metallic multilayers usually produces columnar growth, with columns ranging in 'diameter' from 10 nm to fractions of a micron. Because the direction of growth tends to be perpendicular to the close-packed planes ([111] for (fcc);

[011] for bcc), the layers within a given column are usually reasonably well defined. However, the spaces between the columns may not be so well defined. Occasionally, a thin filament of Nb might penetrate into a space between columns, perhaps even completely through the multilayer. To check for such penetration, the CPP-R is first measured at H = 0 using currents of 0.01, 0.1, 0.5 and 1 times the intended measuring current (50 mA or 100mA). Occasionally the CPP-R for the 0.01 setting is much less than for the 1 setting. Usually applying a field H reduces or even eliminates the difference. Such a 'current-dependent' sample is usually still measured to see if its CPP-MR is consistent with those for samples with no difference in CPP-R for the different current levels. But data from current-dependent samples are not included in publications.

Lastly, from low angle x-ray spectra, Chiang et al. [40] found that sputtered Co/Ag interfaces were rougher when deposited onto Nb than onto a bare Si substrate, presumably due to the roughness of the surface of the Nb base layer. They also found that increasing the sputtering pressure increased interface roughness, the effects of which will be examined in section 8.13.

## 6.2. Lithographed Micro- and Nano-pillars.

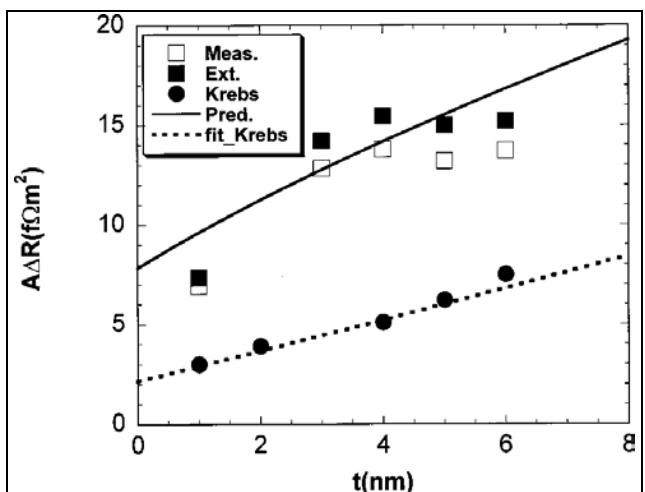

Fig. 12. A$\Delta$R vs t for [Co(t)/Cu(20)/NiFeCo(t)/Cu(20)]$_{10}$ hybrid SVs. The open squares are AR(Pk) and the filled squares are extrapolations to H = 0 (see ref. [115] for details). The solid curve was calculated without adjustment using parameters for NiFeCo found from EBSVs. The dotted line is a fit to the solid circles from ref. [116] using the parameters for the solid curve, but with an added contact resistance of 33 f$\Omega$m$^2$. Reproduced with permission from Vila et al. [115]. Copyright 2000, AIP Publishing LLC.

### 6.2.1. The technique and its advantages and disadvantages.

The first measurements of the CPP-MR up to room temperature were made with micropillars of diameter d ~ μm (radius r ~ μm; area A ~ μm$^2$, made by optical lithography [117, 118]. Nanopillars with diameters d $\lesssim$ 100 nm made by electron-beam lithography are now being used for device testing as we'll discuss in section 10. Fig. 11B shows schematically the geometry of a nanopillar with its leads.

Micro- or nanopillars have the advantages that: (1) they can involve complex multilayers with multiple components; (2) they can be measured from 4.2K to above room temperature; (3) their resistances are large enough to measure with commercial milli- or micro-voltmeters, and increase as their areas A decrease. Areas A = 10$^{-3}$ (μm)$^2$ can give resistances R ~ 1-10 $\Omega$, large enough for devices (e.g. magnetic field sensors). (4) a single chip can have many pillars with a wide range of areas.

Their disadvantages are: (1) that they can require sophisticated optical and electron-beam lithography to produce; (2) care must be taken to avoid or correct for: (a) too-large lead sheet resistance; and (b) too large lead/sample contact resistance. Too-large lead sheet resistance has two effects, first adding an additional resistance to the circuit, and second causing the current flow through the sample to be non-uniform. We consider contact resistance, lead resistance, and non-uniform current flow in that order below.

### 6.2.2. Experimental Details.

Micropillars can be made with optical lithography alone. Nanopillars are usually made with a combination of optical lithography for large scale features and electron-beam lithography for nm scale features. Simple symmetric micropillars with d >> L will give non-uniform current flow, as explained below [119, 120]. In principle, this non-uniformity can be corrected for by depositing at the same time, as-closely-identical-as-possible, samples with a range of values of d and extrapolating the data to d = 0 [117, 120]. A more complex geometry can minimize the contribution of sheet resistance to the total resistance [119]. Since, in the CPP geometry, the interfaces between the leads and the sample are in series with the sample, quantitative analysis requires making these interface contact resistances small.

(a) Contact Resistance: Since the bottom contact and the sample are deposited sequentially without intermediate steps, the main difficulty in lead/sample interfaces occurs at the top interface. We illustrate the problem with three coupled micropillar studies of $Ni_{65}Fe_{15}Co_{20}$ = NiFeCo [116, 121, 122].chosen as a potential alternative to Py = $Ni_{80}Fe_{20}$. All three studies used Cu to Cu top contacts. The first two by Vavra et al.,[121] and Krebs et al., [116] used reactive ion etching to open the contact, and then backsputter cleaning before depositing the top Cu lead. The third, by Bussman et al., [122] used chemical mechanical polishing, which the authors estimated reduced the contact resistance from ~ 100 $f\Omega m^2$ to ~ 0.006 $f\Omega m^2$. This conclusion about the large earlier contact resistances is supported by the data of Krebs et al. [116] for hybrid SVs with A = 1.2μmx1.2μm micropillars composed of a bottom electrode of 250 nm of Cu, the sample $[NiFeCo(t)/Cu(10-t)/Co(t)/Cu(10-t)]_{10}$ with t = 1,2,4,5, or 6 nm, capped with 75 nm of Cu and a top electrode of 150 nm of Cu covered by 150 nm of $Al_{0.98}Cu_{0.02}$ for bonding. Fig. 12 [115] compares the Krebs et al. data (filled circles) with data for A = $mm^2$ samples with similar layering (filled squares) between crossed superconducting strips. The dashed curve shows that the filled circles can be fit with the same parameters derived for the filled squares, by just adding a contact resistance = 33 $f\Omega m^2$. This comparison is oversimplified, as Krebs et al. recognized that their samples also had non-uniform current flow. But it shows how large contact resistances can be.

(b) Effects of too-large lead sheet resistance. The sheet resistance, $R_{sh}$, of a film of uniform thickness t composed of a metal of uniform resistivity ρ is defined as

$$R_{sh} = \rho/t. \tag{10}$$

$R_{sh}$ is the quantity measured with the van der Pauw technique [30]. It is also called the resistance/square, since it is what one would measure with either the van der Pauw technique with small contacts on the corners of a square sample, or with low resistance contacts on two sides of a square sample. On a rectangular sample, the van der Pauw technique requires corrections to be made to give $R_{sh}$. Measurements of a rectangular sample using low resistance contacts should give $N_{sq}R_{sh}$, where $N_{sq}$ is the number of squares in the rectangle. In four-probe measurements of a small pillar sandwiched between much larger area contacts, the current density will increase (current crowding) as the current shrinks down to the size of the pillar. To simplify the mathematics in analyzing a thin pillar of radius r and resistance R, we assume that the much larger area contacts (leads) consist of two identical films of resistivity ρ, area $W^2$, and thickness t, that sandwich the sample between them. As noted above, unless $R_{sh} << R$, the leads will add an additional resistance to the circuit and will also cause non-uniform current flow through the sample pillar. We start with the lead resistance, $R_{ld}$.

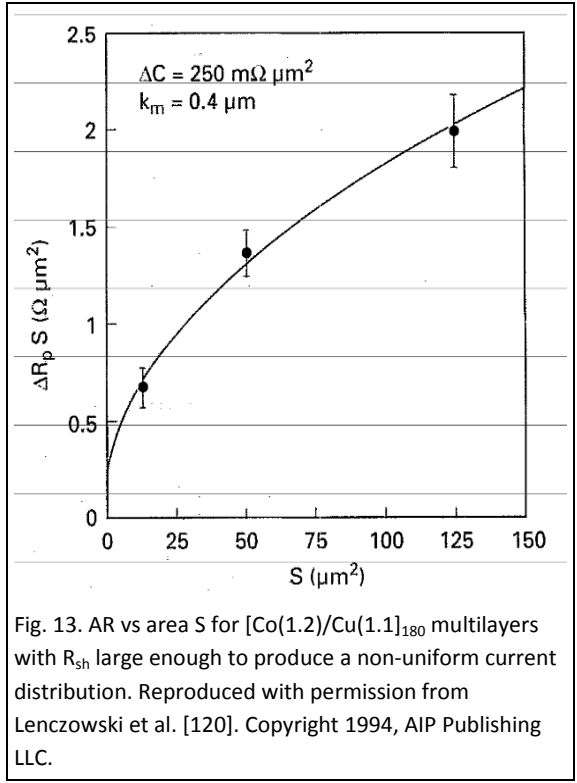

Fig. 13. AR vs area S for [Co(1.2)/Cu(1.1)]₁₈₀ multilayers with $R_{sh}$ large enough to produce a non-uniform current distribution. Reproduced with permission from Lenczowski et al. [120]. Copyright 1994, AIP Publishing LLC.

(b1) Current Crowding in the leads: To treat the lead resistance $R_{ld}$ properly, we must account for current crowding. When both lead films have the same $R_{sh}$, Chen et al. [123] gave an approximate analytical expression for a four-terminal measurement of their sum, $R_{ld}$:

$$R_{ld} = 0.2R_{sh}\ln(W/r). \quad\quad (11)$$

As examples, W = 20 μm and r = 0.5 μm give $R_{ld}$ = 0.7$R_{sh}$. and W = 20 μm and r = 3 μm give $R_{ld}$ = 0.4$R_{sh}$. $R_{ld}$ should be closely the same for AR(AP) and AR(P), thus not much affecting A$\Delta$R, but reducing the CPP-MR. For rectangular leads, a two-terminal measurement requires addition of a term 2N$_{sq}R_{sh}$ from the current terminal to the pillar.

(b2) Non-uniform current in the pillar. The problem of non-uniform current flow arises because, for resistances in parallel, current wants to flow along the path of least resistance. In Fig.11B, current injected into the top lead will flow uniformly through the sample micropillar (MP) only if $R_{sh} \ll R_{MP}$, since only then will the potential across each top or bottom face of the micropillar be approximately constant. In the other limit, $R_{sh} \gg R_{MP}$, current coming to the outer radius of the micropillar will mostly flow down a thin outer annulus of the micropillar of thickness ξ (defined below). Consider our assumed cylindrical pillar of radius r = 0.5 μm sandwiched between two, much wider (W >> r) Cu films with resistivity at 300K of 20 nΩm and thickness 200 nm. Take either a spin-valve with $AR_{SV}$ ~ 10 fΩm² or a multilayer with $AR_{ML}$ ~ 100 fΩm². The sheet resistance of each lead is $R_{sh}$ ~ 0.1 Ω and the micropillar resistances are $R_{MP}$ ~ 0.01 Ω for the SV and ~ 0.1 Ω for the multilayer. Simply comparing $R_{sh}$ with $R_{SV}$ or $R_{ML}$ says that current through the SV will be very non-uniform and that through the multilayer will be less non-uniform. For a pillar with radius 3 μm, the current will be strongly non-uniform in both cases.

As the ratio of the multilayer resistance to the lead resistance, $R_{ML}/R_{ld}$, decreases, the current density becomes more non-uniform, flowing within the pillar only inside a smaller and smaller annulus of thickness ξ in from the radius r. Interestingly, this shrinkage of ξ results in measured quantities, $AR_m$ and A$\Delta R_m$, that increase, as shown in Fig. 13. Lenczowski et al. [120] used a two-dimensional analysis to derive an approximate expression for the measured $AR_m$ = C(x/2)I₀(x)/I₁(x), where I₀ and I₁ are modified Bessel functions of zeroth and 1st order. x, which determines the non-uniformity of the current through the pillar, is given by x = r/ξ, where $\xi = \sqrt{C/R_{sh}}$ and C = $\rho_{sh}t_{sh}$ + $AR_o$ + $AR_c$. Here $AR_o$ is the desired AR of the pillar, $AR_c$ is any unwanted contact specific resistance, $\rho_{sh}$ is the resistivity of the top lead, $t_{sh}$ is its thickness, and to simplify we've assumed that the top and bottom leads are identical. As noted above, the length ξ determines the 'thickness' of the current flow down through the pillar. We give three examples of how $AR_m$ varies with x. In the limit of low lead resistances, x → 0 (ξ → ∞), I₀(x) → 1, I₁(x) → x/2, and $AR_m$ → C. In the limit of high lead resistances, x → ∞ (ξ → 0), I₀(x)/I₁(x) → 1 and $AR_m$ → Cx/2 which grows as $\sqrt{A}$ . For x = 1, $I_o(1)/I_1(1)$ = 2.24, and $AR_m$ = 1.12C. Note that C is larger than the desired $AR_o$ by $\rho_{sh}t_{sh} + R_c$, and thus $AR_m$ must be corrected for both $\rho_{sh}t_{th}$ and $R_C$ if they are not << $AR_o$. As an example of small x at 293K, an EBSV nanopillar with r = 50 nm, 200 nm thick Cu leads, and AR = 10 f$\Omega$m$^2$, would have A = 0.8x10$^{-14}$m$^2$, $\rho_{sh}$ = 20 n$\Omega$m, R = 1.25 $\Omega$, $R_{sh}$ = 0.1 $\Omega$, C = 10 f$\Omega$m$^2$, $\xi$ = 330 nm, and x = 0.17.

Another problem can arise when the pillar radius becomes smaller than the mean-free-path(s) in the multilayer (e.g., $\lambda$ ~ 100 nm for Cu or Ag and less for most other metals). Scattering from a perfectly diffuse pillar boundary should increase the CPP resistance and decrease the CPP-MR, as well as cause non-uniform currents if $\lambda$ varies from layer to layer [124].

To conclude, values of $AR_{AP}$, $AR_P$, A$\Delta$R, and CPP-MR directly measured on pillars are reliable only if the total sheet resistance of the contact films is much less than the resistances $R_{AP}$ and $R_P$ of the pillar, and if there is no significant contact resistance, $R_c$. In general this means that the smaller the pillar diameter, and the greater the thickness and the lower the resistivity of the lead films, the better. The one caveat is that too small a pillar can introduce boundary scattering. If any of the problems described above occur, corrections are needed to obtain reliable values of $AR_{AP}$, $AR_P$, A$\Delta$R, and CPP-MR.

Gijs et al. [119, 120] describe an alternative geometry that eliminates current crowding, but still leaves problems of contact resistance and non-uniform current flow if the lead sheet resistances are too large.

Leung et al. [125] used an FIB microscope to make a cross-bridge Kelvin structure that lets current be input through various leads and voltage be measured across other leads. With small samples, and thick, low resistance (e.g. Cu) leads, appropriate pairs can give correct AR and A$\Delta$R. Their data had large fluctuations; it was unclear if these were due to the sample or the measuring system.

Han et al. [126] described a way to reach diameters < 50 nm; but it gave low CPP-MR.

The most reliable way to correct for non-uniform current is to make a series of as closely identical as possible micropillars with a range of values of r, and to extrapolate to r = 0 using formulae such as Eq. 11 for the leads and those in [119, 120] for the pillar. If the lead effects are relatively modest, one can plot R vs 1/A and take the slope to get AR.

### 6.3. Electrodeposited Nanowires.

### 6.3.1. The technique and its advantages and disadvantages.

The third technique involves nanowires electrodeposited into long, thin, cylindrical channels in commercially available track-etched polymer [127, 128] [129-131] or nanoporous anodic aluminum oxide (Al$_2$O$_3$)[132] membranes. Typically, the membrane disks are cm in diameter and microns thick. Nominal pore diameters can range from as small as 10-20 nm to well over 100 nm. Fig. 11C shows schematically a disk with a few nanowires. A disk can contain 10$^8$ or more per cm$^2$ [133].

The advantages of the technique are: (1) the disks and electrodeposition equipment are relatively cheap; (2) the long, thin nanowire geometry gives a uniform CPP current through each wire; (3) the resistances are large enough to measure with standard electronics; (4) measurements can be made from 4.2K to above 293K; (5) the wires are so long and thin that contact resistances can often be neglected; (6) a single wire can be measured, and gave evidence of switching of finite fractions of the F-layers [134]. Single wires must be handled carefully to avoid destruction by stray voltages.

The disadvantages are: (1) most samples are made in a single bath, limiting the composition to just one F-metal and one N-metal—so far, mostly Co/Cu and to a lesser extent Py/Cu; (2) most measurements involve an unknown number of wires, giving only $\Delta R/R$, and not $A\Delta R$; (3) interface thicknesses appear to be larger than those made by sputtering or MBE, thus limiting the minimum layer thickness that can give reliable data [135]; (4) the average pore diameters in polymers are often larger than the nominal values [136] and the diameters vary with a Gaussian distribution [137]; (5) in polymers, the wire orientations deviate from the vertical, so that H is not always perpendicular to the wire axis; (6) in $Al_2O_3$, some channels can coalesce over at least part of their length, giving an unknown fraction of much larger channels [132]; (7) simple $[F(t_F)/N(t_N)]$ nanowires rarely give complete AP states, so an AP fraction '$p$' must be separately measured or inferred; (8) while a few multiple bath studies have been reported, it is not yet clear how reliable they are (see below);

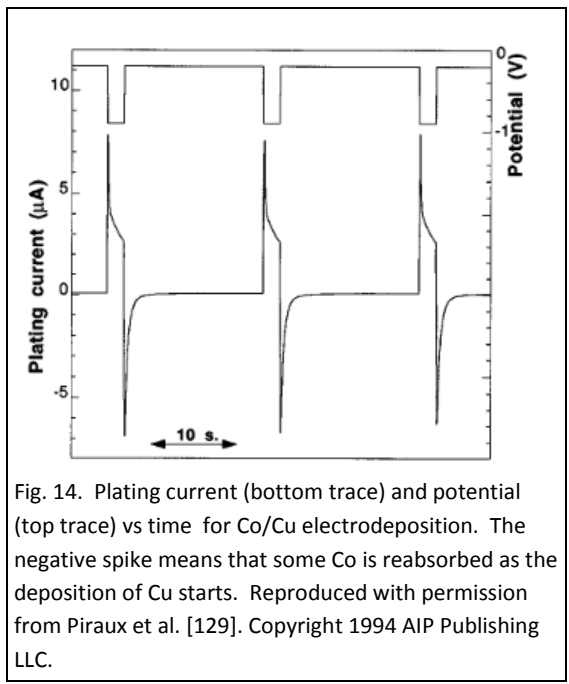

Fig. 14. Plating current (bottom trace) and potential (top trace) vs time for Co/Cu electrodeposition. The negative spike means that some Co is reabsorbed as the deposition of Cu starts. Reproduced with permission from Piraux et al. [129]. Copyright 1994 AIP Publishing LLC.

### 6.3.2. Experimental details.

### 6.3.2.1. Single bath

Pores in commercially available track etched polycarbonate [128-130] or polyester [138] membranes are made by bombarding the polymer with a heavy ion such as Ar and then preferentially etching away the material along the ion track. Alternatively, commercially available anodized alumina templates [132] are made by immersing Al as an anode in an acid bath along with a cathode and a reference electrode. Passing current between the cathode and anode releases oxygen atoms from the electrolyte that convert the surface of the Al into the oxide alumina ($Al_2O_3$). The resulting alumina contains an ordered array of pores of diameter that can be controlled by the current and can be extended all the way through the original Al. Nominal pore diameters can range from as little as 10 nm to over 400 nm. A test of pore diameters in polycarbonate found a Gaussian distribution around a 'best average' larger than the nominal pore diameter: e.g., nominal pore diameters of 10 nm and 50 nm were measured to span 36 ± 3 nm and 61 ± 2 nm [137]. A test of an anodized template 60 μm thick found that within the top ~ 1 μm of the surface many nominally 20 nm pores joined to form one much larger pore of diam. ~ 0.3 μm [132].

The pores are filled by electrodeposition. Typically a few hundred nm thick Au or Cu layer is sputtered or evaporated onto the bottom of the sample disk to serve as the working electrode in a three electrode electrochemical cell and as the bottom contact for resistance measurements. The disk is then submerged in the electrolyte, composed of a mixture of chemicals containing the metals of interest (usually Cu with either Co or Py [139, 140]). A third electrode is used as a reference. At small negative voltage, mostly Cu is deposited. At larger negative voltage, both the F-metal and Cu are deposited. To keep Cu contamination of the F-metal down, the Cu concentration in the electrolyte is kept low so that Cu deposition is diffusion limited, and the F-metal is deposited faster than the Cu. Contamination of the Cu is claimed to be 2% or less. Contamination of the F-metal is typically reported as 7-10%. During deposition, layer thicknesses are set by switching the potential by computer when a preset integrated amount of charge has flowed between the counter and working electrodes. The deposition is stopped when a sudden increase of the plating current indicates that the nanowires have begun to emerge from the membrane to form three-dimensional caps [129]. Fig 14 [129] shows a time trace of the plating current during deposition of Co(10)/Cu(10) multilayers. Pulses of -0.2V give nearly pure Cu and pulses of -0.9 V give Co with up to 10% Cu. The negative current peak indicates desolving of some Co at the start of the Cu deposition. Fig. 15 [129] shows a TEM Image of a nanowire multilayer removed from its membrane.

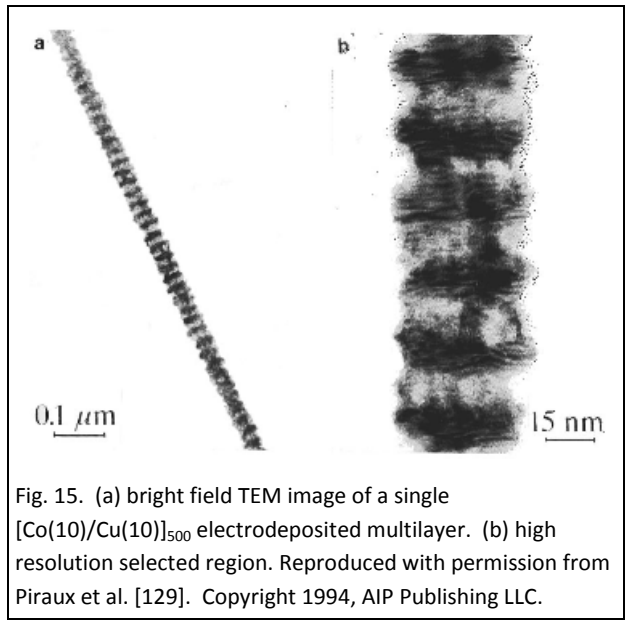

Fig. 15. (a) bright field TEM image of a single [Co(10)/Cu(10)]$_{500}$ electrodeposited multilayer. (b) high resolution selected region. Reproduced with permission from Piraux et al. [129]. Copyright 1994, AIP Publishing LLC.

Top contacts have been made in a variety of ways. (a) Using simply an evaporated Cu layer [129]. (b) Gluing two Au wires with Ag epoxy onto a small area (~ 0.1 mm$^2$) [141]. (c) Pressing a rounded contact into the top of the membrane [142]. Measured total resistances range from 0.1 Ω to 500 Ω. The number of wires contacted is both unknown (although it can be at least roughly estimated from the measured resistance) and variable from membrane to membrane.

Single bath multilayers of the form [F(t$_F$)/N(t$_N$)]$_n$ rarely achieve an AP state, making it necessary to estimate an antiferromagnetic polarization fraction 'p' (typically ~ 50%) and to infer from linear variations with t$_F$ or t$_N$ that 'p' is constant for a given set of multilayers. Better success was achieved using multilayers with the F-layers separated by two different N-layer thicknesses, one short (say t$_s$ ~ 10 nm) and the other long (say t$_L$ ~ 100nm) [99]. The resulting F/N(t$_s$)/F trilayers, separated by N(t$_L$)layers, gave MR(H) curves closer to those for AF-coupling (Fig.1) and estimated values of 'p' as large as 89% [99].

**6.3.2.2. Multiple baths.** Two groups have studied electrodeposited multilayers made with multiple baths. In 1997, Blondel et al. [133] tried double baths with Co/Cu to avoid the Cu contamination of Co that occurs in a single bath. They mounted a polycarbonate membrane on a rotating disc that moved it between the Cu and Co baths with cleaning stations in between. They were able to make dual bath multilayers with t$_{Co}$ ≥ 10 nm. They compared data for two samples, a single-bath multilayer with t$_{Co}$ = t$_{Cu}$ = 8 nm and a dual bath multilayer with t$_{Co}$ = t$_{Cu}$ = 10 nm. The dual bath CPP-MR was smaller than the single bath one, which they ascribed mainly to a smaller interface resistance, perhaps due to the interface being more diffuse.

In 2006, Wang and co-workers published three papers [143-145] on multiple bath electrodeposition into anodized alumina templates, using a separate bath for each of three or four components. As seed layers and bottom contacts they sputtered onto the template either 50 nm of Ta [143, 145] or 100 nm of Au [144]. To prevent oxidation of the layers they added to their solutions an inhibitor such as citric or ascorbic acid and they coated the top of the template with either 4 nm [144] or 20 nm of Cu [143, 145]. Their first samples were EBSVs of the form FeMn(10)/Py(7)/Cu(t$_{Cu}$)/Py(10)/Cu(20) with 1.5 nm ≤ t$_{Cu}$ ≤ 5 nm. Their second samples were hybrid SVs of the form NiFe(6)/Cu(t$_{Cu}$)/Co(4)/Cu(4) also with 1.5 nm ≤ $t_{Cu} \leq 5$ nm. Their third samples were EBSVs of the form FeMn(10)/Py(5)/Co(2)/Cu(2)/Co(2)/Py(10)/Cu(20), with the thin Co layers added to improve thermal stability. The hysteresis curve for an EBSV with $t_{Cu} = 2$ nm [143], or for a hybrid SV with $t_{Cu} = 4$ nm [144], each looked as expected for an uncoupled SV. However, the variations of the CPP-MRs with $t_{Cu}$ in [143] and [144] differ from expectations in opposite ways. In [143], the EBSV CPP-MR drops from 6% at $t_{Cu} = 2$ nm to 0% at 4.5 nm. If the Cu resistivity is much lower than the F-resistivities, VF theory would predict little change. In contrast, in [144], the hybrid SV CPP-MR grows from 0.5% at $t_{Cu} = 2$ nm to 1.7% at 4 nm. Again, VF theory would predict little change. The authors tried to associate these opposite changes with ones seen in different CIP-MRs, but the physics there is different. Unfortunately, these surprisingly opposite variations of CPP-MR with $t_{Cu}$ make it unclear whether their multiple bath technique is yet under control.

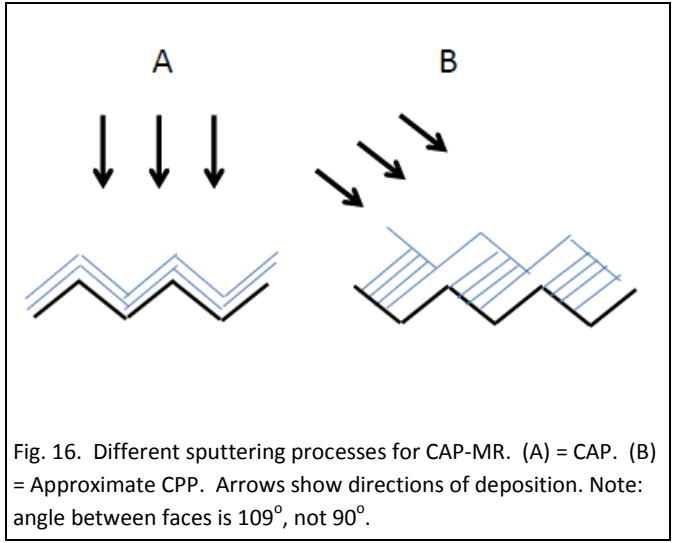

Fig. 16. Different sputtering processes for CAP-MR. (A) = CAP. (B) = Approximate CPP. Arrows show directions of deposition. Note: angle between faces is 109°, not 90°.

### 6.4. V-Grooves (CAP) and Extrapolation to CPP.

The last topic in this section involves preparation and evaporation into V-Grooves, giving Current-at-an-Angle-to-the-Plane (CAP) MRs. Fig. 16 illustrates the two different procedures used. First, a pattern of parallel V-grooves having (111) sides is produced by either: (a) masking plus photolithographic patterning and then wet etching of a (100) oriented Si substrate [146-148]; or (b) using holographic laser interference lithography to pattern and then to wet etch a (100) InP substrate [149-151]. A multilayer is then deposited onto the substrate, giving a layer geometry that lies between CIP and CPP. Ono and Shinjo [146, 147] evaporated [Co/Cu/Py/Cu] hybrid spin-valves normal to the original substrate surface to well beyond the V-groove depth (Fig. 16A). In this geometry, the CAP current flows at an angle of 54.7° to the layer planes and a CIP current flows along the grooves. Levy et al. [152] proposed how to extrapolate from such CIP and CAP data to CPP. In contrast, Gijs et al. [149-151] evaporated [Co/Cu] multilayers at an angle normal to one of the (111) surfaces of the V-groves (Fig. 16B), starting and ending with a moderately thick (20 nm) Cu layer. If they could end the multilayer just at the end of the V-groove, the top and bottom thick Cu layers would connect the multilayers in sequential V-grooves and, if the resistances of the thick Cu layers are low enough, would approximate a CPP geometry.

### 7. Historical Timeline.

This section contains a list of first publications on topics of particular interest.

1991. 2CSR model derived. [27]

1991. First CPP-MR data: [Co/Ag]$_n$ at 4.2K; superconducting cross-strips; CPP-MR > CIP-MR. [26]

1993. 2CSR model analysis of [Co/Ag]$_n$ & [Co/AgSn]$_n$. Parameters $\beta_{Co}$, $\rho^*_{Co}$, $2AR^*_{Co/Ag}$, $\gamma_{Co/Ag}$. [28]

1993. Valet-Fert (VF) model derived. [29]

1993. CPP-MR oscillations of [Co(1.5)/Cu(t$_{Cu}$)]$_n$ with $t_{Cu}$ at 4.2K. [39]

1993. Micropillar CPP-MR of [Fe/Cr]$_n$ from T = 4.2K to 300K. CPP-MR usually > CIP-MR at all T. [117]

1994. Micropillar CPP-MR of [Co/Cu]$_n$ from 4K to 300K. CPP-MR > CIP-MR at all T. [118]

1994. Nanowire CPP-MR of [Co/Cu]$_n$ [128, 129] and [Py/Cu]$_n$ [128].

1994. $l_{sf}^N$ for Cu(Pt), Cu(Mn), Ag(Pt), Ag(Mn). Values from VF model agree with those from CESR. [53]

1995. Correctly predict AR(AP) and AR(P) for [Py/Cu/Co/Cu] from Py/Cu and Co/Cu parameters. [93]

1995. CAP measurements of [Co/Cu/Py/Cu]$_n$ hybrid SVs [146, 152] and [Co/Cu]$_n$ multilayers [149].

1996. Temperature variation of 2CSR model parameters for [Co/Cu]$_n$ from 4.2K to 300K vla CAP. [151]

1996. 2AR$_{N1/N2}$ for N1/N2 = Ag/Cu, Ag/Au, Au/Cu multilayers. [32].

1996. $l_{sf}^{Co} \sim$ 50 nm at 77K from VF model and nanowires. [153, 154]

1997. $l_{sf}^{Py}$ = 5.5 ± 1 nm from VF model plus EBSV. [73]

1997 Interleaved vs Separated [Co/Ag/Py/Ag]$_n$ vs [Co/Ag]$_n$[Py/Ag]$_n$. [155]

1997 $AR_{F/N}^{\uparrow}$ and $AR_{F/N}^{\downarrow}$ formulae & no-free-parameter calculation for Co/Cu with perfect interfaces. [87].

1997. Dual Bath electrodeposition. [133]

1998. Inverse CPP-MRs. [156].

1998 Interface Roughness effects on CPP-MR in Co/Ag. [40]; 2000 in Fe/Cr [42]; 2002 in Fe/Cr. [41]

1999. CPP-MR of Heusler alloy NiMnSb (nominal half-metal). [157]

1999. Polarized neutron and SEMPA confirmation of ≈AP 'virgin' state for [Co(6nm)/Cu(6nm)]$_{60}$. [101]

1999 Interleaved vs Separated [Fe/Cu/Co/Cu]$_n$ vs [Fe/Cu]$_n$[Co/Cu]$_n$. [158]

1999. $\delta_{N1/N2}$: general way to measure and values for several N1/Cu pairs. [159]; 2000 [67]

2000. mfp effect?: Interleaved vs Separated [Co(10)/Cu/Co(6)/Cu)]$_n$. vs [Co(8)/Cu]$_n$[Co(1)/Cu]$_n$. [79].

2000, 2001. Nanooxide Layers(NOL) w/pinholes = Current-Confined Paths (CCP). [160] [161]

2001. No-free-parameter calculations of $AR_{F/N}^{\uparrow}$ and $AR_{F/N}^{\downarrow}$ with disordered interfaces. [89]

2005. Double Blind agreement of experiment with no-free-parameter calculation of 2AR$_{Pt/Pd}$: [86]

2006. Multiple bath electrodeposition. [143]

2006. $\delta_{F/N}$: fit to combined CPP-MR and Spin-Torque data of others: uncertainties unclear [162].

2010. $\delta_{F/N}$: general procedure for measuring, and derived $\delta_{Co/Cu} \cong 0.35$. [163]

2014. Measured large $\delta_{Co/Pt} \cong 0.9$. [164]

## 8. Results Organized by Historical Timeline.

In early studies of a new phenomenon, such as CPP-MR, there are many questions to answer. Some examples are: (a) how large is it?; (b) What do typical data look like?; (c) How should data be analyzed to extract underlying physics?; (d) Is the CPP-MR dominated by 'bulk' or 'interface' scattering?; (d) What techniques reliably produce AR(AP)?; (e) Are the characteristic lengths λ or $l_{sf}$?; (f) How long are $l_{sf}$s for both F- and N-metals?; (g) How do VF parameters vary for different F/N pairs?; (g) Do VF parameters contain all bulk and interface information for real metal pairs?; (h) Are VF parameters 'universal'—i.e., do they agree with equivalent parameters found by completely different techniques, or from no-free-parameter calculations?; (i) Is spin-flipping at interfaces significant? In this chapter we examine studies that address questions such as these. We begin each topic with its first study, but then continue as needed until that topic is done.

**8.1. CPP-MR vs CIP-MR.**

An obvious early question is: How do the CPP-MR and the CIP-MR compare for the same sample? Just before the first CPP-MR data were published, Zhang and Levy (ZL) [27] used a 2CSR model to predict CPP-MR $\geq$ CIP-MR for simple multilayers with no spin-flipping and no lead resistances

**8.1.1. Co/Ag at 4.2K.**

The first CPP-MR measurements, by Pratt et al., were made with superconducting cross-strips and published in 1991 [26]. Values of CPP-MR and CIP-MR were compared at 4.2K for Co/Ag multilayers with total thicknesses $\approx$ 720 nm and the forms: (a) [Co(6)/Ag($t_{Ag}$)]$_n$ with $t_{Ag} \geq 6$ nm, or (b) [Co($t_{Co}$)/Ag($t_{Ag}$)]$_n$ with $t_{Co} = t_{Ag} \geq 6$ nm, Fig. 3 [26] compares hysteresis curves for CPP-MR, CIP-MR, and magnetization $M$ for the multilayer at the intersection of (a) and (b), [Co(6)/Ag(6)]$_{60}$, along with the CPP-MR for a single Co(9) layer. Fig. 3 contains three important results. (1) The CPP-MR is several times larger than the CIP-MR. The ratios for other samples ranged from 2.5 to 13 [26]. (2) The variations of the CPP-MR and CIP-MR with H are similar in form, with the largest values in the as-prepared ('virgin') state, MR(0), before any field has been applied. Subsequent peaks, MR(Pk), after the sample is taken to beyond its saturation field $H_s$, occur at H a bit above the coercive field, $H_c$. (3) The multilayer CIP-MR and CPP-MR are both much larger than the CPP-MR of a single Co layer, which is zero to within the measuring uncertainty of ± 0.5%.

**8.1.2. Fe/Cr and Co/Cu from 4.2K to 300K.**

In 1993 to 1995 [117, 118, 165] micropillar studies by Gijs et al., showed in Figs. 17 and 18 that at 4.2K the CPP-MRs of Fe/Cr and Co/Cu multilayers were also larger than the CIP-MRs, and usually (but not always) stayed larger all the way up to 300K. To correct for non-uniform current flow through the 'short-wide' pillars, the CPP-MRs in Figs.

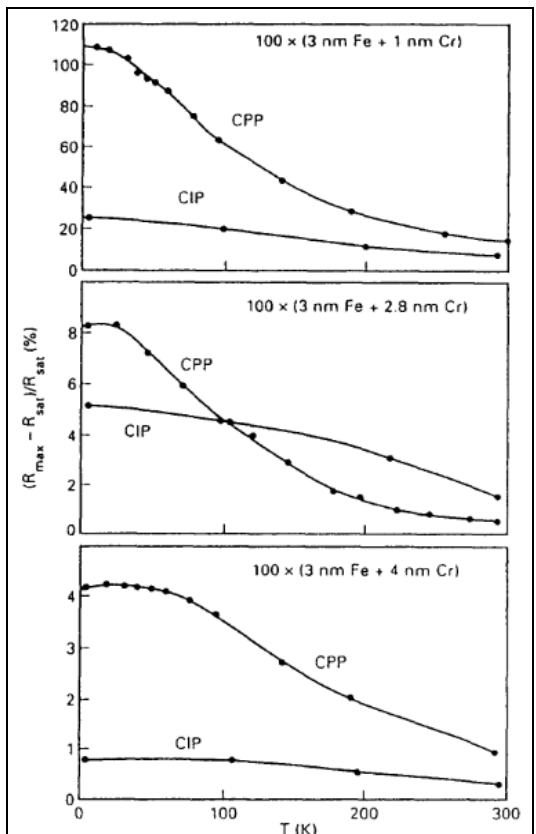

Fig. 17. CPP-MR and CIP-MR vs T(K) for [Fe(3)/Cr(t)]$_{100}$ multilayers with t = 1 nm, 2.8 nm, or 4 nm. Reproduced with permission from Gijs et al. [117]. Copyright 1993 by the American Physical Society.

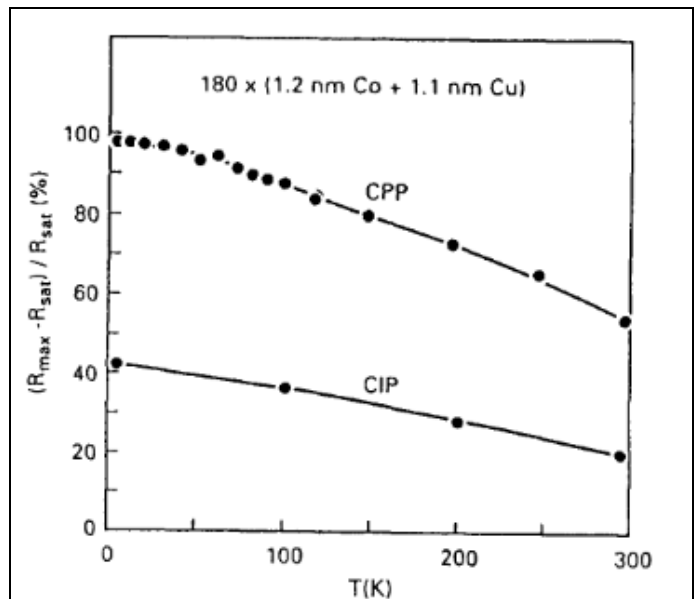

Fig. 18. CPP-MR and CIP-MR vs T(K) for a [Co(1.2)/Cu(1.1)]$_{180}$ multilayer. Reproduced with permission from Gijs et al. [118]. Copyright 1994, AIP Publishing LLC.

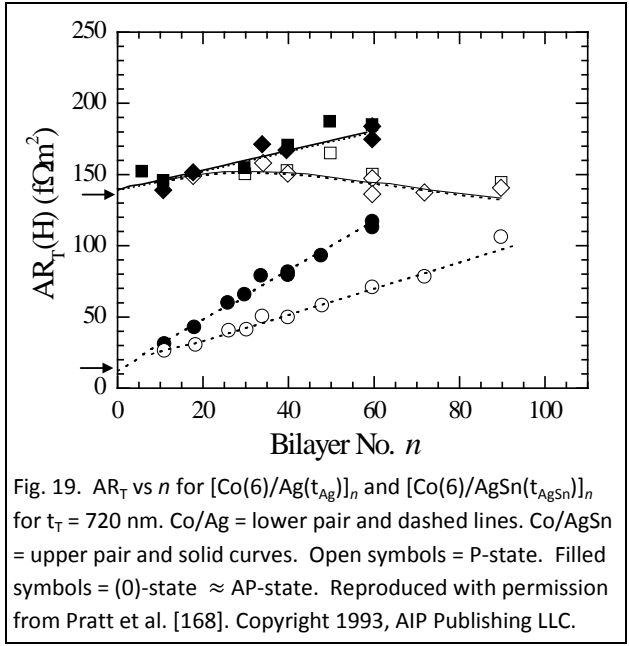

Fig. 19. $AR_T$ vs $n$ for [Co(6)/Ag($t_{Ag}$)]$_n$ and [Co(6)/AgSn($t_{AgSn}$)]$_n$ for $t_T$ = 720 nm. Co/Ag = lower pair and dashed lines. Co/AgSn = upper pair and solid curves. Open symbols = P-state. Filled symbols = (0)-state $\approx$ AP-state. Reproduced with permission from Pratt et al. [168]. Copyright 1993, AIP Publishing LLC.

17 and 18 were found by extrapolating to pillar diameter d → 0. Fig. 17 includes a case of CPP-MR < CIP-MR at higher temperatures, the source of which is not clear.

### 8.2. Analyses of CPP-MR Assuming No Spin-Flipping.

The next important step in CPP-MR studies was to establish a model to analyze CPP-MR data. As noted in section 1.4, when early CPP studies were made, values of $l_{sf}^F$ were unknown, and both $l_{sf}^F$ and $l_{sf}^N$ were expected to be long compared to the F- and N-layer thicknesses, $t_F$ and $t_N$. This expectation led to neglect of moment flipping, in which case, ZL [27] had predicted data describable by a 2CSR model (see sections 1.4.1 and 4.1).

### 8.2.1. Simple SR model with [Co/Ag]$_n$

In the first attempt at data analysis, in 1992 Lee et al [166] measured AR vs $n$ at 4.2K for a series of [Co($t_{Co}$)/Ag($t_{Ag}$)]$_n$ multilayers with fixed total thickness as close as possible to $t_T$ = 720 nm and: (a) fixed $t_{Co}$ = 2 nm; (b) fixed $t_{Co}$ = 6 nm; or (c) $t_{Co}$ = $t_{Ag}$. Co/Ag was chosen for two reasons. First, Co and Ag are immiscible [167], hopefully giving minimal interface interdiffusion. Second, coupling-induced oscillations had not been seen for Co/Ag. So it was hoped that studying Co/Ag with $t_{Ag} \geq 4$ nm might avoid complications of $t_{Ag}$-dependent changes in coupling. To avoid complications of complex magnetic structures in Co, all samples analyzed also had $t_{Co} \lesssim 18$ nm.

For each Co/Ag multilayer, AR(H) varied qualitatively as in Fig. 3. Importantly, AR(0), AR(Pk), and AR(P) all increased linearly with increasing $n$ (see, e.g., Fig. 19, and Figs. 29 and 30 below). Straight lines could be fit with a one-current series resistor (1CSR) model that gave ordinate intercepts consistent with the expected value of $AR_{Nb/Co}$ = 6 ± 1 f$\Omega$m$^2$ (see section 3.5) and an independently measured Ag resistivity, $\rho_{Ag}$ = 10 ± 2 n$\Omega$m. However, unlike the 2CSR model, where all parameters are independent of H, to produce a CPP-MR, the 1CSR model required both $\rho_{Co}$ and $AR_{Co/Ag}$ to be H-dependent, with $\rho_{Co}$(H) growing weakly with H and $AR_{Co/Ag}$(H) growing more strongly.

### 8.2.2. Test of 2CSR model with [Co/Ag]$_n$ and [Co/AgSn]$_n$.

The need for a 2CSR model was established in 1993, when Lee et al. [28] extended CPP-MR measurements to [Co/Ag(4at.%Sn)]$_n$ multilayers—which we simplify to just [Co/AgSn]$_n$. Because Sn has three more conduction electrons than Ag, it gives strong scattering (i.e., a large resistivity, $\rho_{AgSn} \sim 20\rho_{Ag}$ for 4 at.% Sn [37, 38]). But since it is in the same row as Ag in the periodic table, spin-orbit scattering leading to spin-flipping should be weak. We'll see that the Co/AgSn data were inconsistent with the 1CSR model, but the Co/Ag and Co/AgSn data together supported a 2CSR model. The importance of the 2CSR model for later CPP-MR studies motivates a detailed analysis of this paper.

The analysis involves a specific application of Eqs. 4 - 7 in section 1.4.1. The samples all had fixed total thickness $t_T \cong 720$ nm and fixed Co thickness $t_{Co} = 6$ nm. Each multilayer had a top Ag or AgSn layer, to protect the top Co layer from oxidation. Proximity effect with the top superconducting Nb layer presumably turned this top Ag or AgSn layer superconducting. We will note where we need to correct for this effect. To use Eqs. $4 - 7$ with superconducting contacts, one must add to Eq. 4 a constant $2AR_{Nb/Co} = 6$ f$\Omega m^2$ for the two Co/Nb interfaces (see section 3.5), and then use $t_T = nt_N + nt_F$ to eliminate the variable $t_N$. Since AR(AP) should be the largest AR(H) in data such as those in Fig. 3, the authors took AR(0) as the best estimate of AR(AP), a choice we will examine in section 8.2.3.

For multilayers of fixed $t_T$, Eqn. 4 becomes

$$AR(AP) = 2AR_{Nb/F} + \rho_N t_T + n[(\rho_F^* - \rho_N)t_F + 2AR_{F/N}^*]. \quad (4b)$$

Eq. 4b still predicts straight line variations of AR(AP) with $n$ for both Ag and AgSn. But, since $\rho_{AgSn} \sim 20\rho_{Ag}$, the intercept for AgSn (the sum of the two constant terms) should be much larger than for Ag and the slope should be smaller. For large enough $\rho_N > \rho_F^*$, the slope could even become negative.

Eq. 5 is unchanged in form. However, since AR(AP) is now given by Eq. (4b), when N = AgSn and $n$ is small, AR(AP) should be approximately constant, making AR(P) non-linear in $n$. The combination of the term $(\rho_F^* - \rho_N)$ in Eq. 4b with Eq. 5 also allows AR(P) for AgSn to decrease in magnitude with increasing $n$.

Eq. 6 is also unchanged in form. However, with AR(AP) given by Eq. (4b), when N = AgSn and $n$ is small, A$\Delta$R should vary as $n^2$, a variation incompatible with the simple 1CSR model. Lastly, Eq. 7 is also unchanged in form, even with the constant $2AR_{Nb/F}$ included. It, thus, allows a direct test of whether N's mean-free-path, $\lambda_N$, is important, since the right-hand-side (rhs) of Eq. 7 does not depend on either 2AR$_{Nb/Co}$ or any properties of N. That is, even with $\lambda_{Ag} \sim 20\lambda_{AgSn}$, Eq. 7 predicts that the square root for Ag and AgSn should fall on the same straight line passing through the origin.

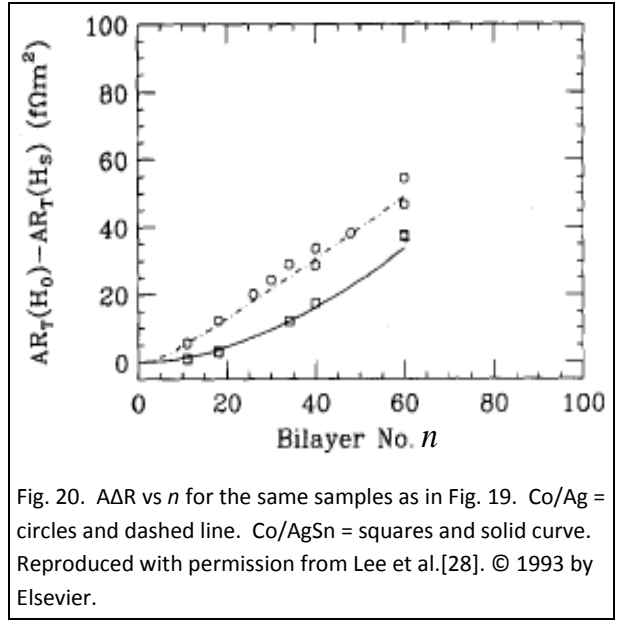

Fig. 20. A$\Delta$R vs $n$ for the same samples as in Fig. 19. Co/Ag = circles and dashed line. Co/AgSn = squares and solid curve. Reproduced with permission from Lee et al.[28]. © 1993 by Elsevier.

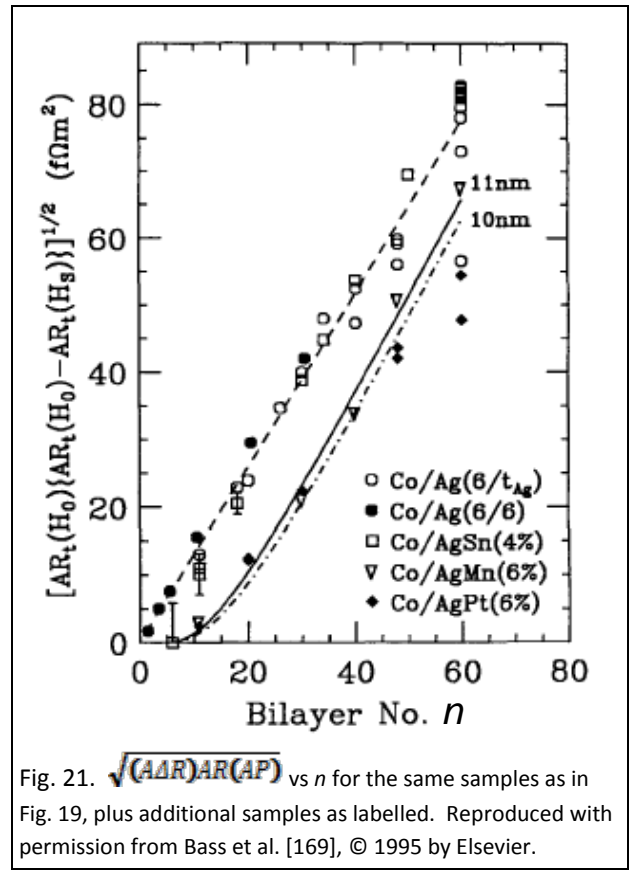

Fig. 21. $\sqrt{(A\Delta R) AR(AP)}$ vs $n$ for the same samples as in Fig. 19, plus additional samples as labelled. Reproduced with permission from Bass et al. [169], © 1995 by Elsevier.

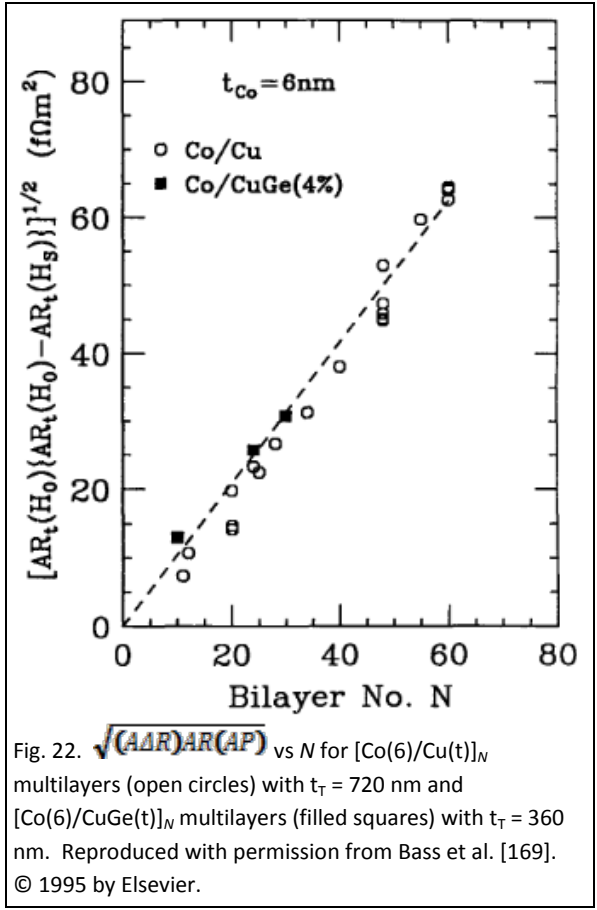

Fig. 22. $\sqrt{(A\Delta R)AR(AP)}$ vs $N$ for [Co(6)/Cu(t)]$_N$ multilayers (open circles) with t$_T$ = 720 nm and [Co(6)/CuGe(t)]$_N$ multilayers (filled squares) with t$_T$ = 360 nm. Reproduced with permission from Bass et al. [169]. © 1995 by Elsevier.

For superconducting leads, the proximity effect changes Eq. (4) differently if the multilayer starts with an F-layer and ends with an N-layer, [F/N]$_n$, or starts and ends with F-layers, [F/N]$_n$F. [F/N]$_n$F adds one extra F-layer. [F/N]$_n$ subtracts one N-layer and two F/N interfaces. The multilayers in [28] ended with an N-layer.

With these equations in hand, we now turn to the data. Figs. 19-21 [28, 168] [169] show data for both [Co(6)/Ag(t$_{Ag}$)]$_n$ and [Co(6)/AgSn(t$_{AgSn}$)]$_n$ multilayers in the forms appropriate to Eqs. 4b, 5, 6, and 7, with AR(0) chosen as the best estimate of AR(AP). Fig.19 shows AR(0) and AR(P) vs $n$ and fits with Eqs. 4b and 5. To within experimental uncertainties the data are consistent with the predictions, especially the large difference in ordinate intercepts and the smaller slopes for AgSn. The data for AgSn have been corrected for the missing AgSn-layer and the missing two Co/Ag interfaces just noted (since AgSn has only 4%Sn, it is assumed that $AR^*_{Co/AgSn} = AR^*_{Co/Ag}$). Fig. 20 shows uncorrected A$\Delta$R vs $n$ data and fits with Eq. 6. As predicted, for small $n$ the Ag data are consistent with a straight line but the AgSn data grow approximately as $n^2$. Finally, Fig. 21 shows uncorrected $\sqrt{(A\Delta R)AR(AP)}$ data vs $n$. Despite the large differences for Ag and AgSn in Figs. 19 and 20, the data in Fig. 21 overlap to within uncertainties. Provided that AR(0) $\cong$ AR(AP), the data for Co/Ag and Co/AgSn in Figs. 19-21 agree well with the predictions of the 2CSR model. The other data in Fig. 21 will be discussed in section 8.5.1.

Section 8.2.3 will describe four different tests of AR(0) $\cong$ AR(AP). Two of these will use [Co/Cu]$_n$ instead of [Co/Ag]$_n$. Fig. 22 [169] shows that $\sqrt{(A\Delta R)AR(AP)}$ vs $n$ for [Co/Cu]$_n$ and [Co/Cu(4%Ge)]$_n$ behaves just like $\sqrt{(A\Delta R)AR(AP)}$ vs $n$ for [Co/Ag]$_n$ and [Co/AgSn]$_n$ in Fig. 21. As for Sn in Ag, Ge in Cu gives a large resistivity, but weak spin-orbit scattering and thus a long $l_{sf}$. For values, see section 8.5.1.

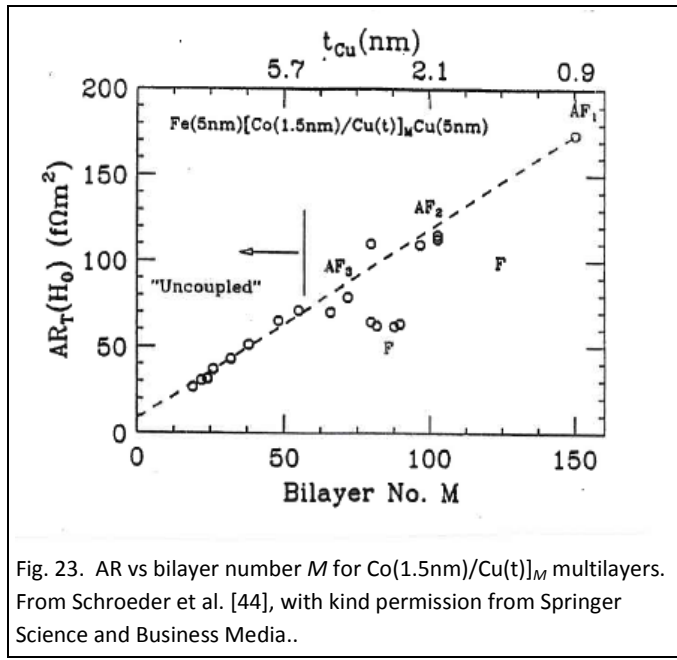

Fig. 23. AR vs bilayer number $M$ for Co(1.5nm)/Cu(t)]$_M$ multilayers. From Schroeder et al. [44], with kind permission from Springer Science and Business Media..

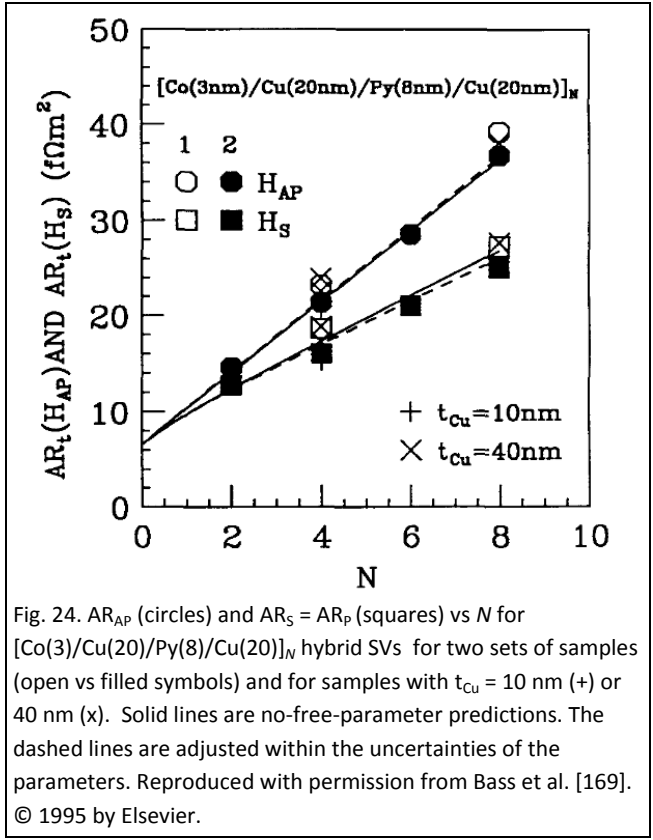

Fig. 24. $AR_{AP}$ (circles) and $AR_S = AR_P$ (squares) vs $N$ for [Co(3)/Cu(20)/Py(8)/Cu(20)]$_N$ hybrid SVs for two sets of samples (open vs filled symbols) and for samples with $t_{Cu} = 10$ nm (+) or 40 nm (x). Solid lines are no-free-parameter predictions. The dashed lines are adjusted within the uncertainties of the parameters. Reproduced with permission from Bass et al. [169]. © 1995 by Elsevier.

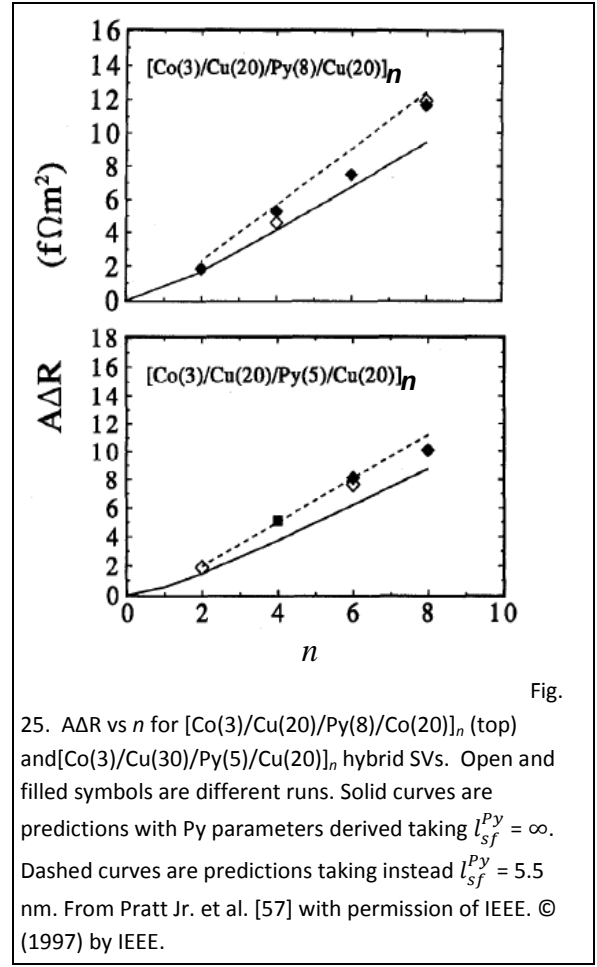

Fig. 25. $A\Delta R$ vs $n$ for [Co(3)/Cu(20)/Py(8)/Co(20)]$_n$ (top) and [Co(3)/Cu(30)/Py(5)/Cu(20)]$_n$ hybrid SVs. Open and filled symbols are different runs. Solid curves are predictions with Py parameters derived taking $l_{sf}^{Py} = \infty$. Dashed curves are predictions taking instead $l_{sf}^{Py} = 5.5$ nm. From Pratt Jr. et al. [57] with permission of IEEE. © (1997) by IEEE.

### 8.2.3. Tests of AR(0) vs AR(AP) using Superconducting Cross-strips.

Strictly, AR(0) ≤ AR(AP). The question is, how close is AR(0) to AR(AP)? Because of the importance of the assumption AR(0) ≅ AR(AP) in the analyses of Co/Ag and Co/Cu data just described, we step forward in time to present four later tests of this 'near equality'. A plausible model for the physics underlying AR(0) ≅ AR(AP) was given in section 5.4.2. Each of the first three tests separately yields AR(0) ≅ AR(AP) for a particular set of samples. Each, by itself, has limits. But combining all three with the fourth test, supports AR(0) ≅ AR(AP) for sputtered Co/Cu and Co/Ag multilayers. We'll see later that such approximate agreement does not hold for all F/N metal pairs.

### 8.2.3.1. Test using AP coupled data.

The first test, in 1993, starts from Fig. 4, which shows that sputtered [Co(1.5)/Cu($t_{Cu}$)]$_n$ multilayers with $t_T = 360$ nm give the very large CPP-MR at $t_{Cu} \cong 0.9$ nm expected for AF-coupling and an AP state, followed by oscillations of the CPP-MR with increasing $t_{Cu}$, with weaker AF coupling for $t_{Cu} \cong 2.1$ nm, still weaker AF coupling for $t_{Cu} \cong 3.5$ nm, and then a smooth decrease in MR beyond $t_{Cu} \cong 6$ nm. The test by Schroeder et al. [44] combined the data in Fig. 4 with Eq. 4b to extrapolate the expected behavior of AR(AP) to large (uncoupled) N-layer thicknesses. As the multilayers in Fig. 4 have fixed $t_T = 360$ nm and fixed $t_{Co} = 1.5$ nm, they should be described by Eq. 4b, which predicts AR(AP) = {(10 ± 2) + $n$[Slope]} fΩm². Taking the $t_{Cu} = 0.9$ point in Fig. 4 as a true AP state, and connecting $AR_T$ for that point to an ordinate intercept = 10 fΩm², gives the straight line in Fig. 23 [44] To within uncertainties, this line passes through the values of AR(0) for $t_{Cu} \gtrsim$ 6 nm, supporting that these values lie close to AR(AP). However, these data are for only a single $t_{co}$ = 1.5 nm, and since the AR(P) data should go to the same intercept, it is hard to set an uncertainty on the approximate relation AR(0) ≅ AR(AP).

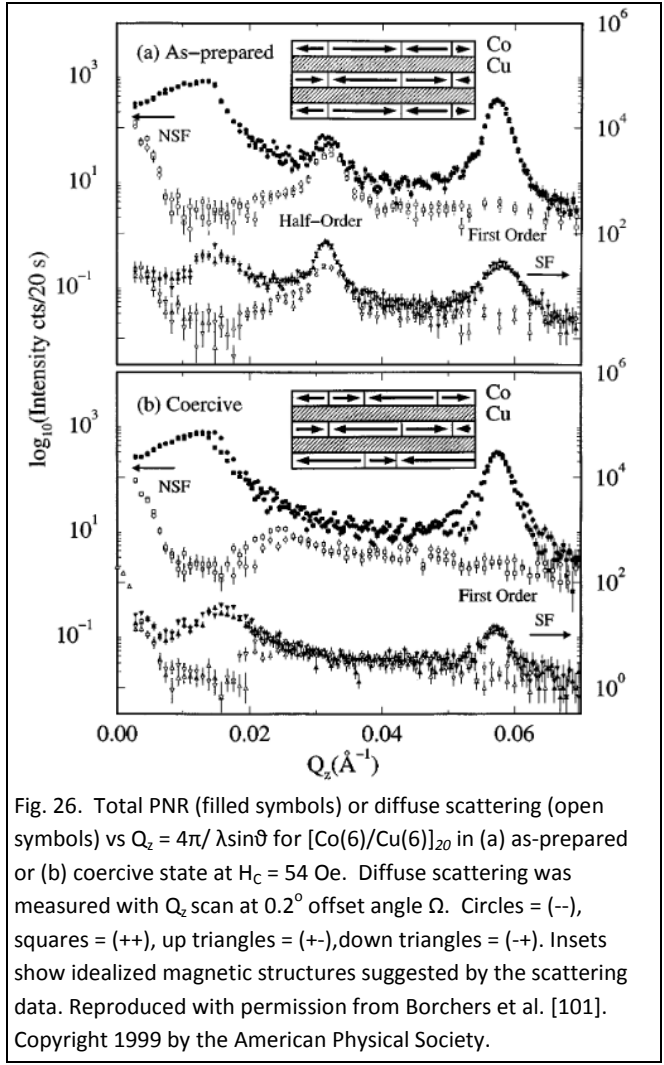

Fig. 26. Total PNR (filled symbols) or diffuse scattering (open symbols) vs $Q_z = 4\pi/\lambda\sin\vartheta$ for [Co(6)/Cu(6)]$_{20}$ in (a) as-prepared or (b) coercive state at $H_C = 54$ Oe. Diffuse scattering was measured with $Q_x$ scan at 0.2° offset angle Ω. Circles = (--), squares = (++), up triangles = (+-),down triangles = (-+). Insets show idealized magnetic structures suggested by the scattering data. Reproduced with permission from Borchers et al. [101]. Copyright 1999 by the American Physical Society.

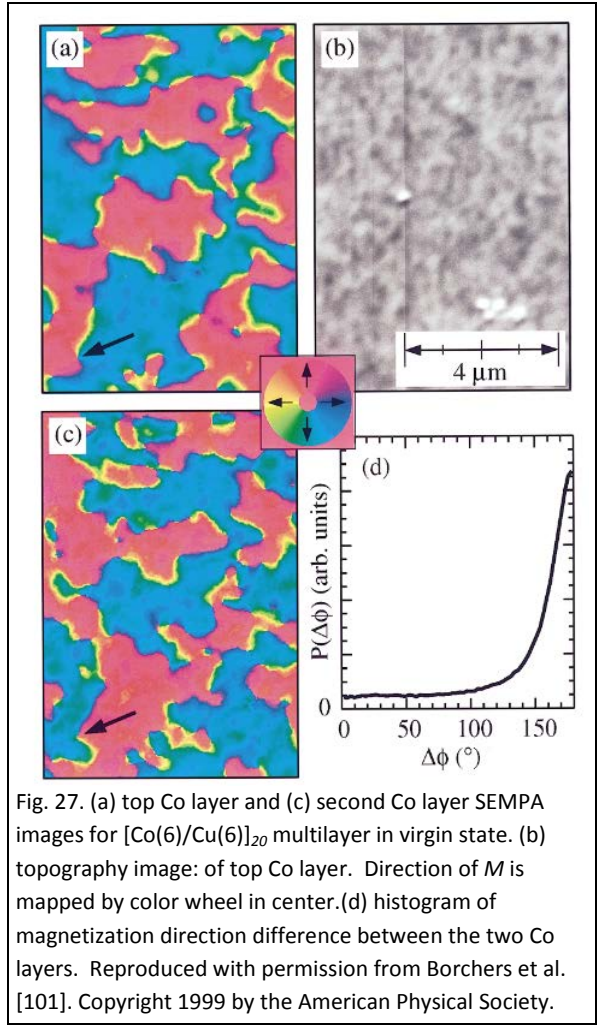

Fig. 27. (a) top Co layer and (c) second Co layer SEMPA images for [Co(6)/Cu(6)]$_{20}$ multilayer in virgin state. (b) topography image: of top Co layer. Direction of *M* is mapped by color wheel in center.(d) histogram of magnetization direction difference between the two Co layers. Reproduced with permission from Borchers et al. [101]. Copyright 1999 by the American Physical Society.

### 8.2.3.2. Test by predicting [Co/Cu/Py/Cu]$_n$ data with no adjustment.

In 1995, the second test was started by Yang et al. [93], who used a hybrid SV to test the parameters that had been derived for [Co/Cu]$_n$ and [Py/Cu]$_n$ assuming that AR(0) = AR(AP). Because $H_c$ is much larger for a 3 nm thick Co layer ($H_c$ ~ 200 Oe) than for an 8 nm thick Py layer ($H_c$ ~ 20 Oe), a [Co(3)/Cu(20)/Py(8)/Cu(20)]$_n$ hybrid SV should display a well-defined AP state as part of a complete hysteresis cycle as in Fig. 8. If AR(0) ≅ AR(AP) for Co/Cu and Py/Cu, then the parameters found for [Co/Cu]$_n$ and [Py/Cu]$_n$ multilayers should correctly predict values of AR(AP) for the [Co/Cu/Py/Cu]$_n$ hybrid SVs without adjustment. Fig. 24 [169] compares the measured values of AR(AP) and AR(P) vs $n$ for the [Co/Cu/Py/Cu]$_n$ multilayers with the non-adjustable predictions (solid lines). To show that $t_{cu}$ is not crucial, Fig. 24 also contains data for samples with $t_{Cu}$ = 40 nm (x) or $t_{Cu}$ = 10 nm (+). The agreements in Fig. 24 are okay for AR(AP) and AR(P), but the solid lines in Fig. 25 [57] show that they are less good for AΔR.

The parameters for Py used in Fig. 24 were derived assuming a 2CSR model for Py/Cu—i.e., assuming that $l_{sf}^{Py}$ = ∞. Measurements made two years after the data in Fig. 24 found an unexpectedly short $l_{sf}^{Py}$ = 5.5 ± 1 nm [73]. This new information led to recalculation of the Py/Cu parameters with $l_{sf}^{Py}$ = 5.5 nm. The dashed vs solid curves in Fig. 25 show that the more stringent test with AΔR favors the new parameters with $l_{sf}^{Py}$ = 5.5 nm. The ability to correctly predict the data of Fig. 24 with no adjustment gives additional support for the assumption AR(0) ≅ AR(AP) for sputtered Co/Cu and Py/Cu multilayers. However, we'll see in section 8.7.1 that the parameters for Py/Cu are less sure than those for Co/Cu.

**8.2.3.3**. **Test with PNS and SEMPA.**

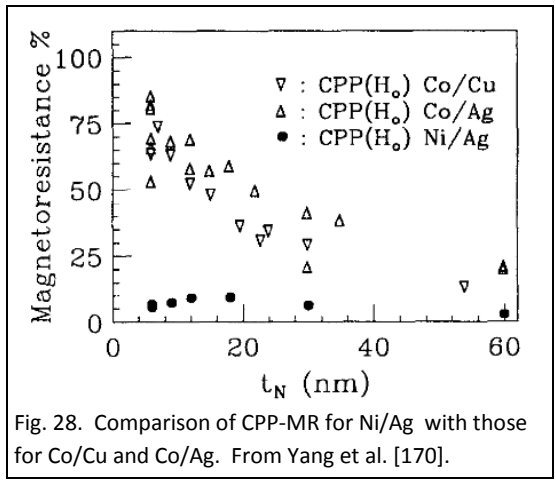

Fig. 28. Comparison of CPP-MR for Ni/Ag with those for Co/Cu and Co/Ag. From Yang et al. [170].

The third test, by Borchers et al. in 1999 [[101-103]] combined two different techniques to directly determine the magnetic ordering of a [Co(6)/Cu(6)]$_{60}$ multilayer in its as-prepared (virgin = AR(0)) and coercive = AR(Pk) states. Polarized Neutron Scattering (PNS) measurements in Fig. 26 [101] show a clear antiferromagnetic (half-order) peak in the as-prepared (virgin) state, indicating strong AP ordering, with magnetic moments in a given layer randomly distributed in the layer plane. The peak width indicates that the domain widths are ~ μm. This peak is gone in the coercive state, indicating a random distribution of moments. Fig. 27 [101] compares Scanning Electron Microscopy with Polarization Analysis (SEMPA) data for the top Co layer and the next Co layer below it (obtained by carefully eroding away the top Co layer and the intermediate Cu layer) in the virgin state. Each Co layer contains micron-sized magnetic domains, with 0.2 μm size Neel domain walls, and the alignments of both the domains and the walls closely reverse between the top and next Co layers. Quantitative analysis of the AF correlation gives roughly 60% for the small domains. Separate topographical images (not shown) give column sizes ~ 0.1 μm. In Transverse Q$_x$ scans (also not shown), coexistence of diffuse and specular peaks implies that the μm size domains are mixed with larger domains (in-plane dimensions ≥ 100 μm), also aligned AP in the virgin state. Together the PNS and SEMPA results show directly that AR(0) is a large fraction of AR(AP) for this sample.

**8.2.3.4. Test by deriving 2$AR_{Co/Cu}^*$. with no free parameters.**

The fourth test, in 1997, compared the experimental 2$AR_{Co/Cu}^*$ derived assuming AR(0) = AR(AP) with a value calculated using no adjustable parameters. We'll see in section 8.14 that the measured value lies between two calculated ones, one for a perfect, 'ballistic' interface and one for a disordered 'diffusive' interface of a 2ML thick 50%-50% random alloy.

To summarize, taken together, these four tests suggest that AR(0) is close to AR(AP) for most of the sputtered multilayers used to determine the Co/Cu and Co/Ag 2CSR parameters. However, it is difficult to specify just how close it is for a given set of multilayers.

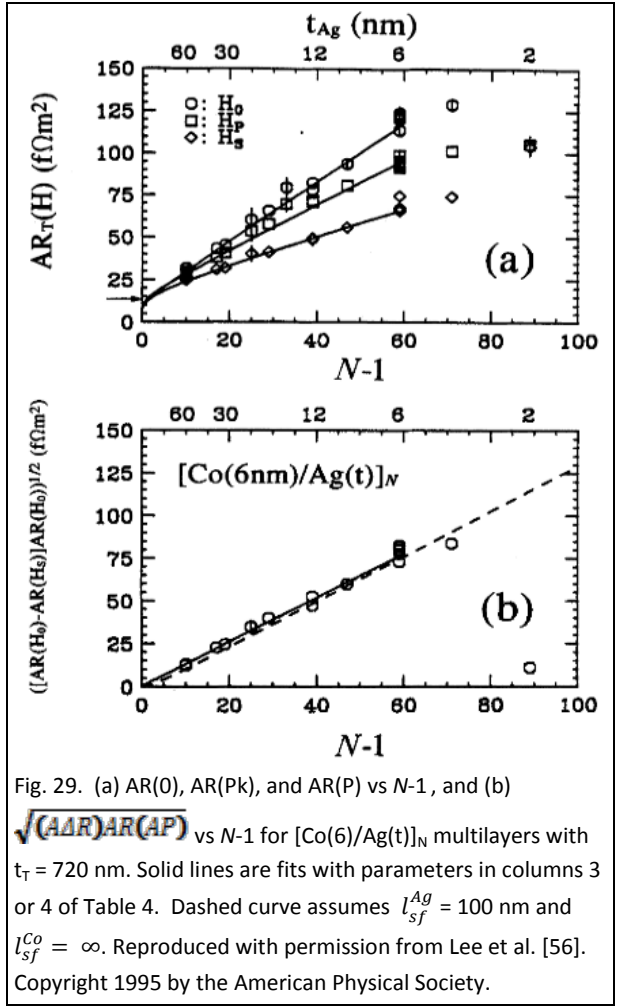

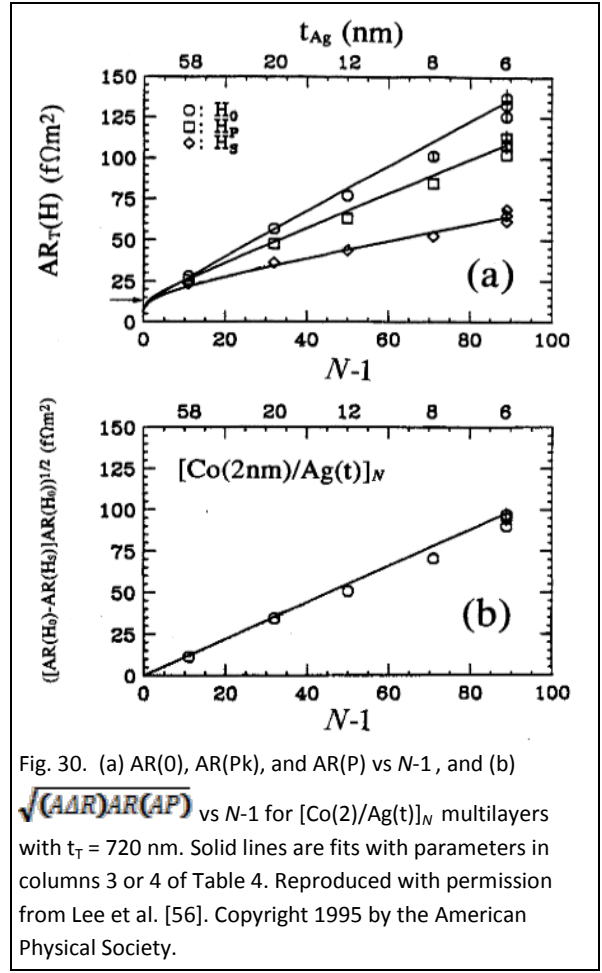

Fig. 29. (a) AR(0), AR(Pk), and AR(P) vs $N$-1, and (b) $\sqrt{(A\Delta R)AR(AP)}$ vs $N$-1 for [Co(6)/Ag(t)]$_N$ multilayers with $t_T$ = 720 nm. Solid lines are fits with parameters in columns 3 or 4 of Table 4. Dashed curve assumes $l_{sf}^{Ag}$ = 100 nm and $l_{sf}^{Co} = \infty$. Reproduced with permission from Lee et al. [56]. Copyright 1995 by the American Physical Society.

Fig. 30. (a) AR(0), AR(Pk), and AR(P) vs $N$-1, and (b) $\sqrt{(A\Delta R)AR(AP)}$ vs $N$-1 for [Co(2)/Ag(t)]$_N$ multilayers with $t_T$ = 720 nm. Solid lines are fits with parameters in columns 3 or 4 of Table 4. Reproduced with permission from Lee et al. [56]. Copyright 1995 by the American Physical Society.

### 8.3. CPP-MR of Ni/Ag and Ni/Cu.

In 1994, with CPP-MR data for Co/Ag, Fe/Cr, and Co/Cu published, interest moved to the other standard ferromagnet, Ni. In 1994, Yang et al. [170] found that the Ni hysteresis curves display AR(0) $\lesssim$ AR(Pk), meaning that the physics producing AR(0) $\approx$ AR(AP) did not apply to Ni/Ag. Fig. 28 shows that the CPP-MRs of Ni/Ag are much less than those of Co/Ag or Co/Cu, A later (2007) study of Ni/Cu by Moreau et al. [171] confirmed small Ni and Ni/Cu parameters: $\beta_{Ni}$ = 0.14 ± 0.02; $2AR^*_{Ni/Cu}$ = 0.36 ± 0.06 f$\Omega$m$^2$, and $\gamma_{Ni/Cu}$ = 0.29 ± 0.05 (consistent with a calculated value of $\gamma_{Ni/Cu}$ = 0.25 by Stiles and Penn [88]). Such weak CPP-MR led interest to move from 'pure' Ni to Ni-based alloys such as Py and NiCr, with results that we describe below.

### 8.4. 2CSR model parameters for Co/Ag and Co/Cu.

The evidence given above that the 2CSR model seems to work well for sputtered Co/Cu and Co/Ag multilayers led to the belief in a long $l_{sf}^{Co}$, and stimulated more complete studies to: (a) measure the parameters of this model for Co/Cu and Co/Ag at cryogenic temperatures to determine the relative importances of bulk and interface scattering, (b) test the limits of the 2CSR model, and (c) see how the parameters vary with temperature. For Co/Ag we analyze the most comprehensive study of simple multilayers, asking especially whether the 2CSR model can reproduce independently measured parameters such as 2AR$_{F/Nb}$, $\rho_N$, and $\rho_F$, and correctly (or not) predict the behavior of data beyond those used in initial fits. For Co/Cu, we first compare cryogenic temperature parameters determined by different groups using the four different CPP-MR techniques: crossed superconductors; micropillars; nanowires; and CAP, to see if the parameters found with different techniques agree. We don't expect close agreement, both because the 'pure metal' Co parameters should not be 'intrinsic' (see section 2.2), and because of limitations on some measuring techniques (see section 6). We then examine what has been learned about the temperature dependences of the 2CSR parameters of Co/Cu, the only F/N pair for which detailed studies have been made.

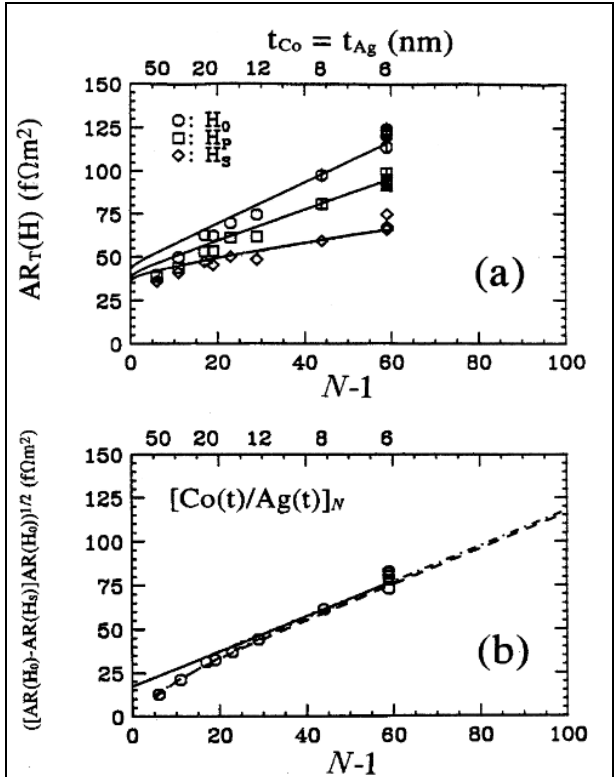

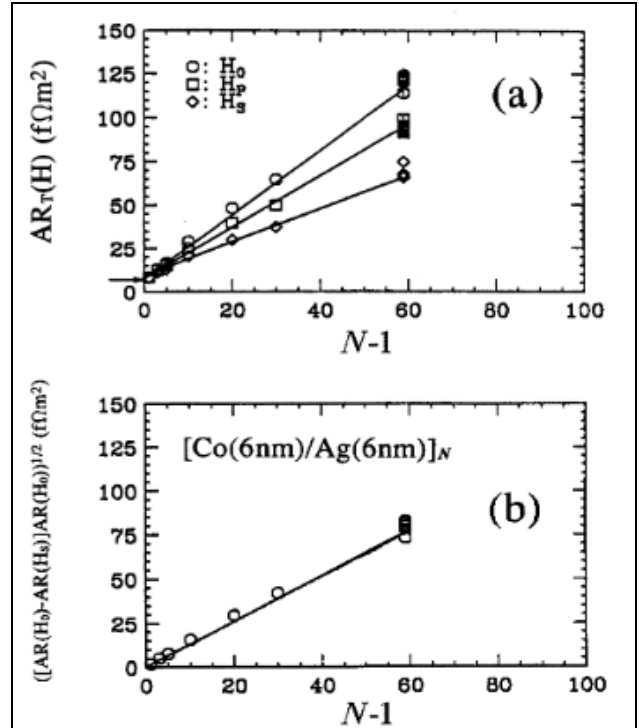

Fig. 31. (a) AR(0), AR(Pk), and AR(P) vs $N$-1, and (b) $\sqrt{(A\Delta R)AR(AP)}$ vs $N$-1 for [Co(t)/Ag(t)]$_N$ multilayers with $t_T$ = 720 nm. Solid lines are fits with parameters in columns 3 or 4 of Table 4. Dashed and broken lines are VF theory with $l_{sf}^{Ag}$ = 100 nm or $l_{sf}^{Co}$ = 6 nm and the other = ∞. Reproduced with permission from Lee et al. [56]. Copyright 1995 by the American Physical Society.

Fig. 32. (a) AR(0), AR(Pk), and AR(P) vs $N$-1, and (b) $\sqrt{(A\Delta R)AR(AP)}$ vs $N$-1 for [Co(6)/Ag(6)]$_N$ multilayers. Solid lines are fits with parameters in columns 3 or 4 of Table 4. Reproduced with permission from Lee et al. [56]. Copyright 1995 by the American Physical Society.

### 8.4.1. 2CSR model parameters for Co/Ag at 4.2K.

We use a 1995 comprehensive crossed-superconductor study of [Co/Ag]$_N$ multilayers by Lee et al. [56] to show the work needed to check for internal consistency (or inconsistency) in the CPP-MR. The multilayers were made with exactly $N$ bilayers, leaving a Ag layer on top to contact the upper superconducting Nb lead and protect the top Co layer in the CIP part of the sample from oxidation (The Nb top strip protects the CPP-MR part of the sample). As explained in section 8.2.2, this choice removes from the CPP-MR both the upper Ag layer and two Co/Ag interfaces. To partly correct for these removals, data with variable $N$ are plotted vs $N$–1. Earlier studies had used the three sets of data in Figs.

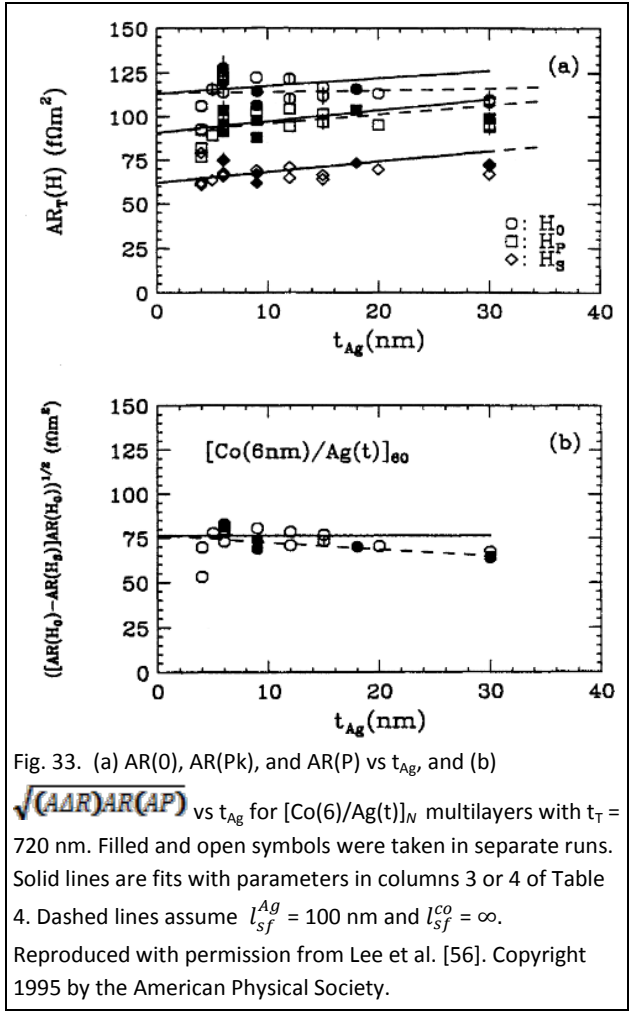

Fig. 33. (a) AR(0), AR(Pk), and AR(P) vs $t_{Ag}$, and (b) $\sqrt{(A\Delta R)AR(AP)}$ vs $t_{Ag}$ for [Co(6)/Ag(t)]$_N$ multilayers with $t_T$ = 720 nm. Filled and open symbols were taken in separate runs. Solid lines are fits with parameters in columns 3 or 4 of Table 4. Dashed lines assume $l_{sf}^{Ag}$ = 100 nm and $l_{sf}^{co}$ = ∞. Reproduced with permission from Lee et al. [56]. Copyright 1995 by the American Physical Society.

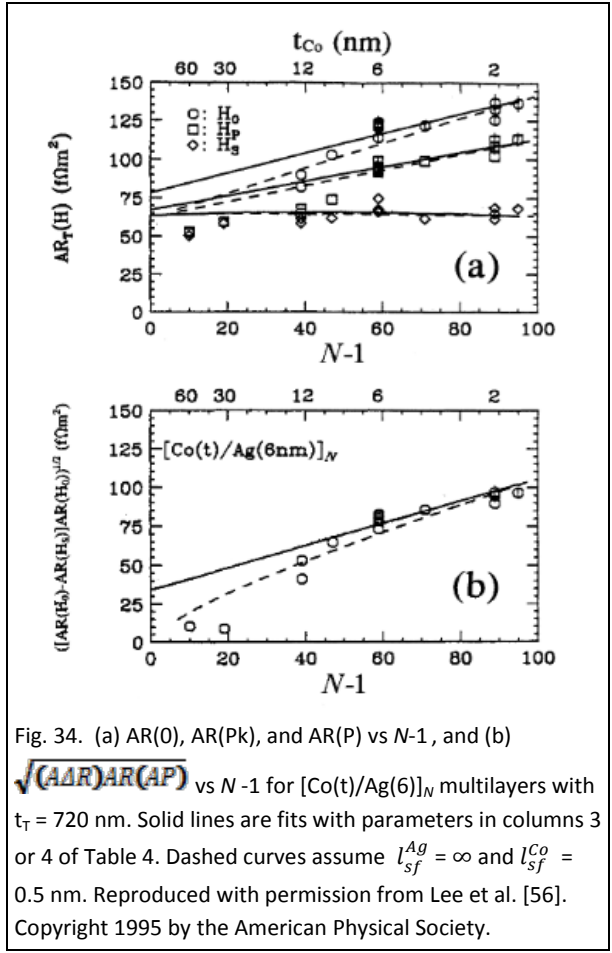

Fig. 34. (a) AR(0), AR(Pk), and AR(P) vs $N$-1, and (b) $\sqrt{(A\Delta R)AR(AP)}$ vs $N$-1 for [Co(t)/Ag(6)]$_N$ multilayers with $t_T$ = 720 nm. Solid lines are fits with parameters in columns 3 or 4 of Table 4. Dashed curves assume $l_{sf}^{Ag}$ = ∞ and $l_{sf}^{Co}$ = 0.5 nm. Reproduced with permission from Lee et al. [56]. Copyright 1995 by the American Physical Society.

29-31 to produce the parameters in Column 3 of Table 4, assuming that AR(0) = AR(AP) and AR(H$_s$) = AR(P). This more complete study added an additional set of data (Fig. 32) to produce the parameters in column 4 of Table 4, and then asked whether these parameters could correctly describe three additional data sets. In each figure in this section we include plots of both A∆R and the square root of Eq. 7 versus the variable of interest.

We start with the earliest three sets of data. Fig. 29 shows the fit to samples of the form [Co(6)/Ag(t)]$_N$ with fixed $t_T$ = 720 nm and $N$ up to 60. Fig. 30 shows the fit to samples of the form [Co(2)/Ag(t)]$_N$ with fixed $t_T$ = 720 nm and $N$ up to 90. In both Figs 29a and 30a, the data for AR(H$_o$), AR(H$_{Pk}$), and AR(H$_s$) are consistent with straight lines going to the predicted ordinate intercepts, except for the downturns as $N \rightarrow 0$, due to proximity effect elimination of one Ag-layer and two Co/Ag interfaces. Also, as predicted, the square-root plots in Figs. 29 and 30 are consistent with straight lines through the origin, assuming AR(0) = AR(AP). Fig. 31 shows a fit to data of the form [Co(t)/Ag(t)]$_N$ and $t_T$ = 720 nm, but with the common t $\lesssim$ 18 nm. This limitation was adopted for five reasons. (1) Bulk Co has hcp structure. While thin Co grown with Ag will adopt to Ag's fcc structure, thicker Co will want to convert to hcp, which might have somewhat different properties [172] . (2) The magnetization of thick Co layers was also found to differ from that of thin Co. (3) Thicker Co layers might couple magnetically across Ag thicknesses at which thin Co layers do not couple. (4) Too thick Co layers might reach the then unknown $l_{sf}^{Co}$. (5) Initial studies showed that data for thick $t_{Co} = t_{Ag}$ fell below expectation, both in plots of $A\Delta R$ and plots of the square root (see Fig. 31). Importantly, beyond about $t_{Co} = t_{Ag} = 20$ nm, even the data for AR(H$_s$) fall below the predicted line. There is, thus, something different about those data. The parameters from a fit to Figs. 29-31 are given in column 3 of Table 4. Note that $\rho_{Co}^*$ and $\beta_{Co}$ are fitting parameters, whereas $\rho_{Co} = \rho_{Co}^*(1-\beta_{Co}^2)$ is derived from their values. The overlap, to within mutual uncertainties, of the 2CSR model derived values of $\rho_{Ag}$ and $\rho_{Co}$ with the values measured on separate thin films supports both the fit and the use of the separately measured values as a check on the analysis.

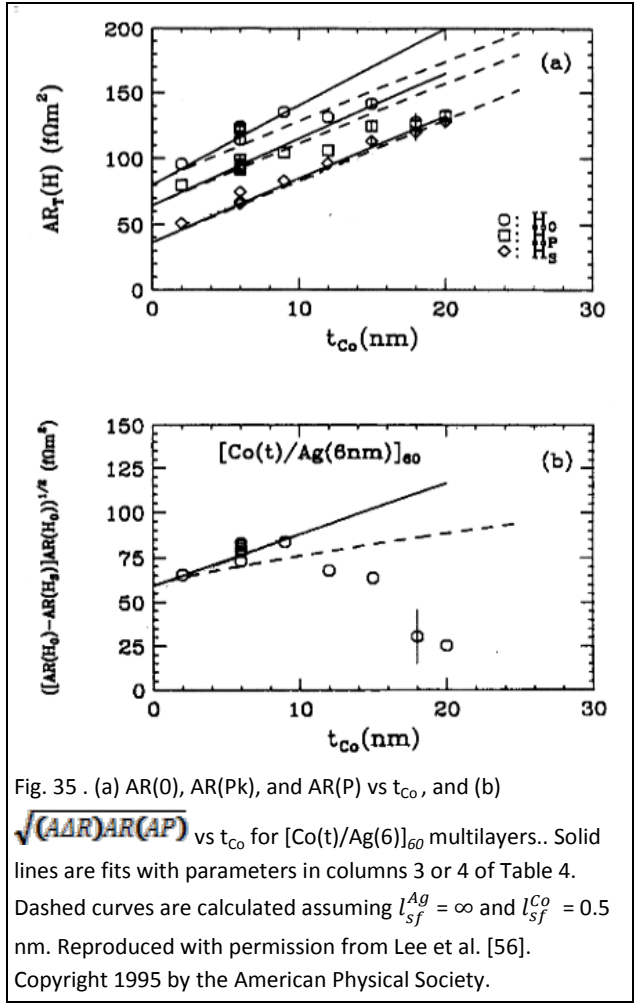

Fig. 35 . (a) AR(0), AR(Pk), and AR(P) vs $t_{Co}$, and (b) $\sqrt{(A\Delta R)AR(AP)}$ vs $t_{Co}$ for [Co(t)/Ag(6)]$_{60}$ multilayers.. Solid lines are fits with parameters in columns 3 or 4 of Table 4. Dashed curves are calculated assuming $l_{sf}^{Ag} = \infty$ and $l_{sf}^{Co} = 0.5$ nm. Reproduced with permission from Lee et al. [56]. Copyright 1995 by the American Physical Society.

This new study [56] added a fourth data set, Fig. 32 for fixed $t_{Co} = t_{Ag} = 6$ nm and variable $N$. This addition let the parameter 2AR$_{Co/Nb}$ also be treated as an unknown. All of the data in Fig. 32 are consistent with the new fit to the four sets, which gave the values in column 4. Note that these values overlap with both the old three-set values and the independent measurements to within mutual uncertainties. The fits with the two slightly different sets of parameters are close enough that we do not distinguish them in the graphs.

Column 5 of Table 4 shows how the alternative choice of AR(Pk) as an estimate of AR(AP) would change the parameters. The values for the three independently measured quantities still overlap to within mutual uncertainties. But, not surprisingly, $\rho_{Co}^*$, $\beta_{Co}$, and 2AR$_{Co}^*$ are now smaller. Interestingly $\gamma_{Co/Ag}$ is essentially unchanged.

With the parameters in column 4 of Table 4 fixed from these 4 sets of data, Lee et al. asked whether these parameters could describe, without adjustment, additional Co/Ag data, some of which fell outside the previously fixed limits ($t_{Ag} \geq 6$ nm, $t_{Co} \leq 18$ nm). They prepared and measured three different sets, shown in Figs. 33-36.

Fig. 33 shows data for multilayers of the form [Co(6)Ag(t$_{Ag}$)]$_{60}$ with fixed $t_{Co}$ and fixed $N = 60$, but variable $t_{Ag}$. The solid lines are no-free-parameter predictions with the parameters in column 4 of Table 4. The open symbols were taken at the start of the study. The data for AR(0) and AR(P) are consistent with predictions up to $t_{Ag} \approx 20$ nm, but fall below the predictions at $t_{ag} = 30$, even for AR(P). To check this discrepancy, new data (filled symbols) were taken. They confirmed the previous results. The data for AR(0) start to fall systematically below the prediction from $t_{Ag} = 15$ nm. The dashed lines show that the data can be fit by VF theory by including a finite spin-diffusion length in Ag, $l_{sf}^{Ag}$ = 100 nm. Ref. [114] suggests a value 4 ± 2 times longer for Ag with $\rho_{Ag} \sim$ 10 nΩm.

Figs. 34 and 35 show data for multilayers with fixed $t_{Ag}$ = 6 nm and variable $t_{Co}$. Fig. 34 shows [Co(t)/Ag(6)]$_N$ with fixed $t_T$ = 720 nm, and Fig. 35 shows [Co(t)/Ag(6)]$_{60}$. In both cases, all of the data fit the predictions for $t_{Co} \lesssim$ 10 nm. But for $t_{Co}$ > 10 nm, the situation becomes more complex, with all of both the AR(0) and AR(Pk) data falling further below the fits as $t_{Co}$ increases. In Fig. 35, the data for AR(P) follow the fit to the thickest $t_{Co}$ = 20 nm measured. In Fig. 34, in contrast, even the AR(P) data fall below the predictions for $t_{Co} \geq$ 30 nm. This discrepancy indicates that something is changing for $t_{Co} \geq$ 30 nm. And for $t_{Co}$ > 10 nm, the deviations from predictions show that AR(0) ≠ AR(AP). The dashed lines show that VF theory with $l_{sf}^{Co}$ = 0.5 nm can describe the data in Fig. 34, but not in Fig. 35. We'll see in section 8.6 that such a value for $l_{sf}^{Co}$ is much too short.

To summarize, taking AR(0) = AR(AP), a single set of 2CSR model parameters can fit all of the data of Figs. 29-32, 34, and 35 for which $t_{Co} \leq$ 10 nm, and also the data of Fig. 33 for which $t_{Ag} \leq$ 14 nm. The rest of the data of Fig. 33 can be fit with VF theory and $l_{sf}^{Ag}$ = 100 nm, which is, however, probably too short. The AR(0) and AR(Pk) data for $t_{Co}$ > 10 nm in Figs. 31, 34, and 35 seem to require a change in behavior of thick Co. Dassonville et al. [172] later found an apparent change in behavior, in the failure of AΔR to grow as expected with increasing $t_{Co}$ > 20 nm in Py-based Double EBSVs with thick Co central inserts. They attributed this failure to thick Co taking its equilibrium hcp structure, giving perpendicular magnetic anisotropy that caused the Co M to tilt out of the plane (locally, both up and down) and mix the two spin-currents. Such mixing might explain the behaviors for thick $t_{Co}$ in Figs. 31, 34, and 35. In addition, waviness at the interfaces of thick Co layers with Ag might give orange peel magnetic coupling [173], which could reduce AΔR. It is not clear why the AR(H$_s$) data deviate from predicted AR(P) for $t_{Ag}$ > 14 nm in Fig. 32 (but see section 8.9.3.2). A measured spin-flip probability at the Co/Ag interface (see section 8.15), neglected in all of the above analysis, does not resolve these issues.

**Table 4. 2CSR Model parameters for Co/Ag at 4.2K.** Parameters are from [56]. The quantities in column 2 were independently measured. Those in column 3 were derived previously from the 3 data sets in Figs.29-31. Those in column 4 were derived by adding the set of Fig. 32. Those in column 5 were derived assuming AR)0) = AR(Pk). $\rho_{Co}^*$ is a VF parameter; $\rho_{Co} = \rho_{Co}^* (1-\beta_{Co}^2)$ is a derived quantity.

| Parameters | Indep. Msmts | H$_o$ (3 sets) | H$_o$ (4 sets) | H$_{pk}$ (4 sets) |
|---|---|---|---|---|
| 2AR$_{Nb/Co}$(fΩm$^2$) | 6 ± 1 | | 6.9 ± 0.6 | 7.3 ± 0.6 |
| $\rho_{Ag}$ (nΩm) | 10 ± 1 | 10 ± 3 | 7.3 ± 1.9 | 10.9 ± 1.9 |
| $\rho_{Co}$(nΩm) | 68 ± 10 | 82 ± 13 | 77 ± 12 | 77 ± 3 |
| $\rho_{Co}^*$ (nΩm) | | 107 ± 10 | 100 ± 6 | 84 ± 6 |
| $\beta_{Co}$ | | 0.48 ± 0.05 | 0.48 ± 0.06 | 0.29 ± 0.06 |
| $2AR_{Co/Ag}^*$ (fΩm$^2$) | | 1.12 ± 0.06 | 1.20 ± 0.04 | 0.90 ± 0.04 |
| $\gamma_{Co/Ag}$ | | 0.85 ± 0.03 | 0.84 ±0.04 | 0.82 ± 0.05 |

### 8.4.2. 2CSR Model Parameters for Co/Cu at Cryogenic Temperatures.

Unlike for Co/Ag, 2CSR model parameters for Co/Cu have been estimated by several groups using various experimental techniques and different choices for AR(AP). In this section we ask how well they agree. The comparison is given in Table 5 [6, 58, 98, 135, 147, 151]. To put the results in context, we briefly describe the conditions and assumptions associated with each set of data taken by a given group.

The MSU group's parameters are from measurements like those in section 8.4.1 for Co/Ag, with AR(0) = AR(AP). Their published parameters have varied slightly over time. But except for Cu, where $\rho_{Cu}$ = 6 ± 1 nΩm is small enough to be almost negligible, their varying values overlap to within their specified uncertainties [39, 44, 174]. We list the values from a 1999 review [6].

The Leeds group's parameters were derived from measurements similar to those for Co/Ag assuming AR(0) = AR(AP), except that they used MBE deposited samples. We list unpublished values from [58] that correct those for List et al. in [6]. Fig. 10 shows that the normalized values of MR(0), but not MR(Pk), are similar for nominally identical MBE and sputtered samples .

The Lausanne (Laus) group [135] used nanowires on two separate sets of samples with slightly different amounts of Co in the Cu layers. They took AR(AP) = AR(Pk) from hysteresis data with no measurements of AR(0). This choice should give some parameters smaller than those in columns 2 and 3 of Table 5. In this section, we limit ourselves to the Lausanne data at 20K. Fig. 36 shows a plot of CPP-MR vs t for the two sets of equal values of t = $t_{Cu}$ = $t_{Co}$. along with the two fits. The authors derived three parameters from this graph—$\beta_{Co}$, $\gamma_{Co/Cu}$, and $2AR^*_{Co/Cu}$. Values of $\rho_{Cu}$, $\rho_{Co}$, $l^{Cu}_{sf}$, and $l^{Co}_{sf}$ were estimated separately, the ρs from separate measurements and Matthiessen's rule. $l^{Cu}_{sf}$ and $l^{Co}_{sf}$ will be considered later. Three parameters are a lot to derive from one graph, especially with uncertainties in the other four. The authors had difficulty constraining $2AR^*_{Co/Cu}$.

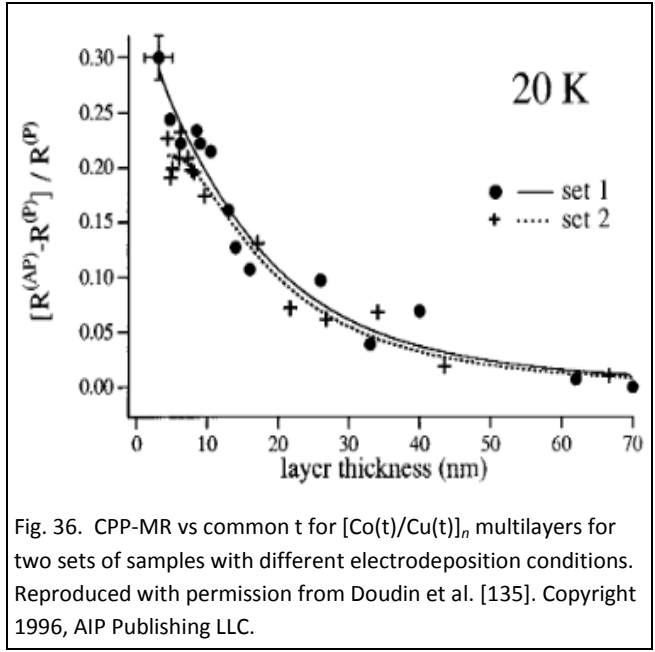

Fig. 36. CPP-MR vs common t for [Co(t)/Cu(t)]$_n$ multilayers for two sets of samples with different electrodeposition conditions. Reproduced with permission from Doudin et al. [135]. Copyright 1996, AIP Publishing LLC.

The Louvain/Orsay (L/O) group's parameters were derived using ≅ 90 nm diameter nanowires [98], using AR(Pk) to fit 2CSR or VF equations. The hysteresis example they gave had AR(0) only about 15% larger. As for Lausanne, the choice of AR(Pk) should make some parameters smaller. In this section, we consider only their data at 77K, which they said were within ~ 10% of data at 4.2K. To estimate $\rho_{Cu}$, they measured the ratio of the 77K and 300K resistances of multilayers with $t_{Cu}$ >> $t_{Co}$. Taking the resulting ratio of 1.47 to be due almost completely to Cu, and assuming Matthiessen's Rule to get ρ(300K) − ρ(77K) = 14.5 nΩm, they estimated $\rho_{cu}$ (300K) = 45.5 nΩm and $\rho_{Cu}$(77K) = 31 nΩm. They then derived $\beta_{Co}$ at 77K using the data of Fig. 37 [153]and the 2CSR model Eq. 6, rewritten as

$$[\Delta R/R(AP)]^{-1/2} = [\rho^*_F t_F + 2AR^*_{F/N}]/[\beta_F\, \rho^*_F t_F + 2AR^*_{F/N}] + \rho_N t_N/[\beta_F\, \rho^*_F t_F + 2AR^*_{F/N}] \qquad (6b)$$

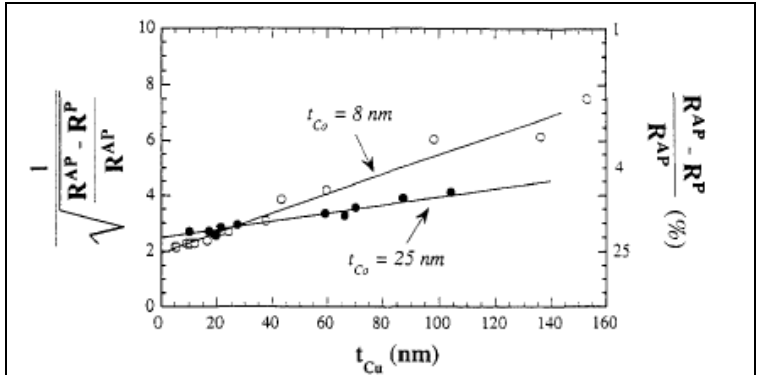

Fig. 37. $1/\sqrt{\frac{(R(AP)-R(P))}{R(AP)}}$ vs $t_{Cu}$ for nanowire Co/Cu multilayers with fixed $t_{Co}$ = 8 nm or 25 nm. From Eq. 6b, the ordinate at the crossing point should equal $(1/\beta_{Co})$. This figure was published in Piraux et al. [153]. © 1996 by Elsevier (http://www.sciencedirect.com/science/journal/03048853).

Realizing that, formally, this equation requires true AP states, the authors argued that it could also apply to random states with $M = 0$. This is true if: (a) the moments are collinear; (b) $t_{Cu}$ < 160 nm is less than their $l_{sf}^{Cu}$ ~ 140 nm found from a fit to the extended data in Fig. 38 (which we consider in section 8.5.1); (c) $t_{Co}$ = 25 nm is less than their estimated $l_{sf}^{Co}(77K)$ ~ 44 nm; and (d) spin-flipping at the Co/Cu interface is negligible. In Fig. 37, conditions (b) and (c) are approximately satisfied. But the moments need not all be collinear, and we'll see in section 8.9 that effects of spin-flipping at the Co/Cu interfaces might not be negligible for a random distribution of non-collinear Co moments. Eq. 6b predicts that a plot of $[\Delta R/R(AP)]^{-1/2}$ vs $t_{Cu}$ should give a straight line, and that the lines for different values of $t_F$ should cross at $[\Delta R/R(AP)]^{-1/2}$ = $1/\beta_F$. The crossing point in Fig. 37 gives $\beta_{Co}$ = 0.36 ± 0.02. The authors derived the other three parameters in Table 5, $\rho_{Co}^*$, $\gamma_{Co/Cu}$, and $2AR_{Co/Cu}^*$, by fitting the two straight lines in Fig. 37.

The Eindhoven (Eind) group [151] used a CAP technique that approximates CPP, as explained in section 6.4. They chose AR(AP) = AR(0), which was almost 50% larger than AR(Pk) in the hysteresis curve that they showed. The values listed are those for H(0) in Table I in [151].

`The Kyoto group [147] extrapolated CAP data to CPP. To get AR(AP), they used Co/Cu/Py/Cu hybrid spin-valves for Co/Cu. We list values from Table I in [147], with their value of $2AR_{Co/Cu}^*$.estimated from their values for $2AR_{Co/Cu}$ and $\gamma_{Co/Cu}$ Due to the large uncertainty in $\beta_{Co}$, we couldn't estimate $\rho_{Co}^*$.

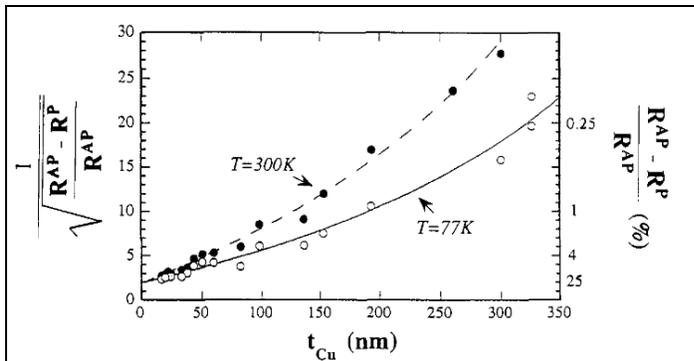

Fig. 38. $1/\sqrt{\frac{(R(AP)-R(P))}{R(AP)}}$ vs $t_{Cu}$ for both T = 77K and T = 300K, and extended to larger $t_{Cu}$ than in Fig. 37. This figure was published in Piraux et al. [153]. © 1996 by Elsevier (http://www.sciencedirect.com/science/journal/03048853).

We conclude by examining the parameters in Table 5. Note first that the resistivities $\rho_{Co}$ and $\rho_{Cu}$ are larger for the electrodeposited samples, by factors of 2 to 6. Thus 'defects and dirt' look to be larger in electrodeposited samples. Given the differences in resistivities, preparation techniques (sputtering, MBE, electrodeposition), and use of AR(0) vs AR(Pk), the various parameters are mostly surprisingly similar. While one cannot specify 'reliable values' for Co and Cu layers with unknown impurities, the values in the first two columns, taken with superconducting cross-strips, seem closest to reliable, since they were taken assuming AR(0) = AR(AP), and agree to within mutual uncertainties on $\beta_{Co}$ and $\gamma_{Co/Cu}$.

**Table 5:  2CSR model parameters for Co/Cu at cryogenic temperatures.**

Each column gives the group that derived the parameters (MSU, Leeds, Lausanne (Laus.), Louvain/Orsay (LO), Eindhoven (Eind), Kyoto), the reference(s), the method used (Superconducting Strips, Nanowires, CAP), the deposition method (Sputtering, MBE, Evaporation, or Electrodeposition), the measuring T, and whether the analysis used AR(0), AR(Pk), or a SV to estimate AR(AP). For Laus. we list values from two separate fits in Fig. 36. Values for Leeds, and the reference, correct erroneous ones in [6].  $2AR_{Co/Cu}^*$ for [58] is ambiguous.

| | MSU [6] Sup. Str. Sputt. 4.2K;H$_o$ | Leeds [58] Sup. Str. MBE. 4.2K, H$_o$ | Laus.[135] Nanow. Electro. 20K; H$_{Pk}$ | L/O [98] Nanow. Electro. 77K; H$_{Pk}$ | Eind. [151] CAP MBE 4.2K (300K);H$_o$ | Kyoto [147] CAP Evap 5K; SV |
|---|---|---|---|---|---|---|
| $\rho_{Cu}$ (nΩm) | 6 ± 1 | 6 ± 0.3 | 13-33 (18-38) | 31 | 3.6±0.6 | 5.6±0.21 |
| $\rho_{Co}$ (nΩm) | 59±10 | 25±1 | ~ 425 | 157±17 | 53±6 | 76±5 |
| $\rho_{Co}^*$(nΩm) | 75 ± 5 | 31±6 | 510-570 (510-570) | 180 ± 20 | 57±7 | |
| $\beta_{Co}$ | 0.46±0.05 | 0.5±0.04 | 0.46±0.05 (0.52±0.06) | 0.36±0.02 | 0.27±0.05 | 0.18-0.45 |
| $\gamma_{Co/Cu}$ | 0.77± 0.05 | 0.71±0.02 | 0.55±0.07 (0.33±0.11) | 0.85±0.15 | 0.52±0.1 | 0.64-0.77 |
| $2AR_{Co/Cu}^*$ (fΩm$^2$) | 1.02± 0.04 | | 0.6-2.2 (0.6-2.2) | 0.6±0.3 | 0.54±10 | 0.7±0.35(est) |

### 8.4.3. Temperature Dependence of 2CSR Model Parameters for Co/Cu.

Although, as shown earlier in this review (see, e.g., Fig. 18), a fair number of studies of the temperature (T) dependence of the CPP-MR have been published, few have examined how the VF parameters change with T. $l_{sf}$ is expected to decrease with increasing T due to increased scattering by phonons and magnons, and $\rho_F$ and $\rho_N$ are expected to increase for the same reason. However, the behaviours of $\beta_F$, $2AR^*_{F/N}$, $\gamma_{F/N}$, and $\delta_{F/N}$ must be determined experimentally. A complication in determining how these parameters vary with increasing T is the growth of a 'spin-mixing' term expected as electron-phonon scattering, and perhaps also electron-magnon scattering [175], grow. VF theory adds a bulk spin-mixing resistivity $\rho\uparrow\downarrow$ [175], and also an interface spin-mixing resistance to allow for scattering by interfacial spin-fluctuations [98] The two papers that we discuss next neglected $\rho\uparrow\downarrow$ in their derivations. For some estimates of $\rho\uparrow\downarrow$ and interfacial spin-mixing see Ono et al. [147].

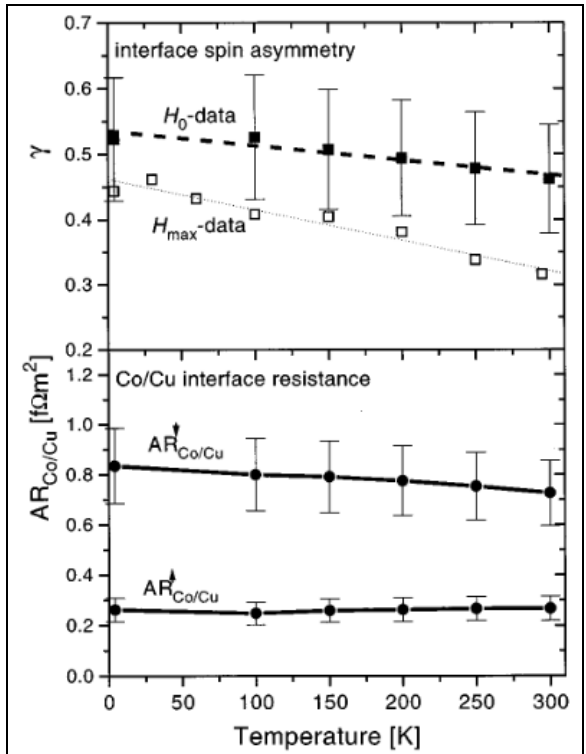

Fig. 39.  $\gamma_{Co/Cu}$ (top), and $AR_{Co/Cu}^{\uparrow}$, $AR_{Co/Cu}^{\downarrow}$, (bottom) vs T(K).  Reproduced with permission from Oepts et al. [151]. Copyright 1996 by the American Physical Society.

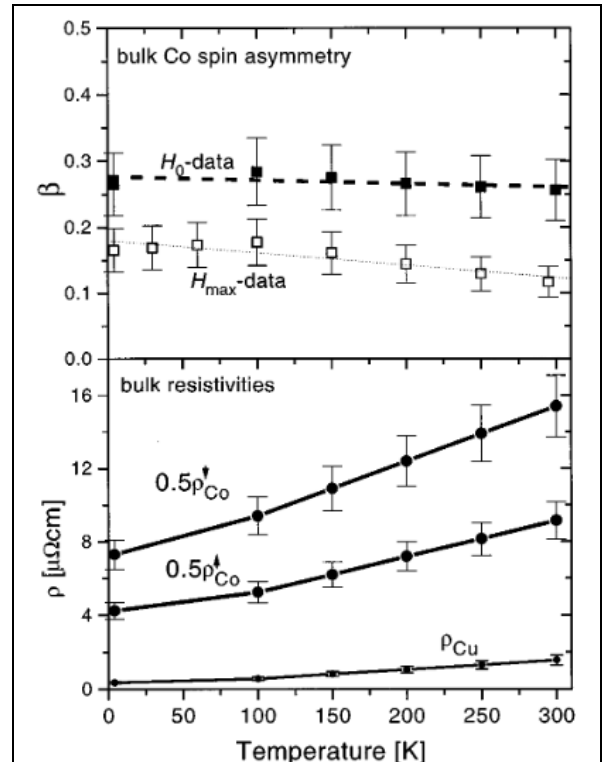

Fig. 40.  $\beta_{Co}$ (top) and $\rho_{Cu}$, $0.5\rho_{Co}^{\downarrow}$ and $0.5\rho_{Co}^{\uparrow}$ (bottom) vs T(K).  Reproduced with permission from Oepts et al. [151]. Copyright 1996 by the American Physical Society.

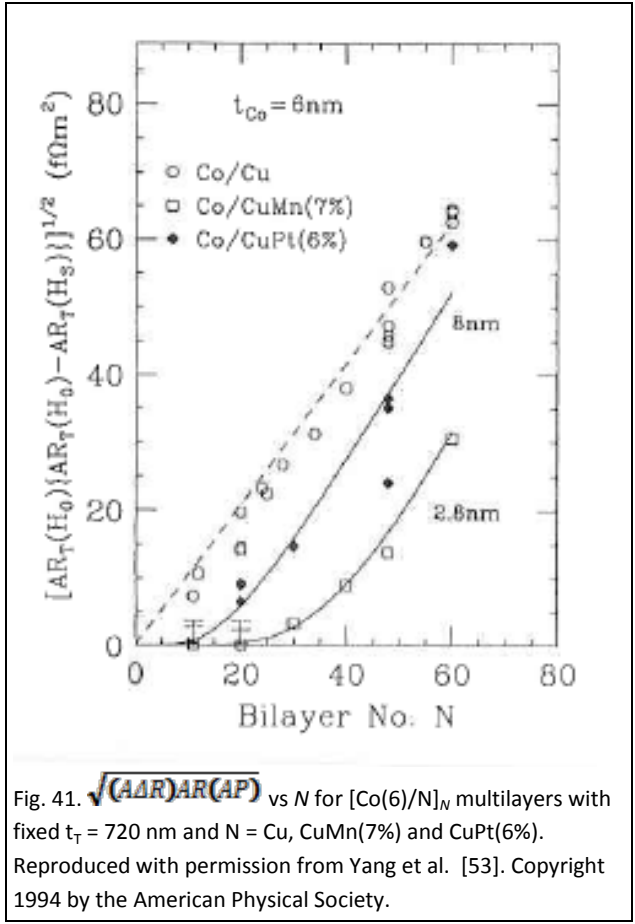

Fig. 41. $\sqrt{(A\Delta R)AR(AP)}$ vs $N$ for [Co(6)/N]$_N$ multilayers with fixed $t_T$ = 720 nm and N = Cu, CuMn(7%) and CuPt(6%). Reproduced with permission from Yang et al. [53]. Copyright 1994 by the American Physical Society.

The most complete study of the changes in VF parameters with T was made in 1996 by Oepts et. [151] using CAP $\approx$ CPP data. Their derived parameters are shown in Figs. 39 and 40. These figures give $\beta_{Co}$ and $2AR^*_{Co/Cu}$ independent of temperature T to within experimental uncertainties. $\gamma_{Co/Cu}$ may decrease slightly with increasing T, but any decrease is on the border of uncertainty. $\rho_{Cu}$ and $\rho^*_{Co}$ grow with increasing T about as expected for additive phonon scattering.

In 1998, Piraux et al. [98] combined the data of Fig. 37 at 77K with equivalent data at 300K, applied the 2CSR model Eq. 6b above, and inferred only a small reduction in $\beta_{Co}$ from 0.36 ± 0.02 at 77K to $\beta_{eff}$ = 0.31 ± 0.02 at 300K, consistent with the little or no reduction inferred by Oepts et al.

The main conclusion from these studies is that any T dependences of $\beta_{Co}$, $\gamma_{Co/Cu}$, and $2AR^*_{Co/Cu}$ are probably weak, and that $\rho_{Cu}$ and $\rho^*_{Co}$ grow with increasing T about as expected. Estimates for some Heusler alloys at T = 14K and 290K will be noted in section 10.5.

### 8.5. $l_{sf}^N$, spin-diffusion lengths in N-metals.

Given the apparent successes of the 2CSR model studies described in sections 8.1-8.4, in 1995-96 interest began to shift to measuring spin-diffusion lengths, starting with $l_{sf}^N$. As explained in section 2.2, $l_{sf}^N$ for any given N-metal is not intrinsic, but is determined by the unknown defects in the N-metal. The best that one can hope is that $l_{sf}^N$ for a given N-metal will be approximately proportional to the mean-free-path $\lambda_N$, which is inversely proportional to the residual resistivity, $\lambda_N \propto 1/\rho_{oN}$. It is, thus, essential to associate any measured value of $l_{sf}^N$ with a corresponding residual resistivity, $\rho_{oN}$.

Given the complications due to the expected variation of $l_{sf}^N$ with both the unknown type and unknown number of scatterers in nominally pure metals, we begin with values of $l_{sf}^N$ for dilute Cu and Ag-based alloys, where $l_{sf}^N$ should be unique for a known concentration of a chosen impurity. We then describe early estimates of $l_{sf}^N$ for Cu, which is moderately long. Lastly, we describe a general CPP-MR technique for measuring values of $l_{sf}^N$ that are not too long, and provide a table of values for several different metals along with the associated resistivities. Most of these values were collected in a review [54] along with values obtained by techniques other than CPP-MR. For a complete picture of the broader data, we refer the interested reader to that review.

### 8.5.1. $l_{sf}^N$ in Cu- and Ag-based alloys.

The first use of VF equations to measure $l_{sf}$ was by Yang et al. in 1994 [53] (see also [176]), who determined $l_{sf}^N$ for dilute alloys of both Ag and Cu with either Pt or Mn. Figs. 21 and 22 show that plots of $\sqrt{(A\Delta R)AR(AP)}$ vs $n$ for [Co(6)/N($t_N$)]$_n$ multilayers with fixed $t_T$ = 720 nm and N = Ag or AgSn (Fig. 21), or N = Cu or CuGe (Fig. 22), fall closely along the same straight lines passing through the origin. As explained in section 8.2.2, these overlaps are predicted by the 2CSR model for dilute alloys that have long $l_{sf}^N$. In contrast, similar data for dilute alloys with either Pt or Mn deviate from the straight line, by larger amounts the smaller $n$ (i.e. the longer $t_N$). The curves in Figs. 21 and 41 for these alloys are VF fits for the values of $l_{sf}^N$ listed in Table 6. For Pt, which is heavy enough to expect substantial spin-orbit scattering, the values of $l_{sf}$ are compared with non-adjustable calculations that used independent measurements of Conduction Electron Spin-Resonance (CESR) cross-sections by Monod and Schultz [46], as explained in section 3.3. The agreements are surprisingly good. The values for AgMn and CuMn are compared with separate calculations by Fert et al. [175] for spin-spin scattering, again with good agreement. Table 6 shows similarly good agreement for extentions to CuGe [169] , and CuNi [178], and with independent support by a different SV-based technique (see section 8.5.2.2 and [67, 177, 179]).

**Table 6.** $l_{sf}^N$ **for Ag- and Cu-based alloys found with Multilayers (ML) or Spin-Valves (SV)**. For simplicity, numbers are rounded. For precise values and uncertainties, see the original papers. The third column lists the experimental values of $l_{sf}^N$. The fourth column lists CESR or spin-spin (SS) calculations. The fifth column lists residual resistivites. The sixth column lists references.

| Alloy | Method | $l_{sf}^N$ (nm) (exp) | $l_{sf}^N$ (nm) CESR or Spin-Spin(SS) | $\rho_o$(n$\Omega$m) | Ref. |
|---|---|---|---|---|---|
| Ag(4%Sn) | ML | $\geq 26$ | | 200 | [53] |
| Ag(3.6%Sn) | SV | 39 | | 150 | [177] |
| Ag(6%Pt) | ML | $\approx 10$ | $\approx 7$ | 110 | [53] |
| Ag(6%Mn) | ML | $\approx 11$ | $\approx 12$(SS) | 110 | [53] |
| Ag(9%Mn) | ML | $\approx 7$ | $\approx 9$ (SS) | 155 | [53] |
| Cu(4%Ge) | ML | $\geq 50$ | $\approx 50$ | 182 | [169] |
| Cu(2.1%Ge) | SV | 117 | $\approx 100$ | 90 | [177]. |
| Cu(6%Pt) | ML | $\approx 8$ | $\approx 7$ | 130 | [53] |
| Cu(6%Pt) | SV | $\approx 11$ | $\approx 7$ | 160 | [67] |
| Cu(7%Mn) | ML | $\approx 2.8$ | 3±1.5 (SS) | 270 | [53] |
| Cu(7%Ni) | ML | $\approx 23$ | $\approx 22$ | 110 | [178] |
| Cu(10%Ni) | ML | $\approx 14$ | $\approx 15$ | 175 | [178] |
| Cu(14%Ni) | ML | $\approx 10$ | $\approx 12$ | 191 | [178] |
| Cu(23%Ni) | ML | $\approx 7.5$ | $\approx 7$ | 355 | [178] |
| Cu(23%Ni) | SV | $\approx 8$ | $\approx 7$ | 310 | [179] |

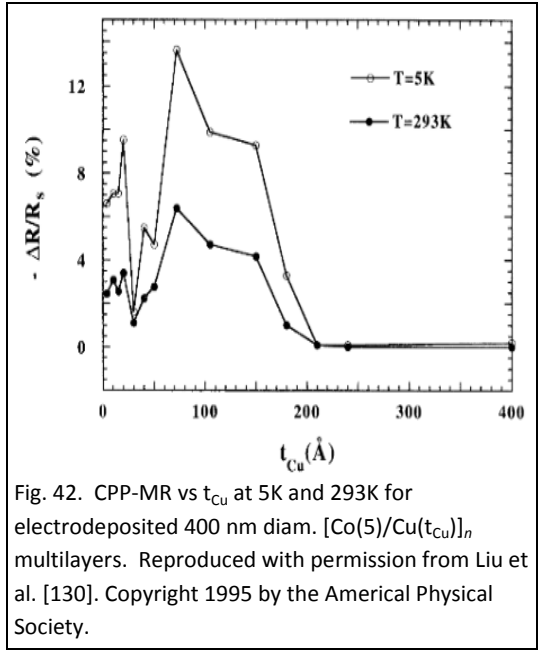

Fig. 42. CPP-MR vs $t_{Cu}$ at 5K and 293K for electrodeposited 400 nm diam. [Co(5)/Cu(t_{Cu})]$_n$ multilayers. Reproduced with permission from Liu et al. [130]. Copyright 1995 by the Americal Physical Society.

The good agreements between values of $l_{sf}^N$ found with multilayers and SVs, as well as between the CPP and CESR values, support both the experimental techniques and the VF model. In contrast, mfp effects look unable to explain the data, since the deviations from the straight line through the origin do not correlate with alloy residual resistivities. As examples, $\rho_{Ag(4\%Sn)}$ and $\rho_{Cu(4\%Ge)}$ are larger than $\rho_{Ag(6\%Pt)}$ or $\rho_{Cu(6\%Pt)}$, but AgPt and CuPt deviate strongly from the square root line, while AgSn and CuGe do not.

### 8.5.2. $l_{sf}^N$ in nominally pure N-metals.
### 8.5.2.1. $l_{sf}^N$, estimates for Cu.

Co/Cu is the most studied F/N metal pair. So there is interest in $l_{sf}^{Cu}$. The simple estimate, $l_{sf}^{Cu} \sim 10\lambda_{cu}$, (see Appendix 1) predicts values from ~ 1300 nm for $\rho_{Cu}$ = 5 nΩm to ~ 220 nm for $\rho_{Cu}$ = 30 nΩm. These are long enough to be unimportant for most CPP-MR data. In this section we summarize results of early efforts to deduce $l_{sf}^{Cu}$ from nanowire measurements.

In 1995, Liu et al. [130] used measurements on d = 400 nm [Co(5)/Cu(t_{Cu})]$_n$ nanowires to estimate $l_{sf}^{Cu}$ = 200 nm at both T = 5K and 293 K from the disappearance of CPP-MR shown in Fig. 42 [130]. No value of $\rho_{Cu}$ was given.

Also in 1995, from measurements of fixed total thickness multilayers with $t_{co}$ = $t_{cu}$, Voegeli et al. [141] estimated $l_{sf}^{Cu} \sim$ 50-85 nm at 20K for $\rho_{Cu}$ = 13-33 nΩm.

In 1996, Doudin et al. [135] extended the Voegli et al. measurements and analysis. Assuming a fixed spin-flip mean-free-path (as opposed to values inversely proportional to the resistivity) they estimated that at 20K $l_{sf}^{Cu}$ increased from roughly 22 to 60 nm as $\rho_{Cu}$ decreased from 100 to 10 nΩm, and that at 300K, $l_{sf}^{Cu}$ increased from roughly 20 to 52 nm as $\rho_{Cu}$ decreased from 100 to 10 nΩm.

Also in 1996, , Piraux et al. [153] used the data of Fig. 38 for multilayers of the form [Co(8)/Cu(t_{Cu})]$_n$ to derive $l_{sf}^{Cu}$ = 140 ± 15 nm for $\rho_{Cu}$ = 31 nΩm.

To summarize, for the relatively dirty Cu-layers of nanowires, these authors found values of $l_{sf}^{Cu}$ ranging from 50 to 200 nm, somewhat smaller than expected from the simple estimate $l_{sf}^N \sim 10\lambda_N$. They also found little variation with T. Values for cleaner Cu should be longer and vary more with T.

### 8.5.2.2. $l_{sf}^N$, General technique and application to several N-metals.

In 2000, Park et al. [67] described a general way to measure $l_{sf}^N$ by inserting an N-layer of variable thickness $t_N$ into the middle Cu layer of a Py-based EBSV to give an EBSV of the form [FeMn(8)/Py(24)/Cu(10)/N(t_N)/Cu(10)/Py(24)] sandwiched between 200 nm thick superconducting Nb cross-strips. Fig. 43 shows the decrease of log (AΔR) with increasing $t_N$ for six different inserts. The maximum Nb thickness of 20 nm is small enough that the Nb in the EBSV does not-superconduct.

The initial rapid decreases in Fig. 43 for N = V, Nb, and W are attributed to formation of the two Cu/N interfaces, including possible spin-flipping therein (see section 8.11). Completion of these rapid decreases gives estimates of the interface thicknesses of 0.6 ± 0.1 nm for V and Nb and 0.9 ± 0.1 for W. Attribution of the decreases to interfaces is supported by the absence of Interface contributions for Cu(6%Pt), where the interface is essentially just Cu/Cu, and for Cu/Ag, where the Cu/Ag interface has very small $2AR_{Cu/Ag} = 0.088 ± 0.006$ (see sections 8.8 and 8.14). The decrease of log ($A\Delta R$) for FeMn is so fast that it is attributed to just the FeMn/Cu interfaces.

The slower decreases after the interfaces are complete is attributed to a combination of spin-relaxation, $l_{sf}^N$, and the additional $AR_N$ added by the growing insert. VF theory predicts an approximate form for this region of

$$A\Delta R \propto \exp[-t_N/l_{sf}^N]/(AR_o + AR_N). \qquad (12)$$

Here, $AR_o$ is the contribution due to the EBSV without the insert, $AR_N$ is the extra AR due to the insert, and the proportionality constant depends upon the properties of Py. Eq. 12 predicts that a plot of log ($A\Delta R$) vs $t_N$ should give a straight line modified slightly by the growing contribution of $AR_N$ so long as $t_N < l_{sf}^N$. Once $t_N > l_{sf}^N$, the contribution from $AR_N$ saturates at $\rho_N l_{sf}^N + 2AR_{Cu/N}$. The fits shown in Fig. 43 were calculated numerically from VF theory taking account of the previously measured properties for Py, (including $l_{sf}^{Py} = 5.5$ nm) and for Py/Cu interfaces [57]. The Cu/N interfaces (I) were treated as additional thin layers of fixed $l_{sf}^I$ and growing t until t = $t_I$.

This same technique was later used with other metals. The resulting estimates of $l_{sf}^N$ are given in Table 7 with references and separately measured values of $\rho_N$. The values of $l_{sf}^N$ are ordered from longest (left) to shortest (right). $l_{sf}^N$ tends to decrease with increasing atomic number Z, as would be expected for spin-orbit scattering, but also with increasing $\rho_N$ as expected from Eq. A.6.c. The result is not simple, since the two effects can conflict (e.g., Au is heavy, but has low $\rho_N$; V is light, but has high $\rho_N$). Tsymbal and Pettifor [8] proposed that the decays in Fig. 43 might measure not $l_{sf}^N$, but rather mean-free-paths $\lambda$, since layer-thickness-dependent interface resistances should approach their asymptotic values exponentially with $\lambda$. But $A\Delta R$ for N-inserts in an EBSV should not be sensitive to layer thickness dependent $\lambda$, which don't flip moments. Moreover, from Eq. A.3, if the values in Table 7 were λ, they should scale with inverse $\rho_N$, which they do not (compare, e.g., V and W).

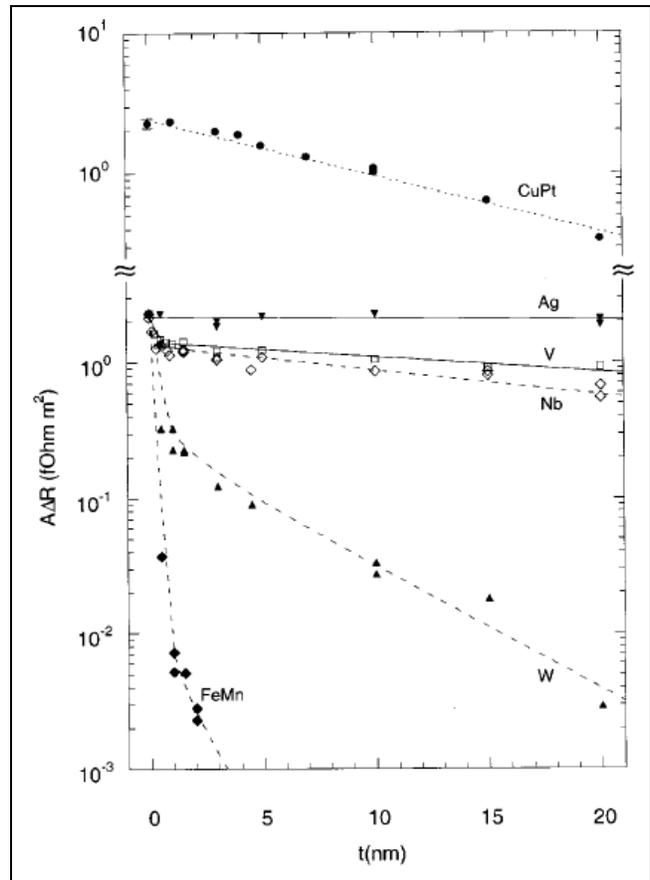

Fig. 43. Log ($A\Delta R$) vs layer thickness t for layers of metals Ag, V, Nb, W, the alloy CuPt(6%), or the antiferromagnet FeMn, inserted into the middle of sputtered Py-based EBSVs. The slopes of the lines at 'large t' give the spin-diffusion-lengths in the inserted metals. Reproduced with permission from Park et al. [67]. Copyright 2000 by the American Physical Society.

**Table 7.** $l_{sf}^N$ and $\rho_N$ for several N-metals measured by the EBSV insert technique.

| Metal | Ag | V | Nb(1) | Nb(2) | Au | Pd | Ru | Pt(1) | Pt(2) | W |
|---|---|---|---|---|---|---|---|---|---|---|
| $l_{sf}^N$ (nm) | >40 | >40 | $25_{.5}{}^{+\infty}$ | 48±3 | $35_{.5}{}^{+65}$ | $25_{.5}{}^{+10}$ | 14 | 14±6 | 9.6±1.1 | 4.8±1 |
| $\rho_N$ (nΩm) | 7±2 | 105±20 | 78±15 | 60±10 | 19±6 | 40±3 | 95 | 42±6 | 75±10 | 92±10 |
| Ref. | [67] | [67] | [67] | [180] | [181] | [182] | [183] | [182] | [164] | [67] |

### 8.6. $l_{sf}^F$, Spin-diffusion lengths in nominally pure F-metals.

The successes of the 2CSR model for Co/Ag and Co/Cu multilayers described above seemed to confirm the general expectation that $l_{sf}^F$ for nominally pure F-metals should be long at cryogenic temperatures. However, the use of relatively thin Co layers (2 nm or 6 nm) for the most clearcut 2CSR model fits means that lower bounds on $l_{sf}^{Co}$ from these studies are weak. And for a number of years, no measurements existed of $l_{sf}^F$ for the other two 'simple' pure F-metals, Fe and Ni. In this section, we examine what is known about $l_{sf}^F$ for nominally pure F = Co, Fe, or Ni. We emphasize again that values are only approximate for a given $\rho_F$, since the dominant scatterers are unknown.

For Co, two very different studies of $l_{sf}^{Co}$ give the best quantitative information. However, each has problems. $l_{sf}^{Co}$ is, thus, not yet fully determined. In Fig. 80, section 8.10.2, we will see that $l_{sf}^{Co}$ is anomalously long for a given Co residual resistivity, $\rho_{Co}$. The reason why is not yet clear; but the scattering in Co might be dominated by stacking faults, which could give weak spin-flipping.

Fe and Ni have only one study each of $l_{sf}^F$. We'll see in sections 8.6.1 - 8.6.4, and 8.10.2, that both values are much shorter than for Co. The Ni value is consistent with those for F-alloys, but the Fe value is low.

### 8.6.1. $l_{sf}^{Co}$ at 77K from L/O.

In 1996, the Louvain/Orsay (L/O group [153] used nanowires in the first study designed to find $l_{sf}^{Co}$. Their Co resistivity at 77K was $\rho_{Co} \sim 160$ nΩm. Fig. 44 [98] plots $R_P/\Delta R$ at 77K and 300K for 90 nm diam. nanowires of the form [Co($t_{Co}$)/Cu(8)]$_n$ for 60 nm ≤ $t_{Co}$ ≤ 950 nm. In the limits $t_F \gg l_{sf}^F$; $t_N \ll l_{sf}^N$; and $\rho_N t_N$, $AR_{F/N}^* \ll \rho_F^* l_{sf}^F$, VF theory gives for a full AP state:

$$R_P/\Delta R = (1-\beta_F^2)t_f/[2\beta_F^2 l_{sf}^F]. \tag{13}$$

The authors assumed that a non-AP state could be described by a fractional antiferromagnetic (AF) order parameter $p$ ($p$ = 1 for an AP state), with Eq. 13 rewritten as

$$R_P/\Delta R = (1-\beta_F^2)t_f/p[2\beta_F^2 l_{sf}^F]. \tag{14}$$

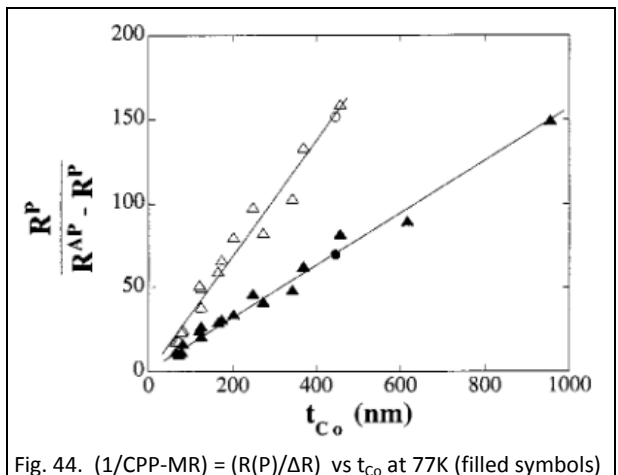

Fig. 44. (1/CPP-MR) = (R(P)/ΔR) vs $t_{Co}$ at 77K (filled symbols) and 300K (open symbols) for [Co($t_{Co}$)/Cu(8)]$_n$ multilayers with unspecified $n$. The two circles are for a single Co(435)/Cu(15)$_n$ sample. It is presumed that $n$ is large enough so that each sample is in a constant CPP-MR limit. From Piraux et a. [98]. With kind permission from Springer Science and Business Media.

Settling upon a value for $p$ involved difficulties. First, it is not clear that $p$ will be independent of $t_{Co}$ in real nanowires. As $t_{Co}$ grows from less than the wire diameter d = 90 nm to much greater than d, shape anisotropy should change the preferred direction of layer magnetization at H = 0 from at least partly in plane to fully along the nanowire axis. Second, for $t_{Co}$ >> d, shape anisotropy should tend to produce a P-state at H = 0, not an AP state. In early studies, the authors simply took $p$ = 0.5, corresponding to an assumed random orientation of moments at H = 0. This value gave $l_{sf}^{Co}$ = 45 nm [153] at 77K. In the latest study [98], Magnetic Force Microscopy (MFM) measurements on a [Co(170)/Cu(8)]$_n$ nanowire (Fig 45) showed an admixture of AP and P orderings with an average AP ordering of $p \sim$ 39%. Taking this average to apply to the whole range of $t_{Co}$, along with an assumed range of 0.33 < p < 0.49, they estimated $l_{sf}^{Co}$ = 59 ± 18 nm at 77K. They associated with this $l_{sf}^{Co}$ an independently measured value of $\rho_{Co} \sim$ 150 nΩm. In 1998 [98] they extended their analysis to 300K, estimating $l_{sf}^{Co}$(300K) = 38 ± 12 nm. Lastly, the value of $\beta_{Co}$ = 0.36 used in this analysis is less than the $\beta_{Co} \cong$ 0.46 found by the most reliable techniques in Table 5. Inserting this larger $\beta_{Co}$ would reduce $l_{sf}^{Co}$ at 77K to ~ 36 nm. However, Cu, the dominant impurity in electrodeposited Co, is unlikely to dominate sputtered or MBE deposited Co. Since Campbell and Fert [36] give no independent estimate of $\beta_{Co}$ for Co(Cu), a smaller $\beta_{Co}$ for electrodeposited Co cannot be ruled out.

### 8.6.2. $l_{sf}^{Co}$ at 4.2K from MSU.

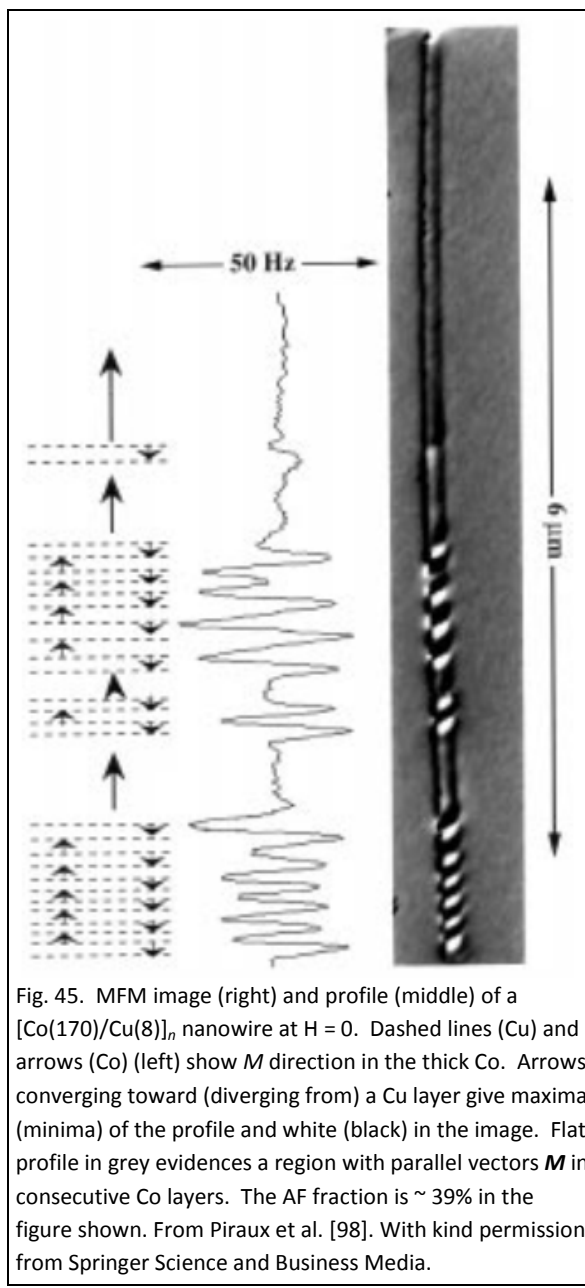

Fig. 45. MFM image (right) and profile (middle) of a [Co(170)/Cu(8)]$_n$ nanowire at H = 0. Dashed lines (Cu) and arrows (Co) (left) show $M$ direction in the thick Co. Arrows converging toward (diverging from) a Cu layer give maxima (minima) of the profile and white (black) in the image. Flat profile in grey evidences a region with parallel vectors $M$ in consecutive Co layers. The AF fraction is ~ 39% in the figure shown. From Piraux et al. [98]. With kind permission from Springer Science and Business Media.

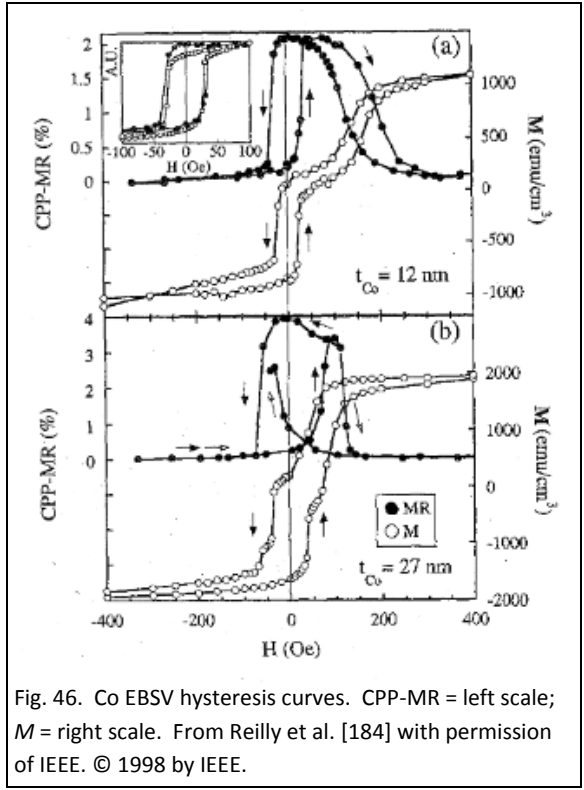

Fig. 46. Co EBSV hysteresis curves. CPP-MR = left scale; $M$ = right scale. From Reilly et al. [184] with permission of IEEE. © 1998 by IEEE.

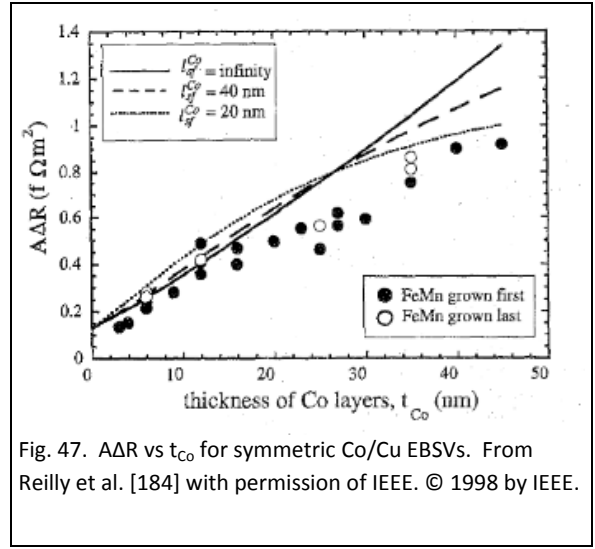

Fig. 47. A$\Delta$R vs $t_{Co}$ for symmetric Co/Cu EBSVs. From Reilly et al. [184] with permission of IEEE. © 1998 by IEEE.

In 1998, the MSU group of Reilly et al. [184] tried to estimate $l_{sf}^{Co}$ at 4.2K using Co-based EBSVs and the crossed-superconductor technique. Their Co resistivity, $\rho_{Co} \sim 60$ n$\Omega$m, is about 2.5 times smaller than that of Pireaux et. al [153]. Fig. 46 shows that the hysteresis curves grew worse as $t_{co}$ increased from 12 nm to 27 nm. Fig. 47 shows A$\Delta$R vs $t_{co}$ for symmetric EBSVs along with calculated values using prior parameters for Co/Cu plus assumed values of $l_{sf}^{Co}$ = 20 nm, 40 nm and $\infty$. None of the calculations fits the data well. To avoid the possible unpinning of the pinned Co layer with increasing $t_{co}$ that might have occurred in Fig. 47, another set of EBSVs was made with fixed pinned layer thickness $t_{co}$ = 6 nm. Fig. 48 shows that the VF fit is no better than in Fig. 47. The authors concluded that the continuous rise in A$\Delta$R to $t_{co}$ > 35 nm, plus the inability to fit the data satisfactorily, made them unable to extract a reliable value for $l_{sf}^{Co}$. Taking account of the earlier L/O data, they concluded that they could not rule out an $l_{sf}^{Co}$ as short as 40 nm.

To at least partially resolve the fit problems in Figs. 47 and 48, we must jump forward to 2002, where Eid et al. [185] proposed a value of $\delta_{Co/Cu} \sim 0.25$. Fig. 49 shows that just adding this $\delta_{Co/Cu}$ gives approximate fits to both sets of EBSV data using the parameters of [184] without adjustment, including a bulk $l_{sf}^{Co} = \infty$. Given the substantially lower resistivity Co in the Reilly samples, such a Reilly result agrees with the L/O conclusion that $l_{sf}^{Co}$ is long, for the Reilly samples possibly substantially longer than 60 nm. Using, instead, a later, more direct estimate of $\delta_{Co/Cu} = 0.35 \pm 0.1$ [163] in Fig. 49 would allow for non-infinite values of $l_{sf}^{Co}$. The uncertainties are large enough for $l_{sf}^{Co}$ to be as short as 60 nm for the lower resistivity Co of Reilly et al., or substantially longer.

We conclude that the available evidence favors a 'long' $l_{sf}^{Co}$, but with large uncertainties in its value for a given $\rho_{Co}$. We'll see in section 8.10.2 that the values of $l_{sf}^{Co}$ just described are unusually long for the measured values of $\rho_{Co}$.

### 8.6.3. $l_{sf}^{Fe}$ at 4.2K.

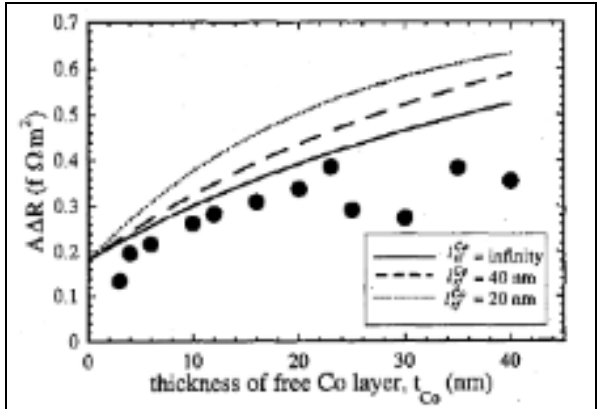

Fig.48. AΔR vs $t_{Co}$ for asymmetric EBSVs (pinned Co thickness = 6 nm). From Reilly et al. [184] with permission of IEEE. © 1998 by IEEE.

In 2000, Bozec [60] derived the value $l_{sf}^{Fe}$ = 8.5 ± 1.5 nm at 4.2K *for* $\rho_{Fe}$ = 38 ± 5 nΩm, along with other parameters for Fe/Cu. He used the crossed-superconductor technique on a combination of three different sets of Fe/Cu multilayers and three different EBSVs, all prepared by MBE. The multilayers had the form $[Fe(t_{Fe})/Cu(8)/Fe(t_{Fe})]_n$, with fixed Fe thicknesses of $t_{Fe}$ = 2 nm, 3 nm, or 6 nm and *n* varying from 5 to 35. The EBSVs had the forms: (1) [FeMn(8)/Fe($t_{Fe}$)/Cu(8)/Fe($t_{Fe}$)]; (2) [FeMn(8)/Fe($t_{Fe}$)/Cu(8)/Py(24)]; and (3) [FeMn(8)/Py(6)/Cu(8)/Fe($t_{Fe}$)]. To oversimplify a bit, values of $\gamma_{Fe/Cu}$ = 0.55 ±0.07 and $2AR^*_{Fe/Cu}$ =1.48 ± 014 were found from the multilayers (with the usual uncertainty in reaching full AP states). Then $\beta_{Fe}$ = 0.78 ± 0.05 and $l_{sf}^{Fe}$ = 8.5 ± 1 nm were found from the form (1) EBSVs and the parameters were cross-checked by correctly predicting, without adjustment, the values of AΔR for the form (2) and (3) EBSVs.

### 8.6.4. $l_{sf}^{Ni}$. at 4.2K.

In 2007, the value $l_{sf}^{Ni}$.= 21 ± 2 nm at 4.2K was found for $\rho_{Ni}$ = 33 ± 3 nΩm by Moreau et al. [171] along with other parameters for Ni/Cu. They used the crossed superconductor technique on sputtered samples involving a combination of: (1) modified [FeMn(8)/Py(8)/Cu(15)/Ni($t_{Ni}$)/Cu(10)/FeMn(2)] EBSVs; (2) [Ni(8)/Cu($t_{Cu}$)]$_n$ multilayers with fixed total thickness = 360 nm, and (3) special 'spoiler' EBSVs of the form FeMn(8)/Py(8)/Ni($t_{Ni}$)/Cu(15)/Py(24) with a Ni layer of variable thickness inserted between the pinned Py layer and the adjacent Cu layer. Adding the FeMn(2) layer to a standard EBSV disrupts spin-memory, thereby enhancing the otherwise small AΔR by removing the $AR_{Ni/Nb}$ contribution to the denominator from the nearby Nb contact (see, e.g., the difference between Eqs. 6 and 8). To slightly oversimplify, the multilayer gave $2AR^*_{Ni/Cu}$ = 0.36 ± 0.06 fΩm², and then the EBSV with extra FeMn(2) layer gave $\gamma_{Ni/Cu}$ = 0.29 ±0.05, $\beta_{Ni}$ = 0.14 ± 0.02, and $l_{sf}^{Ni}$ ≈ 21 nm. However, the weak variation with $t_{Ni}$ of an already small AΔR shown in Fig. 50 left $l_{sf}^{Ni}$ poorly constrained. This limitation led to use of the 'spoiler geometry' to enhance the effect of $l_{sf}^{Ni}$. With $\beta_{Ni}$ << $\beta_{Py}$, and $t_{Ni}$ > $l_{sf}^{Ni}$, VF theory predicts that such a Ni 'spoiler' insert should cause AΔR to decrease approximately as exp(-$t_{Ni}$/$l_{sf}^{Ni}$). Starting from a Py-based EBSV gives a much larger signal than that in Fig. 50 , and the spoiler Ni gives the much larger variation of AΔR shown in Fig. 51 . The result is a better constrained $l_{sf}^{Ni}$ = 21 ± 2 nm.

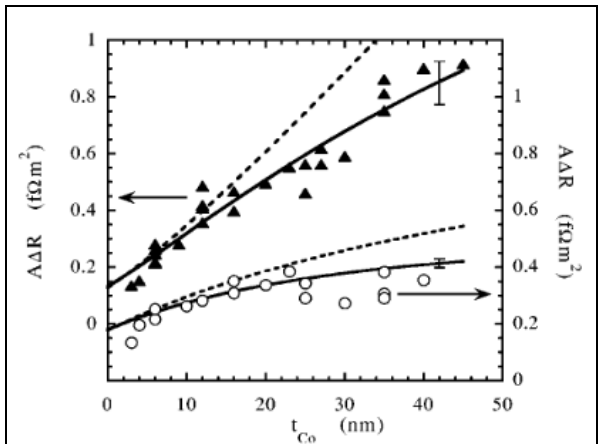

Fig. 49. AΔR vs $t_{Co}$ for the same symmetric (filled triangles) and asymmetric (open circles) EBSVs in Figs. 47 and 48. The dashed curves are the prior fits in Figs. 47 and 48 assuming $l_{sf}^{Co}$ = ∞. The solid curves just add to these curves an assumed $\delta_{Co/Cu}$ = 0.25. Reproduced with permission from Eid et al. [185]. Copyright 2002 by the American Physical Society.

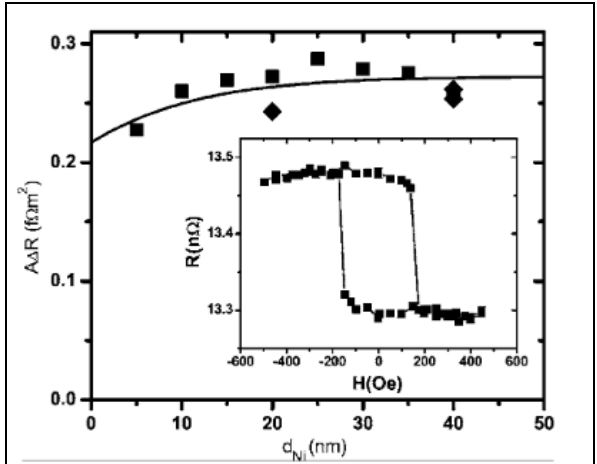

Fig. 50. A∆R vs t$_{Ni}$ for a symmetric Ni/Cu EBSV  The small changes with d$_{Ni}$ leave a large uncertainty in $l_{sf}^{Ni}$. Reproduced with permission from Moreau et al. [171]. Copyright 2007, AIP Publishing LLC.

### 8.7. CPP parameters for Py/Cu.

In section 8.7.1, we explain how sequential derivations of VF parameters for Py/Cu at 4.2K, extending from 1993 to 2000, illustrate rule #1 of section 1.1.  That is, parameters changed as new data accumulated and systematic errors were found and fixed.  The first two analyses of Py/Cu multilayer data assumed a 2CSR model, first with one set of data and then with two.  When EBSV data revealed a short $l_{sf}^{Py}$ ~ 5.5 nm, the multilayer data were reanalyzed with VF theory, first alone and then including the EBSV data.  Lastly, a study of NiCoFe [115] led to depositing of new Py EBSVs that gave a final change in one parameter (β$_{Py}$).  We trace out this history, labelling sequential sets of MSU values as columns (A) - (E) in

Table 8.  For comparison, column (F) lists nanowire values.

In section 8.7.2 we cover the two sets of data and analyses that gave an unexpectedly short $l_{sf}^{Py}$: (1) a 1997 result by Steenwyk et al. of $l_{sf}^{Py}$ = 5.5 ± 1 nm found with EBSVs and superconducting cross-strips, and (2) 1999 measurements on nanopillars by Dubois et al. that gave an overlapping value of $l_{sf}^{Py}$ = 4.3 ± 1 nm. Lastly, we discuss a 2010 ab-initio calculation by Starikov et al. that gave $l_{sf}^{Py}$ = 5 ± 1 nm.

### 8.7.1. VF model parameters for Py/Cu.

The first problem in analyzing data for Py/Cu was that the Py/Cu data did not display the large ratios AR(0)/AR(Pk) ≥ 1.5 typically seen with Co/Ag and Co/Cu.  Instead, as illustrated in Fig.52 [59], most ratios were closer to AR(0)/AR(Pk) ~ 1.1, with some < 1.  The assumption AR(0) ≅ AR(AP) was, thus, not assured, and the initial analysis used whichever of AR(0) or AR(Pk) was larger.   Moreover, as shown in

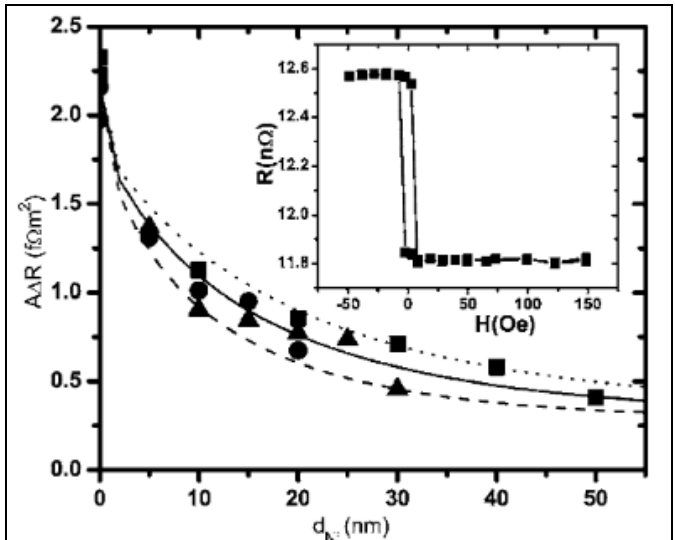

Fig. 51.  A∆R vs Ni thickness d$_{Ni}$. A larger signal than in Fig. 50 was achieved with a spoiler Ni/Cu structure as described in the text. Reproduced with permission from Moreau et al. [171]. Copyright 2007, AIP Publishing LLC.

Figs. 53 and 54, for fixed t$_{py}$ =1.5 nm or 6 nm, both AR(P) and AR(0) increased linearly with bilayer number N over only a limited range of N: in Fig. 53, N ≤ 22 nm, and in Fig. 54, N ≤ 17 nm. .  An early preliminary 2CSR model fit to just the linear t$_{Py}$ = 1.5 nm multilayer data of Fig. 55 gave the parameters in column (A): β$_{Py}$ ~ 0.52 and γ$_{Py/Cu}$ ~ 0.69, assuming from independent measurements ρ$_{Py}$ = 122 nΩm, 2AR$_{Py/Nb}$ = 6.5 fΩm$^2$, and ρ$_{Cu}$ = 5 nΩm [44].  A second 2CSR fit to the linear data in both Figs. 53 and 54 gave the parameters in column (B) [93]. The parameters in column (C) are from a reanalysis by Hollody et al. [59], using VF theory with the newly derived value $l_{sf}^{Py}$= 5.5 nm, which mainly affected  the analysis of the data with t$_{Py}$ = 6 nm. Note that the parameter $2AR^*_{Py/Cu}$ does not appear in that paper. Since a fit without it makes no sense, it was presumably taken to be 1.0 fΩm$^2$. The fit in column (D) was made to a combination of the two multilayer sets already mentioned plus the new EBSV data from [57, 73] in Fig. 55. Lastly, in column (E), newer EBSV data for Py in Fig. 56 [115], were used simply to slightly increase β$_{Py}$. These last data will be discussed in section 8.10.1. The values in (E) became the standards for the MSU group. For comparison, the three values in column (F) are from nanowire data by the L/O group [99]. All three values agree with the latest MSU ones to within mutual uncertainties.

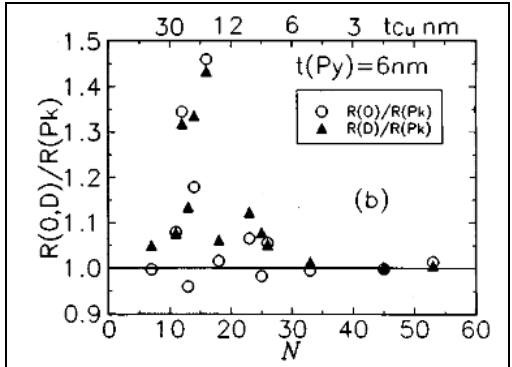

Fig. 52. R(0)/R(Pk) (open circles) and R(D)/R(Pk) (filled triangles) vs N for [Py(6)/Cu(t)]$_N$ multilayers with fixed t$_T$ = 360 nm. D = demagnetized. Reproduced with permission from Holody et al. [59]. Copyright 1998 by the American Physical Society.

**Table 8. Parameters for Py/Cu from different Analyses**.

Items marked (I) were independently measured or assumed--e.g, $l^{Py}_{sf}$ in items (A)-(C) and (E). Columns (A)-(E) are MSU group measurements with superconducting cross-strips. Column (F) is from the Louvain/Orsay group on nanowires.

| | (A)[39] | (B)[93] | (C)[59] | (D)[73] [57] | (E)[115] | | (F)[99] |
|---|---|---|---|---|---|---|---|
| ρ$_{Cu}$(nΩm) | 5 (I) | 4.5±0.5 (I) | 5 (I) | 4.5±0.5 (I) | 4.5±0.5 (I) | | |
| ρ$_{Py}$(nΩm)) | 122 (I) | 123±40 (I) | 122±20 (I) | 123±40 (I) | 123±40 (I) | | |
| β$_{Py}$ | 0.52 | 0.50±0.16 | 0.65± 0.1 | 0.73±0.07 | 0.76±0.07 | | 0.8±0.1 |
| γ$_{Py/Cu}$ | 0.69 | 0.81±0.12 | 0.76±0.1 | 0.7±0.1 | 0.7±0.1 (I) | | 0.8±0.1 |
| 2AR$^*$ $_{Py/Cu}$ (fΩm$^2$) | | 1.00±0.08 | | 1.00±0.08 | 1.00±0.08 (I) | | |
| $l^{Py}_{sf}$ (nm) | ∞ (I) | ∞ (I) | 5.5 (I) | 5.5 ± 1 | 5.5 ± 1 (I) | | 4.3±1 |

### 8.7.2. $l^{Py}_{sf}$, Spin-diffusion length in Py, at cryogenic temperatures.

As already noted, 1997 brought a major surprise, the discovery of an unexpectedly short Py spin-diffusion length, $l^{Py}_{sf}$ = 5.5 ± 1 nm, by Steenwyk et al.[73] from analysis of the data of Fig. 55. The surprise was not only that this value is short, but also that it is shorter even than the longer Py-mean-free-path, $\lambda^f_{Py} \approx 8$ nm, estimated from ρ$_{Py}$ = 124 nΩm and ρ$_b\lambda_b \sim 1$ fΩm$^2$ (see Appendix 1). Given the importance of this result, we discuss in detail the cross checks made, using figures from [57] that contain more details than those in [73].

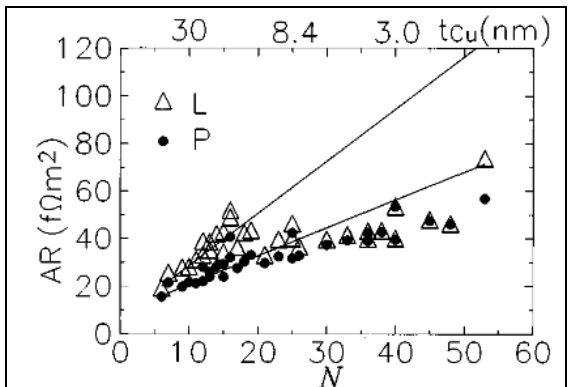

Fig. 53. AR vs N for [Py(6)/Cu(6)]$_N$ multilayers with t$_T$ = 360 nm. Open triangles are largest (L) of AR(0) or AR(Pk). Filled circles are AR(P). Reproduced with permission from Holody et al. [59]. Copyright 1998 by the American Physical Society.

In 1999, this short $l^{Py}_{sf}$ was confirmed by nanowire measurements that yielded $l^{Py}_{sf}$ = 4.3 ± 1 nm [99].

#### 8.7.2.1. $l^{Py}_{sf}$ at 4.2K by the MSU group.

The measurements in [73] involved symmetric Py-based EBSVs of the form [FeMn(8)/Py(t$_{Py}$)/Cu(20)/Py(t$_{Py}$)] with Py thicknesses extending to t$_{Py}$ = 30 nm as shown in Fig. 55 [57]. The hysteresis curves for MR and $M$ in Fig. 9 [73] show that such EBSVs display well-defined AP and P states for both t$_{Py}$ = 3 nm and t$_{py}$ = 24 nm. To check for systematic errors in Fig. 55, data were taken with the FeMn deposited both before the first Py layer (filled circles) and after the second Py layer (filled triangles). Importantly, according to VF theory, for t$_{Py}$ >> $l_{sf}^{Py}$. AΔR should approach a constant limit—the horizontal dashed line in Fig. 55) given by Eq. 9 applied to a Py/Cu EBSV. This limit should be the same for any EBSV where both Py layers have thicknesses t$_{Py}$ >> 5.5 nm This prediction is confirmed by the open square, which is for an asymmetric EBSV with pinned t$_{Py}$ = 15 nm and free t$_{Py}$ = 45 nm. The solid curve in Fig. 55 is a best fit to both the data of Fig. 55 and prior data on Py/Cu multilayers. The history of this curve is a bit complex, because parameters previously derived for Py/Cu had assumed $l_{sf}^{Py}$ = ∞. For internal consistency, it was necessary to refit those old data along with the new data in Fig.55. The details are given in [73]. To simplify, all parameters except β$_{Py}$, γ$_{Py/Cu}$, and l$_{sf}^{Py}$ were fixed as described in [57], and β$_{Py}$, γ$_{Py/Cu}$, and $l_{sf}^{Py}$ = 5.5 nm were determined self-consistently for both the data of Fig. 55 and the prior data for Py/Cu multilayers. The result was β$_{py}$ = 0.73 ± 0.07, γ$_{Py/Cu}$ = 0.7 ± 0.1, and and $l_{sf}^{Py}$ = 5.5 ± 1 nm, giving the solid curve in Fig. 55. For comparison, the dotted curve is the prediction with all of the same parameters, except with l$_{sf}^{Py}$ = ∞. Independent of the detailed parameters, VF theory predicts that $l_{sf}^{Py}$ should be approximately given by the Py thickness at which the solid curve departs from the dashed one. Indeed, this departure occurs at 5 - 6 nm. As a last check, the parameters were successfully used to calculate AR(AP) for the same sample.

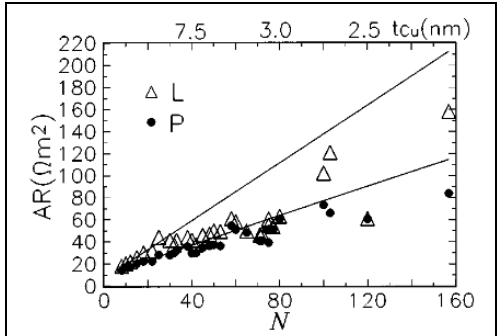

Fig. 54. AR vs $N$ for [Py(1.5/Cu(t)$_N$ multilayers with t$_T$ = 360 nm. Open triangles are largest (L) of AR(0) or AR(Pk). Filled circles are AR(P). Reproduced with permission from Holody et al. [59]. Copyright 1998 by the American Physical Society.

To check if AΔR and $l_{sf}^{Py}$ are sensitive to interfacial details, data were also taken for samples with N =

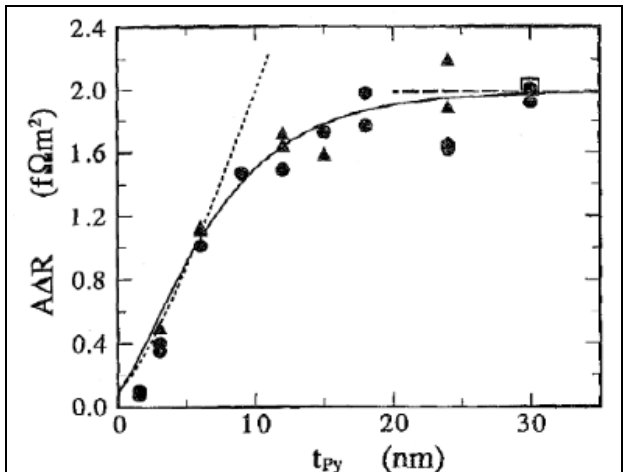

Fig. 55. AΔR vs t for symmetric [FeMn(8)/Py(t)/Cu(20)/Py(t)] EBSVs. Circles and triangles are different runs (see text). The open square is for [FeMn(8)/Py(15)/Cu(20)/Py(45)]. From Pratt et al. [57] with permission of IEEE. © 1997 by IEEE.

Ag(20) instead of N = Cu(20), and for samples with N = Cu dusted with 0.6 nm of Co at each Py/Cu interface. Fig. 56 [57] compares the results with the best fit solid curve from Fig. 55. For large t$_{Py}$, the data lie a little above the curve, but show the same form and are consistent with the same $l_{sf}^{Py}$ = 5.5 nm. As yet another check for systematic errors, Fig. 56 shows a set of data points (open symbols) for t$_{Py}$ = 24 nm using different thicknesses of Cu (10 nm ≤ t$_{Cu}$ ≤ 100 nm). Except for the higher datum for t$_{Cu}$ = 10 nm, the open symbols are consistent with the other data to within mutual uncertainties. There are, however, two unexpected features of the open symbols. First, they show an apparent small systematic increase in AΔR with decreasing t$_{Cu}$. While the low value for t$_{Cu}$ = 100 nm can be attributed to the extra resistance of 100 nm of Cu, the other differences, especially the highest one for $t_{Cu}$ = 10 nm, are larger than expected. Second, the value of AΔR for $t_{Cu}$ = 20 nm, and the average of the open symbols, both fall a bit above the curve of Fig. 55. This slight shift is consistent with the higher values for Py shown in Fig. 57 [115], which were used to slightly increase $\beta_{Py}$ in Table 8. As still another cross-check on the parameters, the authors compared with predictions made without adjustment, new data for symmetric double EBSVs of the alternative forms [FeMn(8)/Py(6 ,12)/Cu(20)/Py(6,12)/Cu(20)/Py(6,12)/FeMn(8)] or [Py(6)/Cu(10)/Py(6)/FeMn(8)/Py(6)/Cu(10)/Py(6)]. As shown in Fig. 58 [57], the new data are consistent with the predictions. The authors concluded that a large selection of data self-consistently support the initially surprisingly low value of $l_{sf}^{Py}$ = 5.5 ± 1 nm, along with the other parameters for Py/Cu given in [57]. The values in column [E] of Table 8 became the standard ones used by the MSU group thereafter for VF fitting.

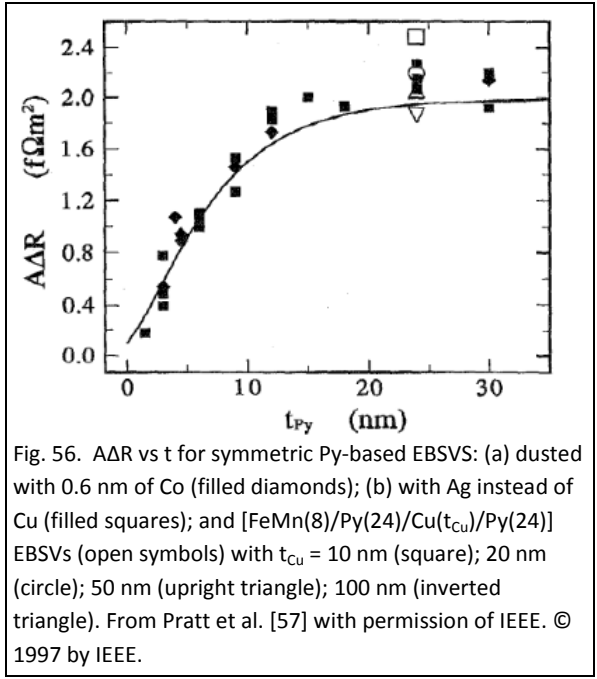

Fig. 56. AΔR vs t for symmetric Py-based EBSVS: (a) dusted with 0.6 nm of Co (filled diamonds); (b) with Ag instead of Cu (filled squares); and [FeMn(8)/Py(24)/Cu(t_{Cu})/Py(24)] EBSVs (open symbols) with t_{Cu} = 10 nm (square); 20 nm (circle); 50 nm (upright triangle); 100 nm (inverted triangle). From Pratt et al. [57] with permission of IEEE. © 1997 by IEEE.

### 8.7.2.2. $l_{sf}^{Py}$ at 77K by the L/O group.

In 1999, Dubois et al. [99] confirmed such a short $l_{sf}^{Py}$ with the nanowire measurements at 77K shown in Fig. 59. To improve control of the AP magnetic state, they constructed their nanowire of [(Py(t_{Py})/Cu(10)/Py(t_{Py})] trilayers separated by 100 nm (open circles) or 500 nm (filled squares) thick layers of Cu to magnetically decouple the trilayers from each other. In the limit $t_F \gg l_{sf}^F$, the inverse of Eq. 6b above then becomes

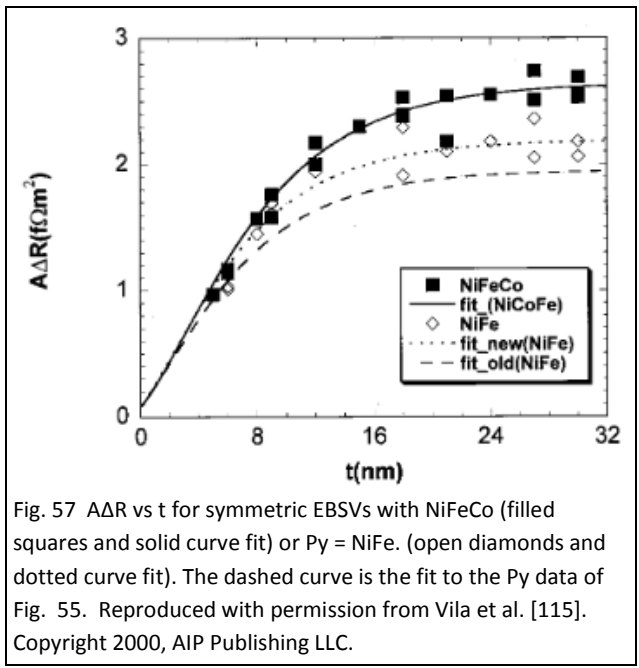

Fig. 57 AΔR vs t for symmetric EBSVs with NiFeCo (filled squares and solid curve fit) or Py = NiFe. (open diamonds and dotted curve fit). The dashed curve is the fit to the Py data of Fig. 55. Reproduced with permission from Vila et al. [115]. Copyright 2000, AIP Publishing LLC.

$$\Delta R/R_P = p*[\beta_F^2 l_{sf}^F]/(1-\beta_F^2)t_F, \qquad (15)$$

where $p*$ is the fraction of trilayers with AP orientation. The solid curve in Fig. 59 is a fit to Eq. 15 with $p*$ = 0.85 and $l_{sf}^{Py}$ = 4.3 nm. The authors concluded that their data yielded $l_{sf}^{Py}$ = 4.3 ± 1 nm, which overlapped with the $l_{sf}^{Py}$ = 5.5 ± 1 nm derived in the previous paragraph. The authors also gave a plausibility argument for a short $l_{sf}^{Py}$, obtaining ~ 9.2nm, within a factor of two of the derived value.

### 8.7.2.3. First principles calculation of $l_{sf}^{Py}$ and $\beta_{Py}$.

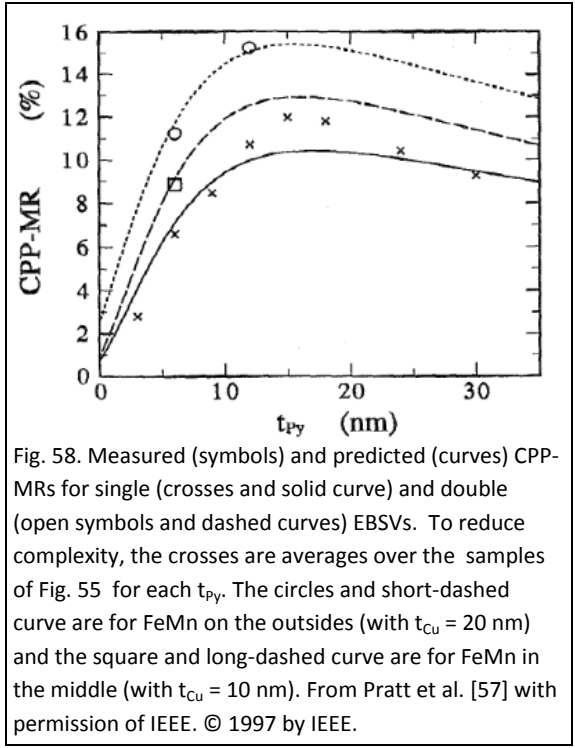

Fig. 58. Measured (symbols) and predicted (curves) CPP-MRs for single (crosses and solid curve) and double (open symbols and dashed curves) EBSVs. To reduce complexity, the crosses are averages over the samples of Fig. 55 for each $t_{Py}$. The circles and short-dashed curve are for FeMn on the outsides (with $t_{Cu}$ = 20 nm) and the square and long-dashed curve are for FeMn in the middle (with $t_{Cu}$ = 10 nm). From Pratt et al. [57] with permission of IEEE. © 1997 by IEEE.

We end this section with a first principles calculation of $l_{sf}^{Py}$ and $\beta_{Py}$. In 2010, Starikov et al. [92] calculated $l_{sf}^{F}$ and $\beta_{F}$ for F = Ni$_{1-x}$Fe$_x$ alloys as shown in Fig. 60. For Py (x = 0.2), they found $l_{sf}^{Py}$ = 5.3 ± 0.3 nm and $\beta_{Py}$ = 0.678 ± 0.003, both in surprisingly good agreement with the experimental values in columns 6 and 7 of Table 8. Their calculated resistivity, $\rho_{Py}$ = 35 nΩm is comparable to expectation for bulk samples, but lower than either sputtered $\rho_{Py} \cong$ 100-125 nΩm [57] or electrodeposited $\rho_{Py}$ = 123 nΩm [99] values, both of which contain some 'dirt'. Note in Fig. 60 the rapid change in $l_{sf}^{Py}$ for x < 0.2 and the coupled decrease in $\beta_{Py}$. This variation confuses the issue a bit, since the original MSU estimate of the content of their early sputtered Py was Ni$_{84}$Fe$_{16}$ [93]. But, overall, the good agreements suggest that the VF parameters for Py are mostly reasonably well understood.

### 8.8. 2AR$_{N1/N2}$: Interface Specific Resistances for N1/N2 metal pairs:

Analysis of complex multilayers often requires knowledge of values of 2AR$_{N1/N2}$ between two non-magnetic metals, N1 and N2. In 1996, Henry et al. [32] introduced a simple way to measure 2AR$_{N1/N2}$. In 2000, Park [67] described a more complex alternative to measure not only AR$_{N1/N2}$ but also $\delta_{N1/N2}$. In this section, we describe both methods, but for the second we give values only for 2AR$_{N1/N2}$, reserving $\delta_{N1/N2}$ for section 8.11.

### 8.8.1. A simple way to find 2AR$_{N1/N2}$.

A simple way to determine 2AR$_{N1/N2}$ is to deposit a series of [N1($t_{N1}$)/N2($t_{N1}$)]$_n$ multilayers with fixed total thickness $t_T$ = 2$nt_{N1}$, and equal layer thicknesses, $t_{N1}$ = $t_{N2}$, of N1 and N2, and then to measure the total AR, AR$_T$, versus $n$. For the crossed superconductor geometry, the procedure requires depositing a thin Co layer between each superconducting Nb strip and the adjacent N-layer, but with the Co thick enough ($t_{Co}$ = 6-10 nm) to eliminate the superconducting proximity effect. If one neglects any interface thickness, then for any $n$, each metal N1 or N2 occupies half ($t_T$/2) of the multilayer, and the multilayer also contains (2$n$-1)AR$_{N1/N2}$ interfaces. The total AR$_T$ of a multilayer with $n$ layers should then be:

$$AR_T = 2AR_{Nb/Co} + 2\rho_{Co}t_{Co} + \rho_{N1}(t_T/2) + \rho_{N2}(t_T/2) - AR_{N1/N2} + n(2AR_{N1/N2}). \qquad (16)$$

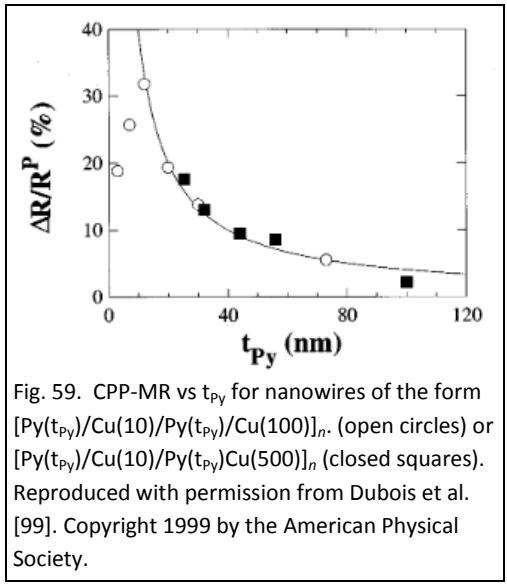

Fig. 59. CPP-MR vs $t_{Py}$ for nanowires of the form [Py($t_{Py}$)/Cu(10)/Py($t_{Py}$)/Cu(100)]$_n$ (open circles) or [Py($t_{Py}$)/Cu(10)/Py($t_{Py}$)Cu(500)]$_n$ (closed squares). Reproduced with permission from Dubois et al. [99]. Copyright 1999 by the American Physical Society.

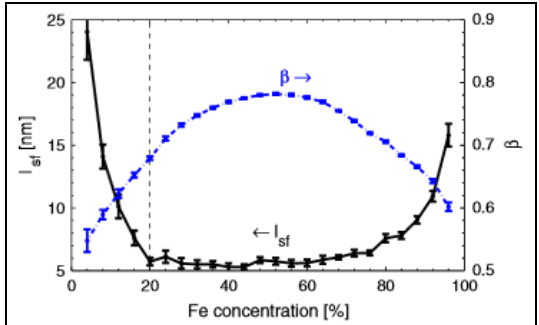

Fig. 60. Calculated $l_{sf}$ (filled squares, solid curve, and left scale) and β (xs, dashed curve, and right scale) for Ni(Fe%). Reproduced with permission from Starikov et al. [92]. Copyright 2010 by the American Physical Society.

A plot of $AR_T$ vs $n$ should give a straight line with intercept equal to the sum of the 5 constant terms and slope = $2AR_{N1/N2}$. Assuming, instead, a finite interface thickness $t_i$, gives two alternatives. First, if the total growth in $AR_T$ as $n$ increases is associated with the interfaces, then Eq. 16 should still apply, until the common layer thickness decreases to the interface thickness, $t_i$, after which the multilayer should become a uniform alloy with constant limiting AR. This is the assumption made in [32] and the behavior shown in Fig. 61, with the resulting values of $AR_{N1/N2}$ listed in Table 9. Alternatively, one can subtract an assumed $(\rho_{N1} + \rho_{N2})(t_i/2)$ for each interface, giving an effective interface resistance $AR'_{N1/N2} = [AR_{N1/N2} - (\rho_{N1} + \rho_{N2})(t_i/2)]$ with the same form as just described (i.e., linear in $n$ until each layer becomes $t_i$ and then constant). To estimate the reduction due to the second term, take $t_i \sim 0.5$ nm, $\rho_{Ag} = 7$ nΩm and $\rho_{Au} = 13$ nΩm, giving $\sim 0.005$ fΩm². This value would be a 10% reduction for Ag/Au and similar estimates would give a 10% reduction for Ag/Cu and a 3% reduction for Au/Cu.

To check for possible systematic errors, two different total thicknesses were used in Fig. 61, $t_T = 360$ nm and 540 nm, and also two sputtering rates differing by a factor of two. No systematic differences were found. To within local fluctuations; AR increased linearly with $n$ for all three metal pairs and the data for different total thicknesses and sputtering rates overlapped.

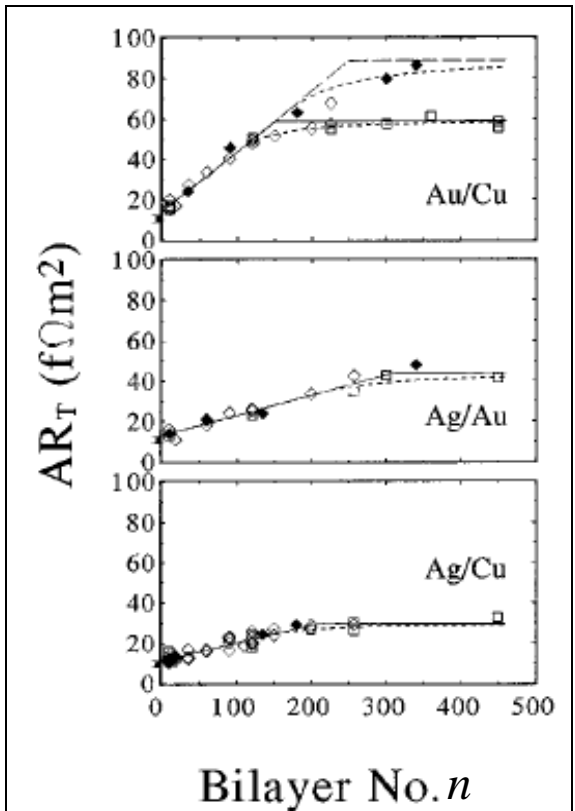

Fig. 61. $AR_T$ vs $n$ for [N1(t)/N2(t)]$_n$ multilayers with equal thickness (t) N1 and N2 layers and fixed $t_T$ = 360 nm (open symbols) or 540 nm (filled symbols). Diamonds are for samples sputtered at standard rates; squares are for samples sputtered at half rates. Reproduced with permission from Henry et al. [32]. Copyright 1996 by the American Physical Society.

For each metal, the mean-free-path $\lambda \sim 100$ nm corresponds to layer thicknesses with $n \sim 2$ to 3. Thus, most of the data in Fig. 61 lie in the range $t_{N1}$, $t_{N2} << \lambda$, where mfp effects would be predicted to appear if they exist. Clearly no mfp effects are required. But limits upon such effects are less clear. The data 'fluctuations' are so large that deviations of $\sim 10$-20% from $AR_T \propto n$ cannot be ruled out, although they could equally involve a decrease in $AR_{N1/N2}$ with $n$ as the mfp-effect-predicted increase. In section 8.14 we'll see that the absolute value of $2AR_{Ag/Au}$ falls between no-free-parameter calculations for a perfect, specular interface and a 2ML thick, disordered 50%-50% alloy interface. The uncertainties there are also $\sim 10$-20%.

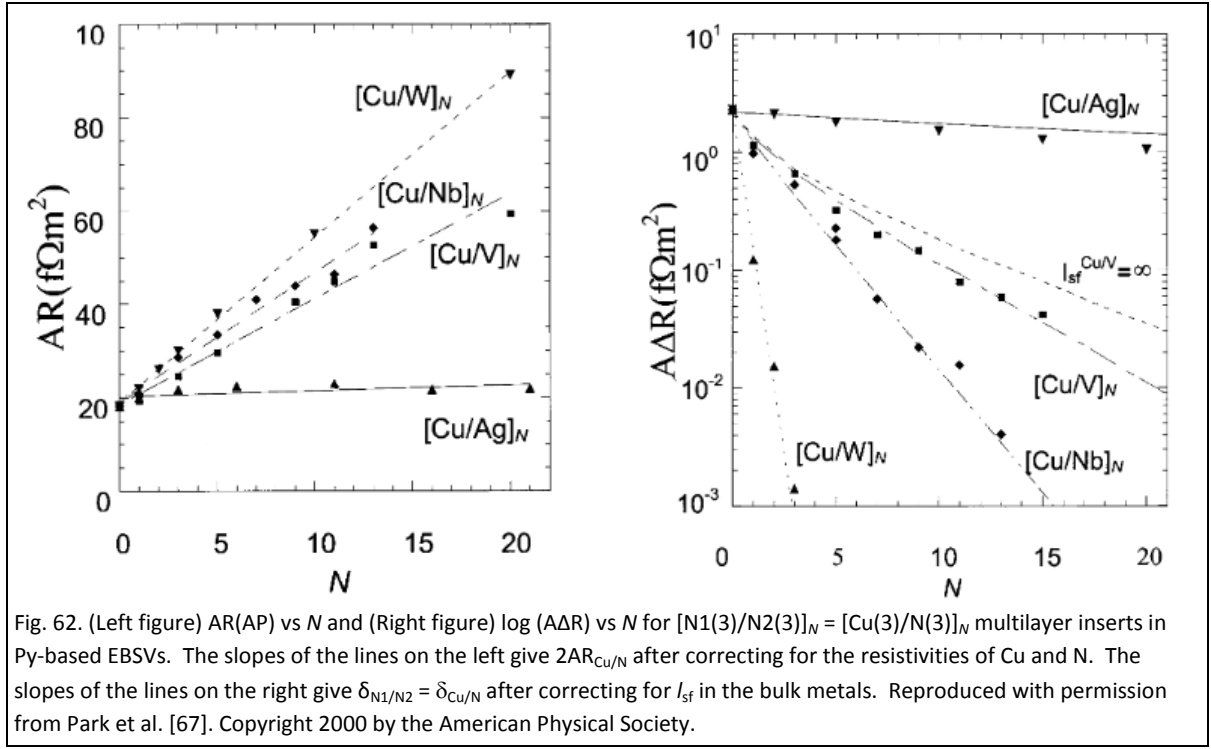

Fig. 62. (Left figure) AR(AP) vs $N$ and (Right figure) log (A$\Delta$R) vs $N$ for [N1(3)/N2(3)]$_N$ = [Cu(3)/N(3)]$_N$ multilayer inserts in Py-based EBSVs. The slopes of the lines on the left give 2AR$_{Cu/N}$ after correcting for the resistivities of Cu and N. The slopes of the lines on the right give $\delta_{N1/N2} = \delta_{Cu/N}$ after correcting for $l_{sf}$ in the bulk metals. Reproduced with permission from Park et al. [67]. Copyright 2000 by the American Physical Society.

Interface thicknesses in each sample set were estimated in two different ways, by seeing where the low angle x-ray multilayer peak disappeared as t$_{N1}$ = t$_{N2}$ was reduced, and by seeing where the data in Fig. 61 become constant. Lastly, the authors estimated the constant limits of AR in Fig. 61 in different ways, finding rough agreement with the experimental values. They concluded that interfacial alloying, giving a 50%-50% alloy, could 'explain' 40-100% of AR$_{N1/N2}$ for different pairs (see [32]).

Table 9 contains the values of 2AR$_{N1/N2}$ for Ag/Cu, Ag/Au, and Au/Cu, along with information about relative solubilities (ranging from 0% to 100%), resistivities per atomic percent impurity ($\Delta\rho_o/\Delta c$) for each metal in the other, resistivities for each metal, and estimates of interface thicknesses from x-rays and from where the linear regions in Fig.61 end. The interface thicknesses are typically ~ 0.6-0.8 nm, or 3-4 ML, consistent with x-ray or NMR studies by others of generally similar multilayers [31, 33-35].

**Table 9. Properties of Sputtered Ag/Cu, Ag/Au, and Au/Cu.** Values from [32] of 2AR$_{N1/N2}$, solubilities, independent measurements of $\rho_{N1}$ and $\rho_{N2}$, resistivities per atomic % impurity (N2 in N1; N1 in N2) in dilute alloys [37, 38]—(? = highly uncertain), and estimates of t$_i$ from x-rays and from Fig. 61.

| Metals (Properties) | Ag/Cu | Ag/Au | Au/Cu |
|---|---|---|---|
| 2AR$_{N1/N2}$ (f$\Omega$m$^2$) | 0.09 ± 0.01 | 0.10 ± 0.01 | 0.30 ± 0.01 |
| Solubilities | Small | 0% | Large |
| $\Delta\rho_o/\Delta c$ (n$\Omega$m/at.%) | 0.7; 1.4 | 3.6; 3.5 | 4.3; 5.3 |
| $\rho_{N1},\rho_{N2}$ (n$\Omega$m) | 7 ± 2; 6 ± 2 | 7 ± 2; 13 ± 3 | 13 ±3; 6 ± 2 |
| t$_i$ (nm) (x-rays) | 0.5 | 0.5 | 0.4 |
| t$_i$ (nm) (From Fig. 61) . | 0.9 ± 0.2 | 0.6 ± 0.1 | 1.2 ± 0.2 |

An intriguing feature of the data in Table 9 is the large difference between 2AR$_{Ag/Cu}$ and 2AR$_{Au/Cu}$, since the lattice parameters and Fermi surfaces of Ag and Au are very close. However, the solubilities of Ag/Cu and Au/Cu are opposite [167] and the resistivities of Ag(Cu) and Cu(Ag) alloys are very different from those for Au(Cu) and Cu(Au) [37]. So something important must be involved in the difference.

### 8.8.2. A more complex way to find $2AR_{N1/N2}$, that also gives $\delta_{N1/N2}$.

The second method, by Baxter et al. [159] and Park et al. [67], also with crossed-superconductors, involves depositing an $[N1(3)/N2(3)]_N$ multilayer in the middle of a Py-based symmetric EBSV, giving samples of the form $[FeMn(8)/Py(6)/Cu(10)/[N1(3)/N2(3)]_N/Cu(10)/Py(6)]$. Plotting AR vs $N$ should give a straight line, as shown in the left-hand figure in Fig. 62, with slope = $2AR_{N1/N2} + 3\rho_{N1} + 3\rho_{N2}$. Measuring the slope should give $2AR_{N1/N2}$ after correcting for the two $3\rho$ terms and (if needed) the one missing N1/N2 interface. 3 nm thick layers were chosen to be much thicker than the interfaces in Table 9, but thin enough that the $3\rho$ terms should not dominate the slopes. Table 10 contains values of $2AR_{N1/N2}$ found by both methods. In the two cases of overlap, Ag/Cu and Au/Cu, the two methods agree.

**Table 10. Experimental Values of $2AR_{N1/N2}$. (a) = method #1. (b) = method #2. [ ] = reference.** The pairs are listed in order of increasing 2AR.

| Metals | $2AR(f\Omega m^2)$ | Metals | $2AR(f\Omega m^2)$ | Metals | $2AR(f\Omega m^2)$ |
|---|---|---|---|---|---|
| Ag/Cu (a) | 0.09±0.01[ [32] | Pd/Au (b) | 0.45±0.15 [186] | Cu/Al (b) | 2.15±0.4 [187] |
| Ag/Cu(b) | 0.09 [67] | Pd/Ag (b) | 0.7±0.15[ [186] | Cu/Nb (b) | 2.2±03 [67] |
| Ag/Au (a) | 0.10±0.01 [32] | Cu/Pd (b) | 0.9±0.1 [182] | Cu/Ru (b) | ~ 2.2 [183] |
| Pd/Pt (a) | 0.28±0.06 [86] | Cu/Pd (b) | 0.85 [186] | Cu/V (b) | 2.3±0.3 [67] |
| Au/Cu (a) | 0.30±0.01 [32] | Pd/Ir (a) | 1.02±0.06 [85] | Cu/W (b) | 3.1±0.2 [67] |
| Au/Cu (b) | $0.35^{+0.1}_{-0.05}$ [181] | Cu/Pt (b) | 1.5±0.1 [182] | | |

The data in the right-hand picture in Fig. 62 will be analyzed in section 8.11.

### 8.9. Interleaved (I) vs Separated (S) Multilayers: Mean-Free-Path effects?

The only lengths in the 2CSR model are the layer thicknesses $t_F$ and $t_N$. Thus, if a 2CSR model is applicable to a hybrid SV of the interleaved (I) form $[F1/N/F2/N]_n$ where F1 and F2 have very different values of $H_c$, then $A\Delta R$ should be the same for the separated (S) form: $[F1/N]_n[F2/N]_n$. Several studies have found large deviations from this predicted equality. These deviations stimulated debate over whether they arise from spin-flipping (relaxation) or mean-free-path (mfp) effects.

### 8.9.1. (I) vs (S) for $[Co/Ag/Py/Ag]_N$

In 1997, Chiang et al. [155], compared AR(H) for I = $[Co(3)/Ag(20)/Py(8)/Ag(20)]_N$ and S = $[Co(3)/Ag(20)]_N[Py(8)/Ag(20)]_N$ hybrid SVs with $N$ = 2 to 8. Fig. 63 shows that AR(H) differs for I and S samples for all $N$ from $N$ = 2 and 8. Fig. 64 shows that the S data for $N$ = 8 are close to just the sum of data for two independent multilayers. Chiang et al. tentatively attributed the differences in Figs.63 and 64 to a presumed short spin-diffusion length in Py. Five months later a derivation of $l_{sf}^{Py}$ = 5.5 ± 1 nm [73] supported this attribution and provided an explanation for most of the differences—the short $l_{sf}^{Py}$ effectively isolated all but adjacent Py layers from each other. However, a short $l_{sf}^{Py}$ didn't explain why the Co layers also appeared to be isolated from each other. We'll address this issue below.

### 8.9.2. (I) vs (S) for $[Fe/Cu/Co/Cu]_n$

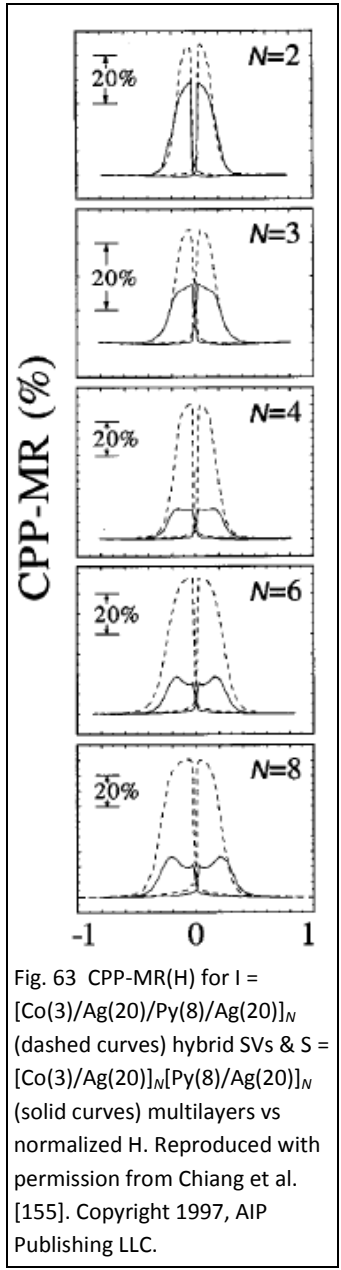

Fig. 63 CPP-MR(H) for I = [Co(3)/Ag(20)/Py(8)/Ag(20)]$_N$ (dashed curves) hybrid SVs & S = [Co(3)/Ag(20)]$_N$[Py(8)/Ag(20)]$_N$ (solid curves) multilayers vs normalized H. Reproduced with permission from Chiang et al. [155]. Copyright 1997, AIP Publishing LLC.

In 1999, Bozec et al. [158], found similar differences in CPP-MRs for *n* = 2 *to* 8 with I = [Fe(2)/Cu(20)/Co(6)/Cu(20)]$_n$, and S = [Fe(2)/Cu(20)]$_n$[Co(6)/Cu(20)]$_n$.(Fig. 65). Arguing that the known $l_{sf}^{Co}$ was long, and that the unknown $l_{sf}^{Fe}$ was unlikely to be as short as $l_{sf}^{Py}$, Bozec et al. attributed the differences in AR(H) to differences in the number of 'boundary layers'—e.g. in the AP state for *n* = 8, the S multilayer has 7 adjacent Fe layers and 7 adjacent Co layers all oriented P to each other, and only one adjacent Fe and Co 'boundary' pair oriented AP. They argued that such an ordering could explain the observed 'nearly simple sum' AR(H) of Fig. 64, if all that was important was the relative orientation of adjacent F-layers. In 2000, Bozec [60] found $l_{sf}^{Fe}$ = 8.5 ± 1.5 nm, short enough to weaken, but not eliminate, the need for a 'new idea', especially for the Co isolation.

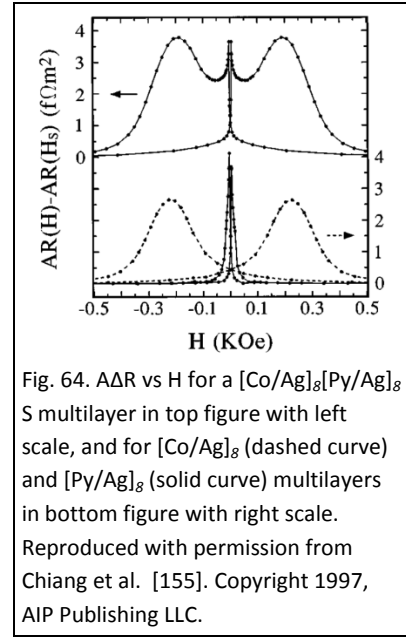

Fig. 64. AΔR vs H for a [Co/Ag]$_8$[Py/Ag]$_8$ S multilayer in top figure with left scale, and for [Co/Ag]$_8$ (dashed curve) and [Py/Ag]$_8$ (solid curve) multilayers in bottom figure with right scale. Reproduced with permission from Chiang et al. [155]. Copyright 1997, AIP Publishing LLC.

### 8.9.3. (I) vs (S) for [Co(1)/Cu/Co(6)/Cu]$_n$

Later, in 2000, Bozec et al. [79] showed similar differences in AR(H) for [Co(1)/Cu(20)/Co(6)/Cu(20)]$_n$ and [Co(1)/Cu(20)]$_n$[Co(6)/Cu(20)]$_n$ hybrid SVs, and argued correctly that the accepted spin-diffusion lengths in both Cu (≥ 100 nm) and Co (> 60 nm) [98] were too long to explain the differences. Similar examples of these differences are shown in Figs. 66 and 67 from [185], which we use because they include data for AR(0). These data show that AR(0) for both I and S samples approximates AR(AP), especially for S samples, where all adjacent Co layers have the same thickness, except for two. Bozec et al. gave two different models to explain their data, both of which they called 'mean-free-path' (mfp) effects [79, 188]. The first model was that when the mfp, λ, becomes longer than the total thickness of the 4-layer unit of the I hybrid SV, the 2CSR model is no longer valid, and (as argued above) all that matters is the relative magnetic orientations of adjacent F-layers [79]. Although we will call this model mfp#1, it is really just a nearest-neighbor (nn) model, since neither the λs of individual layers, nor an effective λ$_T$ including scattering at interfaces, appear as observable parameters. The authors presented as supporting evidence for mfp#1 the ability to fit the complete AR(H) curves with only 3 adjustable parameters. The second model, a calculation based on prior work by Tsymbal [71, 76], attributed the observed behavior to quantum well states in the multilayer. We'll call this second argument 'mfp#2'. Both models agree that the difference between I and S multilayers should disappear, and the 2CSR model should apply, when disorder grows to where λ becomes much less than the layer thicknesses. But 'mfp#1' requires a 'total λ' longer than the total thickness t$_r$ of a 'repeating unit' in I, whereas mfp#2 requires only a λ comparable to t$_r$. Also the physics in the two cases differs.

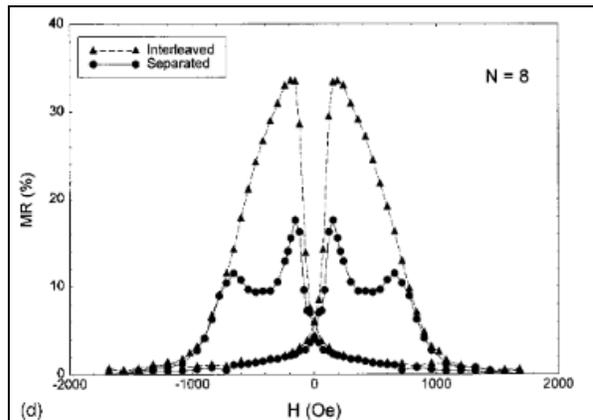

Fig. 65. CPP-MR(H) vs H for I = [Fe/Cu/Co/Cu]$_8$ (triangles*)* and S= [Fe/Cu]$_8$[Co/Cu]$_8$ (circles) hybrid SVs. Reproduced with permission from Bozec et al. [158]. Copyright 1999 by the American Physical Society.

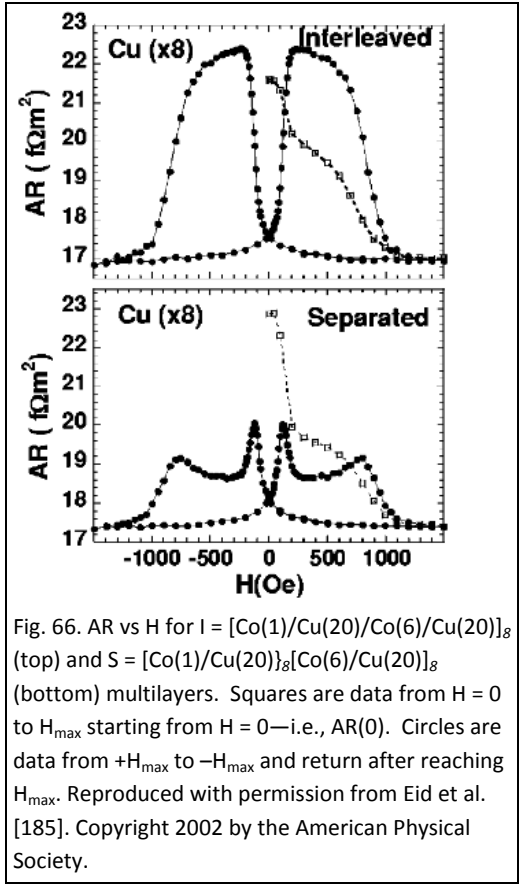

Fig. 66. AR vs H for I = [Co(1)/Cu(20)/Co(6)/Cu(20)]$_8$ (top) and S = [Co(1)/Cu(20)]$_8$[Co(6)/Cu(20)]$_8$ (bottom) multilayers. Squares are data from H = 0 to H$_{max}$ starting from H = 0—i.e., AR(0). Circles are data from +H$_{max}$ to −H$_{max}$ and return after reaching H$_{max}$. Reproduced with permission from Eid et al. [185]. Copyright 2002 by the American Physical Society.

Before examining the arguments of spin-flipping vs mfp effects, we list what is agreed upon. Everyone agrees upon all of the data. The issue Is only interpretation.

(1) The original data in [79] are correct; they have been independently repeated.

(2) Magnetizations of both I and S multilayers are essentially identical [185].

(3) After samples are taken to saturation above H$_s$, 'AP' magnetic orderings of both I and S multilayers are as expected [185]—standard AP for I SVs and half up and half down for S SVs.

(4) The 2CSR model alone cannot explain the differences between I and S multilayers.

(5) If one starts from the 2CSR model, where the relative orientations of the magnetizations of all of the F-layers matter (i.e., all of the F-layers 'communicate'), then explaining the S data requires a way to reduce the ability of F-layers to communicate over long-distances. 'mfp#1' simply postulates that only nearest-neighbor F-pairs communicate. 'mfp#2' incorporates coherent scattering between interfaces. The VF model requires sources of spin-flipping to weaken the communication, either finite spin-diffusion lengths in the F- or N-layers, or spin-flipping at F/N interfaces.

So three basic questions are: (1) Are plausible 'spin-flipping' mechanisms available? (2) Can mfp#1 and/or mfp#2 explain all of the other CPP-MR data that have been explained with the 2CSR or VF models? (3) Are the claimed physics of the mfp#1 or mfp#2 models plausible? We'll answer these questions at the end of section 8.9.3.2.

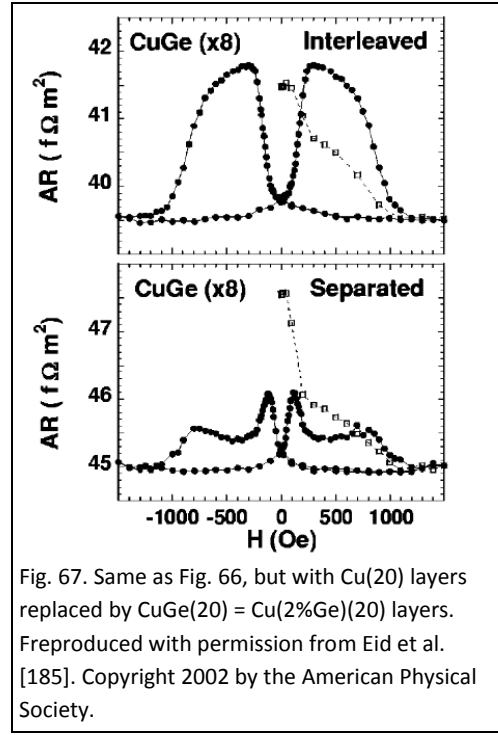

Fig. 67. Same as Fig. 66, but with Cu(20) layers replaced by CuGe(20) = Cu(2%Ge)(20) layers. Freproduced with permission from Eid et al. [185]. Copyright 2002 by the American Physical Society.

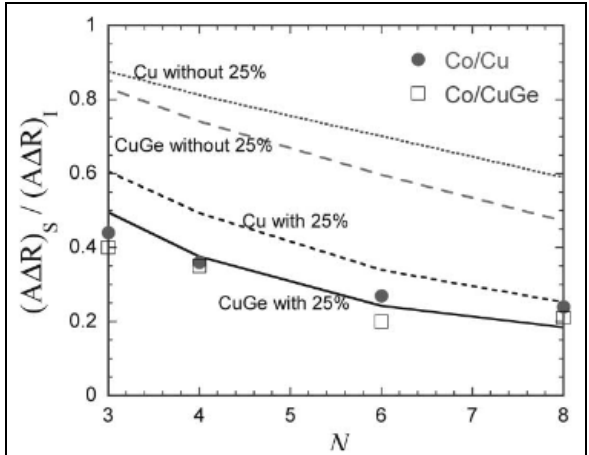

Fig. 68. $(A\Delta R)_S/(A\Delta R)_I$ for separated (S) and interleaved (I) = $[Co(1)/Cu(20)/Co(7)/Cu(20)]_N$ (data = filled circles) or I = $[Co(1)/CuGe(20)Co(7)/CuGe(20)]_N$ (data = open squares). The dotted (Cu) and long-dashed (CoGe) curves are VFcalculations with spin-diffusion lengths $l_{sf}^{Co}$ = 60 nm, $l_{sf}^{Cu}$ = 500 nm, and $l_{sf}^{CuGe}$ = 130 nm and $\delta_{Co/Cu}$ = 0. The short dashed (Cu) and solid (CuGe) curves are the VF calculations plus $\delta_{Co/Cu}$ = 0.25. Reproduced with permission from Eid et al. [189]. © 2001 by Elsevier. [Note: The larger $\delta_{Co/Cu}$ = 0.35 derived in section 8.15 would bring the short-dashed curve down to about the Cu data and do more than enough for the CoGe data.

We now turn to the debate, starting with 'mfp#1'.

### 8.9.3.1. (I) vs (S) for $[Co(1)/Cu/Co(6)/Cu]_N$, 'mfp#1'.

In 2001, Eid et al. [189] showed that the differences between AR(H) for I and S multilayers were unchanged when Cu(20nm) was replaced by Cu(2%Ge)(20nm) (Fig. 67) or even by Cu(2%Ge)(40nm) [185], thereby reducing the mean-free-path to layer thickness ratio $\lambda_{Cu}/t_{Cu}$, by factors ~ 12 and 25 (i.e., from ~ 5 to ~ 1/5) respectively, while still leaving $l_{sf}^N$ long (see section 8.5.1). In Fig. 68 they also showed that VF theory could partly explain the observed reduction in the ratio $(A\Delta R)_S/(A\Delta R)_I$ for separated (S) and interleaved (I) multilayers using reasonable spin-diffusion lengths for Co, Cu, and/or CuGe, and that postulating a spin-flip parameter $\delta_{Co/Cu}$ ~ 0.25 at Co/Cu interfaces gave a more complete explanation. Prior evidence of non-zero δs at N1/N2 interfaces [67] gave a rationale for such a value, but no direct measurement existed. By giving a way to greatly weaken spin-polarization transfer between adjacent Co layers, $\delta_{Co/Cu}$ ~ 0.25 would also complete the explanations for the differences in A$\Delta$R for I and S multilayers of Py and Co and of Fe and Co

In 2002, Eid et al. [190] added two new experimental arguments in favor of spin-flipping. First, to show that the differences in AR(H) for I and S SVs did not require long ratios of $\lambda/t$ for any of the component layers, they substantially reduced the ratios for all three SV layers by replacing Cu(20) by CuGe(20 or 30), Co(1) by Py(30), and Co(6) by CoZr(15 or 30). Fig. 69 shows that the differences between AR(H) for S and I remained. Second, to demonstrate the importance of spin-flipping, they inserted into the central Cu layer of an S multilayer a thin (1 nm) layer of FeMn, which previous work had shown caused very strong 'spin-flipping' [67] (see Fig. 43 and section 8.5.2.2). If inserting the FeMn only reduced the $\lambda$ in that Cu layer, the 'mfp' model would predict a slight decrease in the difference between I and S multilayers.

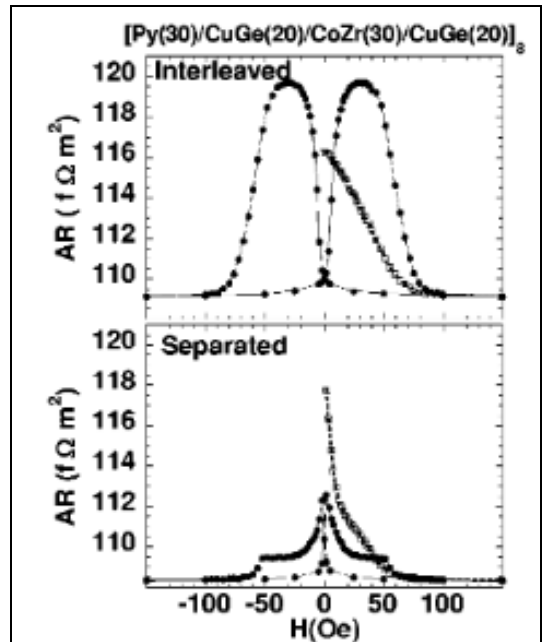

Fig. 69. AR vs H for I (top) and S (bottom) hybrid SVs made of CoZr, CuGe, and Py. Reproduced with permission from Eid et al. [185]. Copyright 2002 by the American Physical Society.

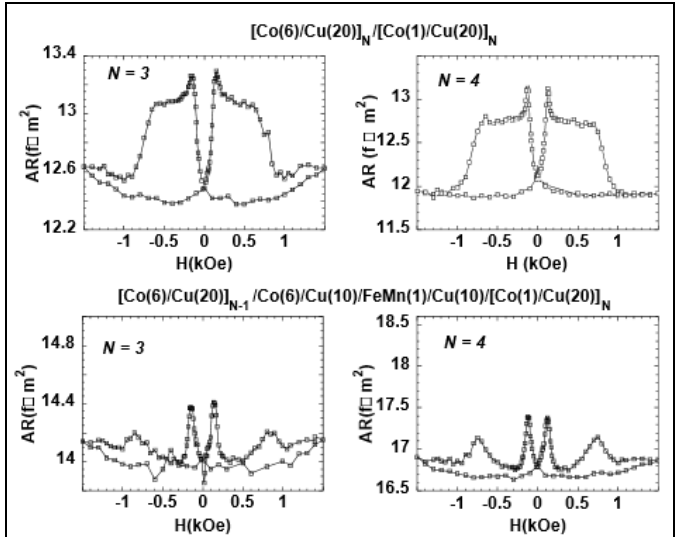

Fig. 70. AR vs H for standard Co/Cu S samples (top) and similar samples with 1 nm of FeMn inserted into the middle of a central Cu layer (bottom). Reproduced with permission from Eid et al. [190]. © 2002 by Elsevier.

But Fig. 70 shows that inserting the FeMn greatly increased the difference in the S sample, completely separating the peaks for the two different Co layer thicknesses. Eid et al. argued that this behavior showed the dominance of spin-flipping. In a longer paper, Eid et al. also showed [185] via Fig. 71 that their postulated $\delta_{Co/Cu} = 0.25$ would have only minor effects on A$\Delta$R for simple I-multilayers, and via Fig. 49 that it largely resolved the discrepancies between predictions and data for Co/Cu EBSVs in Figs. 47 and 48. It would also 'resolve' the 'isolation' of Co layers in Py plus Co and Fe plus Co hybrid SVs above.

In 2003, Michez et al. [191] returned to the fray with data for S and I samples with 3 different F-layers. They argued that the data could be qualitatively understood considering only the relative orientations of the moments of adjacent F-layer pairs. While this argument looks okay, it is not unique. Similar behavior can be produced by spin-flipping at the F/N interfaces. They also made two statements that we believe are incorrect.

(1) They disputed the Eid claim that going from Cu(20) to CuGe(40) reduced λ/t from 5 to 1/5. Based on their own calculations [191], they argued that the 'neck' deviations of the Fermi surface of Cu from sphericity should lead to a much larger value of $\rho_b\lambda_b$ and thus larger λ. Choosing, without details, a factor of 5 in increase in $\rho_b\lambda_b$ gave an inferred λ/t = 1 for CuGe(40), which they claimed was large enough to sustain their mean-free-path argument. Unfortunately, two sets of data contradict this choice. First, Table A in Appendix A shows that Eid et al's 'shorter' $\rho_b\lambda_b$ for Cu is supported by independent measurements of both size effects and the anomalous skin effect. Second, the agreements between measured and calculated values of $l_{sf}^N$ in Table 6 are based upon the same value of $\rho_b\lambda_b$ for Cu.

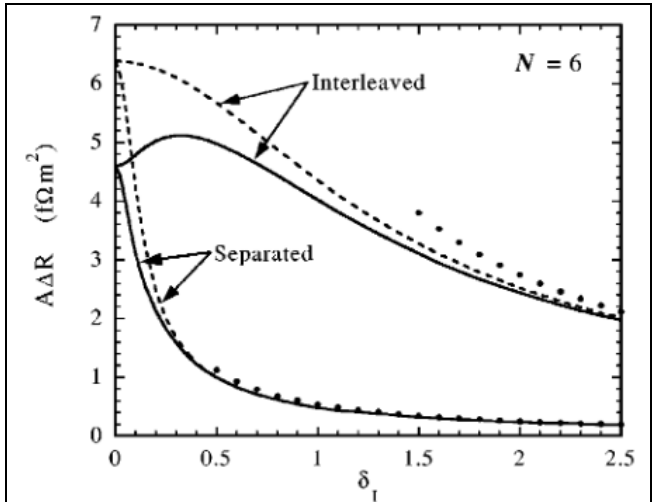

Fig. 71. Calculated A$\Delta$R vs $\delta_I$ (interface $\delta$) for I and S Co/Cu multilayers with $N$ = 6. The solid curves include the Co/Nb contacts; the dashed curves omit them. The filled circles show an analytical form that the calculated solid and dashed curves should approach in the limit $\delta_I \gg 1$. Reproduced with permission from Eid et al. [185]. Copyright 2002 by the American Physical Society.

(2) They also stated that Gijs and Bauer, in their early comprehensive review of CPP-MR [4], had 'pointed out that the requirement for applicability of the 2CSR model is that λ << t', referring the reader especially to pages 301-302 of [4]. Unfortunately, this statement is a misunderstanding. Gijs and Bauer said only that 'the limit λ << t is called the 'local limit', in which the inhomogeneity of the sample must be taken into account locally and where quantum size effects may be disregarded'. At that point, they said nothing about the 2CSR model. But elsewhere it is clear that they are comfortable with use of the VF 'semi-classical' model, and its 2CSR limit for long spin-diffusion lengths, to fit data on present day multilayers with their disordered interfaces.

Then in 2004, Michez et al. [192] showed that they could produce clearer two-peak behavior in S multilayers by inserting into the central Cu layer either 5 nm of Ta or 2 nm of Ru, in each of which large spin-orbit interactions should 'erase spin-memory'. They argued that this behavior supported their 'mfp' model. But the changes that they saw are similar to those in Fig. 70. And their argument attributes the changes to spin-flipping, which is just what Eid et al. had done. They seem to confirm that spin-flipping can produce effects similar to those that they had ascribed to 'mfp' effects.

More recently, Michez et al. published three further papers [193-195], two involving fits to CPP data with adjustable parameters, and the other giving only a qualitative discussion. In the first one, [193], they neglected mfp effects to present a simple phenomenological model for the ratio CPP-MR($n$)/CPP-MR(1) vs $n$ for I and S hybrid SVs containing Py(8) and CoFe(3) F-layers. Using one adjustable parameter each for I and S, they obtained results for $n$ = 2 to 6 that overlapped with a VF calculation by Strelkov et al. [68] that had assumed $l_{sf}^{Py}$ = 5 nm and $l_{sf}^{CoFe}$ = 15 nm. This Michez paper showed that fits to data are often not unique, but did not advance the case for mfp effects. The other two papers involved mfp effects. In [194], their arguments were only qualitative. They showed that AΔR for I = [Py(8)/Cu(20)/Co(3)/Cu(20)]$_n$ hybrid SVs grew linearly with $n$ and that AΔR for S = [Py(8)/Cu(20)]$_n$[Co(3)/Cu(20)]$_n$ hybrid SVs grew linearly or even as a slightly higher power of $n$. In contrast, they noted that AΔR for [Py(t)/Cu(20)/Py(t)] EBSVs saturated with increasing $t$. They attributed this difference in behavior with growth of total Py thickness to mfp effects in the Py. However, the difference between linear growth with $n$ in multilayers, and saturation in EBSVs with t, both occur naturally for I layers in VF theory. Both data (Figs. 24, 29, 32), and extensions of Eq 6 to the I form of hybrid SV multilayers with $n$ layers, give AΔR linear with $n$ for both $t_F < l_{sf}^F$ or $t_F > l_{sf}^F$. And, as in Eq. 9, VF theory gives saturation for a simple single symmetric EBSV when $t_f >> l_{sf}^F$. So their qualitative discussion does not give strong evidence for mfp effects. Lastly, in [195] they show CPP-MRs vs t for [Py(t)/Cu(20)/Co(3)/Cu(20)]$_4$ or [Co(t)/Cu(20)/Co(3)/Cu(20)]$_4$ hybrid SVs in either I-form as listed or in S-form with the first two layers separated from the last two. They fit the data using a mfp#1 model with an unspecified number of adjustable parameters for each sample set. No equations or values of the parameters are given. Since the data vary smoothly, only a few adjustable parameters should be enough with any simple model. Thus, this paper also doesn't strengthen the case for mfp effects.

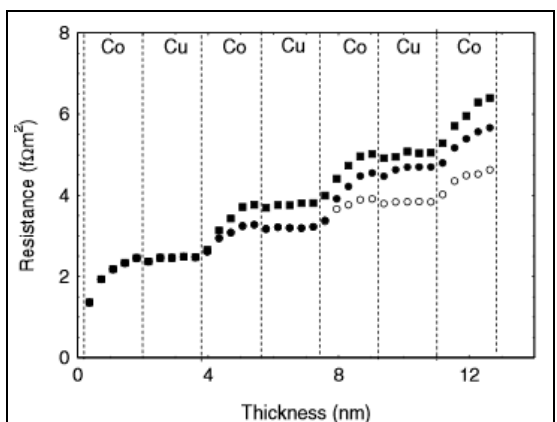

Fig. 72. R vs t, where t = total thickness of a sample composed of 2 nm thick alternating layers of Co and Cu. The bottom curve is for the P state. The top curve is for the AP state. The middle curve is for the S state, which only becomes distinct with the third Co layer. Reproduced with permission from Bozec et al. [79]. Copyright 2000 by the American Physical Society.

To summarize, the differences between I and S Co/Cu multilayers first seen by Bozec et al. have been confirmed. What is in dispute is their interpretation. The 'mfp#1' interpretation by Bozec et al. [79, 188, 191-195], is that all that matters in CPP-MR is the relative orientation of the moments of adjacent F-layers. Eid et al. [185, 189, 190] have argued that, while this model can fit some data sets, it cannot fit others with very short local $\lambda$s. The other interpretation, by Eid et al. [185, 189, 190] is that mean-free-paths are not involved, and that the differences between I and S data for different sets of multilayers are due to spin-flipping, either in bulk F-metals for finite spin-diffusion lengths, $l_{sf}^{F}$, or at interfaces, $\delta_{Co/Cu}$. Section 8.15 gives a direct measurement of $\delta_{Co/Cu}$ consistent with this latter source.

**8.9.3.2. (I) vs (S) for [Co(1)/Cu/Co(6)/Cu]$_n$ 'mfp#2'.** The mfp#2 model presented by Bozec et al. [79, 188] showed differences between I and S multilayers using a simplified system of alternating Co and Cu layers with $t_{co} = t_{cu} = 2$ nm. Disorder was introduced by a scaling parameter that produced random distributions of a scattering potential in both Cu and Co that gave resistivities $\rho_{Cu} = 46$ n$\Omega$m and $\rho_{Co} = 143$ n$\Omega$m, both larger than the experimental ones ($\rho_{Cu} \sim 6$ n$\Omega$m and $\rho_{Co} \sim 70$ n$\Omega$m). From tabulated resistivities per atomic percent impurity [37] the larger resistivities correspond to $\sim 0.8$ at.% Co in Cu and (more uncertain) $\sim 16\%$ Cu in Co, and give mean-free-paths of $\lambda_{Cu} = 14$ nm and $\lambda_{Co} \sim 7$ nm, both shorter than those for the experimental layers. The model contains no intermixing at the interfaces, which are characterized just by their potential steps. The resulting effects are shown in Fig. 72 [79] as a plot of AR vs total sample thickness $t_T$ as layers of Co and Cu are sequentially added. In the 2CSR model, AR would increase linearly as $\rho t$ grows in each layer and then more rapidly at (or near) an interface, for either a perfectly planar or an intermixed interface. In Fig. 72, in contrast, AR grows most rapidly as Co is first added on Cu, the rate of increase slows as the interface is approached, AR actually decreases slightly as the first Cu layer is added and then stays essentially constant until the next interface is reached, after which the process repeats. Adding the third Co layer lets one begin to distinguish between I and S order, with AR for S falling below AR for I. Although the calculation is for an (001) multilayer, and the experimental data are for (111), the increase in AR from the middle of one Cu layer to the middle of the next is almost the same as estimated from VF Co/Cu bulk and interface parameters plus the defined $\rho_{Cu} = 46$ n$\Omega$m and $\rho_{Co} = 143$ n$\Omega$m. According to Bozec et al., the differences between I and S in Fig. 72, are due to the presence of quantum well states associated with the change in electron potential at the Co/Cu interfaces. So the crucial physics question is: Do coherent quantum well states remain between interfaces with both local disorder and randomly varying separations over the mm$^2$ area of the real CPP multilayers? An intriguing feature of Fig. 72, apparently a characteristic of the calculation, is that interfaces only appear when Co is deposited on Cu, but not when Cu is deposited on Co.

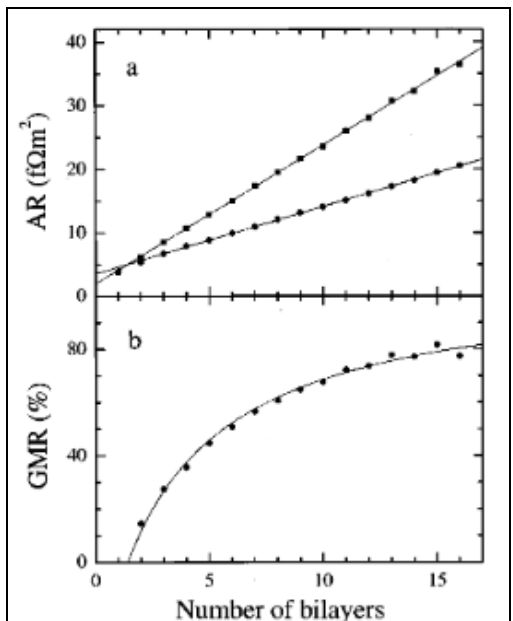

Fig. 73. Calculated (a) AR and (b) CPP-MR vs n of a Co(10ML)/Cu(10ML)$_n$ multilayer for AP(squares) and P (circles) order. Solid lines in (a) are linear fits; solid curve in (b) is resulting CPP-MR. Reproduced with permission from Tsymbal [76]. Copyright 2000 by the American Physical Society.

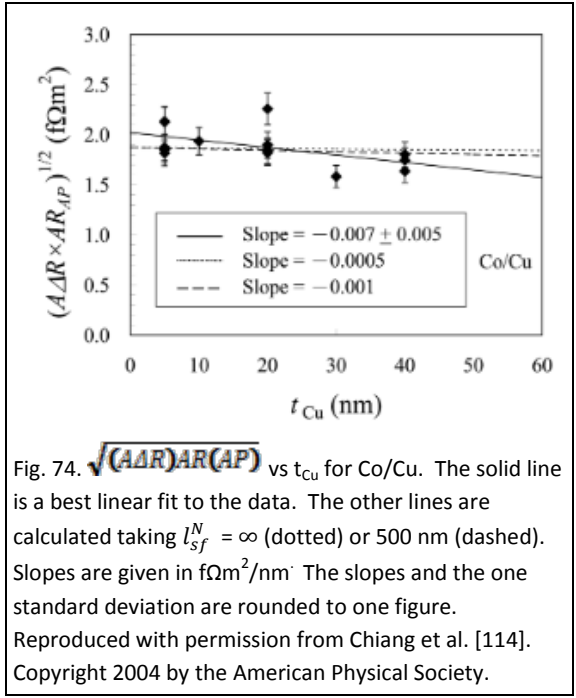

Fig. 74. $\sqrt{(A\Delta R)AR(AP)}$ vs $t_{Cu}$ for Co/Cu. The solid line is a best linear fit to the data. The other lines are calculated taking $l_{sf}^N$ = ∞ (dotted) or 500 nm (dashed). Slopes are given in fΩm²/nm. The slopes and the one standard deviation are rounded to one figure. Reproduced with permission from Chiang et al. [114]. Copyright 2004 by the American Physical Society.

To address the issue of quantum well states, we examine a prediction of an earlier paper on Co/Cu multilayers by Tsymbal using the same model [8]. In Fig. 73 [76], for a [Co(10ML)/Cu(10ML)]$_n$ multilayer with $n$ repeats of 10 monolayers (ML) each of Co and Cu, this paper predicts that extrapolating AR(AP) and AR(P) to $n$ = 0 will give different intercepts that would yield a 'negative' A$\Delta$R at $n$= 0 and A$\Delta$R = 0 at a non-zero value of $n$ = 1.5 ML.

The prediction that A$\Delta$R → 0 at finite $n$ means that the $\sqrt{(A\Delta R)AR(AP)}$ vs $n$ should also go to 0 at a positive value of $n$. Examples of $\sqrt{(A\Delta R)AR(AP)}$ vs $n$ are given in Fig. 21 for Co/Ag, in Fig. 22 for Co/Cu and Co/CuGe, and in Fig. 89 below for Fe/Cr. The data for Co/Ag in Fig. 21, and for Co/CuGe in Fig. 22, and the solid symbols for Fe/Cr in Fig. 89 below all go through 0,0. The data for Co/Cu in Fig. 21 may have an offset, and the open symbols for Fe/Cr in Fig. 89 prefer an offset. The most reliable of these results do not support mfp#2 effects.

To further test mfp #2 with $\sqrt{(A\Delta R)AR(AP)}$ data, Chiang et al. [114] made new measurements on [FeMn(8)/Co(6)/N($t_N$)/Co(6)] EBSVs with fixed $t_{Co}$ = 6 nm and variable $t_N$ for N = Cu, Ag, and Au. EBSVs were chosen to give reliable AP states. As shown by Eq. 7, the 2CSR model for a multilayer or EBSV with only $t_N$ varied predicts that a plot of $\sqrt{(A\Delta R)AR(AP)}$ vs $t_N$ should give a horizontal line. Either a finite spin-diffusion length, or mfp#2 effects, could cause this line to have a finite downward slope, in the former case with the slope growing with increasing t and in the latter case with it decreasing. A slight downward slope for [Co/Ag]$_n$ multilayers was already seen in Fig. 32. Figs. 74-76 show the results of this new test with EBSVs. For all three metals, best fit straight lines slope downward, with comparable slopes for Cu and Ag and a larger slope for Au. For both Ag and Au, the observed slopes can be explained by expected spin-diffusion lengths for the given metal resistivities. However, for Cu, the slope is too large to explain this way. The slope is consistent in size with the ballistic model (mfp#2) with the disorder parameter given above. However, mfp #2 would predict a decreasing slope with increasing t, whereas the data of Figs. 74 and 75 are also consistent with no slope for t ≤ 20 nm, and then a slope. The conclusion of Chiang et al. was as follows. Fitting AR(AP) and AR(P) for these samples (see [114]) did not need mfp #2 effects. Understanding AΔR for Ag and Au also did not need mfp#2 effects. The best fit to AΔR for Cu was consistent with mfp#2 effects, but uncertainty in the slope was just about a standard deviation from zero (and even less if a plausible $l_{sf}^{Cu}$ is included). Unfortunately, the calculation could not be extended to Ag or Au. We conclude that the need for a ballistic model for all of these results is marginal. While the model cannot be ruled out, any contributions seem small enough to not greatly perturb 2CSR or VF model analyses that neglect it.

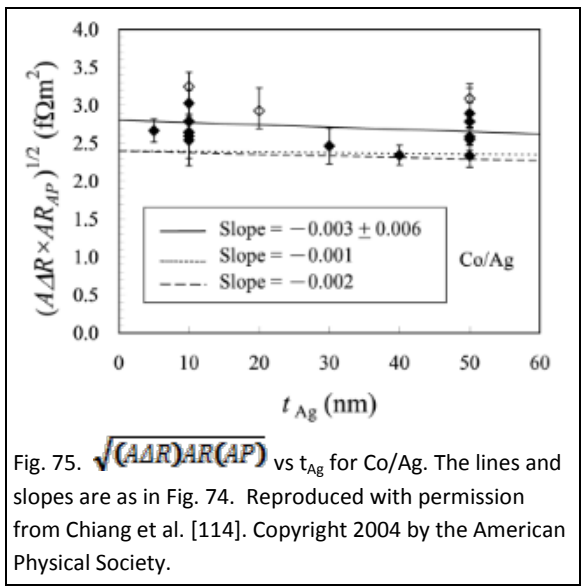

Fig. 75. $\sqrt{(A\Delta R)AR(AP)}$ vs $t_{Ag}$ for Co/Ag. The lines and slopes are as in Fig. 74. Reproduced with permission from Chiang et al. [114]. Copyright 2004 by the American Physical Society.

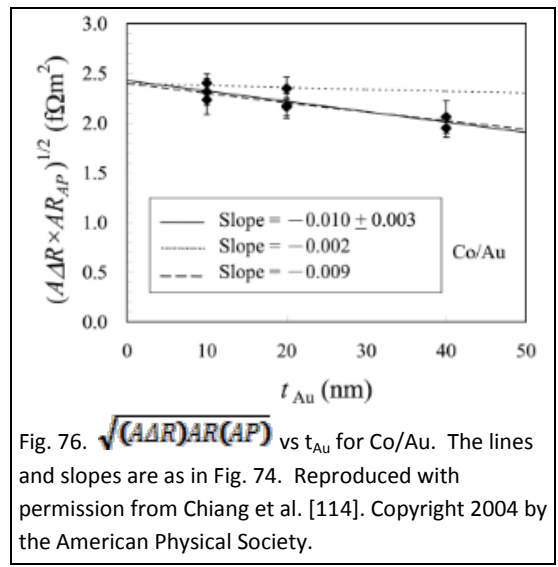

Fig. 76. $\sqrt{(A\Delta R)AR(AP)}$ vs $t_{Au}$ for Co/Au. The lines and slopes are as in Fig. 74. Reproduced with permission from Chiang et al. [114]. Copyright 2004 by the American Physical Society.

To finish, we return to the three questions listed above: (1) Are plausible 'spin-flipping' mechanisms available? (2) Can a nearest neighbor only, no-spin-flips, model explain all of the other CPP-MR data that have been explained with VF models? (3) Are the physics of mfp#1 or mfp#2 plausible?

We answer the first question 'yes'. Independent evidence has been presented for sufficiently large spin-flipping at the Co/Cu interfaces (section 8.15) to resolve all issues with Co in the hybrid structures.

We answer the second question 'no'. In addition to the problems indicated above, proponents of 'mfp' or 'ballistic' effects have not explained how mean-free-paths alone can explain the differences in $\sqrt{(A\Delta R)AR(AP)}$ data for AgSn vs AgPt in Fig. 21 or for CuGe vs CuPt in Fig. 41, since the resistivities (and thus mfps) are similar for each pair. The claim [8] that the data in Fig. 43 might also be due to mfp #2 effects was answered similarly in section 8.5.2.2.

We answer the third question 'unlikely' for 'mfp#1', but less clear for mfp#2. However, any practical effects of the latter seem to be small in multilayers with present interfaces.

### 8.10. $\beta_F$ (tests of 'Unversality') and $l_{sf}^F$ for F-alloys.
### 8.10.1. "Universality" of $\beta_F$ for F-Alloys?

An important question raised in section 1.5 is whether or not VF CPP parameters are 'universal'—i.e., whether their values are similar to those obtained by very different techniques. In this section we examine this relationship for $\beta_F$ in F-alloys where the dominant scatterer is well defined. We compare CPP-MR values of $\beta_F$ for F-alloys with Deviations from Matthiessen's Rule (DMR) values of $\beta_F$ for the same F-alloys, found in the 1970s [36]. Because the CPP-MR values are for finite impurity concentrations, while the DMR values are for the dilute limit, we need not expect exact agreement. Moreover, because a few percent of many impurities is enough to cause significant spin-flipping (relaxation) as conduction electrons traverse F-layers, correctly determining the CPP $\beta_F$ usually requires simultaneously determining $l_{sf}^F$. In this section we consider these two parameters, $\beta_F$ and $l_{sf}^F$, together.

The first direct tests of 'universality' were made in 1997 by Hsu et al. [156] and Vouille et al. [196], who focused upon F-alloys where the DMR values of $\beta_F$ are negative, on the basis that 'universality' should certainly extend to 'sign'. Their method was to produce an 'inverse' CPP-MR —i.e., AR(AP) < AR(P)--as follows. If $\beta_F$ and $\gamma_{F/N}$ both have the same sign for adjacent F-layers, the CPP-MR will always be 'normal'—i.e., AR(AP) > AR(P), independent of whether $\beta_F$ and $\gamma_{F/N}$ are both positive or negative.

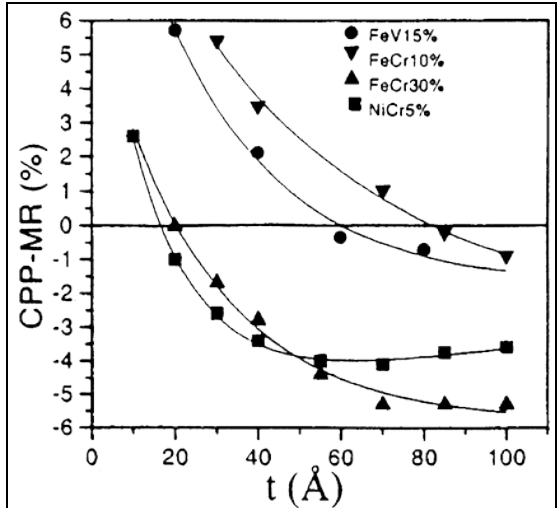

Fig. 77. CPP-MR vs t for [Co(0.4)/Cu(2.3)/X(t)/Cu(2.3)]$_{20}$ hybrid SVs showing inverse MR with sufficiently thick X = FeV(15%); FeCr(10%); FeCr(30%); or NiCr(5%) layers. Reproduced with permission from Vouille et al.[196]. Copyright 1997, AIP Publishing LLC.

The reason is that, when the signs of the parameters in adjacent F-layers are the same, in the P-state, one electron moment direction will always short out the sample. In contrast, if one makes an [F1/N/F2/N]$_n$ multilayer with either $\beta_F$ or $\gamma_{F/N}$ having opposite signs for F1 and F2, then the shorting will occur in the AP-state, giving AR(AP) < AR(P)—i.e. 'inverse' CPP-MR.

To look for 'inverse' CPP-MRs, Hsu et al. [156] and Vouille et al. [196] (see also [197]) used hybrid SVs of the form [A/Cu/Co/Cu]$_n$ with A = Nl(Cr), Fe(V), or Fe(Cr), in all of which negative $\beta$s were found from DMR [36] (see Table 1). They fixed t$_{Co}$ and measured CPP-MR as t$_A$ increased. The 2CSR model gives:

$$A\Delta R = n^2[(\beta_{Co}\rho_{Co}^* t_{Co} + 2\gamma_{Co/Cu} AR_{Co/Cu}^*)(\beta_A \rho_A^* t_A + 2\gamma_{A/Cu} AR_{A/Cu}^*)]/AR(AP) \qquad (17)$$

From prior studies, they expected $\beta_{Co}$, $\gamma_{Co/Cu}$, and $\gamma_{A/Cu}$ to all be positive ($\gamma_{A/Cu}$ because in the dilute limit A just becomes Ni or Fe, which have $\gamma_{F/Cu} > 0$), but $\beta_A$ to be negative. If so, then A$\Delta$R and CPP-MR should be normal for small t$_A$, when the positive interface term in the right-hand parenthesis is dominant, but become inverse for large t$_A$, when the negative bulk term in the right-hand parenthesis becomes dominant. That is, increasing t$_A$ should cause AR to 'invert' from 'normal' AR(AP) > AR(P) for small t$_A$ to 'inverse' AR(P) > AR(AP) for large t$_A$. Fig. 77 [196] shows CPP-MRs with such 'inversion' for all three alloys. The inversions show that all three $\beta_A$s are negative, as expected. Unfortunately, as noted above, the 2CSR model is not valid for large t$_A$, because the maximum thicknesses t$_A$ in Fig.77 are larger than the spin-diffusion lengths in the alloys. Thus, the particular values of $\beta_A$ derived by Hsu, Vouille, et al. are much smaller in magnitude than those from DMR. To obtain reliable values of $\beta_A$ requires fitting the data in Fig. 77 with VF theory, i.e., treating $l_{sf}^A$ as an extra unknown.

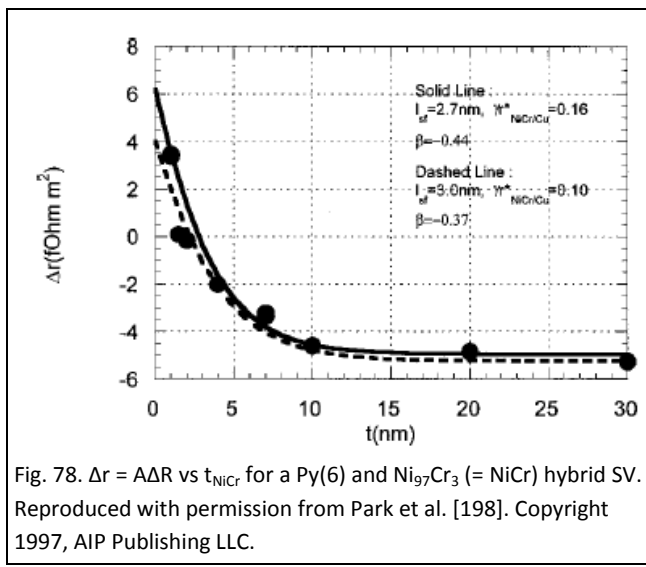

Fig. 78. $\Delta r$ = A$\Delta$R vs t$_{NiCr}$ for a Py(6) and Ni$_{97}$Cr$_3$ (= NiCr) hybrid SV. Reproduced with permission from Park et al. [198]. Copyright 1997, AIP Publishing LLC.

In 1999, Park et al. [198] made the first VF analysis of an inverting system including $l_{sf}^A$ as an additional unknown. They measured A$\Delta$R vs t$_{NiCr}$ for [Py(6)/Cu(20)/Ni$_{97}$Cr$_3$(t$_{NiCr}$)/Cu(20)]$_{10}$ multilayers. The data are shown in Fig. 78, along with two alternative fits with the parameters listed. These fits, plus coupled fits to [FeMn(8)/NiCr(t$_{NiCr}$)/Cu(20)/Py(6)] EBSVs, gave values of $\beta_{NiCr}$ = -0.35 ± 0.1 and $l_{sf}^{NiCr}$ = 3 ± 1 nm. Such a short $l_{sf}^{NiCr}$ shows why simple 2CSR model analysis gave too small values of $\beta_{NiCr}$. Park's value of $\beta_{NiCr}$ is much closer to the DMR dilute limit range of values (-0.38 to -0.67).

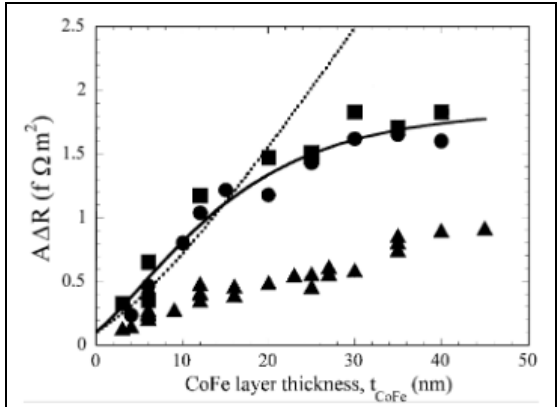

Fig. 79. AΔR vs t for symmetric EBSVs of CoFe (squares and circles) and Co (triangles). The squares and circles are from independent sputtering runs. The solid curve has $\beta_{CoFe}$ = 0.66 and $l_{sf}^{CoFe}$ = 12 nm; the dashed curve has $\beta_{CoFe}$ = 0.66 and $l_{sf}^{CoFe}$ = ∞. Reproduced with permission from Reilly et al. [199]. © 1999 by Elsevier.

Also in 1999, as a further quantitative check on 'universality', Reilly [199] measured AΔR for both symmetric [FeMn(8)/CoFe($t_{CoFe}$)/Cu(10/CoFe($t_{CoFe}$)]] and asymmetric [FeMn(8)/CoFe(6)/Cu/10/CoFe($t_{CoFe}$)]] EBSVs with $Co_{91}Fe_9$. The symmetric EBSV data are given in Fig. 79 The fit shown uses $\beta_{CoFe}$ = 0.65 ± 0.05 and $l_{sf}^{CoFe}$ = 12 ± 1 nm. This $\beta_{CoFe}$ is comparable to the dilute limit DMR value of $\beta_{CoFe}$ = 0.85 ± 0.1. The CoFe parameters were cross-checked by using them to correctly predict (within experimental uncertainties), without adjustment, AΔR for both asymmetric CoFe-based EBSVs with $t_{CoFe}$ = 6 nm for the pinned CoFe layer and for CoFe($t_{CoFe}$)/Cu(10)/Py(6)/Cu(10) hybrid SVs.

As an additional check on universality, Fig. 56 shows AΔR vs $t_{NiFeCr}$ for $Ni_{65}Fe_{15}Co_{20}$ = NiFeCo EBSVs [115]. The fit in Fig. 56 gives $\beta_{NiFeCo}$ = 0.82, slightly larger than $\beta_{Py}$ = 0.76. This value is consistent with the best DMR values of $\beta_{NiCo}$ = 0.9 and $\beta_{NiFe}$ = 0.88 [36].

The values of $\beta_F$ for these four F-alloys are closely enough consistent with dilute limit DMR values to support approximate 'universality'.

**Table 11: $\beta_F$(CPP) vs limit $\beta_F$(DMR).**

β at low temperatures (mostly 4.2K). The top four alloys were analyzed with the listed finite values of $l_{sf}^F$. Uncertainties from CPP and DMR are each typically 10-20%. The CPP values given below the break were found neglecting spin-flipping.

| F-alloy | $\rho_o$ (nΩm) | $\beta_F$(CPP) | $\beta_F$(DMR) | $l_{sf}^F$ (nm) | Ref. |
|---|---|---|---|---|---|
| $Ni_{80}Fe_{20}$ = Py | 123 | 0.76 | 0.88 | 5.5 ± 1 | [115]. |
| $Co_{91}Fe_9$ | 70 | 0.65 | 0.85 | 12 ± 1 | [199]. |
| $Ni_{97}Cr_3$ | 230 | - 0.35 | - 0.54 | 3 ± 1 | [198] |
| $Ni_{65}Fe_{15}Co_{20}$ | 90 | 0.82 | 0.88-0.90 | 5.5 ± 1 (assumed) | [115]. |
| | | | | | |
| $Fe_{85}V_{15}$ | | - 0.11 | - 0.78 | | [197]. |
| $Fe_{90}Cr_{10}$ | | - 0.16 | - 0.63 | | [197]. |
| $Ni_{95}Cr_5$ | | - 0.13 | - 0.54 | | [197]. |

### 8.10.2. $l_{sf}^F$ vs $1/\rho_{oF}$ for F-Alloys.

In addition to values of $\beta_F$ for the four F alloys just described, Table 11 also contains values of $l_{sf}^F$ for the first three alloys (the $l_{sf}^F$ for NiFeCo was simply assumed). To check the values of $l_{sf}$ in Table 11 for internal consistency, in Fig. 80 we plot $l_{sf}^F$ vs the inverse residual resistivity, $1/\rho_{oF}$ for the three alloys other than NiFeCo in Table 11, and the three 'nominally pure' F-metals Co, Fe, and Ni, for which values were given in section 8.6. The straight line is a unweighted fit to just the four points shown, forced to go to (0,0). The inset shows that values of $l_{sf}^F$ for Co fall well above the line and $l_{sf}^F$ for Fe falls below the line.

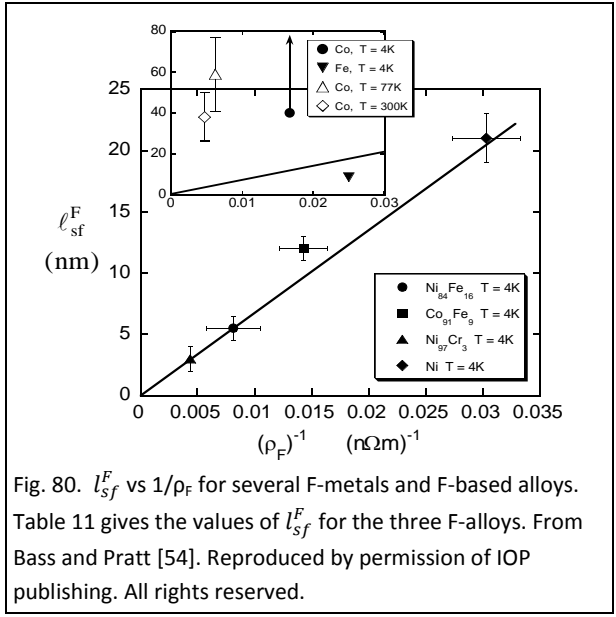

Fig. 80. $l_{sf}^F$ vs $1/\rho_F$ for several F-metals and F-based alloys. Table 11 gives the values of $l_{sf}^F$ for the three F-alloys. From Bass and Pratt [54]. 

## 8.11. $\delta_{N1/N2}$, Spin-Flipping at N1/N2 interfaces.

Most published CPP-MR studies neglect any spin-flipping at N1/N2 or F/N interfaces (I)—i.e., they assume that $\delta_I = 0$. In this section we describe measurements of $\delta_{N1/N2}$ that show that this assumption is invalid for some N1/N2 interfaces, thereby raising a red flag about its validity for F/N interfaces, a topic that we address in section 8.15.

In 1999, Baxter et al. [159] showed how to determine $\delta_{N1/N2}$, the spin-flipping parameter, for N1/N2 interfaces. The method involved inserting an $[N1/N2]_N$ multilayer into the center Cu layer of a symmetric Py-based EBSV and measuring A$\Delta$R vs $N$. They studied N1 = Nb and N2 = Cu, with EBSVs of the form FeMn(8)/Py(24)/Cu(10)/[Nb($t_{Nb}$)/Cu(10)]$_N$/Py(24)].

They confirmed independently the values of both $\rho_{Nb} = 78$ n$\Omega$m and AR$_{Nb/Cu} = 1$ f$\Omega$m$^2$ [180]. for non-superconducting Nb. Fig. 81 plots log (A$\Delta$R) vs $N$ for inserted multilayers with $t_{Nb} = 1.5$, 3.0, and 4.5 nm. For general N1/N2, VF theory predicts:

$$A\Delta R \propto \exp(-N[2\delta_{N1/N2} + t_{N1}/l_{sf}^{N1} + t_{N2}/l_{sf}^{N2}])/(AR_o + AR_{N1/N2}). \quad (18)$$

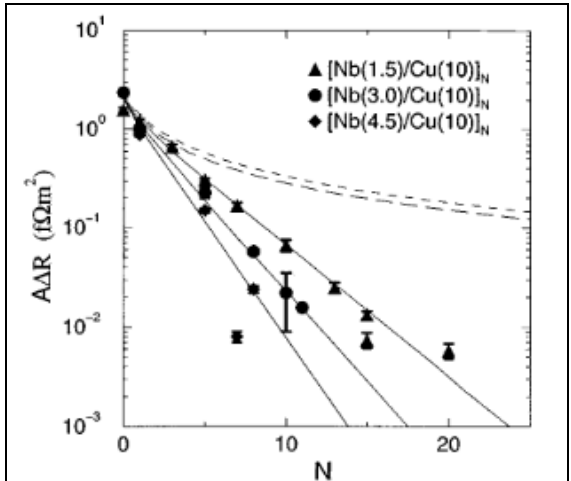

Fig. 81 Log (A$\Delta$R) vs $N$ for Nb(t)/Cu(10)$_N$ inserts with Nb thicknesses t = 1.5 nm, 3.0 nm, and 4.5 nm. The short and long dashed curves show the expected results for Nb = 1.5 nm and 4.5 nm with no spin-relaxation. The solid lines are fits with VF theory assuming spin-relaxation only in thin Nb/Cu interfaces. 

Here AR$_o$ is from the 'bare' EBSV, AR$_{N1/N2}$ is from the [Cu(10)/Nb($t_{Nb}$)]$_N$ insert, and the exponential contains $N$ times the contributions from the two N1/N2 interfaces, the $t_{N1} = t_{Nb}$ thick Nb layer, and the $t_{N2} = t_{Cu} = 10$ nm thick Cu layer. When the Baxter paper was submitted, the only published estimate of $l_{sf}^{Nb}$ was 0.8 $\mu$m from [200]. That value, combined with the expected long $l_{sf}^{Cu}$, led to expectation that the slopes of the lines predicted by Eq. 18 would be nearly independent of $t_{Nb}$. Instead, Fig. 81 shows slopes that grow linearly in magnitude as $t_{Nb}$ increases. Taking the slopes as representing $\delta_{Cu/Nb}$, gave values of $\delta_{Cu/Nb}$, = 0.15, 0.20, and 0.25. From these, the authors estimated $\delta_{Cu/Nb} = 0.20 \pm 0.05$.

In 2000, Park et al. [67] retested Nb and extended studies to Ag, V, and W, plus the antiferromagnet FeMn, all paired with Cu. To correct for the finite values of $l_{sf}^N$ in Eq. 18, they used the related technique described in section 8.5.2.2 to measure $l_{sf}^N$ for N = Ag,

V, Nb, and W.  For Ag and V they could set only lower bounds.  For Nb and W they obtained the values in Table 7.  Their Nb estimate of $l_{sf}^{Nb}$ = 25 nm is much shorter than the $l_{sf}^{Nb}$ = 0.8 µm from [200], but about what is needed in Eq. 18 to remove the differences between the three lines in  Fig. 81.

With these values of $l_{sf}^{N}$ in hand, Park et al. found $2AR_{Cu/N}$ and $\delta_{Cu/N}$ for N = Ag, V, Nb, and W by measuring both AR and log(AΔR) vs $N$ for Py-based EBSVs with [N(3)/Cu(3)]$_N$ inserts.  The 3 nm thicknesses were chosen as small enough that the corrections for $l_{sf}^{N}$ and $l_{sf}^{Cu}$ in Eq. 18 would not overwhelm the spin-loss contribution of $\delta_{N/Cu}$, but several times larger than the expected interface thicknesses ~ 0.5 nm.  Fig. 62 (left side) shows AR vs $N$, and gives the values of $2AR_{Cu/N}$ listed in Table 10. Fig. 62 (right side) shows log (AΔR) vs $N$.  For Cu/Ag, $\delta_{Cu/Ag}$ = 0 to within uncertainties.  For Cu/V, $\delta_{Cu/V}$ = 0.07 ± 0.04 is very small and highly uncertain.  The dashed curve in Fig. 62 (right side) shows how close the calculation for $l_{sf}^{V}$ = ∞ is to the data.  The best common fit to the Nb data of Figs. 62 (right side) and 81 gave $l_{sf}^{Nb}$ = $25_{-5}^{+\infty}$ and $\delta_{Cu/Nb}$ = 0.19 ±0.05.  The best value of $l_{sf}^{Nb}$ = 25 nm supplies what is needed to produce a single $\delta_{Cu/Nb}$ in Fig. 81, but the upward uncertainty is large.  Finally, the derived value of $\delta_{Cu/W}$ = 0.96 ±0.1 is much larger than the other values, consistent with large spin-orbit scattering for W as the heaviest metal.

Table 12 contains references and the measured values of $\delta_{N1/N2}$ for a variety of metal pairs, listed in order of increasing $\delta_{N1/N2}$.  $\delta_{N1/N2}$ generally increases with increasing difference in atomic number between N1 and N2, but the progression is not perfect.  There are not yet calculations to show if these values can come from spin-orbit differences at perfect interfaces, or if they need interfacial alloying.

**Table 12. Experimental values of $\delta_{N1/N2}$.** [  ] =  reference.

| Metals | $\delta_{N1/N2}$  [ref] | Metals | $\delta_{N1/N2}$  [ref] |
|--------|------------------------|--------|------------------------|
| Cu/Ag | 0            [67] | Ag/Pd | 0.15±0.08 [186] |
| Cu/Al | $0.05_{-0.05}^{+0.02}$  [187] | Cu/Nb | 0.19±0.05[[67] |
| Cu/V | 0.07± 0.04 [67] | Cu/Pd | $0.24_{-0.03}^{+0.06}$ [182] |
| Au/Pd | 0.08±0.08 [186] | Cu/Ru | ~ 0.35       [183] |
| Pd/Pt | 0.13±0.08 [86] | Cu/Pt | 0.9±0.1      [182] |
| Cu/Au | $0.13_{-0.05}^{+0.08}$  [181] | Cu/W | 0.96±0.1   [67] |

### 8.12. $\gamma_{F/N}$, $2AR_{F/N}^{*}$ and $2\gamma_{F/N}AR_{F/N}^{*}$: Interface Parameters for F/N Systems.

To be competitive for devices that grow ever smaller requires both large F/N interface resistances, $2AR_{F/N}^{*}$, and large bulk and interfacial scattering asymmetries, $\beta_F$ and $\gamma_{F/N}$.  Values of the bulk asymmetry, $\beta_F$, were discussed in section 8.10.  In this section we cover $2AR_{F/N}^{*}$ and $\gamma_{F/N}$, mostly for nominally pure F-metals, but also including Py and CoFe ($\approx Co_{90}Fe_{10}$) alloys.

Eq. 6 shows that the main contribution to AΔR from an F/N interface comes from the product $2\gamma_{F/N}AR_{F/N}^{*}$.  In a search for promising pairs, investigators have measured a wide range of F/N combinations.  Table 13 collects the values of $2AR_{F/N}^{*}$, $\gamma_{F?N}$, and $2\gamma_{F/N}AR_{F/N}^{*}$ for all of the nominally 'pure metal' F/N pairs studied so far, and for the F-alloys Py and $Co_{90}Fe_{10}$ .  For convenience of searching, the entries are ordered by the F-metal.  The values in Table 13 let one easily calculate $AR^{\downarrow}$ and $AR^{\uparrow}$  as $AR^{\downarrow}$ = $(2AR_{F/N}^{*} + 2\gamma_{F/N}AR_{F/N}^{*})$, and $AR^{\uparrow}$ = $(2AR_{F/N}^{*} - 2\gamma_{F/N}AR_{F/N}^{*})$.

With the possible exception of Co/Al, where two separate analyses disagree on $\gamma_{Co/Al}$, the largest values of $2\gamma_{F/N}AR_{F/N}^{*}$ are for the standard pairs, Co/Cu and Fe/Cr, and don't greatly exceed 1 fΩm$^2$.

These results suggest that obtaining a very large CPP-MR is likely to require an F-metal with high bulk resistivity and large $\beta_F \cong 1$—i.e., something approaching a 'half-metal'.

**Table 13.** $2AR^*_{F/N}$, $\gamma_{F/N}$, and $2\gamma_{F/N}AR^*_{F/N}$ at 4.2K for 'pure metal' F and N and some alloy F (Py and CoFe).

$\Delta a/a_o$ = fractional difference in lattice parameters of F and N.

| Metals(structures) | $\Delta a/a_o$ | $2AR^*_{F/N}$ = $(AR^{\downarrow}+AR^{\uparrow})/2$ (f$\Omega$m$^2$) | $\gamma_{F/N}$ | $2\gamma_{F/N}AR^*_{F/N}$ = $(AR^{\downarrow}-AR^{\uparrow})/2$ (f$\Omega$m$^2$) | Ref. |
|---|---|---|---|---|---|
| Co/Cu(fcc) | 1.8 | 1.02±0.1 | 0.77±0.04 | 0.79 | [6] |
| Co/Ag(fcc) | | 1.12±0.06 | 0.85±0.03 | 0.95 | [6]] |
| Co/Au(fcc) | | | | ~ 0.9 | [181] |
| Co/Pt(fcc) | 10 | 1.7±0.25 | 0.38±0.06 | 0.65 | [201]. |
| Co/Ru (hcp)(0001) | 5 | 1 (~1,2) | -0.2 (-0.16) | -0.2 | [183, 202] |
| Co/Al(a)(fcc)(111)) | 13 | 11.1±0.02 | 0.05±0.01 | 0.56 | [203] |
| Co/Al(b)(fcc)(111) | 13 | 11.6±0.2 | 0.18±0.02 | 2.1 | [204] |
| CoFe/Cu(fcc) | | 1.1±0.2 | 0.8±0.1 | 0.88 | [205] |
| CoFe/Al(fcc)(111) | 13 | 10.6±0.6 | 0.10±0.01 | 1.06 | [203, 204] |
| Co/Ni(fcc) | | 0.52 | 0.94 | 0.49 | [91]] |
| Py/Cu(fcc)(111) | 2.5 | 1.0±0.1 | 0.7±0.1 | 0.7 | [93];[174] |
| Py/Al (fcc)(111) | 13 | 8.5±1 | $0.025^{+0.045}_{-0.01}$ | 0.21 | [206] |
| Py/Pd(fcc)(111) | 10 | 0.4±0.2 | 0.41±0.14 | 0.2 | [201]] |
| Fe/Cr(bcc)(011) | 0.4 | 1.6±0.15 | -0.7±0.15 | -1.1 | [41]. |
| Fe/Cu(bcc/fcc) | | 1.48±0.14 | 0.55±0.07 | 0.81 | [60]] |
| Fe/Al(bcc/fcc) | | 8.4±0.6 | 0.05±0.02 | 0.42 | [204]] |
| Fe/V(bcc)(001) | 5 | 1.4±0.2 | -0.27±0.08 | -0.4 | [201]] |
| Fe/Nb(a)(bcc)(011) | 13 | 2.6±0.3 | -0.17±0.04 | -0.4 | [201]] |
| Fe/Nb(b)(bcc)(011) | 13 | 2.9±0.3 | 0.11±0.04 | 0.3 | [201]] |
| Ni/Cu(fcc)(111) | | 0.36±0.06 | 0.29±0.05 | 0.1 | [171]] |
| Ni/Ru | | $1.7^{+0.4}_{-0.2}$ | \|0.15±0.03\| | 0.25 | [205]] |

**(b)** Co/Al: assoc. with $\delta_{Co/Al}$ = 1.8 ± 0.5 (see [204]; (b) Fe/Nb: assoc. with $\delta_{Fe/Nb}$ = 0.83±0.08 (see [201].

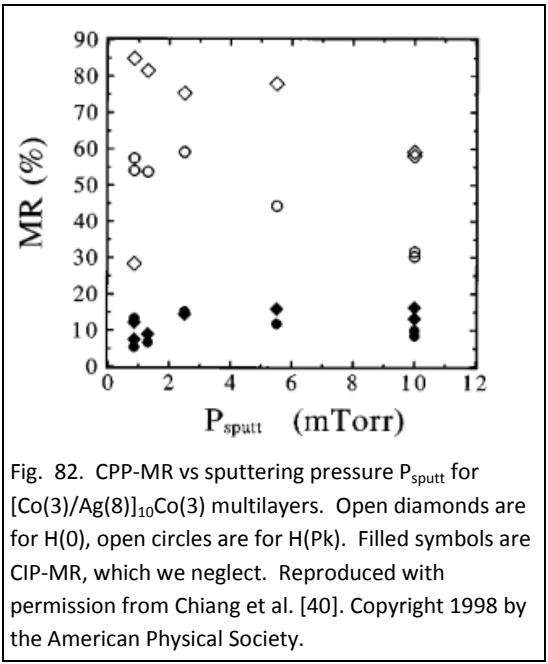

Fig. 82. CPP-MR vs sputtering pressure $P_{sputt}$ for $[Co(3)/Ag(8)]_{10}Co(3)$ multilayers. Open diamonds are for H(0), open circles are for H(Pk). Filled symbols are CIP-MR, which we neglect. Reproduced with permission from Chiang et al. [40]. Copyright 1998 by the American Physical Society.

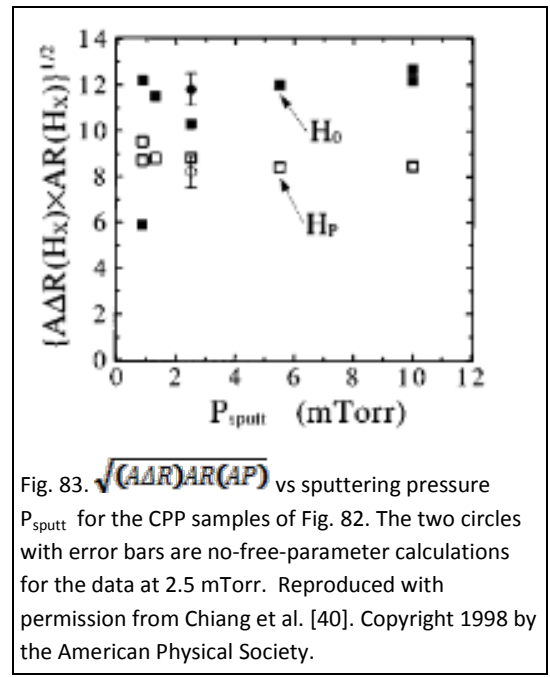

Fig. 83. $\sqrt{(A\Delta R)AR(AP)}$ vs sputtering pressure $P_{sputt}$ for the CPP samples of Fig. 82. The two circles with error bars are no-free-parameter calculations for the data at 2.5 mTorr. Reproduced with permission from Chiang et al. [40]. Copyright 1998 by the American Physical Society.

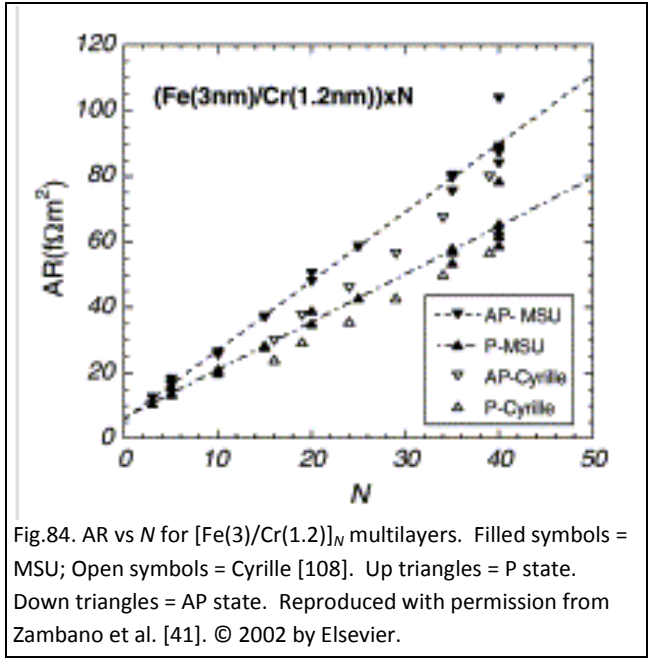

Fig.84. AR vs N for [Fe(3)/Cr(1.2)]$_N$ multilayers. Filled symbols = MSU; Open symbols = Cyrille [108]. Up triangles = P state. Down triangles = AP state. Reproduced with permission from Zambano et al. [41]. © 2002 by Elsevier.

### 8.13. Interface Roughness studies on Co/Ag and Fe/Cr.

In section 2.3 we noted that three different studies of effects of interface roughness gave apparently contradictory results. In this section we examine these studies.

In 1998, Chiang et al. [40] tested the effect of varying the Ar sputtering pressure from 0.86 to 10 mTorr on interface roughness and CPP-MR in [Co(3)/Ag(8)]$_{10}$Co(3) multilayers between 250 nm thick Nb bottom and top cross-strips. As noted in section 6.1.2, low angle x-ray spectra showed that increased sputtering pressure led to increased roughness. Fig. 82 shows that both CPP-MR(0) (with one anomalous exception) and CPP-MR (Pk) decreased as the sputtering pressure increased from 0.86 to 10 mTorr. Using Eq. 7 as a guide, they looked for a change in scattering asymmetry by plotting $\sqrt{(A\Delta R)AR(AP)}$ vs sputtering pressure in Fig.83. They took the absence of significant change to mean that increased roughness had little effect on the scattering asymmetry, but rather increased scattering equally in both spin channels. The decreases in CPP-MR in Fig. 82, then result mostly from increases in the denominator of the CPP-MR.

In 2000, Cyrille et al. [108] published the first of 3 studies of the CPP-MR of sputtered [Fe(3)/Cr(1.2 or 1.3)]$_N$ multilayers, with focus on changing interface roughness with N and with sputtering pressure. To measure the CPP-MR with a microvoltmeter, they connected 100 micron-sized multilayers in series via superconducting Nb layers. To achieve approximately AP states, they chose multilayers with t$_{Cr}$ near the first antiferromagnetic coupling peak. Measuring AR(AP) and AR(P) as functions of N they found data given by the open symbols in Fig. 84. They argued that the P-state data could be extrapolated to AR = 0 at N = 0, thus yielding 2AR$_{Nb/Co}$ = 0. As shown by the open symbols in Fig. 85, they also found the CPP-MR to grow with increasing N. They attributed this growth to systematically increasing interfacial roughness, as evidenced by both x-rays (Fig. 86) and Cr mapping with cross-sectional Electron-Energy-Loss-Spectroscopy (EELS). Followup papers [42, 207] reported that increasing the Ar sputtering pressure, which x-rays also showed increased interface roughness, increased both interfacial disorder and the CPP-MR; their data are open symbols in Fig. 87.

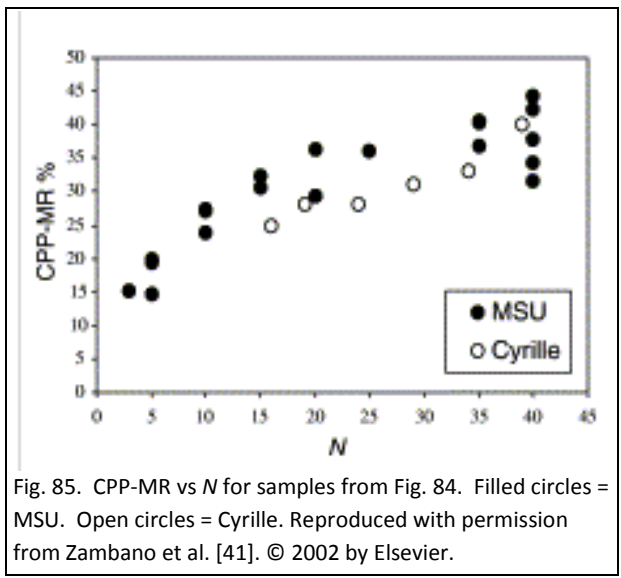

Fig. 85. CPP-MR vs N for samples from Fig. 84. Filled circles = MSU. Open circles = Cyrille. Reproduced with permission from Zambano et al. [41]. © 2002 by Elsevier.

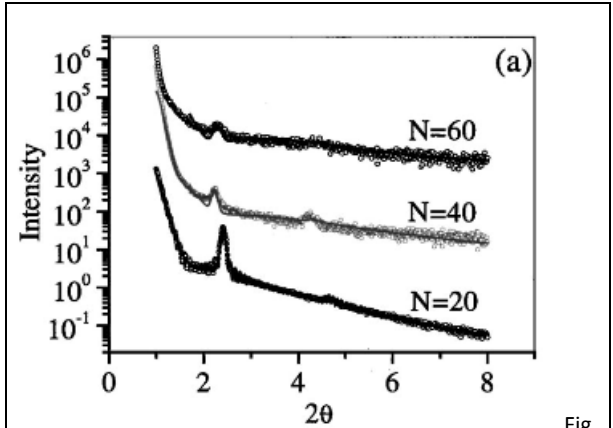

Fig. 86. Low angle Θ-2ϑ x-ray scans of Cyrille [Fe(3)Cr(1.2)]$_N$ multilayers for $N$ = 20, 40, and 60.. Curves shifted for clarity. Reproduced with permission from Cyrille et al. [108]. Copyright 2000 by the American Physical Society.

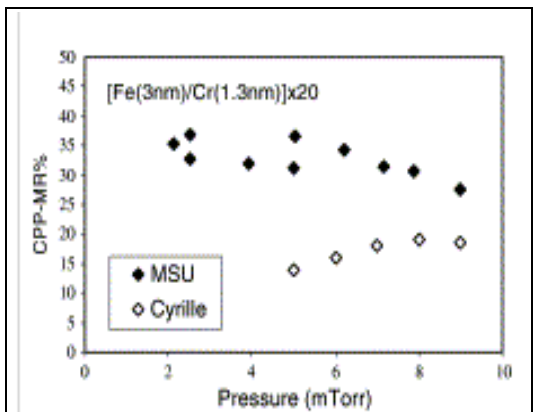

Fig. 87. CPP-MR vs sputtering Pressure for [Fe(3nm)Cr(1.3)]$_{20}$ multilayers. Filled diamonds = MSU. Open diamonds = Cyrille. Reproduced with permission from Zambano et al. [41]. © 2002 by Elsevier..

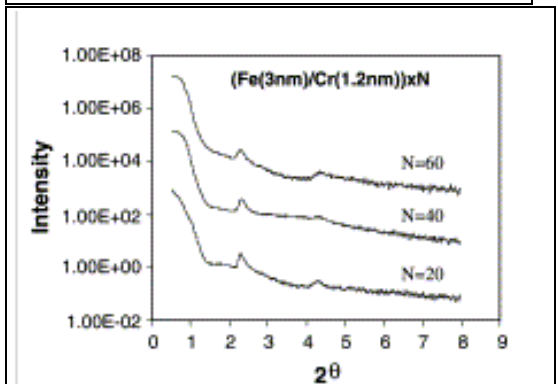

Fig. 88. Low angle ϑ-2ϑ x-ray scans of MSU [Fe(3)/Cr(1.2)]$_N$ multilayers for $N$ = 20, 40, 60. Curves shifted for clarity. Reproduced with permission from Zambano et al. [41]. © 2002 by Elsevier.

Given the approximate agreement on non-zero values of 2AR$_{Nb/Fe}$ in Table 2 for samples prepared by both sputtering and MBE, a value of 2AR$_{Nb/Fe}$ = 0 for sputtered samples was a surprise. This result led Zambano et al. [41] to try to reproduce the Cyrille et al. data. Their results are shown as the filled symbols in Figs. 84, 85, and 87. Figs. 84 and 85 show that Zambano's values of AR(AP), AR(P), and CPP-MR vary similarly, but not identically, to those of Cyrille. Part of the difference may be due to Zambano's use of an extra capping layer of Fe, which, relative to Cyrille, adds to AR(AP) one Cr layer, one Fe layer, and two Fe/Cr interfaces. Together, these changes should increase Cyrille's AR(AP) by about 2 fΩm$^2$. Part of the difference might also be due to systematic differences in areas, which were uncertain by at least 5% in each case. And, because taking antiferromagnetically coupled multilayers to saturation requires high fields (5 – 10 kOe), AR(P) must be corrected for changes in the AR$_{Nb/Fe}$ with increasing high field. Zambano's AR(P) data are described as 'corrected for a modest field dependence of AR for Nb/Fe/Nb sandwiches' that 'does not strongly affect either the data or our conclusions.' Cyrille shows no hysteresis data and makes no comment on this issue, which might affect AR(P). Lastly, Zambano's x-rays (Fig. 88) show less change with $N$ than those of Cyrille in Fig. 86. So the interface structure might change differently with $N$ in the two studies.

The most important difference in Fig. 84 is that Zambano's data extrapolate to an intercept of 2AR$_{Nb/Fe}$ = 6 ± 1 fΩm$^2$, consistent with the MBE result for Nb/Fe by Bozec et al. [60] and with the other data in Table 2. This offset let Zambano et al. fit their data with a 2CSR model. Correcting for the ordinate offset in Fig. 84 then gave a CPP-MR that was constant, independent of $N$. The most intriguing other difference in the two sets of data is that in Fig. 87, where Cyrille's CPP-MR increases with increasing sputtering pressure, but Zambano's shows a slight decrease at their highest pressures. The x-ray spectra for both Cyrille's and Zambano's data behave qualitatively similarly with increased sputtering pressure, both showing reductions in x-ray structure similar to those in Fig. 86. As further support for applying the 2CSR model to their data, Zambano et al. showed in Fig. 89 that their $\sqrt{(A\Delta R)AR(AP)}$ data in the form of Eq. 7 give the predicted straight line through the origin. Lastly, As shown in section 8.14, their $2AR^*_{Fe/Cr}$ = 1.6 ± 0.15 fΩm² agrees with no-free-parameter calculations. However, their $\gamma_{Fe/Cr}$ = - 0.7 ± 0.15 is larger in magnitude than calculated (- 0.3 to - 0.45) [90].

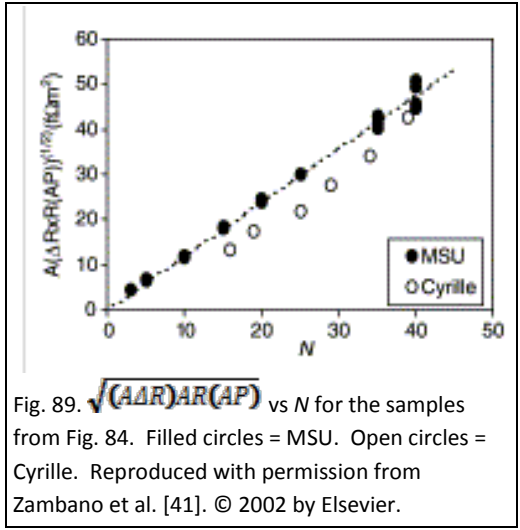

Fig. 89. $\sqrt{(A\Delta R)AR(AP)}$ vs $N$ for the samples from Fig. 84. Filled circles = MSU. Open circles = Cyrille. Reproduced with permission from Zambano et al. [41]. © 2002 by Elsevier.

To summarize, in two experiments, one on Co/Ag [40] and one on Fe/Cr [41], the CPP-MR decreased with increasing sputtering pressure, mostly due to growth of AR(P) in the denominator, while in a third, on Fe/Cr [42, 108] it increased. The reasons for these differences, and others in the Fe/Cr data of Cyrille and Zambano, are not yet known, but may at least partly involve different interfacial structures.

### 8.14. No-free-parameter calculations of $2AR_{N1/N2}$ and $2AR^*_{F/N}$

In section 1 we asked an important question about CPP-MR: Do VF parameters agree with values measured in completely different ways, or with no-free-parameter calculations? Section 8.10.1 answered the first part of this question with a 'qualified yes' for $\beta_F$. In this section we address the second part, $2AR_{N1/N2}$ and $2AR^*_{F/N}$. Section 4.4.1 gave Eq (8), derived by Schep et al. [87] to calculate $2AR_{N1/N2}$, $AR^\downarrow_{F/N}$, and $AR^\uparrow_{F/N}$, assuming diffuse scattering in the bounding metals. In section 4.4.1, we compared its predictions with experimental values for the lattice matched pairs Co/Cu, Fe/Cr and Co/Ni. In this section we expand the discussion to all N1/N2, F/N, and F1/F2 pairs for which both measurements and calculations have been made. To compare values for these different interfaces we focus upon 2AR (for F/N we use 2AR*). Table 14 divides the pairs into two categories. The top six pairs are lattice matched, with closely the same lattice parameters (to within ~ 1%). They give surprisingly good agreement with experiment. The lower six pairs, with greater differences in lattice parameters, give poor agreement. For the six 'good' cases, we include both earlier calculations made with linearized muffin-tin potentials and an 'spd' basis, and later ones made with full muffin-tin potentials and a larger 'spdf' basis. The differences in calculations lie within mutual uncertainties. For completeness, we compare in Table 15 the calculated and experimental values of $\gamma_{Co/Cu}$, $\gamma_{Fe/Cr}$, and $\gamma_{Co/Ni}$. For γ, the best agreement is for Co/Ni.

As the Co/Cu, Fe/Cr, and Ag/Au calculations were all made after the experimental values of 2AR or 2AR* were known, one might worry about calculational bias. The poor agreements obtained for the six pairs below the break argue against bias. But, in the spirit of rules 1 and 2 in section 1.1, it is important to try to remove any bias. To do so, it was decided to measure and calculate $2AR_{N1/N2}$ double blind. That is, one group would measure 2AR, the other would calculate it, and values would be exchanged 'immediately' by e-mail only after each group said that its value was settled.

**Table 14. Experimental 2AR(exp) (2AR\* for F/N or F1/F2) at 4.2K vs calculated 2AR(calc).** Fcc orientation is (111). Bcc orientation is (110). Pairs where exp and calc agree are in bold and separated at the top. The Co/Ni calculations were made with full MTP but only an spd basis. The last column for Co/Ni is for a 4ML(50-50) interface.

| Metals (Struct) | (Δa/a)% | 2AR(exp) (fΩm²) | 2AR(Perf.) (fΩm²) | 2AR(2ML50-50) (fΩm²) | 2AR(Perf.) (fΩm²) | 2AR(2ML50-50) (fΩm²) |
|---|---|---|---|---|---|---|
| *Basis* | | *spd* | *spd* | *spd* | *spdf* | *spdf* |
| **Ag/Au (fcc)** | **0.2** | **0.1±0.01 [32]** | **0.09 [90]** | **0.12 [90]** | **0.09 [85]]** | **0.13 [85]]** |
| **Co/Cu (fcc)** | **1.8** | **1.02±0.1 [6]** | **0.9 [[90]** | **1.1 [90]** | **0.9 [208]** | **1.1 [85]** |
| **Fe/Cr (bcc)** | **0.4** | **1.6±0.15 [41]** | **1.9 [90] 1.5 [88]** | **1.6[90]** | **1.7 [85]** | **1.5 [85]** |
| **Pd/Pt (fcc))** | **0.8** | **0.28± 0.06 [86]** | **0.30±0.04 [86]** | **0.33±0.04 [86]** | $0.40^{+0.03}_{-0.08}$ [85] | $0.42^{+0.02}_{-0.04}$ [85] |
| **Pd/Ir (fcc)** | **1.3** | **1.02±0.06[85]** | **1.21 ± 0.1 [85]** | **1.22 ± 0.1 [85]** | **1.10±0.1 [85]** | **1.13±0.1 [85]** |
| | | **2AR(Exp)** | **2AR(Perf.)** | **2AR(2ML50-50)** | **2AR(4ML50-50)** | |
| **Co/Ni(fcc)** | **0.7** | **0.51±0.05 [91]** | **0.37±0.04 [91]** | **0.44±0.04 [91]]** | **0.61±0.06 [91]** | |
| | | | | | | |
| Ag/Cu (fcc | 12 | 0.09 [32] | 0.45 [186] | 0.7 [186] | | |
| Au/Cu (fcc) | 12 | 0.30 [32] | 0.45 [186] | 0.6 [186] | | |
| Pd/Cu (fcc) | 7 | 0.9 [182] | 1.5 [186] | 1.6 [186] | | |
| Pd/Ag (fcc) | 5 | 0.7 [186] | 1.6 [186] | 2.0 [186] | | |
| Pd/Au (fcc) | 5 | 0.45 [186] | 1.7 [186] | 1.9 [186] | | |
| Ni/Cu (fcc)u | 2.5 | 0.36 [84] | 0.74 [84] | | | |

**Table 15. $\gamma_{F/N}$ (exp) vs $\gamma_{F/N}$(calc) for Co/Cu, Fe/Cr, and Co/Ni.**

| Metals | γ(exp) | γ(perf) | γ(2ML50-50) |
|---|---|---|---|
| Co/Cu | 0.77 ± 0.04 [6] | 0.55 ± 0.05 [208] | 0.67 ± 0.05 [208] |
| Fe/Cr | - (0.70 ± 0.15) [41] | - (0.45) [90] | - (0.30) [90] |
| Co/Ni | 0.94 ± 0.04 [91] | 0.96 [91] | 0.97 [91] |

Pd and Pt were chosen first, because of the closeness of their lattice parameters, their known good sputtering characteristics, their complete miscibility [167]] and the absence of any intermetallics, all of which were expected to give 'simple interfaces'. As an internal cross-check, measurements were made with both of the techniques described in section 8.8. The 'simple' technique of section 8.8.1 gave the data in Fig. 90 [86], with best fit of $2AR_{Pd/Pt}$ = 0.29 ± 0.03 fΩm². The 'more complex' technique of section 8.8.2 gave the more scattered data in Fig.91 [86] and less certain 2AR$_{Pd/Pt}$ = 0.17 ± 0.13 fΩm$^2$. The weighted average 'settled value' was 2R$_{Pd/Pt}$(exp) = 0.28 ± 0.06 fΩm$^2$, as in Table 14. The calculated settled values, found with the same linearized muffin tin potentials (MTP) and spd basis used for the prior calculations, were 2AR$_{Pd/Pt}$(perf) = 0.30 ± 0.04 fΩm$^2$ and 2AR$_{Pd/Pt}$(50-50) = 0.33 ± 0.04 fΩm$^2$ for 2 ML of a 50/50 alloy.

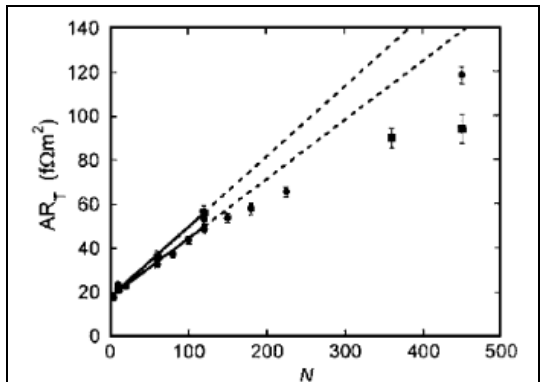

Fig. 90. $AR_T$ vs *N* for [Pd(t)/Pt(t)]$_N$ multilayers with fixed $t_T$ = 360 nm. The lower line is a best fit up to *N* = 130. The upper line is an estimate of maximum slope. Reproduced with permission from Olson et al. [86]. Copyright 2005, AIP Publishing LLC.

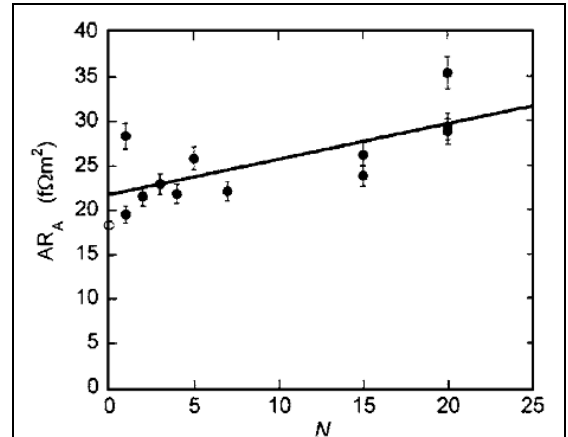

Fig. 91. AR vs *N* for [Pd(3)/Pt(3)]$_N$ multilayer inserts in the middle of symmetrical Py-based EBSVs. Reproduced with permission from Olson et al. [86]. Copyright 2005, AIP Publishing LLC.

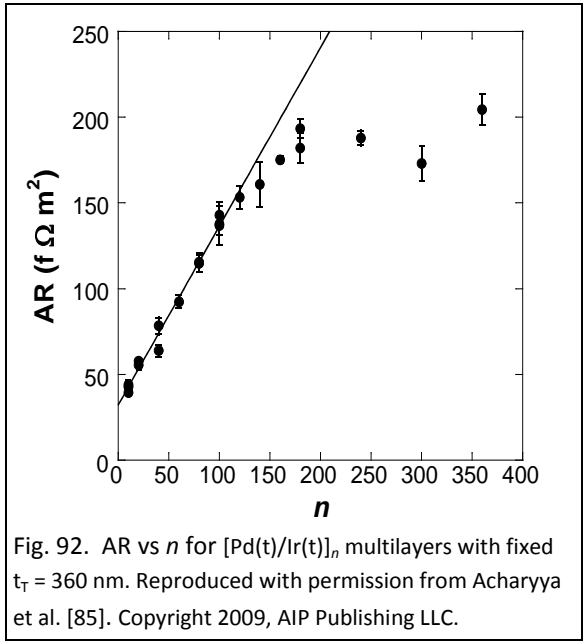

Fig. 92. AR vs $n$ for [Pd(t)/Ir(t)]$_n$ multilayers with fixed $t_T = 360$ nm. Reproduced with permission from Acharyya et al. [85]. Copyright 2009, AIP Publishing LLC.

The good agreement for Pd/Pt stimulated a second double-blind test, with Pd/Ir [85]. For Pd/Ir, the difference in lattice parameters is just over 1% and solubility is opposite to Pt/Pd; the two metals separate except for < 2% Ir solubility in Pd and < 5% Pd solubility in Ir [209]. Table 14 lists the results. The data in Fig. 92 gave 2AR$_{Pd/Ir}$(exp) = 1.02 ± 0.06 fΩm$^2$. The calculation gave 2AR$_{Pd/Ir}$(perf) = 1.21 ± 0.1 fΩm$^2$ and 2AR$_{Pd/Ir}$(50-50)

= 1.22 ± 0.1 fΩm$^2$, not quite overlapping with the measured value. By the time these Pd/Ir data were available, computers had improved to where it was possible to upgrade the calculations to full MTPs and an spdf basis. Such an updated calculation gave 2AR$_{Pd/Ir}$(Perf) = 1.10 ± 0.1 fΩm$^2$ and 2AR$_{Pd/Ir}$(50-50) = 1.13 ± 0.1 fΩm$^2$, now within mutual uncertainties of 2AR$_{Pd/Ir}$(exp).

However, applying the same update to Pd/Pt gave new values listed in Table 14 that moved away from 2AR$_{Pd/Pt}$(exp). Now, 2AR$_{Pd/Pt}$(exp) and 2AR$_{Pd/Pt}$(calc) barely overlap to within uncertainties. Table 14 shows that the updates for the other three metal pairs gave only minor changes.

The last test so far with lattice matched metals was made with Co/Ni [91]. The experimental values of $2AR^*_{Co/Ni}$ and γ$_{Co/Ni}$ in Tables 14 and 15 were derived from a combination of data for multilayers and two different EBSVs. The calculations were done with full MTPs but only an spd basis, as both Co and Ni have atomic numbers too low for f-electrons. The best fit to $2AR^*_{Co/Ni}$ is for an interface thickness between 2 ML and 4 ML.

We conclude from Table 14 that experiments and calculations for 2AR or 2AR* agree reasonably well for six metal pairs with lattice parameters within ~ 1%. However, the agreements are generally poor for metal pairs with larger differences in lattice parameters. Presumably, in these latter cases the calculations are too sensitive to unknown structural details of the interfaces, as found for calculations of dilute alloy resistivities [84].

Table 15 shows that the calculated γ$_{Co/Ni}$ agrees with experiment, but γ$_{Co/Cu}$ and γ$_{Fe/Cr}$ are smaller (for Fe/Cr in magnitude) than the experimental values.

### 8.15. δ$_{F/N}$ or δ$_{F1/F2}$: Spin-Flipping at F/N and F1/F2 interfaces.

The last of the eight VF parameters, and the least studied, is δ$_{F/N}$ (or δ$_{F1/F2}$) [210]. It is hard to study using simple [F/N]$_n$ multilayers since, as illustrated in Fig. 71, unless it is large (δ ≥ 0.5), it does not have a large affect on AΔR. In contrast, even small values can strongly affect separated (S) multilayers (see section 8.9) or SVs. But then it is often non-trivial to distinguish its effects from those of a shorter $l^F_{sf}$.

The first proposal of a non-zero $\delta_{F/N}$ was for Co/Cu, where it was argued (see section 8.9) that $\delta_{Co?Cu} = 0.25$ [185] could 'explain' the observed differences between interleaved (I) and separated (S) Co/Cu multilayers.

The next proposal was by Manchon et al. [162], who derived values of $\delta_{Co/Cu} = \delta_{CoFe/Cu} \sim 0.3$; $\delta_{Co/Ru} \sim 0.3$; $\delta_{CoFe/Ru} \sim 0.37$, and $\delta_{Py/Cu} \sim 0.4$ (Note: these values are converted from $P = [1 − \exp(−\delta)]$ to just $\delta$}, from published data by others for combined CPP-MRs and spin-torques in several different spin-valves. Although their analysis is indirect, and involved assumptions of most parameters from still other studies, we'll see in Table 16 that it gave values consistent with those of later, more direct studies. A study by Delille et al. [211] in the same year will be discussed at the end of this section.

These studies were suggestive, but left the need for a more direct way to measure $\delta_{F/N}$.

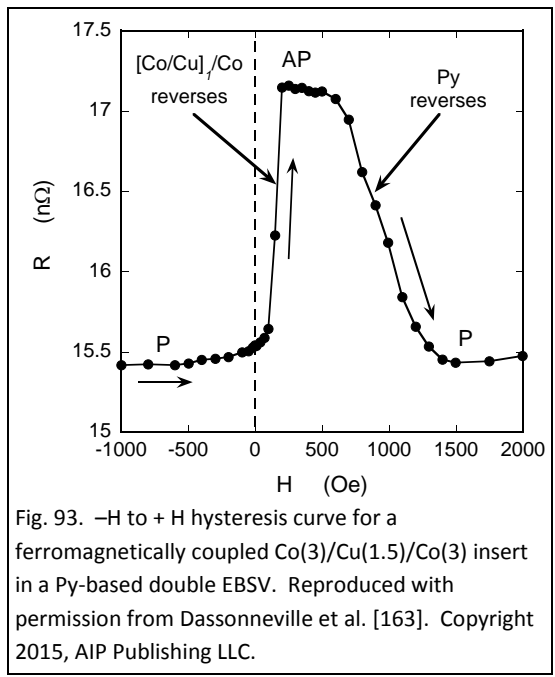

Fig. 93. −H to + H hysteresis curve for a ferromagnetically coupled Co(3)/Cu(1.5)/Co(3) insert in a Py-based EBSV. Reproduced with permission from Dassonneville et al. [163]. Copyright 2015, AIP Publishing LLC.

Section 8.11 showed that measuring $\delta_{N1/N2}$ at N1/N2 interfaces is straightforward, as inserting an $[N1/N2]_n$ multilayer into the middle of a Py-based EBSV doesn't change the magnetic structure of the EBSV. In contrast, a magnetic $[F/N]_nF$ insert changes an EBSV's magnetic structure. In 2010, Dassonneville et al. [163] controlled the magnetic structure by inserting into the central Cu layer of a Py-based double EBSV (DEBSV) a *ferromagnetically coupled* $[F/N]_nF$ multilayer, giving the symmetrical form Nb(150)/Cu(10)/FeMn(8)/Py(6)/Cu(10)/[F(t_F)/N(t_N)]_n/F(t_F)/Cu(10)/Py(6)/FeMn(8)/Cu(10)/Nb(150). The symmetry simplifies the VF quantitative fitting, and the DEBSV nearly doubles the A$\Delta$R signal compared to a single EBSV. In this geometry, both Py layers are pinned, and the CPP-MR results from reversal of the moment of the ferromagnetically coupled $[F(t_F)/N(t_N)]_nF(t_F)]$ insert. So long as the coercive field of the insert is much smaller than the common one of the pinned Py layers, the strong ferromagnetic coupling ensures that the magnetic states with the moment of the insert either P or AP to the common moments of the Py-layers will be well defined. The 10 nm Cu layers bounding the insert are thick enough to eliminate any exchange coupling between the $[F/N]_nF$ multilayer and the Py layers. Fig. 93 shows an example of a hysteresis curve for an inserted F = Co and N = Cu. $t_F$ and $t_N$ must be kept much thinner than their spin-diffusion lengths, so that their bulk contributions to A$\Delta$R with $n = 8$ will be small enough to not overawe that of $\delta_{F/N}$. This requirement led to use of Co for most such studies to date. $t_F$ and $t_N$ should also be kept at least 2-3 times as long as the expected interface thicknesses of ~ 0.5 nm [32-35], so that the F- and N-layers stay well defined. Separate magnetization measurements on simple F/N multilayers with different layer thicknesses are needed to choose values of $t_N$ that give ferromagnetic coupling.

The Dassonneville technique was applied first to Co/Cu, where, as noted above, one interpretation of data for Interleaved (I) and Separated (S) multilayers (section 8.9) led to a prediction of $\delta_{Co/Cu} = 0.25$ [185]. Combining this value for $\delta_{Co/Cu}$ with previously measured parameters by the same group for all of the constituents of the DEBSV with its insert, allowed prediction of how A$\Delta$R should grow with increasing $n$ using no adjustable parameters. With a Co thickness chosen as $t_{Co} = 3$ nm, magnetization studies showed that $t_{Cu}$ = 1.3 nm and 1.5 nm gave the smallest saturation fields. The AΔR vs $n$ measurements were made with $t_{Cu}$ = 1.5 nm, but $n$ = 1 and $n$ = 8 data points for $t_{Cu}$ = 1.3 nm gave similar results as shown in the plot of AΔR vs $n$ in Fig. 94 [163]. Correct prediction of the measured value of AΔR at $n$ = 0 serves as a check on the parameters other than $\delta_{F/N}$, since $n$ = 0 contains no insert. The dotted curve in Fig. 94 is the VF prediction for AΔR with no adjustable parameters taking $\delta_{Co/Cu}$ = 0. It starts correctly at the data for $n$ = 0, but grows much too fast as $n$ increases. The Co layers alone don't give enough moment flipping. The dashed curve is the VF prediction with the previously predicted $\delta_{Co/Cu}$ = 0.25. Now, the moment flipping due to the Co/Cu interfaces causes the curve to approach saturation faster, moving it close to, but still slightly above, the data. The solid curve is the best fit treating only $\delta_{Co/Cu}$ as adjustable. Including uncertainties in both the data and the fixed parameters, the authors estimated $\delta_{Co/Cu}$ = $0.33^{+0.03}_{-0.08}$. This larger $\delta_{Co/Cu}$ gives the faster saturation needed to fit the data. The technique's reliability depends upon the reliability of the assumed fixed parameters and on confidence that the thin F- and N-layers do not intermix enough to approach a uniform alloy for large $n$. The fixed parameters were measured previously in the same laboratory with the same deposition conditions. And the results are not sensitive to most of the parameters. Correct prediction of AΔR(0) gives a check on the totality of parameters. But unwanted errors cannot be completely ruled out. So an alternative way to check the results would be nice. X-ray measurements can rule out complete intermixing [212].

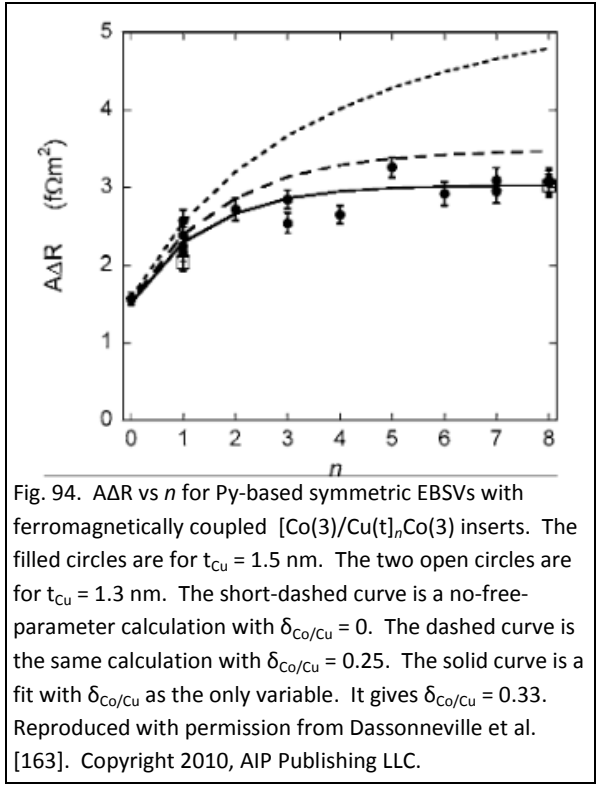

Fig. 94. AΔR vs $n$ for Py-based symmetric EBSVs with ferromagnetically coupled [Co(3)/Cu(t)$_n$Co(3) inserts. The filled circles are for $t_{Cu}$ = 1.5 nm. The two open circles are for $t_{cu}$ = 1.3 nm. The short-dashed curve is a no-free-parameter calculation with $\delta_{Co/Cu}$ = 0. The dashed curve is the same calculation with $\delta_{Co/Cu}$ = 0.25. The solid curve is a fit with $\delta_{Co/Cu}$ as the only variable. It gives $\delta_{Co/Cu}$ = 0.33. Reproduced with permission from Dassonneville et al. [163]. Copyright 2010, AIP Publishing LLC.

Later use of this technique continued mostly with Co, using its unusually long $l_{sf}^{Co}$ (see sections 8.6.1 and 8.6.2) to keep the bulk contribution to AΔR from F from masking that from $\delta_{Co/N}$. Measurements on Co/Ni [91], (Co$_{91}$Fe$_9$)/Cu [213], and Co/Ag [212] all gave moderate values of δ in the range 0.19 to 0.35, as listed in Table 16. Co/Ru, measured two different ways, gave $\delta_{Co/Ru}$ = $0.33^{+0.04}_{-0.02}$ and 0.35±0.08 [214]. The only much larger value is the most recent case of Co/Pt, which gave $\delta_{Co/Pt}$ = $0.9^{+0.5}_{-0.2}$ [164]. Such a large value soon received support from both a different experiment by Royas-Sanchez et al. [215] and an ab-initio calculation for Py/Pt by Liu et al. [216], and also helped to largely resolve a dispute over $l_{sf}^{Pt}$, engendered partly by people forgetting that $l_{sf}^{Pt}$ is approximately ∝ 1/$\rho_{Pt}$ and partly by neglect of $\delta_{Co/Cu}$ in analyzing complex data (For details see [164, 215, 216], especially Fig. 1 in [164] and Fig. 4 in [216]).

To summarize and conclude this discussion, the similarities of the middling values in Table 16 of $\delta_{Co/N}$ ~ 0.2-0.35 for most N from the techniques of Dassonneville, Triplet Superconductivity, and Manchon, provide some support for each value and each technique. The larger N = Pt value is supported by two other studies. Despite these approximate agreements, in the spirit of rules (1) and (2) in section 1.1, it would be nice to see these values checked by additional techniques.

**Table 16. $\delta_{F/N}$ and $\delta_{F1/F2}$ using the Dassonneville technique.** Listed are: the metal pair; the N-layer thickness(es), $t_n$; $\delta_{F/N}$ or $\delta_{F1/F2}$; the technique used (Dass. = Dassonneville; Trip. Sup. = triplet superconductivity; Man. = Manchon); and the reference(s). The F-layer thickness for Dass. is always 3 nm.

| Metals(structure) | $t_N$(nm) | $\delta_{F/N}$ or $\delta_{F1/F2}$ | Technique | Ref. |
|---|---|---|---|---|
| Co/Cu | 1.5;1.3 | $0.33^{+0.5}_{-0.8}$; 0.29 | Dass.; Man. | [163]; [162] |
| Co/Ni | 3 | 0.35±0.05 | Dass. | [91] |
| Co$_{91}$Fe$_9$/Cu | 1.4 | 0.19±0.04 | Dass. | [213] |
| Co/Ru | 1.4 | $0.34^{+0.04}_{-0.02}$; 0.35±0.08; 0.37 | Dass; Trip. Sup.; Man. | [214]; [214]; [162] |
| Co/Ag | 1.8;2.0 | 0.33±0.1 | Dass. | [212] |
| Co/Pt | 1.1 | $0.9^{+0.5}_{-0.2}$ | Dass. | [164] |

Notes: The Manchon values have been converted from the reported $P = [1 - \exp(-\delta)]$ to just $\delta = t_I/l_{sf}^I$ .(see section 4.2.2). The Manchon value listed for Co/Ru is more precisely for CoFe/Ru.

We end this section by examining another F/N pair where a non-zero $\delta_{F/N}$ was claimed-- Co$_{50}$Fe$_{50}$/Cu. Three different sets of measurements and analyses have been published, only one of which reported a non-zero δ. Table 17 compares the three sets of parameters. For simplicity, in the following we write Co$_{50}$Fe$_{50}$ = CoFe. Since Co$_{50}$Fe$_{50}$ is bcc, but Cu is fcc, different growth conditions might give structural differences that could complicate the interpretation that we present.

The first measurements were made in 2002 at 300K by Yuasa et al., [217] on micropillar EBSVs with active elements of the form [PtMn(15)/CoFe(t$_{CoFe}$)/Cu(5)/CoFe(t$_{CoFe}$)] and 2 nm ≤ t$_{CoFe}$ ≤ 7 nm. They found a linear variation of AΔR with t$_{CoFe}$. Since they had only enough data to determine two parameters, they assumed $l_{sf}^{CoFe} = \infty$, $\delta_{CoFe/Cu} = 0$, and 2AR = 0.4 fΩm$^2$ (taken from the value for Co$_{90}$Fe$_{10}$/Cu in ref. [6]) for all of the various interfaces in their samples. They derived β$_{CoFe}$ = 0.62 and γ$_{CoFe/Cu}$ = 0.72, with no uncertainties specified.

The next measurements were made in 2006 from 4.2K to 300K on 500 nm diam. pillars by Delille et al. [211], who were interested in both CoFe itself and its performance when laminated with thin Cu layers. The 'laminated' data will be discussed in section 10.2. Delille et al. studied two SVs, each containing a synthetic ferromagnetic pinned layer (made of CoFe-based alloys) and either a single 5 nm thick CoFe layer ('bare' SV), or a [CoFe(1)/Cu(0.3)]$_4$CoFe ('laminated' SV). Most of their parameters were taken from measurements by others or simply assumed as 'plausible'. We'll argue that some of the choices are likely incorrect, especially an assumed unusually long $l_{sf}^{CoFe}$ = 50 nm, to which the authors say their analysis is 'very sensitive'. At 4.2K, the data to be fit for the 'bare' and 'laminated' SVs were just AR(AP), AR(P), and AΔR. The parameters left as unknowns were (a) $2AR_{CoFe/Cu}^*$; (b) $\delta_{CoFe/Cu}$; and (c) the contact resistances, AR$_{contact}$, with a chosen value of $\gamma_{CoFe/Cu} \approx 0.72$ apparently also adjustable at the margin. Expecting AΔR to be insensitive to contact resistances, they used it to derive $2AR_{CoFe/Cu}^*$ and $\delta_{CoFe/Cu}$, and then used the values of AR(AP) to find the contact resistances. The authors emphasized two unexpected results, a large $2AR_{CoFe/Cu}^*$ = 3.2 fΩm$^2$ and a large polarization $P$ = 52% (giving $\delta_{CoFe/Cu}$ = 0.73).

The last measurements were made in 2010 at 4.2K by Ahn et al. [80] who studied both Co$_{50}$Fe$_{50}$/Cu and Co$_{70}$Fe$_{30}$/Cu. They used the superconducting cross-strip geometry on three sample sets at 4.2K: (1) simple [CoFe(4)/Cu(6)]$_n$CoFe(4) multilayers; (2) symmetric (t$_{CoFe}$(pinned) = t$_{CoFe}$(free)) EBSVs, and (3) asymmetric (t$_{CoFe}$(pinned) = 12 nm and t$_{CoFe}$(free) = variable) EBSVs. They used AR(AP) and the $\sqrt{(A\Delta R)AR(AP)}$ of Eq. 7 for the multilayers, and AΔR for the EBSVs to find four unknowns: (a) $2AR_{CoFe/Cu}^*$; (b) β$_{CoFe}$; (c) γ$_{CoFe/Cu}$.; and (d) $l_{sf}^{CoFe}$. They first used the multilayers to constrain parameters (a), (b), and (c) and then used AΔR for the two EBSVs to fix all four parameters. Finally, they checked their values by predicting (approximately) with no adjustability AR(AP) for the EBSVs. Their best values are given in Table 17, with the uncertainties listed in ref. [80]. We note first that Ahn et al.'s parameters for $Co_{50}Fe_{50}$ and $Co_{70}Fe_{30}$ are similar, but not identical. In contrast, most of their parameters for $Co_{50}Fe_{50}$ differ significantly from those of both Yuasa and Delille. Ahn et al.'s value of $l_{sf}^{CoFe}$= 9 nm lies far below those simply assumed by Yuasa ($\infty$) and Delille (50 nm), but falls close to the 'best fit' line in Fig. 80. Their value of $\beta_{CoFe}$ is larger than Yuasa's (which Delille simply assumed), and their values of $\gamma_{CoFe/Cu}$ and $2AR_{CoFe/Cu}^*$ are smaller, in the case of Delille much smaller. Lastly, Ahn et al. were able to fit their data assuming, with Yuasa, that $\delta_{CoFe/Cu}$ = 0. They concluded that their data are inconsistent with the large $\delta_{CoFe/Cu}$ inferred by Delille et al, but that they could not rule out modest values of $\delta_{CoFe/Cu}$ ~ 0.1.

Comparing the three sets of parameters for $Co_{50}Fe_{50}$ in Table 17, the most complete and reliable look to be those of Ahn et al., which have the advantages of known uniform current and known lead ARs, as well as much more extensive data to be fit. Combining the Ahn and Yuasa data, the large values of $2AR_{CoFe/Cu}^*$ and $\delta_{CoFe/Cu}$ proposed by Delille et al. look unlikely. It seems more likely that their large $\delta_{CoFe/Cu}$ = 0.73 results from having to fit their experimental A$\Delta$R while countercating their assumed long $l_{sf}^{CoFe}$ = 50 nm and their large $2AR_{CoFe/Cu}^*$ = 3.2 f$\Omega$m$^2$. Some independent support for Ahn et al's numbers are the agreements of their two values of $l_{sf}^{CoFe}$ with the 'best fit' straight line in Fig. 80.

**Table 17. Parameters for $Co_{50}Fe_{50}$/Cu and $Co_{70}Fe_{30}$/Cu.** The top row lists the materials, the reference, the temperatures measured, and the kind of sample used.

| | Co(50)Fe(50)/Cu Yuasa et al. [217] T = 300K. Sputt.; Micropill. | Co(50)Fe(50)/Cu Delille et al. [211] T = 4.2K-300K. Sputt.; Micropill. | Co(50)Fe(50)/Cu Ahn et al. [80] T = 4.2K Sputt.; Crossed-Sup. | Co(70)Fe(30)/Cu Ahn et al. [80]. T = 4.2K Sputt.; Crossed-Sup. |
|---|---|---|---|---|
| $\rho_{Cu}$(4.2K) (n$\Omega$m) | | 20(d) | 5±1(d) | 5±1 |
| $P_{Cu}$(300K) (n$\Omega$m) | 65 | 70(d) | 22±2(d) | 22±2 |
| $\rho_{CoFe}$(4.2K) (n$\Omega$m) | | 80 | 75±5 | 61±4 |
| $\rho_{CoFe}$(300K) (n$\Omega$m) | 129 | 191 | 113±8 | 103±6 |
| $\beta_{CoFe}$ | 0.62 | 0.62(a) | 0.89±0.02 | 0.86±0.02 |
| $\gamma_{CoFe/Cu}$ | 0.72 | 0.72 | $0.54^{+0.3}_{-0.13}$ | 0.62±0.1 |
| $2AR^*_{CoFe/Cu}$ (f$\Omega$m$^2$) | 0.83(b) | 3.2 | $0.62^{+0.3}_{-0.13}$ | 0.60±0.1 |
| $l_{sf}^{CoFe}$ (nm) | $\infty$(a) | 50(a) | $9.0^{+0.7}_{-0.4}$ | 10.9±0.5 |
| $\delta_{CoFe}$ | 0(a) | 0.73(c) | 0(a); ≤0.1 | 0(a); ≤0.1 |
| | | | | |

(a) Value simply assumed.

(b) Derived assuming $2AR_{CoFe/Cu}$ = 0.4 f$\Omega$m$^2$ and using the derived $\gamma_{CoFe/Cu}$

(c) The Delille et al. value is changed to conform to the definition of $\delta_{CoFe/Cu}$ used in this review.

(d) The difference between Cu resistivities at 300K and 4.2K in Ahn are consistent with Matthiessen's rule, $\rho_{300K}$(Cu) $-\rho_{4.2K}$(Cu) = 17 n$\Omega$m [37] , whereas those in Delille are not.

## 9. Miscellaneous Topics.

In this section we briefly discuss some miscellaneous topics that didn't fit elsewhere.

### 9.1. Pseudorandom variation of F-layer thickness.

In 1996, Mathon [218, 219] predicted that pseudorandom fluctuations (PRF) in layer thicknesses would lead to Anderson localization of electrons that would greatly enhance the CPP-MR with layer thicknesses smaller than the electron mean-free-paths. This argument was based on the agreement (see Table 14) of the measured value of $2AR^*_{Co/Cu}$ with calculations using a combination of specular interface scattering and diffuse bulk scattering.

In 1997 Chiang, et al. [220] tried to test this prediction using [Co (1.5)/Cu(0.9 or 2.2)]$_n$ multilayers at 4.2K at the first and second AF coupling peaks. The Co and Cu layers were chosen as thin as feasible in hopes of enhancing any PRF effects. With $\lambda_{Co}$ ~ 16 nm and $\lambda_{Cu}$ ~ 130 nm, these samples had $t_{Co}$, $t_{Cu}$ << $\lambda_{Co}$, $\lambda_{Cu}$, the minimum conditions needed for ballistic transport within the layers. Introducing PRF required reprogramming the control computer regularly, and manually opening and closing shutters between sputtering of layers. Checks were made that this 'slower' procedure didn't significantly perturb the data for standard samples. A PRF sequence was obtained by flipping a die and defining 1 or 2 as an increase (+) in $t_{Co}$ by a chosen $\Delta$ = 0.2nm or 0.4nm ($\cong$ 1 or 2 ML ), 3 or 4 as zero change, and 5 or 6 as a decrease (-) in $t_{Co}$ by $\Delta$. Further details of the randomization of PRF are given in [220]. If overall scattering is diffuse, so that the 2CSR model applies, and if randomization of $t_{Co}$ doesn't destroy the AF coupling, the 2CSR model predicts no change in A$\Delta$R from introducing PRF. If, in contrast, overall scattering is ballistic (bulk) and specular (interface), Mathon predicted a large increase in A$\Delta$R. For the first peak data, the CPP-MR with PRF did not display the characteristics of AF coupling. That coupling was apparently disrupted, vitiating the desired test. So the authors focused on the second peak data. There, the hysteresis curves for normal and PRF samples looked similar, and A$\Delta$R for the PRF samples were only ~ 50% of those for the regular samples, independent of $\Delta$ = 1 ML or 2 ML, different substrates, or $n$ = 30 or 40. No evidence was found for a large increase in A$\Delta$R with PRF. But, due to interface roughness, it is unlikely that the samples satisfied the 'ballistic' conditions specified by Mathon.

**9.2. Point Contact CPP-MR**.

In 1995, Schep et al. [221] predicted that a point contact to a Co/Cu magnetic multilayer would show a large GMR in the ballistic regime, even without defect scattering. In 1997, Tsoi et al. [222] tried to test this prediction at 4.2K, using a mechanically controlled Ag tip to measure the point-contact MR of a sputtered [Co(1.5)/Cu(2.0)]$_n$ multilayer that came close to satisfying Schep's requirements. The Cu thickness lies at the second antiferromagnetic exchange-coupling peak, so the 'bulk' multilayer gives a nearly-AP-state at H = 0 and large CIP-MR (~ 30%) and CPP-MR (~ 50%). The combination of Cu and Co thicknesses allows a reorientation to a P-state in a field H $\leq$ 0.1T, low enough to not make the tip unstable, and also lets injected electrons probe at least two bilayers deep for a Sharvin ballistic point-contact resistance ~ 10 $\Omega$. The probed volume should be ~ $10^3$ nm$^3$, much smaller than any previously probed. Strictly, since the current injected normal to the contact interface spreads out inside the multilayer, the Point Contact GMR should lie between CIP and CPP. Although $\lambda_{Cu}$ ~ 130 nm and $\lambda_{Co}$ ~ 16 nm are much longer than $t_{Cu}$ and $t_{Co}$, strong scattering at the rough Co/Cu interface probably makes the point contact resistance non-ballistic. For $R_t$ = 2 $\Omega$ to 50 $\Omega$, they found only small $\Delta R$ that is essentially independent of $R_t$. For even the smallest $R_t$, the effective MR ~ 25% is only comparable to the bulk CIP - MR, and the lack of systematic change of $\Delta R$ with $R_t$ gives one pause in associating the $\Delta R$ with GMR.

In 1999, Wellock et al. [223] tried nanofabricated point contacts to avoid possible local distortions due to a mechanical contact. They tried three geometries: (a) MBE deposition of a magnetically uncoupled Co/Cu multilayer into a preexisting hole in a silicon nitride membrane; (b) sputtering of an AF coupled Co/Cu multilayer before etching the contact hole in the covering membrane, but shunting the contact by a thick Cu layer; and (c) returning to geometry (a), but depositing Co(Cu) granular material into a preexisting hole. At 4.2K, cases (a) and (b) gave small MRs ($\leq 5\%$), which the authors attributed to disorder in the multilayers. Case (c) gave 14% MR, but without a multilayer.

Lastly, in 2000, Theeuwen et al. [224] used a scanning tunneling microscope (STM) at 4.2K to probe multilayers of the form $[Co(2)/Cu(1)]_n$ at the first AF peak, and $[Co(2)/Cu(2)]_n$ at the second AF peak. In both cases they found relatively large MRs, but additional studies led them to attribute most of the MRs to magnetostriction in the STM, which caused an increase in the field to push the tip harder into the sample, reducing the contact resistance. For the first AF peak samples, the field-based corrections were so large that they could not isolate any GMR. For the second peak, with its lower coercive field, they felt able to make field-based corrections, deriving GMRs that ranged up to 10% in one measuring sequence and up to 60% in another.

### 9.3. Enhancing CPP-MR by increasing Spin-Memory Loss.

Comparing the denominators of Eqs. 6 and 9 shows that placing an insert (ins) that gives strong spin-flipping, but has relatively small $AR_{ins}$, between the free-F-layer of a Py-based EBSV and its superconducting contact with $AR_{F/S} > AR_{Ins}$, should enhance the CPP-MR by removing the contribution of the larger $AR_{F/S}$ contact from the denominator of $A\Delta R$. Indeed, at 4.2K, Gu et al. [225] found that inserting 0.5 nm or 1 nm thick FeMn layers into the free-F-layer of a Py-based EBSV increased $A\Delta R$ relative to a comparison standard by amounts ranging from ~ 15% for a 12 nm thick free Py layer to over a factor of two for a 1 nm thick free Py layer.

## 10. CPP-MR for Devices.

### 10.1. Overview and outline.

As explained in section 1, CIP-GMR was discovered in 1988 [20, 21], and CPP-MR was first measured at 4.2K in 1991 [26] and then to 300K in 1993 [117]. In 1997, CIP-MR replaced anisotropic magnetoresistance (AMR) in the read heads of computers, giving an areal density ~ 1.5 Gbit/in$^2$ [Fig. 95].

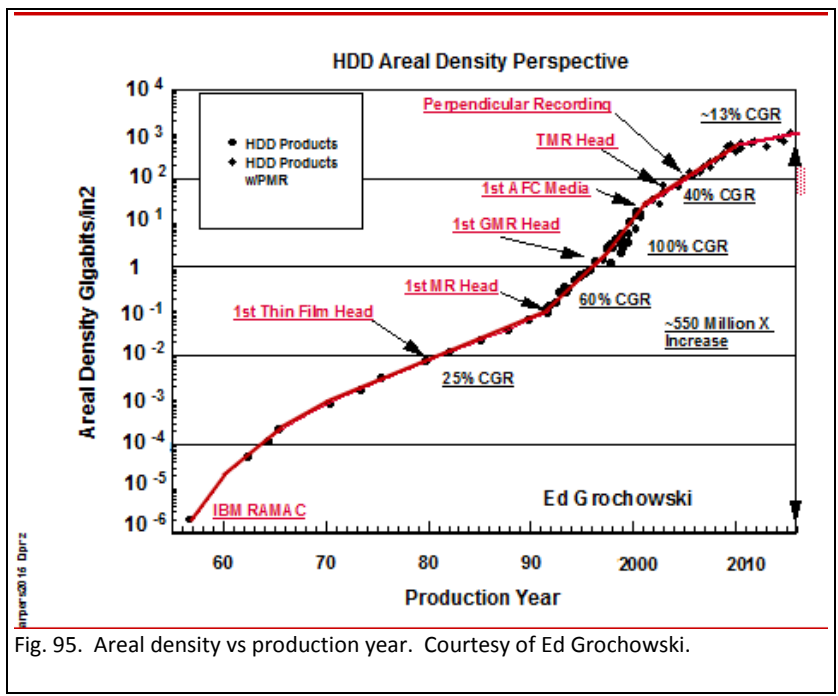

Fig. 95. Areal density vs production year. Courtesy of Ed Grochowski.

Already in 1995, Rottmayer and Zhu [226] argued that the CPP-MR had the advantages of being usually larger than the CIP-MR, having an output voltage that grows as the device area shrinks, with 'the read back voltage amplitude virtually independent of track width', and involving a simpler structure since the 'read and write gaps are coincident'. For a Py-based design, Rottmayer and Zhu calculated a recording areal density potential of 25 Gbit/in$^2$ at a time when the achieved density [Fig. 95] was ~ 0.7 Gbit/in$^2$. In 1996 and 1997,

Spallas et al. tested Co/Cu CPP multilayers at the third [227] and second [228] AF peaks, finding promising CPP-MRs ~ 30%. However, problems of hysteresis and insufficient output voltage or sensitivity precluded their structures reaching 25 Gbit/in$^2$. Below, we'll continue the 'history' of CPP-MR for devices with work after 1997.

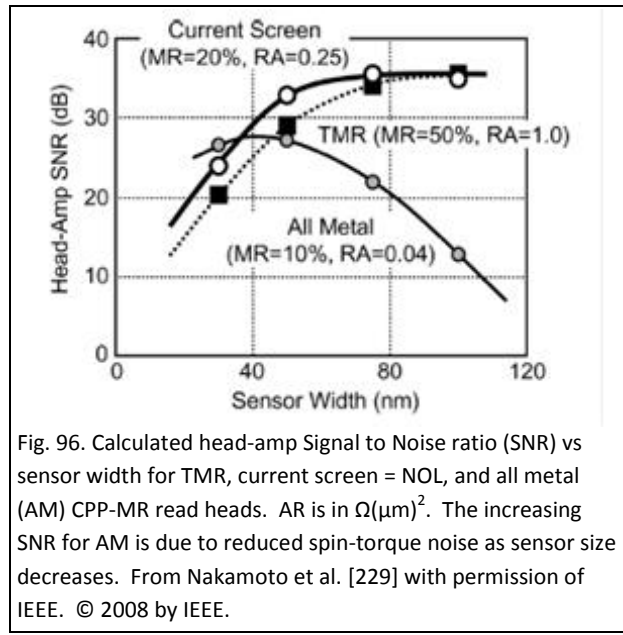

Figure 96. Calculated head-amp Signal to Noise ratio (SNR) vs sensor width for TMR, current screen = NOL, and all metal (AM) CPP-MR read heads. AR is in $\Omega(\mu m)^2$. The increasing SNR for AM is due to reduced spin-torque noise as sensor size decreases. From Nakamoto et al. [229] with permission of IEEE. © 2008 by IEEE.

Had the CIP-MR stayed dominant, the CPP-MR might have replaced it in read heads before now. However, in 2004, CIP-MR was replaced by the higher sensitivity Tunneling MR (TMR) with an Al$_2$O$_3$ tunneling barrier, giving areal density ~ 100 Gbit/in$^2$. Then in 2007, Al$_2$O$_3$ was replaced by MgO, which gave an initial areal density ~ 700 Gbit/in$^2$ and present areal density ~ 1 Tbit/in$^2$ (TBPSI). As this review is being written, the goal is to exceed 1 TBPSI, corresponding to sensor widths ≤ 26 nm. The limit for TMR is set by large AR for large TMR. Unless the AR for large TMR can be reduced, TMR is expected to 'top out' as sensor width decreases, as shown in in Fig. 96 (For assumptions see [229]), leaving a need for a next generation device. A much lower AR than TMR makes CPP-MR a potential competitor, if AR, A$\Delta$R, and the CPP-MR can all be made large enough in a stack of total thickness, $t_T$, small enough for a desired areal density (e.g., 2 TBPSI). Fig. 97, from calculations by Takagishi et al. [230] in 2010 show examples of the ARs and CPP-MRs needed for 2 TBPSI. Appendix B shows that traditional pairs such as Py and Cu cannot meet this need, making the goal of the studies in section 10 to find ways to increase AR, A$\Delta$R, and the CPP-MR, with limited $t_T$.

The rest of Section 10 is divided into four parts, each containing studies ordered chronologically. To motivate each part, we start with the generic EBSV CPP-MR read head sensor, composed of an AF pinning layer, a pinned F-layer, a spacer N-layer, and a free F-layer. For a standard F/N pair, such as Py/Cu, Appendix B shows that a total thickness ~ 26 nm gives AR and A$\Delta$R too small for a 1 TBPSI sensor.

Section 10.2. describes lamination of one or both F-layers, by inserting one or more N layers so thin that they leave the divided F-layers ferromagnetically coupled. Such lamination can increase A$\Delta$R by a factor of two or more, but not enough to be competitive by itself—i.e., without new F/N pairs.

Section 10.3. describes the insertion into the CPP multilayer stack of one or more Nano-oxide Layers (NOL = a non-conducting oxide layer with conducting metallic inclusions extending through it), leading to Current-Confined-Paths (CCP), where the current is confined to a local area much smaller than the area A of the rest of the CPP-MR stack. The NOL inclusions are usually non-magnetic and mostly inserted into the N-layer. But some examples use insertion into the F-layer(s) of NOL(s) with either non-magnetic or magnetic inclusions, and one involves magnetic inclusions replacing the N-layer. CCP studies, mostly by groups at Toshiba, Fujitsu, and Hitachi, have progressed. But problems remain with uniformity and reproducibility, likely current limitations, and the need for more effective F-materials.

Section 10.4. covers a wide variety of techniques for enhancing the CPP-MR, or reducing problems with a simple EBSV, without needing new F/N pairs. It also describes results of tests of trial read heads.

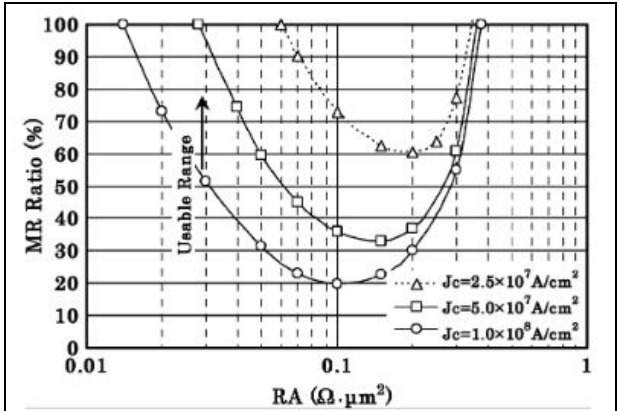

Fig. 97. Useable ranges (above the curves) of MR ratio vs AR for areal density = 2 Tb/in$^2$ and three critical current densities, $j_c$ = 1, 0.5, and 0.25 x 10$^{12}$ A/m$^2$, due to spin transfer torque. From Takagishi et al. [230] with permission of IEEE. © 2010 by IEEE.

Section 10.5. covers the search for more effective F-metals and F/N pairs, with larger resistances and larger spin-scattering asymmetries than standard F/N pairs. The goal is to find either an F-metal with large $\rho_F^*$ and $\beta_F \cong 1$, or an F/N pair with large $2AR_{F/N}^*$ and $\gamma_{F/N} \cong 1$, or both together. An F-metal with $\beta_F = 1$ has electron states of only one spin orientation at the Fermi level, and is called a half-metal. Some alloys, such as Heusler alloys, have been predicted to be half-metallic. But none has yet shown complete half-metallicity in real CPP-MR samples, where structural imperfections can weaken scattering asymmetry.

Each section begins with an overview to explain the topic and outline the main issues. It continues with a chronological history of developments, including both papers and patents. From each work, we try to extract the essential information, including AR and CPP-MR or A$\Delta$R, where given. Where we can, we give the author's estimate of potential areal density, which can be compared in Fig. 95 with the actual density in the same year. When a paper could fit into either section 10.4 or 10.5, we chose the section most convenient for our presentation. For patents, we list both the filing and published dates, using the former for chronology. If two or more patents are related, we list only the latest.

### 10.2. F-layer lamination.

### 10.2.1. Overview.

The rationale for lamination comes from Eq. 16 of the 2CSR model for a hybrid SV with metals F1 and F2 (A and Co in Eq. 16). For either F1 or F2, the contributions to A$\Delta$R from within the F-layer— $\beta_F\rho_F^*t_F$, and from the F/N interfaces— $2\gamma_{F/N}AR_{F/N}^*$, are additive. Thus, if one keeps the total F-layer thickness fixed, while inserting thin N-layers to 'laminate' the F-layer, the added F/N interfaces should increase A$\Delta$R, by larger amounts the larger is $2\gamma_{F/N}AR_{F/N}^*$. The relative effect of lamination is greater in the numerator, which initially contains only two terms, than in the denominator, which contains additional terms from the rest of the SV, including the leads. Laminating both F1 and F2 should enhance the effect, since the numerator is squared. For best effect, the inserted N-layers should be thin enough that the F-sublayers couple ferromagnetically. Their thickness should thus be: (a) less than the ~ 1 nm that gives the first antiferromagnetic exchange coupling peak, but (b) at least the typical interface thickness, $t_I$ ~ 0.5 nm, so that the laminated F-metal is still layered. Similarly, the thickness of each ferromagnetic sublayer should be $\geq t_I$. Several studies of lamination have been published.

To see how large an effect might be expected, consider Co/Cu, Py/Cu, or an assumed Heusler-like F-alloy (H/N), with unknown N. In Table 13, the maximum $2\gamma_{F/N}AR_{F/N}^*$ is ~ 1 f$\Omega$m$^2$. We use that for estimates, while hoping that an appropriate H/N pair might give a (much) larger value. For Co/Cu, from Table 5, $\beta_{Co}\rho_{Co}^*t_{Co}$ ~ 0.4 f$\Omega$m$^2$ for $t_{Co}$ = 10 nm. So inserted interfaces could give a large effect. For Py/Cu, from Table 8, $\beta_{Py}\rho_{Py}l_{sf}^{Py}$ ~ 1.2 f$\Omega$m$^2$ for $t_{Py}$ = $l_{sf}^{Py}$ = 5.5 nm. So interfaces could still be significant. If we assume large values of $\rho_H$ = 50 x 10$^{-8}$ $\Omega$m, $\beta_H$ = 0.9, but a relatively short $l_{sf}^H$ = 2 nm (see Fig.80), we get $\beta_H \rho_H^* l_{sf}^H \approx 5$ fΩm². So, unless $2\gamma_{H/N} AR_{H/N}^*$ is larger than those in Table 13, effects of lamination would be more modest. Lamination effects might also be reduced by spin-flipping at the new interfaces (see [231] and section 8.15).

**10.2.2. History.**

In 2002, Oshima et al. [232] [233], embedded a single 1.5 nm thick Cu layer between two CoFeB layers in a $Co_{88}Fe_{10}B_2$ = (CoFeB) EBSV nanopillar to form a 'laminated' free layer. At T = 300K, AΔR increased by ~ 20% for $t_{CoFeB}$ = 7 nm and by ~ 15% for $t_{CoFeB}$ = 8.5 nm. They confirmed that the increased AΔR was interfacial, by showing that AΔR grew as the number of inserts increased from zero to one to two, and that AΔR was independent of the Cu layer thickness.

Also in 2002, Yuasa et al. [217] found that inserting a half-atomic Cu layer into the middle of the $Co_{50}Fe_{50}$ layers of an EBSV increased AΔR by almost a factor of 2. However, they did not ascribe the increase to lamination. Rather, they argued that the Cu dissolved in the CoFe, and, extrapolating from a large β for Cu in Ni [36], that the Cu scatterers increased β in the CoFe from 0.62 without Cu to 0.77 with Cu. A followup study [234] showed that increasing the Cu insert thickness up to 1 nm gave a small additional increase in AΔR when the Cu was dissolved in the $Co_{50}Fe_{50}$, but a larger decrease when it stayed separated. A third study [235] showed that, upon keeping the total $Co_{50}Fe_{50}$ layer thickness fixed at 5 nm, but laminating with n Cu inserts each 0.13 nm thick, AΔR increased for n = 1 and 3, but then decreased approximately linearly, by ~ 13% from the maximum by n = 10. They associated the decrease with separately observed structural changes, but did not discover a simple explanation.

In 2003, Saito [236] submitted a patent, published in 2005, that proposed laminating the free layer of an EBSV-like structure to give AF/F(pinned)/Cu/F1/Cu/F2/Cu/AF with different thicknesses of F1 and F2 aligned antiparallel to each other to give a ferrimagnetic F1/Cu/F2 free layer. To produce strong coupling between the first AF layer and the adjacent pinned F layer, but weaker and different coupling between the second AF and the separated F2, two separate annealings in magnetic field were required, with further processing of the second AF layer in between. Saito claimed that proper choices of layer materials and thicknesses could enhance AR and AΔR and reduce the demagnetizing field of the free layer. He showed no data and gave no calculations of CPP-MR to support his claims.

Two other studies involved Cu inserts with the total thickness of the F-layer held constant.

In 2003, Eid et al. [231] inserted 0.5 nm thick Cu layers into the Co layer of a [Py/6)/Cu(4)/Co(3 or 6)/Cu(3.5)]₃ hybrid SV to give a laminated {Py(6)/Cu(4)/[Co($t_{co}$/n)/Cu(0.5)]ₙ/Cu(3.5)}₃ hybrid SV with fixed total Co thicknesses $t_{co}$ = 3 nm and n ≤ 3 or $t_{co}$ = 6 nm and n ≤ 6—i.e., minimum $t_{co}$ = 1 nm. $t_{Cu}$ = 0.5 nm was chosen to give ferromagnetic coupling while still being comparable to the expected interface thickness. At 4.2K, AΔR grew with increasing number of interfaces, N = (2n -2), by about 100% for $t_{co}$ = 3 nm and N = 4 and just over 100% for $t_{co}$ = 6 nm and N = 10. Both growths were less than expected from a simple 2CSR model. Given the expected long $l_{sf}^{Co} \cong 60$ nm, plus other results arguing against the reason being too thin Cu giving 'incomplete interfaces', Eid et al. ascribed the slower than expected growths mostly to spin-flipping (relaxation) at the Co/Cu interface (see section 8.15).

In 2006, Delille et al. [211] compared AΔR for an EBSV with a 5 nm thick $Co_{50}Fe_{50}$ free layer with AΔR for a laminated free-layer of the form [CoFe(1)/Cu(0.3)]₄CoFe(1) (see section 8.15). Lamination increased AΔR by ~ 10% and also reduced CoFe magnetostriction.

The conclusion is that lamination can increase A$\Delta$R. But, given the likelihood of some spin-flipping at the new interfaces, whether the increase in a real device will be large enough to justify the extra complexity of fabrication is less clear.

### 10.3. Nano-oxide Layers (NOL) and current-confined-paths (CCP).

### 10.3.1. Overview.

As shown in Appendix B, the AR of a simple EBSV containing standard F- and N-metals is so small, ~ 20 f$\Omega$m$^2$ {= 20 m$\Omega$($\mu$m)$^2$}, that to achieve the R ≥ 10 $\Omega$ needed for impedance matching to the components of a device, the area A must be reduced to ~ 10$^{-3}$ ($\mu$m)$^2$, giving typical dimension ~ 30 nm. In 2000, Fujiwara and Mankey [160] of the University of Alabama filed a patent, published in 2003, proposing a way to increase AR (and perhaps also A$\Delta$R) without needing more effective F-metals and F/N interfaces. Their idea was to make the N-layer a mosaic of thin regions of a conducting metal within an insulating oxide, thereby greatly reducing the area of current flow within the N-layer. The result is equivalent to a set of conducting pinholes giving current confining paths (CCPs). AR and A$\Delta$R increase due to a combination of current crowding into the much smaller area of the set of pinholes, along with the increased local resistance due to the much smaller area of flow. The patent's pictures show an array of conducting wires in parallel, similar to the top picture of nanowires in Fig. 11C. No data were given, and three descriptons of how one might make such structures were generic: (a) co-deposit a metal or alloy and an oxide that are immiscible; (b) make a layered structure of immiscible metal and oxide and heat treat; (c) co-deposit two immiscible metals, one much easier to oxidize, and then oxidize. In 2001 two related patents were filed. In one, published in 2002, Heijden et al. [237] of Seagate, proposed increasing the CPP AR by including in the multilayer stack a partially oxidized, non-magnetic nano-oxide layer (NOL). In the other, published in 2005, Dieny et al. [238] of Headway Technologies, proposed inserting thin, magnetic nano-oxide layers into the middle of each of the two F-layers in an EBSV.

The first data showing that the CCP mechanism can increase both AR and A$\Delta$R were published in 2001 by Nagasaka et al. [161], who found that inserting two very thin (nano-) oxide layers (NOL) into an EBSV substantially increased both AR and A$\Delta$R. We'll see below that their motivation and interpretation did not involve CCP. The effects of NOLs were first attributed to CCP in 2002 by Tanaka et al. [239] and Takagishi et al.[240]. These three papers were published between the submission and publication dates of the Fujiwara-Mankey patent, and the early history in section 10.3.2 strongly suggests that these authors did not know of that patent.

Most NOLs have been inserted into the spacer N-layer rather than into either of the two active F-layers. But we'll see exceptions. Most of the papers and patents describe different ways to make the NOL, with the goal of maximizing AR and A$\Delta$R. But some involve more than one NOL, and additional goals include achieving: (a) a uniform distribution of mono-diameter conducting channels to minimize deleterious effects of local heating and electromigration at the NOL, and (b) reproducible values of AR and A$\Delta$R for the many samples on a chip that will be needed to make CCP samples competitive for devices. As most NOL and CCP studies were made at company laboratories, in section 10.3.2 we list the laboratory for each paper or patent.

### 10.3.2. History.

In 1997, Egelhoff et al. [241] found that oxidizing the upper surface of a CIP-MR multilayer increased the CIP-MR. They ascribed the increase to increased specularity of scattering from the oxidized surface, which reduced the multilayer's sheet resistance. Stimulated by Egelhoff, in 2001, Nagasaka et al. [161]

of Fujitsu found that inserting NOLs both in the middle of the pinned CoFeB layer of an EBSV, and on top of the free CoFeB layer, led to a nearly 8-fold increase in AR to 1120 $f\Omega m^2$ and more than a 30-fold increase in A$\Delta$R to 23 $f\Omega m^2$. For two different devices they achieved MRs of 1.9% and 2.3%. They attributed the increases to scattering between two specular oxidized layers that lengthened the electron mean-free-path in the multilayer. Later studies confirmed that NOLs can increase both AR and A$\Delta$R, but concluded that the physics involved was not specular reflection, but rather conduction through pin-holes in the NOL. Agreeing with the terminology of Fujiwara and Mankey, this picture of strongly confined current flow through the NOL is called CCP.

In 2002, Tanaka et al. [239], also of Fujitsu, confirmed Nagazaka's observation that judicious oxidation of thin CoFeB layers enhanced AR and A$\Delta$R of SVs. However, they found that AR increased linearly with oxidization time, but A$\Delta$R grew for oxidation times up to ~ 100 sec and then decreased. They inferred that the A$\Delta$R enhancement 'was partially due to microscopically inhomogeneous current flow through the oxide layers'. Their maximum A$\Delta$R was 4.5 $f\Omega m^2$ for AR = 2000 $f\Omega m^2$.

Later in 2002, Takagishi et al. [240] of Toshiba, examined the applicability of CCP-CPP-MR heads for magnetic recording. They reference a private communication from H. Fujiwara as suggesting that a NOL produces pinholes that compress the current flow. Intriguingly, as had Fujiwara and Mankey [160], they called this phenomenon Current-Confined Paths (CCP). Their two sets of data with a NOL in the spacer layer were sensitive to details of sample preparation and they found that modeling the effect was not trivial. Their best MR was about 2% with AR about 600 $f\Omega m^2$.

Three patents were filed in 2003. In one published in 2004, Hoshiya et al. [242], of Hitachi, described a variety of different ways to produce NOLs . In one published in 2006, Sugawara [243] of Fujitsu proposed to make one or both of the pinned and free F-layers the NOL, consisting of a granular film with electrically conducting grains of magnetic material penetrating through the oxide. In one published in 2009, Fujiwara et al. [244] of the University of Alabama, asserted that larger AR and A$\Delta$R can be achieved if two separate NOLs are placed either within layers or at interfaces in an EBSV, with the values of AR and A$\Delta$R depending upon where the NOL layers are placed. The best locations depend upon whether $\beta_F$ or $\gamma_{F/N}$ is larger. They supported the claim by simple 2CSR model calculations for two NOL layers placed at various locations within the multilayer, and by some hysteresis curve data for selected placements.

2003 and 2004 also saw two more studies by Fujitsu. In 2003, Oshima et al. [245] asserted that a NOL in an EBSV is most effective at increasing the CPP-MR when it is inserted into the nonmagnetic spacer between the pinned and free F-layers. They examined several materials for the spacer NOL in CoFeB-based EBSVs, finding the largest MRs (~3% to 5%) with oxidized $Co_{50}Fe_{50}$ or oxidized CoFeB. For increasing NOL thickness, the MR grew until $\approx$ 1.2 nm and then saturated. Separate magnetization studies showed that the CoFeB  NOL was non-magnetic below 1 nm, which Oshima et al. ascribed to complete oxidation. Above 1 nm, the magnetization grew linearly with NOL thickness, which Oschima et al. attributed to only partial oxidation. Linear I-V curves led them to conclude that conduction was not tunneling, but metallic through pinholes, giving 'current-confinement'. In 2004, Tanaka et al. [246] constructed a single SV CPP read head based on the CCP multilayers described by Oshima et al. [245]. Finding a four terminal CPP-MR = 3.2% and AR ~ 600 $f\Omega m^2$, they obtained a promising read head output voltage of 0.9 $mV_{p-p}$  at $I_s$ = +4.8 mA. They noted that, if CCP read-heads could surpass Oshima's best 5%

CPP-MR, such 'CPP heads have sufficient potential for ultrahigh density recording over 150 Gb/in$^2$ with moderate resistance.'

In a patent filed in 2004 and published in 2008, Funayama et al. [247], of Toshiba, proposed enhancing both AR and A$\Delta$R using two NOLs, one inserted into the middle of each of the two spacer N-layers in a dual EBSV with two outer pinned F-layers and a single central free F-layer. In a second patent filed in 2004, but published in 2005, Horng and Tong [248] of Headway Technology described a procedure of oxygen doping all of the constituents of an EBSV and also inserting one or more NOLs into the CPP stack. They reported up to a three-fold increase in R and 2%-3% increase in CPP-MR. In a third patent, filed in 2004 but published in 2006, Li et al. [249] of Headway reported improved CPP-MRs upon replacing the usual Ta seed (on a NiFe shield layer) by a bilayer seed of NiCr on Ta, and also replacing the usual Cu spacer layer by a Cu/NOL/Cu spacer.

Extending from 2004 to 2005, Fukuzawa et al. of Toshiba published 3 related papers [250-252]. In the first two [250] [251], they used an Al$_{90}$Cu$_{10}$ starting layer for oxidation within the Cu spacer of a pinned CoFe(4) and free CoFe(1)/NiFe(3.5) EBSV. They found Ion-assisted oxidation (IAO) to improve the CPP-MR over natural oxidation (NO), by lowering the resistivity of the Cu spacer layer, to which the results are sensitive. With a fitted $\rho_{Cu}$ = 65 n$\Omega$cm for the IAO EBSVs, they achieved CPP-MRs of 5.4% for AR = 500 f$\Omega$m$^2$ and 6% for AR = 1400 f$\Omega$m$^2$. As the 'pure' Cu resistivity at room temperature is ~ 17 n$\Omega$m [37], they noted scope for substantial further improvement. In [251] they also used HRTEM images to show that the NOL consisted of an Al rich amorphous oxide with Cu rich metallic channels. Lastly, by replacing Co$_{90}$Fe$_{10}$ (CoFe) with Fe$_{50}$Co$_{50}$ (FeCo) in their EBVS, Fukuzawa et al. [252] improved their IAO produced CCP-CPP-MR to 7.5% at AR = 500 n$\Omega$m$^2$ and 10.2% at AR = 4200 n$\Omega$m$^2$. They noted that their high fitted value of $\rho_{Cu}$ = 75 n$\Omega$m left more scope for improvement than in 2004.

Later in 2005, Nakamoto et al. [253] of Hitachi produced a CPP-GMR reader and wraparound writer using a NOL. With a CPP-MR = 3% and AR = 600 f$\Omega$m$^2$ they achieved an output voltage of 2.2 mV and head-amp SNR of 30 db at an operating voltage of 120 mV.

Also in 2005, Hoshino et al. [254] of Hitachi inserted oxidized CoFe as the NOL in the central Cu layer of an unspecified multilayer. They compared CPP-MR, A$\Delta$R, and AR for samples with different precursor CoFe thicknesses, finding maxima of almost 7% MR and AR ~ 3000 f$\Omega$m$^2$ for CoFe thickness just over 3 nm. Different bottom leads of Ru, NiFe, or Cu gave similar MRs ~ 4-5%, but ARs decreasing from about 20000 f$\Omega$m$^2$ for Ru, to ~ 8000 f$\Omega$m$^2$ for NiFe, and to ~ 1000 f$\Omega$m$^2$ for Cu. The decreasing ARs correlated with decreasing grain size and decreasing roughness height.

2005 also saw two patents filed. In one published in 2007, Carey et al. [255] of Hitachi proposed to achieve larger signals in CCP-CPP-MR samples by using e-beam lithography to make one or more nano-holes in one or two insulating layers, with the hole or holes located near the sensing edge (air-bearing surface {ABS}) of the CPP-multilayer. Since the NOL is where the field to be sensed is largest, they expected substantial increases in AR and A$\Delta$R for one hole (or a set of holes) with total area = 0.1A of the area A of the multilayer and placed as near as feasible to the ABS. In the second, published in 2012, Fukuzawa et al. [256] of Toshiba proposed a CCP type CPP structure that they call a 'spring spin-valve'. Each pair of 3 or more F-layers is separated by a NOL in which the metal penetrating through the oxide is magnetic (e.g., Co), thereby weakly coupling the F-layers magnetically. The moments of the top F-layer and the bottom F-layer are initially pinned at 90$^o$ to each other, with the bottom F-layer moment strongly pinned to stay in its initial direction. If a field H is applied in the direction of the bottom F-moment, all of the moments will align parallel to this direction. If the field is reversed, the top F-layer reverses, but the authors argue that the intermediate F-layers only partly reverse, so that the F-layer moments rotate in stages from the fixed bottom F-layer to the reversed top F-layer. Since the NOLs have only small Co vias, this system seems to give a fairly large AR. But the authors do not prove, either by calculation or by data, that it enhances A$\Delta$R over that for a standard CCP-CPP EBSV.

In 2006, Jogo et al. [257] of Fujitsu showed that a Co-SiO$_2$ granular NOL gave a CPP-MR = 5.7% at AR = 1000 f$\Omega$m$^2$ with the potential for 5% at AR = 500 f$\Omega$m$^2$. They found a voltage limitation problem that they attributed to high current densities in the NOL pin-holes; their MRs dropped at applied voltage ~ 230 mV, much lower than the ~ 600 mV tolerated by TMR films.

2006 also saw three patents filed. In the first, published in 2009, to make a CPP layer more robust against electromigration, Zhang et al. [258] of Headway Technology proposed using Mg in the CCP layered structure. In the simplest case, Al-oxide is simply replaced by Mg-oxide, e.g. by depositing a CuMg alloy and oxidizing it. In other alternatives, Mg subjected to Ion-assisted oxidation (IAO) is included along with Al-oxide. In the second, published in 2010, Hoshiya et al. [259] of Hitachi, described fabricating an EBSV with a bottom, several nm thick, pinned Co (or Co$_{90}$Fe$_{10}$) layer that is first oxidized to a depth of ~ 1 nm and then covered with an N-metal separating layer and a free F-layer. The system is then annealed to convert the ~ 1 nm oxide layer into a somewhat thicker layer composed of insulating oxide surrounding Co holes that penetrate from the remaining Co in the pinned layer up to the N-layer interface. The authors argue that having the NOL located at the Co/N interface and extending into the Co-layer combines the best features of GMR and a NOL. They provide TEM data consistent with their description of the NOL structure and data showing that this NOL enhances the CPP-MR, with Co$_{90}$Fe$_{10}$ up to CPP-MR $\approx$ 6% for AR $\approx$ 400 to 800 f$\Omega$m$^2$. In the third, published in 2011, Nowak et al. [260] of Seagate proposed several different ways to achieve CCP by producing a conducting path or paths through a high resistivity layer (such as an oxide) via nanoconstriction precursers. They listed three examples of methods: (a) make the high resistivity layer thinner in local regions and break through the local regions by applying a voltage to utilize their lower breakdown voltages; (b) use a highly focused electron beam to convert a local region of an insulating layer into a metal, and use a punch current to initiate dielectric breakdown of high resistivity material around the nanoconstriction precursor; (c) use a highly focused reactive ion beam to serve the same purpose as the electron beam in (b).

In 2007, Sato et al. [261] of the Japanese Nanotechnology Research Institute (NRI) and the Institute of Advanced Industrial Science and Technology (AIST) used VF theory to compare the CPP-MR of a standard CoFe/Cu-based EBSV with the CCP-CPP-MR for a CoFe/Cu-based EBSV with a NOL insert. Their CCP-CPP-GMRs were maximum at a pin-hole diameter that varied with Cu resistivity. They found the pinhole diameter that maximized the CCP-CPP-MR to be describable by effective resistance matching.

In 2007, Fukuzawa et al. [262] of Toshiba combined Co$_{50}$Fe$_{50}$ with IAO to reach a CCP-CPP-MR = 8.2% at AR = 580 f$\Omega$m$^2$. To check for the heating problem noted above, they 'stressed' samples at 120 mV for up to 60 hrs, and found no effect.

2007 also saw three patents filed. In one published in 2008, Yuasa and Fukushima [263] of the Japanese AIST claimed that a large CPP-MR should result from a multilayer using as a CCP layer a very thin (< 1 nm) [001] oriented single-crystal or polycrystalline MgO layer with micropores, sandwiched between bcc (001) oriented F-layers. They argued that conduction through the CCP micropores of such an oriented system will occur mainly through a $\Delta$1 Bloch state, which is highly spin-polarized. They also described more complex alternative structures. In a second, published in 2008, Zhang et al. [264] described using a composite spacer involving a metal and a semiconductor, giving examples for Cu and ZnO. In relatively complex structures, a Cu(0.3)/ZnO(1.5)/Cu(0.3) spacer gave AR = 82 f$\Omega$m$^2$ and CPP-MR = 10%, and a ZnO(0.8)/Cu(0.2)/ZnO(0.8) spacer gave AR = 342 f$\Omega$m$^2$ and CPP-MR = 17%. It is not clear if the mechanism is simply CCP NOL, with the ZnO acting as an insulator containing Cu 'inclusions', or if something more is involved. In one published in 2009, Funayama et al. [265] of Toshiba argued that making a NOL with Cu and oxidized Al, leads to truncated cones of Cu embedded in Al-oxide, with more larger bases on one side of the NOL, the side varying with the preparation process. They argued that a device lives longer when the current is sent in the direction that gives lower measured noise.

In 2008, Nakamoto et al. [229] of Hitachi made a CoFe/Cu-based read-head with a NOL (which they called 'current-screened') that had a CCP-CPP-MR = 4-5% with AR = 250 f$\Omega$m$^2$. They also reported reaching a CPP-MR of 18-19% with AR = 200-300 f$\Omega$m$^2$, using an unspecified 'different F-material'. In Fig. 96 they compared signal-to-noise-ratios (SNR) for TMR (assumed MR = 50%, AR = 1000 f$\Omega$m$^2$), CCP-CPP-MR (assumed MR =20%, AR = 250 f$\Omega$m$^2$), and standard all-metal CPP-MR (assumed MR = 10% and AR = 40 f$\Omega$m$^2$). Below a track width of 40 nm, the all-metal CPP-MR became the preferred option.

A patent, filed in 2008 and published in 2012 by Berthold et al. [266], of Hitachi, described two ways to achieve a closely uniform, and reproducible, distribution of small current paths through an insulating layer. One involves depositing an array of ferritin protein molecules with inorganic cores onto an electrically conducting support layer, dissolving the ferritin molecules to leave an array of insulating oxide particles, and then depositing an electrically conducting layer over and between the oxide particles. The other involves depositing ferritin molecules containing inorganic particles on an insulating support, and dissolving the ferritin molecules to leave an array of inorganic particles to function as an etch mask. The insulating support is etched through the mask to form vias down to the layer below the support. The vias are filled by depositing an electrically conducting layer. A patent filed by Dieny et al. [267] in 2008 described a variety of ways to make CCP spin-valves.

In 2009, Wang et al. [268], of the Singapore Agency for Science, Technology and Research (A*Star) used simplified calculations to try to model CCP-CPP-MR data. Given their simplifications and assumed parameters, their calculations unfortunately don't seem to provide much guidance for new studies.

Also in 2009, a patent was filed and published in 2013 by He et al. [269], of Seagate, who described a way to use heat annealing of a mixture of Cu and MgO to obtain CCP samples with narrower ranges of GMR and AR distributions than those typically found for TMR distributions. To demonstrate reproducibility, they showed a graph of MR vs AR for pillars with d = 0.15 $\mu$m in which MR varied only from 13%-19% and AR from 170-250 f$\Omega$m$^2$.

In 2010, Yuasa et al. [270] of Toshiba reported NOL-based MR ratios ranging from 25% to 27.4%, coupled with ARs ranging from 400 to 4000 f$\Omega$m$^2$, after using a hydrogen ion treatment (HIT) on a FeCo-based EBSV CCP multilayer. To form the NOL, they started with a thin AlCu layer, which they first subjected to IAO to fully oxidize the Al to Al$_2$O$_3$ but leave the Cu less oxidized to give Cu pinholes to form the CCPs. They then applied the HIT to reduce any CuO or Cu$_2$O that had been formed by the IAO. They argued that x-ray photoelectron spectroscopy (XPS) confirmed that the Cu-oxides were significantly reduced to Cu, and that this reduction led to the increased CPP-MR.

Lastly 2011 saw a paper and two patents. In the paper, Zeng et al. [271], of the National University of Singapore, numerically analyzed failures of CCP-CPP CoFe/Cu nanopillar SVs involving reduction of both MR and exchange-bias field, and change in interlayer coupling. They concluded that current-induced high temperatures in the CCP region led to Cu mass transport into the F-layer, which roughened the interfaces, thereby inducing more spin-flipping; changing the interlayer coupling; and reducing the pinning. They argued that these unwanted effects could be mitigated by 'tuning the path density, the purity (electrical resistivity) of the Cu, and the uniformity of the pinhole areas'. In the first patent, also published in 2011, Zhang et al. [272], of Headway Technology, argued that they improved the uniformity of the Cu filaments through a Cu/Al-oxide NOL, by adopting a two step (i.e., two adjacent NOL) process. They first deposited a thin Cu layer and a thin AlCu layer, applied a short plasma ion treatment (PIT) and then a short ion-assisted oxidation (IAO) treatment. The NOL from this process, they argued, had conical shaped Cu inclusions in Al-oxide, with larger diameters at the bottom of what had been the Cu layer and a random selection of narrower diameters (down to almost points) at the top of what had been the AlCu layer. They then repeated the same process, finding evidence of more nearly uniform Cu cones from the 'double NOL'. They argued that the wider the diameter of the tops of the Cu cones at the top of the first NOL, the better the base for the cones to grow and extend to the top of the final surface. In contrast, cones with small points after the first step don't propagate through the second NOL. In the second patent, also published in 2011, Zhang et al. [273] of Toshiba, proposed to improve the reproducibility of NOLs by using an amorphous layer, and several alternative multi-step processes, to produce smoother CCP layers, with more uniform metal paths, in the middle of the spacer Cu layer. The simplest version included the following steps. (a) deposit a thin Cu layer. (b) deposit an amorphous metal or oxide layer to give a smooth surface. (c) Use a pre-ion plasma treatment (PIT) and then ion-assisted oxidation (IAO) to transform the amorphous layer and part of the Cu layer into oxides containing segregated Cu paths. (d) deposit the rest of the Cu spacer layer. Other versions involve more complex structures, including a starting Cu layer, an amorphous layer, other oxidizable layers, and PIT and IAO processing.

We conclude that NOL-CCP devices can increase AR and A$\Delta$R. But work is still needed to control the number and spatial uniformity of pinholes, uniformity of their sizes, and the purity of the metal inside them, so as to achieve reproducible output across wafers containing many devices, and to minimize deleterious effects of high-current-density, such as local heating and dielectric breakdown.

**10.4. Issues for CPP-MR devices and ways to improve them.**

**10.4.1. Overview.**

This section covers papers and patents that describe: (a) techniques intended to improve device performance or (b) measurements of device structures. Examples of techniques include different ways: (a) to enhance AR and A$\Delta$R; (b) to pin or self-pin the pinned layer, including more complex pinning structures; (c) to magnetically shield the sensor to strengthen the signal and/or minimize pickup from outside the desired sensor area; (d) to bias the free layer to keep its moment single domain as it rotates; (e) to minimize noise, including deleterious effects of spin-transfer-torque (STT); (f) to reduce electromigration at AF/F interfaces.

As the NOL and CCP systems in section 10.3. are all broadly similar, pursuing the references in each paper and patent should have uncovered almost all appropriate references. In contrast, the techniques in section 10.4 are so varied that this process is less sure. Thus, in this section we specify only that we found particular papers and patents. Both are ordered chronologically, for patents using the filing year and listing the published year in parentheses.

**10.4.2. History**.

In 1996 we found two patents. In the first, Lederman and Kroes [274] (1997) described a flux guide yoke structure to enhance the local magnetic field at a CPP-MR sensor. In the second, Dykes and Kim [275] (1997) proposed simple CPP-MR EBSV structures with: (a) one EBSV, or (b) two EBSVs separated by a conducting lead that allows a differential read head. They described shielding either with magnetic conductor leads or with separate magnetic shields. Magnetic leads simplify the structure, but give higher lead resistance than do lower resistivity metals.

In 1998 we found one paper and one patent. Pohm et al. [276] described a two-leg geometry for CPP-MR devices that increases both the active sensor length and width. In the patent, Barr et al. [277] (2001), described how to reduce the resistances of CPP current leads by extending them outside of the CPP multilayer stack to increase their area.

In 1999 we found one patent. To reduce the current flow through the sensor edges, which are likely more damaged and thus less sensitive than the sensor body, Barr et al. [278] (2000) gave a CPP-MR design with upper and lower contacts having locally smaller areas than the sensor itself.

In 2000 we found three patents. One by Knapp and Barr [279] (2001) described a series of deposition and processing steps to produce a groove-shaped quasi-CPP structure roughly like the CAP structures described in section 6.4. A second by Mao [280] (2002) described a series of modifications of a simple AF/F1/N/F3 EBSV sensor. (a) Replace the AF/F1 layers by an AF1/F1/Ru/F2/ synthetic antiferromagnet (SAF), composed of a pinning AF layer, an adjacent 'pinned' F1-layer, a thin Ru spacer, and a second 'reference' F2-layer. Choose the Ru thickness to couple the two F-layers antiparallel to each other, thus reducing the demagnetization field at the free F-layer relative to a single pinned F-layer. This step also affects the CPP-MR, because the moments of the two 'coupled' F-layers are oriented opposite to each other. If the two F-layers have the same thickness, and no spin-flipping occurs, a simple 2CSR model would predict A$\Delta$R (see Eq. 17) to be reduced from containing the sum ($\beta_F \rho_{F2}^* t_F + \gamma_{F2/Cu} AR_{F2/Cu}^*$) to containing just ($\gamma_{F2/Cu} AR_{F2/Cu}^*$) due to the F2/Cu interface. Different F1 and F2 layer thicknesses, plus likely spin-flipping in the F1/Ru/F2 system (see Table 16 in section 8.15), could increase A$\Delta$R. (b) Pin the moments of the F-layers in the SAF perpendicular to that of the free F3-layer. With this orientation, a small external magnetic field gives a linear MR response as the free F-layer moment oscillates around its stable position. (c) To minimize edge states, and stabilize the free F-layer moment as single domain as it rotates, place an extra AF layer above, and separated from, the free F3-layer by a Cu spacer. This gives AF1/F1/Ru/F2/N/F3/Cu/AF2. Choose the AF2 and Cu spacer layer thicknesses to give the desired exchange bias, but let the free F3-layer moment rotate as needed. (d) Wrap the upper current lead and shield around (but insulate it from) the free F3-layer. Wrapping reduces side reading from adjacent tracks, thus allowing higher track density. No details of layer thicknesses were given, but listed values of CPP-MR ~ 11% were almost independent of the Cu layer thickness for pinning the free F-layer. A third by Li and Araki [281] (2004), incorporated a flux guide into a single or double EBSV, so as to achieve a read gap not limited by the spin valve thickness. The EBSV lies between magnetic shields that function also as current leads. The free F-layer is electrically connected to one of the shields, and magnetically connected to the medium to be read, by a low magnetic moment, soft magnetic material, flux guide that ends at the sensor as a thin NiFeX (X = Cr, Ta, etc.) layer forming part of the free F-layer: e.g. Free layer = [CoFe(1)/NiFeTa(3)/CoFe(1)].

In 2001 we found four patents. The first, by Smith and Yang [282] (2002), stabilized the free F-layer of an EBSV in a single domain state by placing above it three layers: a non-magnetic spacer (such as Ru), a second F-layer, and an AF layer. The AF layer pins the moment of the adjacent second F-layer, the non-magnetic spacer induces anti-parallel coupling between the second F-layer and the free F-layer, and the combination of anti-parallel coupling with magnetostatic coupling between the second F-layer and the free F-layer was claimed to stabilize the free F-layer moment. The second, by Dieny et al. [238] (2005), proposed to enhance both AR and A$\Delta$R by inserting either into the middle of the F-layers of a CPP-MR multilayer, or at the F/N interfaces, thin (0.4-6 nm) layers of very high resistivity ($\rho_F > 10^5$ n$\Omega$m) magnetic oxides that were also claimed to have large values of $\beta_F$. The third and fourth, by Fontana Jr. et al. [283] (2004) and by Khizroev et al. [284] (2004), described magnetic shielding to minimize side reading from adjacent tracks

In 2002 we found two papers and six patents. Furakawa et al. [285] showed, by calculations and measurements, that head efficiency is increased in a configuration where the circular sense-current field cancels the hard bias field at the air-bearing side of the free layer. Matsuzono et al. [286] examined CPP-MR needs for 200 Gbt/in$^2$ areal density. They concluded that the CPP-MR had to be > 2.1% for AR = 1600 f$\Omega$m$^2$. In the first patent, Nishiyama [287] (2003) argued that a CPP geometry with sides tilted from the vertical to form a cone with two different slope angles, was needed to avoid shorting of the CPP stack during fabrication. In the second patent, Hienonen et al. [288] (2004) proposed to replace the simple AF pinning layer in an EBSV with a synthetic AF (SAF), which they claimed should reduce scattering of the majority electrons and thereby enhance the CPP-MR. The discussion above about Mao's 2000 patent filing suggests that the argument is not so simple. The third patent, by Gill [289](2004), described a way to reduce the thickness of a dual EBSV by coupling the two outer pinned layers to self-pinning layers, thus eliminating the need for AF pinning layers. The self-pinning layers are achieved by either: (a) high uniaxial anisotropy (e.g., a self-pinned hcp $Co_{75}Pt_{25}$ or hcp $Co_{80}Sm_{20}$ layer is anti-parallel coupled to the adjacent pinned F-layer by a thin layer of Ru); or (b) high positive magnetostriction (e.g., a self-pinned layer of fcc $Co_{50}Fe_{50}$ or $Ni_{45}Fe_{55}$ is anti-parallel coupled to the adjacent F-layer). The moments of both the self-pinning layer and the pinned layer are oriented perpendicular to that of the common free layer. The fourth patent, by Pinarbasi [290] (2004), described a CPP-sample with an AP pinned layer structure stabilized by intrinsic uniaxial anisotropy and magnetostriction, and a free layer stabilized by a spatially separated AF layer. The fifth patent, by Gill [291] (2004), described a symmetric dual EBSV, with the pinned layers on the outsides and a common, central, current carrying layer in between. Separate currents flow into each EBSV from their opposite ends and out through the common central layer. The patent claims that the current flowing out the central layer biases the free layers of the two EBSVs. The sixth patent, by Mauri and Lin [292] (2006), described a quasi-CPP geometry that they argued has two advantages: (a) its structure is like a DEBSV, and (b) it reduces the sensing area from the usual stripe height (SH) x read width (RW) to just SH x $t_F$, where $t_F$ is the free layer thickness. Concerning claim (b), in principle, the bit size can always be reduced to $t_F$ if one can shield the rest of the multilayer stack.

In 2003 we found two papers and four patents. Jiang et al. [293] found that using an asymmetric synthetic AF (SyAF) = $Co_{90}Fe_{10}(5)/Ru(0.45)/Co_{90}Fe_{10}(3)$ free layer increased the CPP-MR over that for a single $Co_{90}Fe_{10}(3)$ free layer from 0.83% to 3.56%, while doubling AR to 472 f$\Omega$m$^2$. The SyAF also gave single-domain structure and size-independent switching field. Mao et al. [294] designed a CPP-MR read-head for areal densities of 100 Gb/in$^2$. Their stack had the form Ta(5)/AF1(13)/NiFe(3)/Cu(2.5)/NiFe(3)/AF2(13)/Ta(5), with the two 'free' F-layers biased by AF1and AF2 with different blocking temperatures to let the biasing fields be set along different directions. The best result occurred with F-layer moments set at symmetric canting angles of 25$^o$, giving biasing fields at an angle of 130$^o$. In the first patent, Saito [236] (2005) described how a laminated free F-layer can be stabilized by a separated AF layer. In the second patent, Zheng et al. [295] (2006) proposed to reduce the contribution to AR from the pinning AF layer of an EBSV by greatly increasing the area of the AF layer, covering it with a thin, highly conducting F-layer that does not reduce the antiferromagnetic pinning by the AF layer, but largely shorts out its resistance. In the third patent, Hasegawa [296] (2003) described a CPP sensor stack that he called a double, dual EBSV, with dual EBSVs connected in line so that, as a track of a perpendicular recording medium passes under them, they pass sequentially over each bit. The moments of each dual EBSV are oriented so that the sensor puts out no signal when it is over either a 'positive' or 'negative' bit (i.e., the outputs of the two dual EBSVs cancel when both feel fields in the same direction), but a pulsed signal when it passes over the transition region between two bits (when the two dual EBSVs feel fields in opposite directions). Each dual EBSV has a free F-layer sandwiched between two synthetic antiferrimagnetic (SFi) layers, with appropriate N-layers separating each pair of F-layers. The moments of the SFi magnetic layers are oriented perpendicular to the ABS and those of the free layers are oriented in plane. A graph of experimental data showed that ΔR for a given sensor area A grew about 5 times from a single to a dual EBSV and another 5 times from a dual to a double dual EBSV. In the fourth patent, Nishiyama [297] (2003) described in detail a complex DEBSV in which the area of the free F-layer structure is less than that of the rest of the multilayer stack. Nishiyama argued that not reducing the area of the upper pinned F-layer structure to that of the free F-layer structure avoids contaminating the 'side faces' of the free F-layer.

In 2004 we found two papers and nine patents. Saito et al. [298] examined CPP-MR SV read heads with AF = IrMn, F =Co$_{90}$Fe$_{10}$ or Py, and N = Cu, Ru, and Ta. They found that dual SV heads with 50 nm track width and stripe height < 100 nm could potentially give 200-300 Gb/in$^2$. But the output V with these standard F and N metals was still too small for competitive devices. Zhu et al. [299] showed that current-induced noise in CPP-SVs arises from STT driven sense current excitation of coherent spin waves and 1/f noise. The first patent, by Kasiraj and Maat [[300]] (2006), proposed to stabilize the single-domain moment of a free F-layer structure against vortices induced by the sense current, by adding a nearby stabilizing F-layer that is maintained in a 'negative' vortex pattern by an adjacent AF layer. The vortex pattern in the stabilizing layer is set by taking the sample to above the blocking temperature of the AF adjacent to the stabilizing layer, applying the expected sensing current, but in the opposite direction to what will be used for measuring, and then cooling the sample to lock the vortex pattern into the stabilizing layer. They argue that applying the sense current should now make the free F-layer stay close to a single-domain, since the magnetic structure of the neighboring stabilizing layer should approximately cancel the vortex state that the sense current would otherwise produce. The second patent, by Carey et al. [301] (2006), described another way to stabilize the moment of the free F-layer in an EBSV. Their EBSV contains a pinned F-layer, an N-layer, a free-F-layer, another N-layer, and a biasing F-layer on the other side of the free F-layer. Both pinned and biasing F-layers are exchange biased by AF layers. The sensor is designed to work with the moment of the free F-layer oriented perpendicular to the moment of the pinned F-layer (but both in the layer plane). In prior designs, the self-field of the biasing F-layer stabilized the moment of the free F-layer to be anti-parallel to that of the biasing F-layer, with both orthogonal to the moment of the pinned F-layer. In the new design, the moment of the biasing F-layer is parallel to that of the pinned layer, and the moments of the free F-layer and biasing F-layer are set perpendicular to each other by exchange coupling. This geometry has the advantages that the pinning of the pinned and biasing layers can be done in a single step, and the width of the biasing layer and its AF layer can be larger than that of the rest of the sensor, thereby reducing their contribution to AR. The third [302] (2007) and fourth [303] (2009) patents, by Saito et al., proposed different ways to eliminate the AF-pinning layer from the CPP stack. The third gave a geometry in which the AF used to pin the pinned F-layer is located outside the area of the CPP-stack, thereby allowing a shorter stack. The high resistivity of the AF minimizes shunting current through it. The fourth, involved an extended geometry for the pinned layer, which must have a positive magnetostriction constant or a high coercive force. The fifth patent, by Guo and Zhu [304] (2007), described another way to bias the free F-layer moment to stay single domain as it rotates. They argue that connecting the upper contact to the multilayer stack by a strip with width less than the CPP stack's, and thus also less than the space between two hard-biasing magnets that abut the stack, lets the two magnets provide needed biasing. The sixth patent, by Li et al. [305] (2007) described a CPP multilayer stack with a wider, SyAF biasing layer above the free F-layer to more strongly stabilize the free F-layer and allow it to be made thicker to increase $A\Delta R$. The seventh patent, by Gill [306] (2007) described a CPP differential GMR sensor, where the stack consists of a central AF layer with an odd number of antiparallel coupled F-layers above it and an even number below it. The eighth patent, by Gill [307] (2006), described a procedure to hard bias the free F-layer without an AFM. On top of the free F-layer is deposited a thin (1-2 nm) conducting amorphous layer (e.g., NiTa), on top of which is deposited a thin (1-2 nm) Cr layer to break the (fcc) symmetry, and on top of which is deposited a magnetic hard bias layer about 1.5 times as thick as the free F-layer. The bcc hard bias layer has a high (~ 1000 Oe) coercive field, and Gill claims that its hard bias can be set with just a field, no annealing. The ninth patent, by Gill [308] (2005), described a method, involving oxide layers lying outside of the CPP stack, to give strong self-pinning of the top F-layer of a DEBSV, thus eliminating an AF layer.

In 2005 we found two papers by Maat et al. [309] [310] that dealt with pinning, and two patents. [309] looked for strong antiparalled coupling to allow thicker pinned and reference layers to increase $A\Delta R$. AF coupling of $Co_{90}Fe_{10}$ layers by Ir(0.6) layers was ultrastrong, but, unfortunately, thermally unstable. [310] found ultrathin (4 nm) $Co_{82}Pt_{18}$ (CoPt) layers on 2 nm of Cr to give good enough antiparallel coupling in Cr/CoPt/Ru/CoFe structures to replace FeMn or IrMn pinning in small-gap sensors. The first patent, by Freitag et al. [311] (2009) described an EBSV with a spatially extended SyAF that is intended to enhance magnetic anisotropy and thereby enhance pinning of the reference layer. The second patent, by Zhang et al. [312] (2007), argued that changing F2 in an F2/Ru/F1 antiparallel coupled SyAF, from bcc $Co_{50}Fe_{50}$ to an fcc trilayer such as $Co_zFe_{1-z}/Fe_{1-x}Ta_x/Co_zFe_{1-z}$ with z = 0.9 and x = $0.03 - 0.3$: (a) minimizes electromigration; (b) slightly increases the MR; and (c) leaves AR < 500 $f\Omega m^2$.

In 2005 & 2006, Smith and coworkers published 3 papers on effects of spin-transfer-torque (STT) on critical currents for CPP-MR SVs. In [313] they derived analytical expressions for the STT-limited critical current and presented measurements on CPP-devices with synthetic AF pinned layers. They found some unexpected complications concerning the locations of instabilities. In [314], they examined thermal noise (via micromagnetic simulations) and Spin-Transfer-Torque (STT) noise (via analytical modeling plus experiments) in CPP read sensors. They found that symmetric DBSVs ameliorate STT noise effect on the free layer (FL), making the limiting problem the STT induced instability of the reference layer (RL). In [315], they found that analytic spin-transfer torque models in CPP multilayers gave similar forms to numerical calculations that included thermal fluctuations, with only an ~ 25% difference in magnitude. They found both to generally agree with frequency spectral measurements of spin-torque-induced noise in dual EBSVs.

In 2006 we also found three patents. In the first,Gill [316] (2008) described a geometry that lets the free F-layer of an EBSV be biased by an adjacent AF layer that is located behind the CPP stack. In the second, Carey et al. [317] (2009) described a variation on a laminated free F-layer structure, in which a three-component free F-layer consists of two thin (~ 0.1-0.5 nm) (CoFe)M (M = Al or Si) F-alloy layers symmetrically bounding a thicker (> 2 nm) central NiFe alloy layer, and each is separated from the central F-layer by a thin (~ 0.1-0.5 nm) Cu layer  The F-layer contents were $(Co_xFe_{1-x})M_y$ with $0.4 \leq x \leq 0.6$ and $0.2 \leq y \leq 0.3$, and $Ni_{1-z}Fe_z$ with $0.02 \leq z \leq 0.25$. x and y were chosen to keep the free F-layer moment below a target value and magnetostriction $\leq 0$. In the third, Lin [318] (2010) described ways to grow epitaxial, fcc oriented CPP stacks to improve thermal stability.

From 2006 to 2008, a Hitachi group examined the capabilities of EBSVs and DEBSVs for read-heads. In 2006, Childress et al. [319] compared data for Ir-pinned  EBSVs and DEBSVs with standard F-metals, finding that the DEBSV: (a) increased A$\Delta$R by about a factor of 2, from 0.6 f$\Omega$m$^2$ to 1.2 f$\Omega$m$^2$; (b) increased AR from 35 f$\Omega$m$^2$ to 46 f$\Omega$m$^2$; and (c) increased the STT stable bias current by about a factor of 5. Later, examining  DEBSV read heads with sensor height = 50 nm and trackwidths ranging from 60 to 30 nm, they found [320]the CPP-MRs to be'substantially similar' down to 30 nm, indicating that all-metal CPP-MR sensors can be reliably made to such small sizes. Higher MRs should, thus, allow densities well beyond 300 Gb/in$^2$. In 2008, this group [321] extended their studies using CoFeAl. For a single SV with a CoFe pinned layer, 0.8nm thick Ru layer, and Heusler CoFeAl reference and free layers as in [322]][323], they made sensors with track widths from 60 nm to 30 nm and sensor stripe heights of 30-50 nm. They found CPP-MRs = 5.5% for 45 nm shield-to-shield distances and magnetic read widths, presenting results for various parameters, including signal/noise ratio and spin-torque limited maximum currents (maximum current density $\geq 10^{12}$ A/m$^2$). They estimated sensor compatibility with areal densities ~ 400 Gb/in$^2$, but concluded that improvements in $\Delta$R/R were needed to reach 500 Gb/in$^2$.

2007 also saw a patent filed by Gill and Pemsiri [324] (2011), who proposed to reduce the read gap between magnetic shields (and leads) by moving the AF- and capping layers of an EBSV back from the ABS and filling the space with part of the magnetic shield. This process reduced the read gap by the total thickness of the AF and capping layers.

In 2008, the Hitachi group further addressed limitations on the measuring critical current-density due to spin-transfer-torque (STT). Carey et al. [325] and Smith et al. [326] found that replacing a free layer of the form CoFe(0.6)/NiFe(3.8)/CoFe(0.2) with a given magnetization, by a synthetic ferrimagnet (SF) of the form CoFe(0.6)NiFe(4+t)/CoFe(0.2)/Ru(0.55)/CoFe(0.2) having the same magnetization, increased A$\Delta$R by almost 40% (0.68 to 0.93 f$\Omega$m$^2$) and the critical current for spin-torque instability by more than a factor of three, to 'sustainable sense current densities of $J_{max} > 2 \times 10^{12}$ A/m$^{2''}$. The SF thus substantially increased the sensor output voltage A$\Delta$RxJ$_{max}$, but only with current flowing from the free to the reference layer. Maat et al. [327] found that adding a Dy cap layer suppressed STT noise enough to increase the critical current density by a factor ~ 3.

In 2008 we also found one more paper and three patents. Nikolaev et al.[328] described the performance of a read head including a Heusler alloy (not specified) reference layer, Cu spacer, and CoFe/Cu/NiFe based compound free layer. A bias voltage ~ 80 mV (current density ~ 9x10[11] A/m[2]) gave optimal performance with low-bias CPP-MR of 9% for R = 50 fΩm[2] and a potential of 425 Gb/in[2] at bit error rate of 10[-4]. In 2009 they expanded the study and specified the F- and N-alloys (see section 10.5). The first patent, by Liu et al. [329] (2011), described use of a back shield to enhance the CPP-MR by slowing the decay of the media-produced magnetic field with distance from the ABS. The second patent, by Lin et al. [330] (2011), used elongated AF and pinned F-layers to supposedly let the current flow longer through the pinned F-layer to enhance ΔR and bypass part of the AF layer to reduce R. ΔR should be unaffected by F-layer length much beyond $l_{sf}^F$, but R will now include the larger F-length (complicated by non-uniform current flow). The third patent, by Min et al. [331] (2011), proposed using a stabilized vortex magnetization state for the free F-layer. For stack A = 10[-2] (μm)[2], the vortex is produced by applying perpendicular to the free F-layer a 1 Tesla field and a 5 mA current. They cited the advantage of having the moment of the pinned F-layer in the more stable direction parallel to the ABS.

In 2009, Zeng et al. [332] studied theoretically the degrading at high current density of IrMn-pinned CPP-MR read sensors with F = $Co_{80}Fe_{20}$. They concluded that high current density locally heated the IrMn/CoFe boundary, causing Mn atom diffusion into the CoFe that: (a) reduced the pinning, and (b) thermally stressed the CoFe, thus weakening its magnetic ordering. Fortunately, the degrading decreased dramatically as device size decreased to well below 100x100 nm[2].

In 2010, Zeng et al. [333] used 3D thermoelectrical finite element models to numerically simulate high-current-density (j ≥1 x 10[12] A/m[2]) induced failures in CIP- and CPP-MR SVs with IrMn as the AF pinning metal and Py and $Co_{80}Fe_{20}$ (= CoFe) as the F-metals. They found that Mn-impurities driven into the CoFe weakened the pinning at the IrMn/CoFe interface and reduced spin-polarization in the CoFe.

Also in 2010, Takagishi et al. [230] examined, theoretically and experimentally, the 'useable ranges' of CPP-MR for fixed critical current densities due to spin-transfer torque, $j_c$ and for areal densities of 2 Tb/in[2] or 5 Tb/in[2]. For 2 Tb/in[2], Fig. 97 shows the calculated curves of CPP-MR vs AR that separate the useable ranges (above the lines) from the unusable for $j_c$ = 1, 0.5, and 0.25 x 10[12] A/m[2]. For $j_c$ = 1 x 10[12] A/m[2], the lowest useable CPP-MR is ~ 20% at AR ~ 100 fΩm[2], giving AΔR ~ 20 fΩm[2]. For comparison, for 5 Tb/in[2] with $j_c$ = 1 x 10[12] A/m[2], the CPP-MR must be ≥ 40% with AR ~ 100 fΩm[2], giving AΔR ~ 40 fΩm[2]. By 5 Tb/in[2], they noted that thermal magnetic noise may also become critical.

In 2012, Zeng et al. [334] examined numerically the effects of stray fields from longitudinal or perpendicular magnetic recording media on electromigration in CPP-MR devices. The mean-time-to-failure (MTTF) was found to be strongly affected by stray field pulse width (effect differs for longitudinal or perpendicular media), bit length (best to reduce), and head moving velocity (best to increase).

**10.5. Search for improved F-metals (including Heusler alloys) and F/N pairs.**

**10.5.1. Overview**

As noted in section 10.1, to overcome deleterious effects on the CPP-MR of high resistivity, non-active layers (e.g. leads and pinning-AFs) in series with the active F/N layers, the best F-layers should have high resistivity and large $β_F$, and an F/N pair should have high $2AR_{F/N}^*$ and large $γ_{F/N}$. Especially nice would be a half-metallic F-layer, such as some Heusler alloys are calculated to be. We'll see that several groups have studied CPP-MR EBSVs with Heusler alloys. Unfortunately, many of these alloys appear to need annealing to T ≥ 673K to achieve largest CPP-MRs. For read-heads, annealing to T ≥ 573K is probably impractical [335], due to problems with: (a) interdiffusion and interface intermixing; (b) for some F-metals, degrading of magnetic properties, and/or (c) degrading of magnetic shields. Thus of special interest are Heusler (or Heusler-like) alloys that don't require high-T anneals. Also, as the areal density to be read grows, the bit size decreases, and the allowed total thickness of the CPP-multilayer shrinks. Thus, large CPP-MRs with F-layer thicknesses ≤ 10 nm are of special interest.

### 10.5.2. History

In 1998-1999, Caballero et al. [157, 336] first measured the CPP-MR with a Heusler alloy, NiMnSb. Their largest CPP-MR was only 9% at 4.2K.

Later in 1999, Reilly et al. [199], using superconducting cross-strips at 4.2K, and Seyama et al. [337], using a quasi-micropillar geometry at 293K, found that combining $Co_{90}Fe_{10}$ (= CoFe) with Cu outperformed Co with Cu.

In 2002, Hosomi et al. [338] studied hybrid SVs of CoFe with NiFe (we presume $Co_{90}Fe_{10}$ and Py = $Ni_{80}Fe_{20}$, but no details were given) plus a thin CoFe layer in both abutted and pillar type bottom synthetic SSVs (BSSV) and dual SSVs (DSSV). The BSSV had PtMn below active layers of CoFe(2)/Ru(0.9)/CoFe(2)/Cu(2.5)/CoFe(1)/NiFe(4). The DSSV had similar form, but symmetrically doubled. To their surprise, increasing the PtMn pinning layer thickness from 20 to 40 nm increased AΔR by up to 60%, for which they had no clear explanation. Otherwise, AΔR increased as $t_{NiFe}$ increased to 5 nm but then saturated and AΔR was independent of $t_{Cu}$. Their maximum AΔR was 3 fΩm$^2$, found with a DSSV, giving a CPP-MR of only ~ 1-2%.

In 2004, Hoshiya et al. [339] found that laminating free layers of CoFe with either CoMnGe or Fe-added magnetite, increased AΔR to 1.6 -2 fΩm$^2$, well above the AΔR = 0.9 fΩm$^2$ for CoFe alone of the same total thickness. Further details were given in 2005 [340].

In 2005. Aoshima, et al. [341] found that $Co_{75}Fe_{25}$ improved the CPP-MR over $Co_{90}Fe_{10}$ from 2.0% to 2.9%. But element shape, varied by etching depth, also affected their results, giving a largest CPP-MR = 3.3% for the most pillar-like structure. In a patent, filed in 2005 and published in 2006, Hasegawa et al. [342] proposed using as F-layer or layers in an EBSV, alloys of the form $(Co_{0.67}Fe_{0.33})_{1-x}Z_x$, with Z = Al, Ga, Si, Ge, Sn or Sb and 0 < x ≤ 0.3. For an EBSV with Al, as 'x' increased from 0 to 0.3 , the magnetostriction coefficient dropped from 90 ppm to 20 ppm and AΔR grew from ~ 4% at x = 0 to a maximum of ~ 5.7% at x = 0.25.

From 2005-2007, a Fujitsu group published 3 papers describing enhancing the CPP-MR by: (a) doping the pinned layer of an SYF to reduce $l_{sf}^F$ (which they called producing 'spin-blocking'), thereby removing the deleterious effect of the anti-aligned SYF layer on the CPP-MR [343]; (b) inserting a thin Cu layer into the free layer to add interfaces and increase the CPP-MR [344]; and (c) using high resistivity CoFeAl alloys to enhance the CPP-MR [343-345]. Jogo et al. [345] give details of the CoFeAl alloy used in [343]. A high resistivity $(Co_{75}Fe_{25})_{75}Al_{25}$ alloy enhanced the CPP-MR, and a dual EBSV gave maximum AΔR = 8.9 fΩm$^2$, with AΔR = 5.7 fΩm$^2$ for total thickness = 45 nm, and a relatively low AR ~ 1000 fΩm$^2$.

Closely related, in 2007, Maat et al. [322, 323] tested a range of $(Co_xFe_{100-x})_{100-y}Al_y$ alloys, finding $(Co_{50}Fe_{50})_{75}Al_{25}$ to maximize the CPP-MR. Replacing $Co_{50}Fe_{50}$ by this alloy in EBSVs increased the CPP-MR from 1.7% to 3.3%, but lowered the STT threshold to where STT-induced noise partly offset the advantage of the higher CPP-MR.

In 2006, studies began to use Heusler alloys such as $Co_2MnSi$ and $Ci_2FeAl_{0.5}Si_{0.5}$.

In 2006, Singh et al. [346] measured at 15K the CPP-MRs of a sputtered and FIB patterned [$Co_2MnSi(300)/Cu(10)/Co(16)/Cu(300)$] SV and a standardly patterned [$Co_2MnSi(80)/Cu(5)/Co(40)/Cu(4)$] SV, both with $Co_2MnSi$ Heusler alloys. The FIB gave $A\Delta R$ = 2.4 $f\Omega m^2$ and CPP-MR = 0.16%. The standard pattern gave $A\Delta R$ = 5.6 $f\Omega m^2$ and CPP-MR = 0.4%. Since the $A\Delta R$s were comparable to those for standard F-metals, they attributed the low CPP-MRs with FIB to large lead or contact resistances.

In 2006, Yakushiji et al. [347] measured at 300K a large $A\Delta R$ = 19 $f\Omega m^2$, but a small MR = 2.4%, for CMS(50)/Cr(3)/CMS(10) trilayers, with CMS = $Co_2MnSi$. Their samples had a bottom CMS layer epitaxially grown at 573K on a 10 nm thick Cr buffer layer to give an $L2_1$ structure, and a top CMS layer grown at 300K to give $B2$ with partial $L2_1$. Kodama et al. [348] later replaced Cr by Cu as the spacer, annealed a 20 nm thick bottom CMS layer at 673K to get an $L2_1$ structure, found the 5 nm thick top CMS layer to grow epitaxially on the Cu as $B2$ (despite a large lattice mismatch), and used $Fe_{25}Co_{75}/Ir_{28}Mn_{72}$ to pin the top CMS layer. At 300K, for a sample with 200 nm of Ag on the base 10 nm of Cr they found AR = 165 $f\Omega m^2$ and $A\Delta R$ = 14 $f\Omega m^2$, giving a maximum CPP-MR = 8.6%. At 4.2K they found $A\Delta R$ = 35 $f\Omega m^2$ and CPP-MR = 31%. Without the Ag base layer they found AR = 1080 $f\Omega m^2$, $A\Delta R$ = 15 $f\Omega m^2$ and CPP-MR = 1.4%, comparable to the values of Yakushiji et al. This large change in AR for different bottom leads is bothersome, despite the stability of $A\Delta R$. The absence of a thick bottom lead in both the Yakushiji and higher AR Kodama samples should result in current crowding and current non-uniformity (see section 6.2), giving a much larger AR without the Ag bottom layer and likely having some effect on $A\Delta R$. Yakushiji et al.'s simple note that they used Pt (a relatively high resistivity metal) as the top contact, with no thickness specified, and the absence of any information about the top contacts in Kodama et al., make these issues for the samples sputtered directly on Cr even murkier.

In 2008, Furubayashi et al. [349] made samples similar to those that Kodama et al. [348] had sputtered onto Ag, except with $Co_2FeAl_{0.5}Si_{0.5}$ (CFAS) replacing CMS and Ag replacing Cu as the spacer. The lower CFAS layer was annealed to 673K, the upper CFAS layer was unannealed. To minimize current crowding and non-uniformity, Furubayashi et al. used a 200 nm thick Ag base layer and a 200 nm thick top layer of a low resistivity metal, Cu. For CFAS thicknesses of 20 nm and 5 nm they found 300K values of AR = 108 $f\Omega m^2$ and $A\Delta R$ = 7.4 $f\Omega m^2$, giving CPP-MR $\approx$ 6.9%. In 2010, Furubayashi et al. [350] later extended 673K annealing to both layers, finding the 300K CPP-MR to increase to 12%.

In 2008, Tripathy and Adeyeye [351] reported 300K CPP-MRs in pseudo-spin-valves of the form $Fe_3O_4$/Cu/Py including what they called half-metallic $Fe_3O_4$. However, their CPP-MRs were < 1%. In addition, their CPP-MRs showed several strange features. For fixed thicknesses of $Fe_3O_4$ and Py, their CPP-MRs decreased from ~ 0.9% to 0.1% as the Cu layer thickness increased from 5 nm to 30 nm. Although they described such behavior as consistent with VF theory, it is unexpected given an expected Cu resistivity much less than those of $Fe_3O_4$ or Py. From growth of the CPP-MR with increasing t of the $Fe_3O_4$ up to 80 nm they inferred a spin-diffusion length in their $Fe_3O_4$ greater than 80 nm. Such a value seems much too long for an expected high resistivity (no information on this is given) of $Fe_3O_4$. Given their relatively thin bottom Cu lead (only 50 nm) and their unusual top Al lead (since Al oxidizes, it is rarely used for leads), plus no information on how the Py/Al interface was prepared, one wonders if their currents are uniform and their contact and lead resistances are too high compared to those of the pillars (see section 6.2) and possibly variable from sample to sample.

In 2008, Maat et al. [353] examined (CoFe$_{1-x}$Ge$_x$) EBSVs, finding a maximum CPP-MR = 6.3% (after correcting for lead resistances) with A$\Delta$R = 2.6 f$\Omega$m$^2$ for a sample of the form Underlayer/IrMn(6)/CoFe(3)/Ru(0.8)CoFe(7)/(CoFe)$_{74}$Ge$_{26}$(4)/Cu(3.5)/(CoFe)$_{74}$Ge$_{26}$(54)/Cap.

In 2008, Mizuno et al. [354] reported CPP-MR measurements on d = 0.2 µm pillars of pinned SVs involving the alternatives Co$_{51}$Mn$_{25}$Si$_{24}$ (called CMS), and Co$_{48}$Mn$_{21}$Si$_{31}$ (called Si-rich CMS), both in B2-structure after annealing. They gave no details of contacts or pinning AF. The largest CPP-MRs occurred for [AF/CMS(6)/FeCo(1)/Cu/CMS(6)/FeCo(1)] SVs with FeCo layers next to the CMS. For CMS, CPP-MR = 6%. For Si-rich CMS, CPP-MR = 9%. From calculations, they attributed the larger CPP-MR for Si-rich CMS to larger $\gamma_{CMS/FeCo}$.

Except for Mizuno et al., the Heusler alloy papers described so far focused on enhanced spin-scattering asymmetry ($\beta$) within the F-layer. In 2008-9, Nikolaev et al. [328, 355] tried an 'all-Heusler alloy' CPP-MR structure, composed of the F-metal Co$_2$MnGe (CMG) and the N-metal Rh$_2$CuSn (RCS), in hopes of enhancing interface spin-scattering-asymmetry ($\gamma$), to achieve large CPP-MRs with only thin layers. Their idea was to use band matching of N- and F-metals with 'similar Fermi momentum or band slopes near the Fermi surface' for one momentum direction (e.g., majority) but not the other (minority). Their metals gave a 300K CPP-MR = 6.7% with an A$\Delta$R = 4 f$\Omega$m$^2$ with free and reference layers as thin as 4 nm. They concluded that their majority state band matching was good, but that $\gamma$ was limited by disorder-induced minority states at the Fermi level.

In 2009 and 2010, a Tohoku University group published 3 papers [352, 356, 357] on CPP-MRs of epitaxial SVs involving the Heusler alloy Co$_2$MnSi, one with a Cr spacer and two with a Ag spacer. These papers extended the work of Yakushiji, Kodama, and Furubashi [347-349] described above. In [356] they used Cr(5)/Au(40)/Cr(15) bottom leads and a Au cap and Pt upper lead with thicknesses unspecified. Their sample areas ranged from 75x150 nm$^2$ to 300x600 nm$^2$. Annealing these samples with a Cr spacer to 673K promoted $L2_1$ ordering. Samples with active layers [CMS((20)/Cr(3)/CMS(7)] gave $\Delta$R = 6.5 f$\Omega$m$^2$ (less than Yakushiji's 19 f$\Omega$m$^2$ ), but MR = 5.2% (larger than Yakushiji's 2.4%). Iwase et al. [357] compared results with Cr and Ag spacers, using Cr(20)/Ag(40)/Cr(10) bottom leads and a cap of Ag(2)/Au(5) before an unspecified upper lead was deposited. Their active layers were [CMS(8.8)/Ag(5) or Cr(5)/CMS(8.8)]. Their areas ranged from 100x200 nm$^2$ to 300-600 nm$^2$. The one R vs H curve given shows GMR hysteresis with main switchings around 20 Oe and 100 Oe, but also extra structure. Since neither of the two identical thickness CMS layers was pinned, the hysteresis means that their behaviors differ, although TEM studies showed no clear evidence of structural differences between either the two CMS layers or with Ag or Cr. Plots of R vs 1/A gave straight lines for AR and approximately constant values for A$\Delta$R. AR was smaller for Ag, but the average A$\Delta$R = 8.2 fΩm² was larger than for Cr, AΔR = 5.8 fΩm². The maximum CPP-MRs were 17.2% for Ag and 9.7% for Cr. The authors attributed this larger CPP-MR to larger γ for CMS/Ag. They also reported that a CMS/Ag/CMS sample annealed at still higher temperatures, 773K for the lower CMS and 723K for the upper CMS, gave maximum CPP-MR = 28.8% and AΔR = 8.9 fΩm². They attributed this larger CPP-MR to better $L2_1$ ordering giving higher β. But since the largest AΔR among their first set of Ag samples was 8.8 fΩm², the larger new CPP-MR seems to be due mainly to a reduction in AR. Lastly, Sakuraba et al. [352] studied the effects of changing both the annealing temperature and the common CMS layer thickness on [CMS(t)/Ag(5)/CMS(t)] active regions with 3 nm ≤ t ≤ 11 nm. They found AΔR to grow with annealing temperature up to 823K and to also grow with t, albeit more slowly. Assuming a long spin-diffusion length for the CMS, they fit the AΔR vs t data in Fig 98 with straight lines to derive room temperature VF parameters vs annealing temperature $T_A$ ranging from β = 0.46 and γ = 0.82 for $T_A$ = 623K (CPP-MR = 15%), to β = 0.50 and γ = 0.84 for $T_A$ = 773K (CPP-MR = 36%). As illustrated, e.g., for CoFe in Fig. 79 and for NiFe in Fig. 55, the slow rise of AΔR with t in Fig. 98, extrapolating linearly to a large intercept, could also be consistent with the shorter $l_{sf}^{CMS} \approx 2$ nm that would be expected from Fig. 80 for the relatively high CMS resistivity ($\rho_{CMS}$ = 400 nΩm) estimated by the authors [352]. We'll revisit this issue of $l_{sf}^{CMS}$ below.

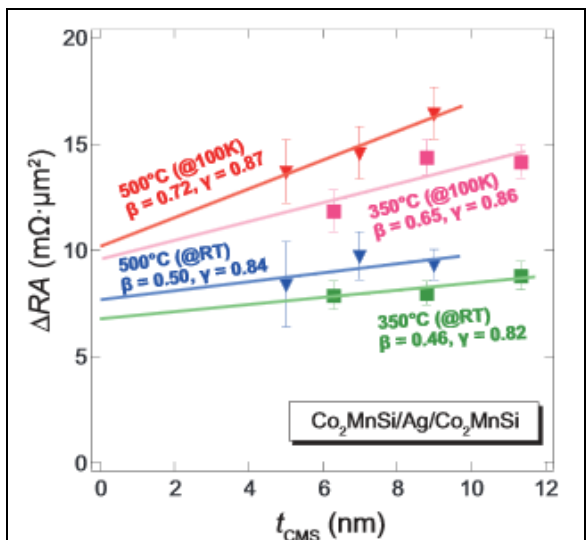

Fig. 98. A$\Delta$R vs t$_{CMS}$ for CMS/Ag/CMS trilayers annealed at 623K or 773K. Solid curves are fits assuming a 2CSR model. Reproduced with permission from Sakuraba et al. [352]. Copyright 2010 by the American Physical Society.

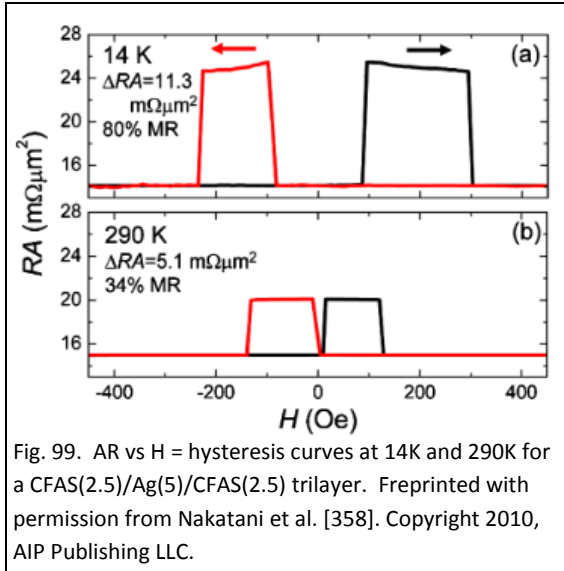

Fig. 99. AR vs H = hysteresis curves at 14K and 290K for a CFAS(2.5)/Ag(5)/CFAS(2.5) trilayer. Freprinted with permission from Nakatani et al. [358]. Copyright 2010, AIP Publishing LLC.

In 2010, 2011, Nakatani et al. [64, 358] published 2 papers on CPP-MR for symmetrical trilayer 'pseudospin-valves' of approximately $Co_2Fe(Al_{0.5}Si_{0.5})$ = CFAS with a Ag spacer. Fig. 99 shows a hysteresis curve for an Ag(lead)/CFAS(2.5)/Ag(5)/CFAS(2.5)/Ag(lead) elliptically shaped trilayer from among ones with eccentricities ~ 2 (e.g., 0.07x0.14 μm² to 0.20x0.40 μm²). At 290K, AΔR = 5.1 fΩm² and the CPP-MR = 34%. X-rays showed B2 order for the CFAS. From the data in Fig. 100 for AΔR vs $t_{CFAS}$ at 14K and 290K, Nakatani derived nominal values for VF parameters $\beta_{CFAS}$, $l_{sf}^{CFAS}$, and $\gamma_{CFAS}$, but, unfortunately, using the VF equations given in [29] for an infinite multilayer. The group soon realized its error and, as noted in section 6.2, Taniguchi et al. [65] corrected the analysis but then made some incorrect statements that they corrected in an errata [66]. Their best VF estimates at 14K are: $\beta_{CFAS}$ = 0.86; $\gamma_{CFAS/Ag}$ =0.93; $l_{sf}^{CFAS}$ = 4.2 nm; $\rho_{CFAS}$ = 624 nΩm; $\rho_{Ag}$ = 6.2 nΩm; and $2AR_{CFAS/Ag}^*$ = 3.7 fΩm². Their corrected value of $\beta_{CFAS}$ was ~ 10% larger than Nakatani's and their $l_{sf}^{CFAS}$ was ~ 40% longer. Their maximum CPP-MR at 290K was 24.6% [64]. An unresolved issue is how equal thickness CFAS layers, and an elliptical sample, could give the hysteresis behavior in Fig. 99. Something must differ between the two CFAS layers that dominates over the dipolar coupling expected for an elliptical sample.

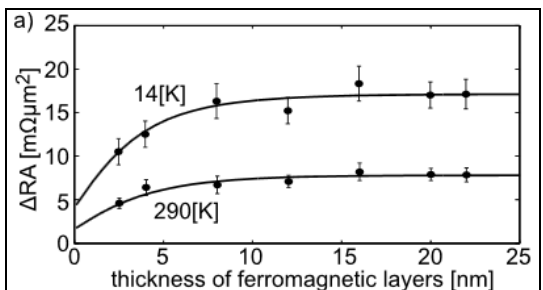

Fig.100. AΔR vs $t_{CFAS}$ for Ag/CFAS/Ag/CFAS/Ag pseudo spin-valves. Reproduced with permission from Taniguchi et al. [65]. Copyright 2011, AIP Publishing LLC.

In 2010, You et al. [359] found that inserting $Co_{75}Fe_{25}(3) = CoFe(3)$ layers between $Co_2Fe_2B(3) = CoFeB(3)$ and $Cu(2.5)$ layers in an EBSV to give [IrMn(10)/CoFeB(3)/CoFe(3)/Cu(2.5)CoFe(3)/CoFeB(3)] gave a higher CPP-MR ~ 1.6 % than the ~ 0.9% for simpler [IrMn(10)/CoFeB(6)/Cu(2.5)/CoFeB(6)] EBSVs.

Also in 2010, Shimazawa et al. [360] reported CPP-MRs ranging from 10% to over 20% for A = $0.2 \times 0.2$ $\mu m^2$ EBSVs with ZnO-based spacers. The pinning layer was IrMn and alternative spacers were $Cu(0.8)/ZnO(1.6)/Cu(0.7)$, $Cu(0.8)/ZnO(1.6)/Zn(0.7)$, or $Zn(0.8)/ZnO(1.6)/Zn(0.7)$. The pinned layer was described as composed of two magnetic layers antiferromagnetically coupled through a Ru layer, with CoFe(1.3) in contact with the spacer. The free layer was described as containing Ni rich NiFe and Co rich CoFe with a CoFe(1.0) layer in contact with the spacer. Initial CPP-MRs for Cu/Cu and Cu/Zn spacers ranged from 8-13%, and initial CPP-MRs for Cu/ZnO spacer ranged from 10-15%. By changing the oxidation condition in a way unspecified, they were able to improve their Cu/ZnO CPP-MRs to 17-24%. They used the growth of AR with bias current and agreement of noise data with Johnson noise alone, not including shot noise, to argue that the ZnO spacer resistance was ohmic.

In 2011, the Hono group published 2 papers on Heusler alloys annealed to 673K. Takahashi et al. [361, 362] reported achieving CPP-MR = 41.7% at 300K with A$\Delta$R = 9.5 f$\Omega m^2$ using ellipsoidal trilayer pillars with dimensions ranging from $0.7 \times 0.14$ to $0.2 \times 0.4$ $\mu m^2$ of $Co_{49}Fe_{23}(Ga_{14}Ge_{14})$ (CFGG $\cong$ $Co_2Fe(Ga_{0.5}Ge_{0.5})$ active layers separated by Ag. Fig. 101 shows AR(AP), AR(P), and A$\Delta$R vs T. Then, Hase et al. [363] found that they could achieve CPP-MR = 20% (A$\Delta$R = 6 f$\Omega m^2$) for B2-ordered $Co_2Mn(Ga_{0.5}Sn_{0.5})$ (= CMGS) based pillars with Ag spacer by inserting $Co_{50}Fe_{50}(1)$ layers at the two CMGS(12)/Ag interfaces.

Later in 2011, Carey et al. [335] studied CPP-MRs from 4K to 300K for EBSVs with $Co_2MnGe$ (CMG) and $Co_{50}Fe_{50}$ = CoFe. For single EBSVs, they achieved values of A$\Delta$R up to 4 f$\Omega m^2$ and CPP-MR up to 12% even with total sensor thicknesses of only 40 nm and annealing T < 523K. These values were gained with both pinned (reference) and free 'CMG' layers of the 'complex' form: CoFe(0.5)/CMG(4.4)CoFe(0.5) [Fig. 102]. DEBSVs, gave even larger A$\Delta$R = 6 f$\Omega m^2$, but CPP-MR only = 9%.

Lastly in 2011, Sato et al. [364] reported a room temperature CPP-MR ~ 75% for an epitaxial L2$_1$ ordered sample of the form Cr(20)/Ag(50)/CFMS(20)/Ag(5)/CFMS(10)/Ag(3)/Ru(3) grown on MgO(001) with the bottom CFMR layer annealed at 773K and the top at 723K. Here the Heusler alloy CFMS = $Co_2Fe_{0.4}Mn_{0.6}Si$. Unfortunately, the ARs and CPP-MRs varied by factors of two or more on a given chip.

In 2012, Sakurba et al. [365] followed Sato et al. [364] by trying CFMR layers with $Fe_xMn_{1-x}$ values of 0, 0.2, 0.3, 0.4, 0.5, and 1. Their maximum CPP-MR ~ 50% occurred with Sato's x = 0.4. Thin active layers (e.g., CFMR(4)/Ag(3)/CFMS(2)) gave A$\Delta$R ~ 12 f$\Omega m^2$. Also in 2012, Hase et al. [366] showed that replacing Ag with NiAl as both underlayer and spacer layer with Co$_2$Mn(Ga$_{0.5}$Sn$_{0.5}$) gave desired structures and better temperature stability, but smaller CPP-MR, presumably due to short $l_{sf}^{NiAl}$.

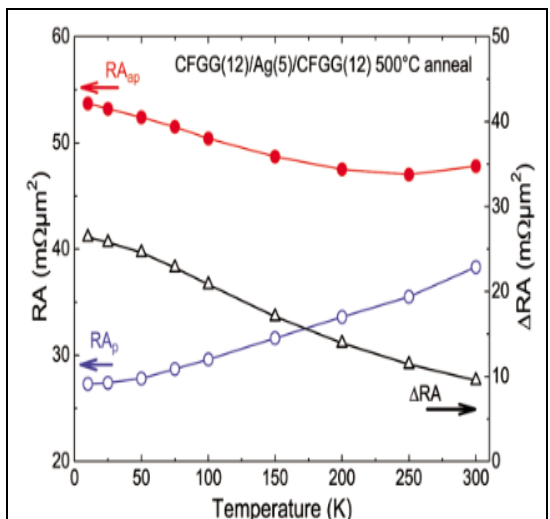

Fig. 101. AR(AP) and AR(P) (left scale), and A$\Delta$R (right scale) for trilayer ellipsoidal pillars with CFGG F-layers and Ag spacer. Reproduced with permission from Takahashi et al. [361]. Copyright 2011, AIP Publishing LLC.

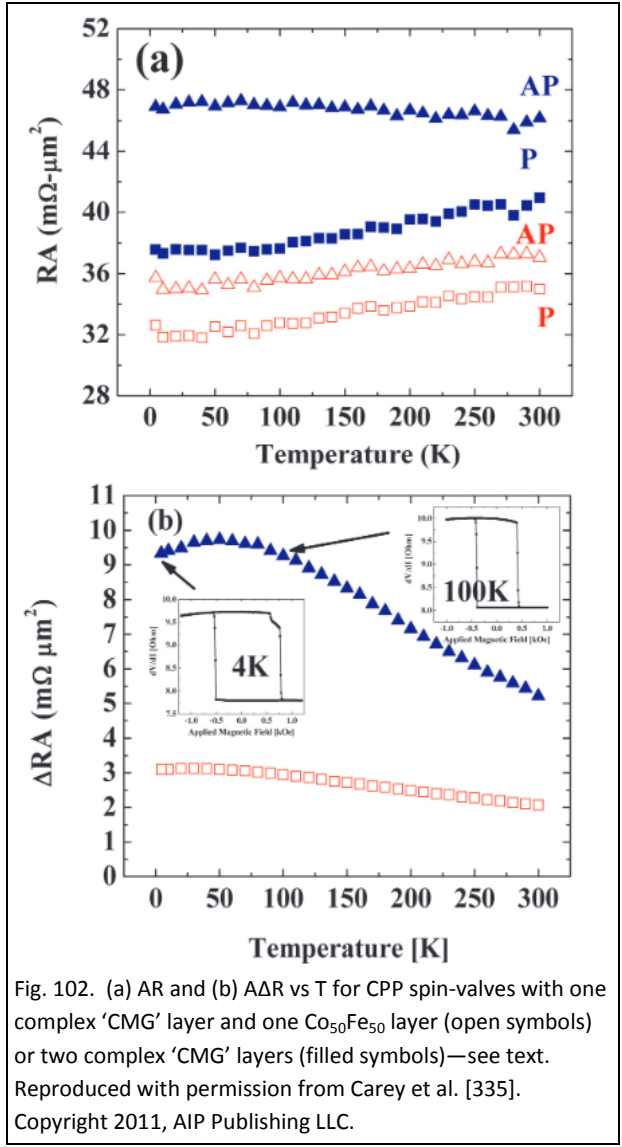

Fig. 102. (a) AR and (b) AΔR vs T for CPP spin-valves with one complex 'CMG' layer and one Co$_{50}$Fe$_{50}$ layer (open symbols) or two complex 'CMG' layers (filled symbols)—see text. Reproduced with permission from Carey et al. [335]. Copyright 2011, AIP Publishing LLC.

2013 saw 5 papers from the Hono group. Goripati et al. [367] found maximum AΔR = 8.7 fΩm$^2$ at 293K with Co$_2$Fe(Ga$_{0.5}$Ge$_{0.5}$) = CFGG based pseudo spin-valves of the form CFGG(10)/Ag(5)/CFGG(10 ) annealed to 773K. Ge = 0 or 1 gave lower AΔR than Ge = 0.5, as did lower temperature annealing. Their VF parameters are not reliable at ≥ 10-20%, as they analyzed their data with the erroneous equation of Nakatani et al. [358] for an infinite multilayer. Their hysteresis curve [Fig. 103] is closer to expectation than Fig. 99, in that the AP state is at H = 0, as expected for dipolar coupled elliptical layers. But asymmetry still indicates that the two CFGG layers differ somehow. Still studying CFGG, Li et al. [368] found that annealing a CFGG(10)/Ag(5)/CFGG(10) pseudo spin-valve in situ to 873K gave strongly L2$_1$ ordering and larger AΔR = 12 fΩm$^2$ and CPP-MR = 57% at 293K. Takahashi et al. [369] studied Co$_2$Mn(Ga$_{0.25}$Ge$_{075}$) = CMGG with a Ag(5) spacer. Annealing of CMGG(5) layers to 673K gave AΔR = 6.1 fΩm$^2$ and CPP-MR = 40.2%. Lastly, Du et al. examined polycrystalline versions of CMGG [370] and CFGG [371], thereby eliminating the high temperature anneals needed to achieve single-crystal-like samples. For CMGG, using CoFe(2) layers both below and above the CMGG(5)/Ag(7)/CMGG(5) active layer to improve the chemical ordering of CMGG(011) and Ag(111), gave larger AΔR = 3.7 fΩm$^2$ and CPP-MR = 12.2% at 293K than samples without the CoFe. For CFGG, [001] polycrystalline texture after 673K annealing gave the best AΔR = 5.8 fΩm$^2$ and CPP-MR = 16%.

2014 and 2015 saw four papers by the Hono group. Chen et al. [372] extended CFGG studies to see how the CPP-MR of epitaxially grown pseudo spin-valves varies with crystal orientation for Ag or Cu spacers. Interfaces with smaller misfits tended to give larger CPP-MRs. Then Chen et al. [373] examined B2 structure CFGG with thin NiAl (2 or 5 nm) spacers, finding good lattice matching and no dependence on (001) or (110) orientation. Their largest AΔR was ~ 4 fΩm$^2$ after annealing at 673–723K. Next, Du et al. [374] found that fully epitaxial pseudo spin valves with 10 nm layers of CFGG separated by 5 nm of AgZn and annealed to 903K gave AΔR = 21.5 fΩm$^2$ and CPP-MR = 200% at 293K. They attributed these large values to enhanced $L2_1$ ordering, induced by atomic diffusion of Zn through the CFGG. For the more practical annealing temperature of 623K, they found A$\Delta$R = 10.9 f$\Omega$m$^2$ and CPP-MR = 25.6%. Lastly, Furubayashi et al. [375] found that annealing CFGG with a CuZn spacer to 623K gave A$\Delta$R up to 8 f$\Omega$m$^2$.

2014 also saw a paper by Diao et al. [376] on both CoFeMnSi (CFMS) Heusler alloy based pseudo-SVs and AF-pinned SVs. 293K data for CFMS(20)/Ag/CFMS(10) pseudo-SVs, grown on MgO and annealed to 723 and 773K, extrapolated to CPP-MR $\geq$ 55% and A$\Delta$R $\geq$ 27.5 f$\Omega$m$^2$ upon correcting for current-non-uniformities. After similar corrections, AF-pinned SV based read-heads, grown on AlTiC, gave CPP-MR ~ 18% and A$\Delta$R ~ 9 f$\Omega$m$^2$ with a track width of 35.6 nm and a potential recording density up to 800 Gb/in$^2$.

Two more papers were published in 2015, Narisawa et al. [377] found that, after post annealing to 773K, $L2_1$ ordered Co$_2$Fe$_{0.4}$Mn$_{0.6}$Si Heusler alloys, with partially L1$_2$ ordered Ag$_{83}$Mg$_{17}$ spacers, gave 293K CPP-MRs up to 40% (48% with lead contributions removed) and A$\Delta$Rs up to 25 f$\Omega$m$^2$. The active part of the sample was CFMS(20)/AgMg(5)/CFMS(7). Read et al. [378] argued that thin (< 3 nm) AgSn (< 20%Sn) spacers with Co-based Heusler-alloys gave similar CPP-MRs (up to 15% at 293K) to Ag spacers, with still fairly long $l_{sf}^{AgSn}$ (see Table 6), but smaller grains, less roughness, better annealing stability with less interdiffusion, and better corrosion resistance.

Some of these more recent results with Heusler alloys, plus techniques in section 10.4 to reduce spin-torque effects, etc., give hope of CPP-MR read heads able to exceed 1 Tbit/in$^2$. The media could then become the limiting factor.

## 11. Summary and Conclusions.

In section 11.1, we summarize and comment on the studies that focused on physics. In section 11.2 we summarize and comment on studies related to devices. In section 11.3 we present our conclusions and suggestions for further work.

### 11.1. Summary and Comments for physics.

In 1991, within three years of the discovery of 'Giant' Current-in-Plane Magnetoresistance (CIP-MR) [20, 21], crossed superconductor measurements [26] showed that the Current-Perpendicular-to-Plane (CPP)-MRs at 4.2K of F/N multilayers (there Co/Ag) could be several times larger than their CIP-MRs. In 1993, this relationship was confirmed for Fe/Cr micropillars all the way up to 300K [117]. Once the CIP-MR was incorporated into devices such as read heads, this larger CPP-MR seemed to promise device applications. Unfortunately, the CPP-MR has two disadvantages compared to the CIP-MR: (1) the resistance of a multilayer with area > $\mu$m$^2$ is too low for devices, and (2) non-active components of devices, such as antiferromagnetic (AF) pinning layers to control AP states, are in series with CPP-MR samples rather than in parallel as with CIP-MR samples. Combined with the subsequent development of Tunneling MR (TMR) devices that had both high resistances and larger MRs than the CIP- and CPP-MRs, these problems weakened the competitiveness of CPP-MR devices until recently. The ability to make area A = 10$^3$ nm$^2$ samples, combined with recent progress on new materials that give higher CPP-MRs, have only now begun to make the CPP-MR again potentially competitive for devices. This delayed development left the focus of research until recently mostly upon the physics of CPP-MR and upon finding new materials for improved devices. For these reasons, this review has focused mainly on the understanding of physics that has grown from CPP-MR studies. Only section 10 describes studies of devices. This summary will also begin with physics and end with device potential.

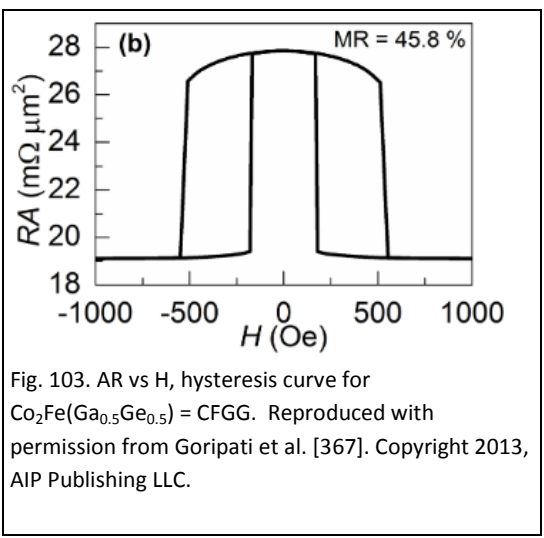

Fig. 103. AR vs H, hysteresis curve for Co$_2$Fe(Ga$_{0.5}$Ge$_{0.5}$) = CFGG. Reproduced with permission from Goripati et al. [367]. Copyright 2013, AIP Publishing LLC.

When CPP-MR studies began, it was not clear how to analyze CPP-MR data, and initial studies focused upon this issue. Early studies used a simple two-current series-resistor (2CSR) model, first derived by Zhang and Levy [27] (section 1.4), to fit data on simple [Co/N]$_n$ multilayers (see, e.g. Figs. 19-22 and 29-32). This model contains no lengths except the layer thicknesses $t_F$ and $t_N$. The model worked because moment flipping was so weak in the Co and N-metals (Ag or Cu) being studied that no other lengths were important. But as research expanded to alloy F-metals and other N-metals, moment-flipping became important, and the more general Valet-Fert (VF) theory was required.

A main conclusion of this review is that many years of data analysis show that VF theory (sections 1.4 and 4.2), which generalizes the 2CSR model to include moment flipping, can 'explain' essentially all of the experimental data covered, with the possible exception of the deviations of the data in Figs. 33 and 74 from horizontal lines (which are on the edge of uncertainty). This ability does not prove that VF theory is always correct. A given set of data can often be fit by different models. However, the wide ranging successes of VF theory puts the onus on those who dispute it to show clear-cut data where it fails. To analyze data on a given F/N pair, the 2CSR model uses 5 parameters (section 1.4). For bulk N, a resistivity, $\rho_N$, which can be separately measured. For bulk F, an enhanced resistivity, $\rho_F^*$, and a scattering asymmetry, $\beta_F$, with $\rho_F^*$ related to the independently measured F-resistivity, $\rho_F$, by $\rho_F^* = \rho_F/(1-\beta_F^2)$. For an F/N interface, twice an enhanced interface specific resistance, $2AR_{F/N}^*$, and an interface scattering asymmetry, $\gamma_{F/N}$. VF theory adds 3 more parameters (section 1.4). For bulk F- and N-metals, spin-diffusion lengths, $l_{sf}^F$ and $l_{sf}^N$. For the F/N interface, an interface spin-flipping parameter, $\delta_{F/N}$. When CPP-MR studies began, only a little was known about $\beta_F$ for F-based alloys [36] (section 3.2), and a little about $l_{sf}^N$ for a few metals and some Cu- and Ag-based alloys [46, 54] (sections 3.3,3.4). Essentially nothing was known about $l_{sf}^F$; $2AR_{F/N}^*$; $\gamma_F$; or $\delta_{F/N}$. Section 8 of the review covers tests of the applicability of the VF model to real data, derivations of the VF parameters, and tests of whether those parameters agree with values from measurements by different techniques or with no-free-parameter calculations.

To apply VF theory correctly requires three abilities:

(1) To produce antiparallel (AP) and parallel (P) states of the magnetic moments of adjacent F-layers. The P state can be achieved by raising the magnetic field (applied in the plane of the layers) to above the larger saturation field, $H_s$, of the two F-layers. Section 5 shows that the AP state can be produced and controlled using Hybrid (Fig. 8) or exchange-biased (EB) (Fig. 9) spin-valves (SV).

(2) To obtain uniform current flow through the area, A, of CPP current-flow. Uniform current flow can be produced by: (a) Crossed superconductors at 4.2K (section 6.1); (b), nanowires to above 293K (section 6.3); and (c) nanopillars of small enough diameter and with low enough resistance contacts to above 293K (section 6.2).

(3) To determine A. Both crossed superconductors (section 6.1) and small diameter nanopillars (section 6.2) can give reliable values of AR(AP), AR(P), AΔR, and CPP-MR. Carefully made nanowires and Current-at-an-Angle-to-the-Plane (CAP), can give reliable values of CPP-MR (sections 6.3, 6.4).

With this background, we now turn to CPP-MR data, focusing especially upon experimental values of the 2CSR and VF parameters and what we have learned about them. We first treat parts of section 3, and then treat sections 8.1 to 8.15, mostly in order. For each section we first explain what it covers and, where appropriate, specify the Table or Tables containing values of the relevant VF parameter.

(1) Sections 3.1 - 3.4 cover important results obtained prior to CIP- and CPP-MRs. In Section 3.2, Table 1 lists values of $\beta_F$ for the dilute limit of selected F-based alloys. Section 3.3 discusses Conduction Electron Spin Resonance (CESR) measurements of spin-flipping in some Cu- and Ag-based alloys due to spin-orbit interactions. Table 6 compares the resulting values of $l_{sf}^N$ with ones derived from CPP-MR. Section 3.5 shows (see Fig. 5) how to measure the contact specific resistances $2AR_{F/S}$ (S = superconducting Nb) that are needed to properly analyze data taken with superconducting Nb contacts. Table 2 lists values of $2AR_{F/S}$ for several F-metals; most values fall within the range $6 \pm 1$ f$\Omega$m$^2$. An explanation for these values does not yet exist.

(2) Section 8.1, and Figs. 3, 17, and 18, show that, as predicted [27], the CPP-MRs for [Co/Ag]$_n$, [Fe/Cr]$_n$, and [Co/Cu]$_n$ multilayers are usually larger (often several times larger) than the CIP-MR.

(3) Sections 8.2 – 8.4 describe mostly successful tests of the 2CSR model, and derivations of 2CSR parameters, for [Co/Ag]$_n$ (see Figs. 29-32 and Table 4) and [Co/Cu]$_n$ multilayers (see Table 5). The [Co/Ag]$_n$ data in Figs. 33-35 are only partly understood. Table 5 compares [Co/Cu]$_n$ parameters from different groups. Differences are at least partly due to different choices of AR(0) vs AR(Pk). Analyses in Section 8.4.3, along with Figs. 39 & 40, show that $\beta_F$, $2AR_{F/N}^*$, and $\gamma_{F/N}$ for Co/Cu vary only weakly with temperature T from 4.2K to 300K.

(4) Section 8.5.1 describes (see Figs. 21,22,41) how applying VF theory to appropriate data gave values of $l_{sf}^N$ for Cu and Ag-based alloys. Table 6 lists values of $l_{sf}^N$ for Cu-alloys with Ge, Pt, Mn, and Ni, and for Ag-alloys with Sn, Pt, and Mn. The derived values agree well with ones calculated independently from CESR data for Ge, Pt, and Ni impurities, or from calculations for Mn. These agreements provide support for both the 2CSR and VF models.

(5) Section 8.5.2 shows how (see fig. 43) to measure $l_{sf}^N$ for nominally pure N-metals. Table 7 shows that values of $l_{sf}^N$ for Cu or Ag are long (> 100 nm), and those for several other metals are > 20 nm, except for the heavy metals Pt and W where spin-orbit scattering should be large. Section 2.2 explains why $l_{sf}^N$ for a 'pure' N-metal is not intrinsic, but should scale roughly with inverse resistivity, $1/\rho_N$.

(6) Section 8.6 explains how values of $l_{sf}^F$ were determined (see Figs. 44, 47-51) for Co at 4.2K and 77K, Fe at 4.2K, and Ni at 4.2K. As just noted in (5), these values are not intrinsic.

(7) Section 8.7.1 describes how the 2CSR model parameters for Py and Py/Cu were obtained. Table 8 shows how the parameters improved as more and better data became available. Here, because the dominant scatterer is known to be Fe, at 4.2K the properties should be close to intrinsic.

(8) Sections 8.7.2.1 and 8.7.2.2 explain how $l_{sf}^{Py}$ was found in two independent ways to be ~ 5 nm (Figs 55-59). Section 8.7.2.3 shows that a first-principles calculation agrees with this experimental value.

(9) Section 8.8 describes two ways, one simple (Figs. 61, 90, 92)) and one more complex (Figs. 62, 91), to measure interface specific resistances, $2AR_{N1/N2}$, for N1/N2 non-magnetic metal pairs, which are often needed for complex multilayers. Tables 9 and 10 give values found by the two techniques. Table 14 in section 8.14 shows that the values for the lattice matched pairs (same crystal structure and lattice parameters the same to within ~ 1%) Ag/Au, Pt/Pd, and Pd/Ir agree surprisingly well with no-free-parameter calculations for both perfect (flat and periodic) or disordered (2 monolayers of a random 50%-50% alloy) interfaces. Values for non-lattice matched pairs do not agree with calculated ones, probably due to sensitivity to details of interfacial structure.

(10) Section 8.9 covers the contentious issue of mean-free-path (mfp) effects. If spin-flipping is completely absent, so that the 2CSR model applies, values of A$\Delta$R should be the same for Interleaved (I = [F1/N/F2/N]$_n$) or separated (S = [F1/N]$_n$[F2/N]$_n$) multilayers. In fact, significant differences have been seen, and grow with increasing $n$ (Figs. 63-67, 69, 70). In our view, these differences are most reasonably explained by spin-flipping within the F-layers or at the F/N interfaces, or a combination of both. But others attribute the differences to 'mean-free-path' (mfp) effects—which do not appear in VF theory. Section 8.9 gives the arguments pro and con, along with data and relevant calculations.

(11) Section 8.10.1 and Tables 1 and 11 show that values of $\beta_F$ from CPP-MR for F-alloys agree well in sign and reasonably well in size with ones from Deviations from Matthiessen's Rule (DMR) [36] for dilute F-alloys, provided that $\beta_F$ is determined concurrently with $l_{sf}^F$ (Figs. 55,56,78,79). These agreements tend to support a claim of 'universality' for $\beta_F$.

(12) Fig. 80 in section 8.10.2 shows that values of $l_{sf}^F$ for F = Py, Co$_{91}$Fe$_9$, Ni$_{97}$Cr$_3$ and Ni fall reasonably well along a single straight line through the origin. In contrast, $l_{sf}^{Co}$ falls well above this line and $l_{sf}^{Fe}$ falls below it. Values for Co$_{50}$Fe$_{50}$ and Co$_{75}$Fe$_{25}$ in columns 4 and 5 of Table 17 in section 8.15 fall close to the same line. Both why so many values lie close to this line, and why $l_{sf}^{Co}$ and $l_{sf}^{Fe}$ deviate so far from it, have not yet been well explained. The long $l_{sf}^{Co}$ might be due to weak spin-flipping at stacking faults.

(13) Section 8.11 explains (see Figs. 62,80) how to measure the spin-flipping parameter, $\delta_{N1/N2}$, at interfaces between non-magnetic metals, N1/N2. Table 12 gives values for twelve N1/N2 pairs. The largest values are for Cu with heavy metals, as expected if spin-orbit interactions are the main source of spin-flipping. No calculations yet show whether these values can occur for perfect interfaces, or if interface alloying is required. The existence of non-zero values of $\delta_{N1/N2}$ suggested the possibility of non-zero values of $\delta_{F/N}$, the topic of section 8.15.

(14) In section 8.12, Table 13 lists values of the interface parameters, $\gamma_{F/N}$ and $2AR_{F/N}^*$, and their product, $2\gamma_{F/N} AR_{F/N}^*$, for various standard F- and N-metals and alloys, Eqn. 5 shows that this product determines their interfacial contribution to A$\Delta$R. The largest value of $\gamma$ is $\gamma_{Co/Ni} = 0.94$. The largest value of $2AR_{F/N}^*$ is $2AR_{Co/Al}^* \cong 11$ f$\Omega$m$^2$. The largest reliable values of $2\gamma_{F/N} AR_{F/N}^*$ are $\cong 1$ f$\Omega$m$^2$.

(15) Section 8.13 covers three studies of how the CPP-MR is affected by enhanced interfacial roughness due to increased sputtering pressure. Two studies, on Co/Ag (Figs. 82,83) and Fe/Cr (Fig. 88), gave a decrease, but the third, on Fe/Cr (Fig. 88), gave an increase. The difference is not yet explained.

(16) Section 8.14 and Tables 3, 14, and 15 compare no-free-parameter calculations with measurements of $2AR_{N1/N}$ and $2AR_{F/N}^*$. Tables 3 and 14 show that the calculations and measurements agree surprisingly well for lattice matched metal pairs. However, they disagree, usually substantially, for non-lattice matched pairs. Tables 3 and 15 show that no-free-parameter calculations of $\gamma_{F/N}$ for Co/Cu, Fe/Cr, and Co/Ni, are reasonable for Co/Cu or Co/Ni, but too small in magnitude for Fe/Cr.

(17) Section 8.15 describes (see Fig. 94) how to measure $\delta_{F/N}$ or $\delta_{F1/F2}$, using ferromagnetically coupled [F/N]$_n$F or [F1/F2]$_n$F1 inserts into the central Cu layer of a symmetric Py-based EBSV. Table 16 shows that values of $\delta_{F/N}$ or $\delta_{Co/Ni}$ are mostly modest ($\delta$ = 0.19-0.33), except for Co with the heavy metal Pt, where $\delta_{Co/Pt} \approx 0.9$ is large. These values, which need to be reproduced, are not yet explained.

(18) In Section 9, studies of pseudorandom variations of Co-layer thickness show little effect, point contact CPP-MR measurements don't show much enhancement over usual values, and judiciously introducing spin-memory loss within the free F-layer of an EBSV is shown to increase the CPP-MR.

**11.2. Summary and comments for devices.**

Section 10.2 shows that laminating an F-layer by inserting thin layers of N can increase A$\Delta$R and the CPP-MR. But it isn't clear that the increase when an F-alloy has high resistivity and large $\beta_F \sim 1$ will be large enough to justify the extra complication of sample fabrication. Spin-flipping at the newly introduced interfaces will also likely reduce the benefit of lamination.

Section 10.3 shows that Nano-Oxide Layers (NOL) leading to current confined paths (CCP) can enhance both A$\Delta$R and CPP-MR. However, uniformity and reproducibility problems are not yet solved. And Fig. 96 shows that CCP becomes relatively less effective as areal densities approach 1 Tbit/in$^2$.

Section 10.4 describes ways to improve CPP-MR performance and also contains examples of outputs of test read heads. Improvements include using Double EBSVs (DEBSVs) to increase A$\Delta$R and CPP-MR and reduce unwanted Spin-Transfer-Torque (STT) effects, but at the expense of greater sample thickness. Using capping or synthetic ferrimagnet (SFi) layers can also reduce unwanted STT effects.

Section 10.5 describes the progress made in increasing both A$\Delta$R and the CPP-MR, within total thickness constraints, by finding new F-metals (e.g. Heusler alloys) and F/N combinations. Nearly half-metallic F-alloys, with high $\rho_F$ and $\beta_F \sim 1$, combined with appropriate N-layers, can provide larger values of A$\Delta$R and CPP-MR than the 'standard' F/N pairs. Unfortunately, some require too high annealing temperatures, and the correlation of short values of $l_{sf}^F$ with high $\rho_F$ (see Fig. 80) will likely limit the useful thicknesses of F. But, with help from items in section 10.4, such alloys may well be able to produce read-heads for areal densities $\geq$ 1 Tbit/in$^2$.

**11.3. Conclusions and Work for the Future.**

Since the first measurements in 1991 [26], much has been learned about the physics underlying the CPP-MR. In almost every case, that physics seems to be surprisingly well described by the simple Valet-Fert (VF) model, which assumes diffusive transport and (if leads are neglected) describes the CPP-MR of an F/N multilayer using 8 parameters, the values of which are independent of layer thicknesses or number. Of the 8, 5 describe bulk properties of the F- and N-metals, and the evidence is good that 2 of these, the bulk resistivities $\rho_F$ and $\rho_N$, can usually be set (at least approximately) by independent measurements on separately prepared thin films. So, once it became clear that the VF model worked, studies focused on the other 3 bulk parameters, $\beta_F$, $l_{sf}^F$, and $l_{sf}^N$, and the 3 interface parameters, $\gamma_F$, $2AR_{F/N}^*$, and $\delta_{F/N}$, plus upon additional interface parameters, $2AR_{N1/N2}$, and $\delta_{N1/N2}$, that arise in multilayers that contain N1/N2 interfaces. This review explains how each parameter is measured and contains both figures and tables for a range of F/N combinations. The physics underlying most of the parameters seems reasonably well described by VF theory. However, in only a few cases, such as $2AR_{F/N}^*$ and $2AR_{N1/N2}$ for lattice-matched metal pairs, $l_{sf}^N$ for Cu- and Ag-based alloys, $l_{sf}^{Py}$, and $\beta_F$ for some F-alloys, is there semi-quantitative or quantitative agreement of CPP-parameters with those measured by a different technique or calculated with no adjustable parameters. Further work is needed to answer questions such as the following.

(a) Why are the values of $2AR_{F/Nb}$ for several F-metals and F-alloys with superconducting Nb all roughly consistent with $2AR_{F/Nb}$ = 6 $\pm$ 1 f$\Omega$m$^2$..

(b) What determines the values of the spin-flipping parameters at F/N or N1/N2 interfaces, $\delta_{F/N}$ or $\delta_{N1/N2}$, and do they require interfacial intermixing?

(c) How does interfacial structure, including physical roughness, effect $2AR_{N1/N2}$, $\gamma_{F/N}$, and $2AR^*_{F/N}$?

(d) Is the failure of calculated values of $2AR_{N1/N2}$ and $2AR^*_{F/N}$ for non-lattice matched metal pairs to agree with measured values due to lack of proper treatment of how the lattices of the two metals adjust across the interface, or to some other limitation on the calculations?

(e) How widespread is the correlation of $l^F_{sf}$ vs $1/\rho_F$ shown by the straight line in Fig. 80, and why is $l^{Co}_{sf}$ so much longer than this correlation would predict.

(f) Does VF theory adequately explain the observed differences in data between interleaved (I) and separated (S) multilayers covered in section 8.9? In our view it does, so long as spin-flipping at interfaces is included. However, not everyone agrees with us.

(g) Is our argument correct that the deviations from VF expectations of the data in Figs. 34 and 35 are due to extra spin-flipping in hcp oriented thick Co layers?

(h) Why do the plots of $\sqrt{(A\Delta R)AR(AP)}$ vs $t_{Ag}$ or $t_{Cu}$ in Figs. 33b and 74 deviate from the VF predicted horizontal lines? The data for $t_{Ag}$ in Fig. 33b are the lesser problem, as similar deviations in better controlled EBSVs in Fig. 75 are consistent with the expected spin-diffusion length in Ag. However, the data for $t_{Cu}$ in Fig. 74 are not so easily explained. Given the different results in Figs. 33b and 75, and that the data in Fig. 74 deviate from a horizontal line by only about a standard deviation, we conclude that the deviations are not far enough above noise to invalidate the VF model. But, as explained in section 8.9, others disagree and attribute the deviations to mean-free-path (mfp) effects. The two models could be distinguished by measuring to much larger values of $t_{Cu}$ and $t_{Ag}$, since for mfp effects the deviations should saturate, whereas for finite spin-diffusion lengths they should confnue to grow. Any mfp effects might also be more visible in multilayers with more nearly perfect interfaces.

For devices, the main tasks are to find: (a) F-layers with both larger $\rho^*_F$ and $\beta_F$ closer to 1, as expected for half-metals, and/or (b) F/N pairs with large values of both $2AR^*_{F/N}$ (see, e.g. large values of F/Al in Table 13) and $\gamma_{FN}$ (unfortunately small for F/Al), thereby enhancing the CPP-MR, without needing annealing temperatures too high for devices. Achieving higher densities in media is also an issue.

**Acknowledgments**

The author thanks W.P. Pratt Jr., N.O. Birge, and E. Grochowski for helpful comments and suggestions, and E. Grochowski for supplying Fig. 95. Some support for research covered in this review was provided by the Division of Materials Research (DMR) of the US National Science Foundation (NSF), by Seagate Inc., and by the Korean Institute of Science and Technology (KIST). Permission to reproduce figures was obtained from either one of the authors, or the copyright owner, or both.

## 12. Appendices.
### 12.1. Appendix A: Mean-free-paths and Spin-diffusion lengths.

For this review we need to distinguish between three scattering times, $\tau$, $\tau_s$, and $\tau_{sf}$, and two lengths, mean-free-paths, $\lambda$, and spin-diffusion lengths, $l_{sf}$. $\tau$ is the momentum relaxation time, $\tau_s$ is the spin-relaxation time, and $\tau_{sf}$ is the spin-flip time. The values of $\lambda$ are coupled to those of $\tau$ by the relation $\lambda_x = v_F\tau_x$, where $v_F$ is the Fermi velocity. $\lambda_x$ is the mean linear distance between events separated by $\tau_x$. That is, each electron moves linearly through the metal between scattering events. In contrast, $l_{sf}$ is the mean distance an electron diffuses between scattering events in which its magnetic moment reverses. Making these simple definitions quantitative can be a bit tricky.

### 12.1.1. Mean-Free-Path.

The Drude theory of free electron transport gives the relation between the transport conductivity, $\sigma_T$, and the momentum relaxation time, $\tau$, as

$$\sigma_T = ne^2\tau/m \qquad\qquad (A.1)$$

Here n is the number of electrons per unit volume, e is the magnitude of the electron charge, and m is the electron's mass. Defining the resistivity $\rho = 1/\sigma_T$, converting $\tau$ to $\lambda$ by $\lambda = v_F\tau$, and rearranging gives

$$\rho_b\lambda_b = (mv_F/ne^2). \qquad\qquad (A.2).$$

In eq. A.2, the subscripts b (for bulk) indicate that the product $\rho_b\lambda_b$ is a constant representative of the metal being studied. Ref. [37] shows that, for most metals, this constant lies between 0.5 and 2 f$\Omega$m$^2$. Thus, $\rho_b\lambda_b = 1$ f$\Omega$m$^2$ is a reasonable rough estimate for metals where a better value is not available. For metals of special interest, Cu and Ag, we'll specify better values just below. From Eqs. A.1 and A.2, the mean-free-path for a sample of measured bulk resistivity $\rho$ is estimated from

$$\lambda = \rho_b\lambda_b/\rho. \qquad\qquad (A.3)$$

Although Eq. A.3 was derived for a free electron metal, it is assumed to be valid in general, given an appropriate value of $\rho_b\lambda_b$ for the metal of interest. Ziman [379] shows that Eq. A.1 can be generalized to a real Fermi surface and to the possibility that $\lambda$ might vary over the Fermi surface by the equation:

$$\sigma_T = (1/2\pi)(e^2/3h)\int \lambda (dS_{F.}). \qquad\qquad (A.4)$$

Here h is Planck's constant and the integral is over the Fermi surface $S_F$. If $\lambda$ is taken to be constant and pulled out of the integral, then integration over a spherical Fermi surface gives Eq. A.1. Inserting the real Fermi surface area gives the best estimate of $\rho_b\lambda_b$. Since Eq. A.3 is only approximate, we refer to it as giving only an 'estimate' of $\lambda$.

$\rho_b\lambda_b$ can also be estimated experimentally, by measuring 'size effects' on thin wires or films [37], or the anomalous skin-effect [37]. In a few metals, e.g. Cu and Ag, free electron calculations are reasonably consistent with such experiments at 4.2K as shown in Table A. For analyses in the text, we use $\rho_b\lambda_b$(Cu) = 0.66 ± 0.05 and $\rho_b\lambda_b$(Ag) = 0.84 ± 0.1.

**Table A.  $\rho_b \lambda_b$ f$\Omega$m$^2$ from  Free Electron Calculation (FEC), Anomalous skin-effect (ASE), and size effect (SE) measurements at 4.2K [37].**

| Metal | FEC | ASE | SE | SE | SE | SE |
|-------|-----|-----|-----|-----|-----|-----|
| Cu | 0.66 | 0.65 | 0.65 | 0.66 | 1.1 | 0.69 |
| Ag | 0.84 | 1.16 | 0.85 | | | |
| Au | 0.84 | 1.19 | 0.9 | 1.7 | 0.82 | |

In an F-metal, there are two mean-free-paths, $\lambda^{\downarrow}$ and $\lambda^{\uparrow}$.  In the two-current model, conductivities add, giving an effective transport mean-free-path of $\lambda = \frac{1}{2}(\lambda^{\downarrow} + \lambda^{\uparrow})$.  For the CPP-MR, $\lambda^{\downarrow}$ and $\lambda^{\uparrow}$ are given by $\lambda^{\downarrow} = [\rho_b \lambda_b / \rho_F](1 - \beta)$ and $\lambda^{\uparrow} = [\rho_b \lambda_b / \rho_F](1 + \beta)$.  If the magnitude of $\beta$ is large, the longer of $\lambda^{\downarrow}$ or $\lambda^{\uparrow}$ (usually $\lambda^{\uparrow}$) dominates the transport mean-free-path.

**12.1.2. Spin-Diffusion Lengths.**

We consider the spin-diffusion lengths for two different cases: (a) a non-magnetic alloy, calculated from separate measurements of the spin-flip cross-section found from Conduction Electron Spin Resonance (CESR) measurements; and (b) a ferromagnet (F).

(a)  To describe spin relaxation in a non-magnetic alloy we need to include the other two relaxation times, $\tau_s$ and $\tau_{sf}$, defined above.  According to [23], $(1/\tau_s) = 2(1/\tau_{sf})$, since each spin-flip equilibrates both spins equally.  For a spin-diffusion length to involve diffusion through a sample, we need $\tau_s$ and $\tau_{sf}$ to both be several times longer than $\tau$.

For each spin-relaxation, the equation giving its spin-diffusion length is [23]

$$l_{sf}^N = \sqrt{D\tau_x} \qquad (A.5)$$

where $D = (1/3)(v_f \lambda)$.

Plugging this expression for $D$ into Eq. A.5 gives the general relation

$$l_{sf}^N = \sqrt{(1/3)\lambda\lambda_x} \qquad (A.6.a)$$

For spin-relaxation we get

$$l_{sf}^N = \sqrt{(1/3)\lambda\lambda_s} \qquad (A.6.b)$$

For spin-flipping (CESR) we get

$$l_{sf}^N = \sqrt{(1/6)\lambda\lambda_{sf}} \qquad (A.6.c)$$

(b)  Finally, a subtlety.  For a ferromagnetic metal, Fert, Duvall, and Valet [175] found that the mean-free-path $\lambda$ that goes into Eq. A.5 should be $(1/\lambda) = (1/\lambda^{\downarrow}) + (1/\lambda^{\uparrow})$, giving.

$$l_{sf}^F = \sqrt{\frac{\lambda_{\uparrow}\,\lambda_{\downarrow}\,\lambda_{sf}^F}{3(\lambda_{\uparrow} + \lambda_{\downarrow})}} \qquad (A.7)$$

Thus, if $\beta$ is large (i.e. near +1 or -1), $\lambda$ will be dominated by the *shorter* of $\lambda^{\uparrow}$ and $\lambda^{\downarrow}$, *not* the longer 'transport' one.  This difference reduces the $l_{sf}$ expected for a ferromagnet.

## 12.2. Appendix B.  Magnetic media and CPP-MR read-head sensor characteristics.

### 12.2.1. Magnetic media.

Since 2005, the magnetic information on a hard disc has been stored in bits of magnetic material with magnetic moments oriented perpendicular to the surface of the disc (see Fig. 95).  The bits are stored along tracks in the media.  The areal density is the product (kBPI)x(kTPI) of the kilobits/inch (kBPI) along a track times the track density in kilotracks/inch (kTPI).  For example, 382 GB/in$^2$ = 1252 KBPI x 305 KTPI [229], illustrating that bits are asymmetric.  Each bit contains a number of smaller single magnetic domains.  These domains are aligned by a write head, which produces a controllable local magnetic field large enough ( > $H_s$, but $\leq$ 2.4 T) to align the domains parallel to each other.

### 12.2.2. CPP-MR read head sensor characteristics.

The minimum features for a CPP-MR sensor are shown in Fig. B.1, as viewed from the surface of the magnetic medium (called the Air Bearing Surface (ABS)).  The sensor consists of a CPP-MR EBSV of the form [AF/pinned F1/N/free F2] sandwiched between two wider and thicker leads (L) that carry current I into, through, and out of the sensor and, if the current leads are a soft magnet, shield the sensor from stray fields from bits other than the one to be read.  Important lengths are: (a) the shield-to-shield ' width W, which is the multilayer total thickness, $t_T$, (b) the length ($L$) of the multilayer, (c) the height (h) of the multilayer, and (d)  the thickness (t) of each of the two conducting lead films.  The write head is presently a separate entity.

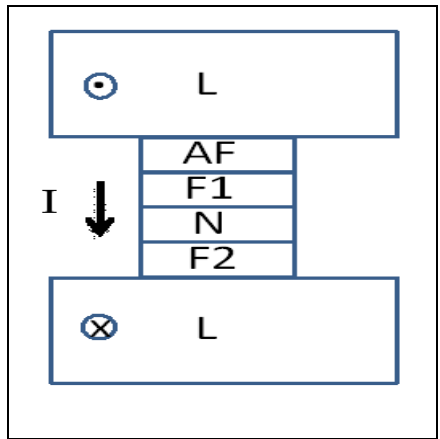

Fig. B.1. Read head schematic of AF/F1/N/F2 EBSV, as seen from the medium (ABS). Symbols on left show current I flows out of the top lead, down through the EBSV and into the bottom lead.  The leads can serve also as shields.

Since a given areal density requires the ability to read separate bits along a track, $t_T$ must not exceed the bit width, which limits the sum $t_T = t_{AF} + t_{F1} + t_N + t_{F2}$.  Separate shields are needed to isolate different tracks.  Note that area A for the CPP-MR is given by $L$xh, which does not involve $t_T$.  The contribution of h to the CPP-MR is limited by the distance that the magnetic field generated by the bit extends 'into' the sensor.

The discussion of device physics in section 10 started with the assumption that an EBSV with the standard F/N pairs, such as Py/Cu, Co/Cu, CoFe/Ag, etc., cannot produce large enough A$\Delta$R and CPP-MR for present day devices.  We test this assumption by estimating values of AR(AP), A$\Delta$R and CPP-MR  at 4.2K and 300K for a Py-based EBSV with Cu that has area A $\approx$ 0.65 f-m$^2$, which would give $\approx$ 1 Tb/in$^2$. We compare the results with the values for an all-metal sample in Fig. 96 or with the estimates for areal densities 2 and 5 Tb/in$^2$ in Takagishi et al. [230] given near the end of section 10.4.2 (see, e.g., Fig. 97).

We take an EBSV of the form FeMn(8)/Py(6)/Cu(6)/Py(6) with total thickness $t_T$ = 26 nm.  We also take the geometry as in Fig. B.1, but with a cubic multilayer having $t_T$ = W = h = L = 26 nm.  We'll see that such an EBSV's resistance R is ~ 20 $\Omega$.  For the resistivities of Cu listed below, two Cu leads, each 200 nm thick, will have total sheet resistance ~ 0.05 $\Omega$ at 4.2K and ~ 0.2 $\Omega$ at 300K, both small enough to neglect. We choose the Py thickness = 6 nm to round up its $l_{sf}$ at 4.2K, so that we can use Eq. 9 and Fig. 55 to estimate AΔR. For simplicity, we assume that this $l_{sf}$ drops by the ratio $\rho_{Py}(4.2K)/\rho_{Py}(300K) \sim 0.55$ from 4.2K to 300K, i.e., from 5.5 nm to 3 nm. At 4.2K, we use the measured parameters for FeMn, Py, and Cu. At 300K we allow for phonon scattering by increasing both $\rho_{FeMn}$ and $\rho_{py}$ by 100 nΩm and $\rho_{Cu}$ by 15 nΩm. Using the data for Co/Cu in section 8.4.3 as a guide, we leave $\gamma_{Py/Cu}$ and $AR^*_{Py/Cu}$ unchanged, and reduce $\beta_{Py}$ slightly from 0.76 to 0.7.

The parameters for our estimates at 4.2K are from column E in Table 8 and [57].

At 4.2K. $\rho_{FeMn}$ = 875 nΩm; $\rho_{py}$ = 123 nΩm; $\beta_{Py}$ = 0.76; $\gamma_{Py/Cu}$ = 0.7; $AR^*_{Py/Cu}$= 0.5 fΩm$^2$); $\rho_{Cu}$ = 5 nΩm; $t_{FeMn}$ = 8 nm, $t_{Py}$ = 6 nm, $t_{Cu}$ = 6 nm, and AR$_{FeMn/Py}$ = 1 fΩm$^2$.

At 300K. $\rho_{FeMn}$ = 975 nΩm; $\rho_{py}$ = 223 nΩm; $\beta_{Py}$ = 0.7; $\gamma_{Py/Cu}$ = 0.7; $AR^*_{Py/Cu}$= 0.5 fΩm$^2$); $\rho_{Cu}$ = 20 nΩm; $t_{FeMn}$ =8 nm, $t_{Py}$ = 11 nm, $t_{Cu}$ = 5 nm, and AR$_{FeMn/Py}$ = 1 fΩm$^2$.

**4.2K.** For AΔR at 4.2K, Eq. 9 correctly gives the thick Py limit of AΔR ($t_{Py} \gg l_{sf}^{Py}$) $\approx$ 2.2 fΩm$^2$ that is associated with the dotted curve in Fig. 57. This limit is larger than that in Fig. 55 due to the larger $\beta_{Py}$ (0.76 vs 0.73). For a Py thickness of 6 nm, we can then use AΔR ~ 1 fΩm$^2$ from the dotted curve in Fig. 57. For AR(AP), adding the contributions from the FeMn layer, the FeMn/Py interface, the two Py layers, the two Py/Cu interfaces, and the one Cu layer, gives AR(AP) ~ 13 fΩm$^2$.

**300K.** With our assumptions, Eq. 9 gives a thick Py limit of AΔR = 1.9 fΩm$^2$, a little less than that at 4.2K. Since now $t_{Py}$ = 6 nm is twice $l_{sf}^{Py}$, we take from Fig. 57 the fraction (~ 0.8) of the thick Py limit corresponding to $2l_{sf}^{Py}$. Together these give AΔR ~ 0.8 x 1.9 ~ 1.5 fΩm$^2$. Adding up the same components to AR(AP) as at 4.2K now gives AR(AP) = 15 fΩm$^2$. These assumptions give rounded rough estimates:

At 4.2K. AR(AP) = 13 fΩm$^2$; AΔR = 1 fΩm$^2$; CPP-MR =100xAΔR/(AR(AP) − AΔR) = 9%

At 300K. AR(AP) = 15 fΩm$^2$; AΔR = 1.5 fΩm$^2$; CPP-MR = 11%.

For 1 Tb/in$^2$, area A = (25.4 nm)$^2$ = 0.65 f-m$^2$.

At 4.2K, R(AP) = 20 Ω and ΔR = 1.5 Ω.

At 300K, R(AP) = 23 Ω and ΔR = 2.3 Ω.

The estimates for 4.2K should be accurate. Those for 300K may overestimate AΔR and the CPP-MR, but should be good enough to compare with estimates of need.

Comparing with Fig. 96, our estimated AR(AP)s is only ~ 40% of the 40 fΩm$^2$ assumed for a competitive all metal device, the CPP-MR is comparable to the assumed 10%, but AΔR is only ~ 40% of what is assumed.

Comparing with the estimates by Takagishi et al. in Fig. 97 [230], the results are even worse. For 2 Tbit/in$^2$, our AR(AP) is only ~ 15% of the best value ~ 100 fΩm$^2$, thus requiring a CPP-MR ~ 90% compared to an estimated 11%. For 5 Tbit/in$^2$ the situation is even less favorable.

These results show the need for higher resistivity F-alloys with βs at least as large as that for Py.

`                                    References

1.      Feynman, R.P., *The Pleasure of Finding Things Out*. 1999, Cambridge: Perseus Book, Cambridge
        Press.
2.      Jeng, M., *A selected history of expectation bias in physics.* American Journal of Physics, 2006. **74**:
        p. 578.
3.      Levy, P.M., *Giant magnetoresistance in magnetic layered and granular materials.*, in *Solid State
        Physics Series*  D.T.H. Ehrenreich, Editor. 1994, Academic Press, : New York. 47, p. 367-462.
4.      Gijs, M.A.M. and G.E.W. Bauer, *Perpendicular giant magnetoresistance of magnetic multilayers.*
        Advances in Physics, 1997. **46**: p. 285.
5.      Bass, J., W.P. Pratt Jr, and P.A. Schroeder, *Current-perpendicular (CPP) giant magnetoresistance-
        -larger and simpler than current-in-plane magnetoresistance.* Comments on Condensed Matter
        Physics, 1998. **18**: p. 223.
6.      Bass, J. and W.P. Pratt Jr, *Current-perpendicular (CPP) magnetoresistance in magnetic metallic
        multilayers.* Journal of Magnetism and Magnetic Materials, 1999. **200**: p. 274.
7.      Fert, A. and L. Piraux, *Magnetic nanowires.* Journal of Magnetism and Magnetic Materials, 1999.
        **200**: p. 338.
8.      Tsymbal, E.Y. and D.G. Pettifor, *Perspectives of giant magnetoresistance.* Solid State Physics
        Series, 2001. **56**: p. 113.
9.      Bass, J., *Giant magnetoresistance: experiment.*, in *Handbook of  spin transport and magnetism*
        E.Y. Tsymbal and I. Zutic, Editors. 2011, CRC Press, Taylor & Francis: Boca Ratonp. 69.
10.     Ansermet, J.-P., *Perpendicular transport of spin-polarized electrons through magnetic
        nanostructures.* Journal of Physics Condensed Matter, 1998. **10**: p. 6027.
11.     Bass, J., *CPP-GMR: Materials and Properties.*, in *Handbook of Spintronics*  D. Awshalom, J. Nitta,
        and Y. Xu, Editors. 2015, Springer: Bristol, UK.
12.     Bass, J. and W.P. Pratt Jr, *Current perpendicular (CPP) magnetoresistance in magnetic metallic
        multilayers (erratum).* Journal of Magnetism and Magnetic Materials, 2006. **296**: p. 65.
13.     Pratt Jr, W.P. and J. Bass, *Perpendicular-current studies of electron transport across metal/metal
        interfaces.* Applied Surface Science, 2009. **256**: p. 399.
14.     Dieny, B., et al., *Anisotropy and angular variation of the giant magnetoresistance in magnetic
        multilayers (invited).* Journal of Applied Physics, 1996. **79**: p. 6370.
15.     Urazhdin, S., R. Loloee, and W.P. Pratt Jr, *Noncollinear spin transport in magnetic multilayers.*
        Physical Review B., 2005. **71**: p. 100401(R).
16.     Slonczewski, *Current-Driven excitation of magnetic multilayers.* Journal of Magnetism and
        Magnetic Materials, 1996. **159**: p. L1-L7.
17.     Berger, L., *Emission of spin waves by a magnetic multilayer traversed by a current.* Physical
        Review B., 1996. **54**: p. 9353.
18.     Tsoi, M., et al., *Excitation of a magnetic multilayer by an electric current.* Physical Review
        Letters, 1998. **80**: p. 4281.
19.     Ralph, D.C. and M.D. Stiles, *Spin transfer torques.* Journal of Magnetism and Magnetic
        Materials., 2008. **320**: p. 1190.
20.     Baibich, M.N., et al., *Giant magnetoresistance of (001)Fe/(001)Cr superlattices.* Physical Review
        Letters, 1988. **61**: p. 2472.
21.     Binasch, G., et al., *Enhanced magnetoresistance in layered magnetic structures with
        antiferromagnetic interlayer exchange.* Physical Review B., 1989. **39**: p. 4828.
22.     Daughton, J.M., *GMR applications.* Journal of Magnetism and Magnetic Materials, 1999. **192**: p.
        334.

23.     Zutic, I., J. Fabian, and S. Das Sarma, *Spintronics: fundamentals and applications.* Reviews of Modern Physics, 2004. **76**: p. 323.

24.     Grunberg, P., et al., *Layered magnetic structures: evidence for antiferromagnetic coupling of Fe layers across Cr interlayers.* Physical Review Letters, 1986. **57**: p. 2442.

25.     Parkin, S.S.P., A. Modak, and D.J. Smith, *Dependence of giant magnetoresistance on Cu-layer thickness in Co/Cu multilayers: A simple dilution effect.* Physical Review B., 1993. **47**: p. 9136(R).

26.     Pratt Jr, W.P., et al., *Perpendicular giant magnetoresistances of Ag/Co multilayers.* Physical Review Letters, 1991. **66**: p. 3060.

27.     Zhang, S. and P.M. Levy, *Conductivity perpendicular to the plane of multilayered structures.* Journal of Applied Physics, 1991. **69**: p. 4786.

28.     Lee, S.F., et al., *Two-channel analysis of CPP-MR data for Ag/Co and AgSn/Co multilayers.* Journal of Magnetism and Magnetic Materials, 1993. **118**: p. L1.

29.     Valet, T. and A. Fert, *Theory of perpendicular magnetoresistance in magnetic multilayers.* Physical Review B., 1993. **48**: p. 7099.

30.     van der Pauw, L., *A method of measuring specific resistivity and Hall effect of discs of arbitrary shape. .* Phillips Research Reports, 1958. **13**: p. 1.

31.     Meny, C., P. Panissod, and R. Loloee, *Structural study of cobalt-copper multilayers.* Physical Review B., 1992. **45**: p. 12269.

32.     Henry, L.L., et al., *Perpendicular interface resistances in sputtered Ag/Cu, Ag/Au, and Au/Cu multilayers.* Physical Review B., 1996. **54**: p. 12336.

33.     Gu, T., A.I. Goldman, and M. Mao, *Characterization of interfacial roughness in Co/Cu multilayers by x-ray scattering.* Physical Review B., 1997. **56**: p. 6474.

34.     Jedryka, E., et al., *Structure of Co layers in Co/Cu multilayers at the first antiferromagnetic msximum studies by nuclear magnetic resonance.* Journal of Applied Physics, 1997. **81**: p. 4776.

35.     Larson, D.J., et al., *Atomic-scale analysis of CoFe/Cu and CoF/NiFe interfaces.* Applied Physics Letters, 2000. **77**: p. 726.

36.     Campbell, I.A. and A. Fert, *Transport properties of ferromagnets*, in *Ferromagnetic Materials* E.P. Wolforth, Editor. 1982, North Holland: Amsterdam. 3, p. 747.

37.     Bass, J., *Electrical resistivity of pure metals and dilute alloys. ,* in *Landolt-Bornstein, New Series, Group 3.  Metals: Electronic Transport Phenomena*  K.-H. Hellwege and J.L. Olsen, Editors. 1982, Springer-Verlag: Berlin. 15a, p. 1-288.

38.     Bass, J., *Electrical Resistivity of pure metals and dilute alloys. ,* in *Landolt-Bornstein New Series, Group 3,* K.H. Hellwege and J.L. Olsen, Editors. 1985, Springer-Verlag: Berlin. 15b, Metals, Electronic Transport Phenomena, p. 1-12.

39.     Schroeder, P.A., et al., *Perpendicular magnetoresistance in Cu/Co and Cu/(NiFe) multilayers. .* Materials Research Society Symposium Proceedings, 1993. **313**: p. 47-61.

40.     Chiang, W.-C., et al., *Effect of sputtering pressure on the structure and current-perpendicular-to-the-plane magnetotransport of Co/Ag multilayered films.* Physical Review B., 1998. **58**: p. 5602.

41.     Zambano, A., et al., *Interfacial properties of Fe/Cr multilayers in the current-perpendicular-to-plane geometry.* Journal of Magnetism and Magnetic Materials, 2002. **253**: p. 51.

42.     Cyrille, M.C., et al., *Effect of sputtering pressure-induced roughness on the microstructure and the perpendicular giant magnetoresistance of Fe/Cr superlattices.* Physical Review B., 2000. **62**: p. 15079.

43.     Parkin, S.S.P., R. Bhadra, and K.P. Roche, *Oscillatory magnetic exchange coupling through thin copper layers.* Physical Review Letters, 1991. **66**: p. 2152.

44.     Schroeder, P.A., et al., *Perpendicular magnetoresistance in Ag/Co and Cu/Co multilayers*, in *NATO ASI series, Magnetism and Structure in Systems of Reduced Dimension.*  R.F.C. Farrow, Editor. 1993, Plenum Press, NYp. 129.

45. Mott, N.F., *The electronic conductivity of transition metals.* Proceedings of the Royal Society, 1936. **A153**: p. 699.

46. Monod, P. and S. Schultz, *Conduction electron spin-flip scattering by impurities in copper.* Journal of Physics, Paris, 1982. **43**: p. 393.

47. Fierz, C., et al., *Superconductor/ferromagnet boundary resistances.* Journal of Physics Condensed Matter, 1990. **2**: p. 9701.

48. Bergmann, G., *Inelastic life-time of the conduction electrons in some noble metal films.* Zeitschrift fur Physik, 1982. **48**: p. 5.

49. Santhanam, P., S. Wind, and D.E. Prober, *One-dimensional electron localization and superconducting fluctuations in narrow aluminum wires.* Physical Review Letters., 1984. **53**: p. 1179.

50. Santhanam, P., S. Wind, and D.E. Prober, *Localization, superconducting fluctuations and superconductivity in thin films and narrow wires of aluminum.* Physical Review B., 1987. **35**: p. 3188.

51. Johnson, M. and R.H. Silsby, *Interfacial charge-spin coupling: injection and detection of spin magnetization in metals.* Physical Review Letters, 1985. **55**: p. 1790.

52. Johnson, M. and R.H. Silsby, *Ferromagnetic-nonferromagnetic interface resistance.* Physical Review Letters, 1988. **60**: p. 377.

53. Yang, Q., et al., *Spin diffusion length and giant magnetoresistance at low temperatures.* Physical Review Letters, 1994. **72**: p. 3274.

54. Bass, J. and W.P. Pratt Jr, *Spin-diffusion lengths in metals and alloys, and spin-flipping at metal/metal interfaces: an experimentalists critical review.* Journal of Physics Condensed Matter, 2007. **19**: p. 183201.

55. Eid, K., et al., *Changes in magnetic scattering anisotropy at a ferromagnetic/superconducting interface.* Physical Review B., 2004. **70**: p. 10411(R).

56. Lee, S.F., et al., *Current perpendicular and parallel giant magnetoresistances in Co/Ag multilayers.* Physical Review B., 1995. **52**: p. 15426.

57. Pratt Jr, W.P., et al., *Perpendicular-current transport in exchange-biased spin-valves.* IEEE Transactions on Magnetics, 1997. **33**: p. 3505.

58. List, M.J., et al., *CPP-magnetoresistance and electron scattering spin asymmetry in epitaxial Co/Cu multilayers*, in *Unpublished*. 1997.

59. Holody, P., et al., *Giant magnetoresistance in copper/permalloy multilayers.* Physical Review B., 1998. **58**: p. 12230.

60. Bozec, D., *Current perpendicular to the plane magnetoresistance of magnetic multilayers.*, in *Physics and Astronomy*. 2000, University of Leeds: Leeds, West Yorkshire, England.

61. Johnson, M., *Analysis of anomalous multilayer magnetoresistance within the thermomagetoelectric system.* Physical Review Letters, 1991. **67**: p. 3594.

62. Fert, A. and S.F. Lee, *Theory of the bipolar spin switch.* Physical Review B., 1996. **53**: p. 6554.

63. Bauer, G.E.W., *Perpendicular transport through magnetic multilayers.* Physical Review Letters, 1992. **69**: p. 1676.

64. Nakatani, T.M., T. Furubayashi, and K. Hono, *Interfacial resistance and spin-dependent scattering in the current-perpendicular-to-plane giant magnetoresistance using $Co_2Fe(Al_{0.5}Si_{0.5})$ Heusler alloy with Ag.* Journal of Applied Physics, 2011. **109**: p. 07B724.

65. Taniguchi, T., et al., *Effect of the number of layers on determinations of spin asymmetries in current-perpendicular-to-plane giant magnetoresistance* Applied Physics Letters, 2011. **98**: p. 042503.

66.    Taniguchi, T., et al., *Effect of the number of layers on determination of spin asymmetries in current-perpendicular-to-plane giant magnetoresistance (Erratum).* Applied Physics Letters, 2011. **99**: p. 019904.

67.    Park, W., et al., *Measurement of resistance and spin-memory loss (spin relaxation) at interfaces using sputtered current perpendicular-to-plane exchange-biased spin valves.* . Physical Review B., 2000. **62**: p. 1178.

68.    Strelkov, N., A. Vedyaev, and B. Dieny, *Extension of the semiclassical theory of current-perpendicular-to-plane giant magnetoresistance including spin flip to any multilayered magnetic structures.* Journal of Applied Physics, 2003. **94**: p. 3278.

69.    Butler, W.H., X.-G. Zhang, and J.M. MacLaren, *Solution to the Boltzmann equation for layered systems for current perpendicular to the planes.* Journal of Applied Physics, 2000. **87**: p. 5173.

70.    Shpiro, A. and P.M. Levy, *Resistance across an interface, and that measured far from it.* Physical Review B., 2000. **63**: p. 014419.

71.    Tsymbal, E.Y. and D.G. Pettifor, *Quantum-well resistivity for perpendicular transport in magnetic layered systems.* Physical Review B., 2000. **61**: p. 506.

72.    Penn, D.R. and M.D. Stiles, *Spin transport for spin diffusion lengths comparable to the mean free path.* Physical Review B., 2005. **72**: p. 212410.

73.    Steenwyk, S., et al., *Perpendicular current exchange biased spin-valve evidence for a short spin diffusion length in Permalloy.* Journal of Magnetism and Magnetic Materials, 1997. **170**: p. L1.

74.    Borlenghi, S., et al., *Multiscale approach to spin transport in magnetic multilayers.* Physical Review B., 2011. **84**: p. 035412.

75.    Levy, P.M. and S. Zhang, *Our current understanding of giant magnetoresistance in transition-metal mutilayers.* Journal of Magnetism and Magnetic Materials, 1995. **151**: p. 315.

76.    Tsymbal, E.Y., *Effect of disorder on perpendicular magnetotransport in Co/Cu multilayers.* Physical Review B., 2000. **62**: p. R3608.

77.    Baxter, R.J., D.G. Pettifor, and E.Y. Tsymbal, *Interface proximity effects in current-perpendicular-to-plane magnetoresistance.* Physical Review B., 2005. **71**: p. 024415.

78.    Sanvito, S., C.J. Lambert, and J.H. Jefferson, *Breakdown of the resistor model of CPP-GMR in magnetic multilayer nanostructures.* Physical Review B., 2000. **61**: p. 14225.

79.    Bozec, D., et al., *Mean free path effects on the CPP magnetoresistance of magnetic multilayers.* Physical Review Letters, 2000. **85**: p. 1314.

80.    Ahn, C., et al., *Current-perpendicular-to-plane transport properties of CoFe alloys: spin-diffusion length and scattering asymmetry.* Journal of Applied Physics, 2010. **108**: p. 023908.

81.    Nicholson, D.M.C., et al., *Magnetic structure of the spin valve interface.* Journal of Applied Physics, 1994. **76**: p. 6805.

82.    Oparin, A.B., et al., *Magnetic structure at copper-permalloy interfaces.* Journal of Applied Physics, 1999. **85**: p. 4548.

83.    Robinson, J.W.A., et al., *Zero to Pi transition in superconductor-ferromagnet-superconductor junctions.* . Physical Review B., 2007. **76**: p. 094522.

84.    Xu, P.X. and K. Xia, *Ab initio calculations of the alloy resistivities of lattice-matched and lattice-mismatched metal pairs: influence of local-impurity-induced distortions.* Physical Review B., 2006. **74**: p. 184206.

85.    Acharyya, R., et al., *Specific resistance of Pd/Ir interfaces.* Applied Physics Letters, 2009. **94**: p. 022112.

86.    Olson, S.K., et al., *Comparison of measured and calculated specific resistances of Pd/Pt interfaces.* Applied Physics Letters, 2005. **87**: p. 252508.

87.    Schep, K.M., et al., *Interface resistances of magnetic multilayers.* Physical Review B., 1997. **56**: p. 10805.

88.     Stiles, M.D. and D.R. Penn, *Calculation of spin-dependent interface resistance.* Physical Review B., 2000. **61**: p. 3200.

89.     Xia, K., et al., *Interface resistance of disordered magnetic multilayers.* Physical Review B., 2001. **63**: p. 064407.

90.     Bauer, G.E.W., et al., *Scattering theory of interface resistance in magnetic multilayers.* Journal of physics D: Applied Physics, 2002. **35**: p. 2410.

91.     Nguyen, H.Y.T., et al., *Conduction electron scattering and spin-flipping at sputtered Co/Ni interfaces. .* Physical Review B., 2010. **82**: p. 220401(R).

92.     Starikov, A.A., et al., *Unified first-principles study of Gilbert damping, spin-flip diffusion, and resistivity in transition metal alloys.* Physical Review Letters, 2010. **105**: p. 236601.

93.     Yang, Q., et al., *Prediction and measurement of perpendicular (CPP) giant magnetoresistance of Co/Cu/Ni$_{84}$Fe$_{16}$/Cu multilayers.* Physical Review B., 1995. **51**: p. 3226.

94.     Steenwyk, S., et al., *A comparison of hysteresis loops from giant magnetoresistance and magnetometry of perpendicular-current exchange-biased spin-valves.* Journal of Applied Physics, 1997. **81**: p. 4011.

95.     Acharyya, R., et al., *Spin-flipping associated with the antiferromagnet IrMn.* IEEE Transactions on Magnetics, 2010. **46**: p. 1454.

96.     Acharyya, R., et al., *A study of spin-flipping in sputtered IrMn using Py-based, exchange-biased spin-valves.* Journal of Applied Physics, 2011. **109**: p. 07C503.

97.     Cowburn, R.P., et al., *Single domain circular magnets.* Physical Review Letters, 1999. **83**: p. 1042.

98.     Piraux, L., et al., *The temperature dependence of the perpendicular giant magnetoresistance in Co/Cu multilayered nanowires.* Europhysics Journal B., 1998. **4**: p. 413.

99.     Dubois, S., et al., *Evidence for a short spin diffusion length in permalloy from the giant magnetoresistance of multilayered nanowires.* Physical Review B., 1999. **60**: p. 477.

100.    List, M.J., et al., *Perpendicular resistance of Co/Cu multilayers prepared by molecular beam epitaxy.* Journal of Magnetism and Magnetic Materials, 1995. **148**: p. 342.

101.    Borchers, J.A., et al., *Observation of antiparallel order in weakly coupled Co/Cu multilayers.* Physical Review Letters., 1999. **82**: p. 2796.

102.    Borchers, J.A., et al., *Polarized neutron reflectivity characterization of weakly coupled Co/Cu multilayers.* Physica B, 2000. **283**: p. 162.

103.    Unguris, J., et al., *Magnetic depth profiling Co/Cu multilayers to investigate magnetoresistance.* Journal of Applied Physics, 2000. **87**: p. 6639.

104.    Schroeder, P.A., et al., *Magnetic states of magnetic multilayers at different fields.* Journal of Applied Physics, 1994. **76**: p. 6610.

105.    Schuller, I.K. and P.A. Schroeder. 1988. p. Unpublished.

106.    Slaughter, J.M., W.P. Pratt Jr, and P.A. Schroeder, *Fabrication of layered metallic systems for perpendicular resistance measurements.* Review of Scientific Instruments, 1989. **60**: p. 127.

107.    Highmore, R.J., et al., *Magnetoresistance of Cu-Ni multilayers.* Journal of Magnetism and Magnetic Materials, 1992. **104-107**: p. 1777.

108.    Cyrille, M.C., et al., *Enhancement of perpendicular and parallel giant magnetoresistance with the number of bilayers in Fe/Cr superlattices.* Physical Review B., 2000. **62**: p. 3361.

109.    Slater, R.D., et al., *Perpendicular-current exchange-biased spin valve structures with micron-size superconducting top contacts.* Journal of Applied Physics, 2001. **90**: p. 5242.

110.    Bell, C., et al., *Fabrication of nanoscale heterostructure devices with a focused ion beam microscope.* Nanotechnology, 2003. **14**: p. 630.

111.    Nallamshetty, K. and M.A. Angadi, *Perpendicular electrical conduction in Cu/Mn multilayer films at low temperatures.* Physics Letters A, 1993. **177**: p. 67.

112. Edmunds, D.L., W.P. Pratt Jr, and J.R. Rowlands, *0.1 ppm four-terminal resistance bridge for use with a dilution refrigerator.* Review of Scientific Instruments, 1980. **51**: p. 1516.

113. Dauguet, P., P. Gandit, and J. Chaussy, *New methods to measure the current perpendicular to the plane magnetoresistance of multilayers.* Journal of Applied Physics, 1996. **79**: p. 5823.

114. Chiang, W.-C., et al., *Search for mean-free-path effects in current perpendicular-to-plane magnetoresistance.* Physical Review B., 2004. **69**: p. 184405.

115. Vila, L., et al., *Current perpendicular magnetoresistances of NiCoFe and NiFe 'permalloys'.* Journal of Applied Physics, 2000. **87**: p. 8610.

116. Krebs, J.J., et al., *Perpendicular transport and magnetic properties in patterned multilayer magnetic microstructures (invited).* Journal of Applied Physics, 1996. **79**: p. 6084.

117. Gijs, M.A.M., S.K.J. Lenczowski, and J.B. Giesbers, *Perpendicular giant magnetoresistance of microstructured Fe/Cr magnetic multilayers from 4.2 to 300K.* Physical Review Letters, 1993. **70**: p. 3343.

118. Gijs, M.A.M., et al., *Perpendicular giant magnetoresistance of microstructures in Fe/Cr and Co/Cu multilayers.* Journal of Applied Physics, 1994. **75**: p. 6709.

119. Gijs, M.A.M., et al., *New contacting technique for thin film resistance measurements perpendicular to the film plane.* Applied Physics Letters, 1993. **63**: p. 111.

120. Lenczowski, S.K.J., et al., *Current-distribution effects in microstructures for perpendicular magnetoresistance measurements.* Journal of Applied Physics, 1994. **75**: p. 5154.

121. Vavra, W., et al., *Perpendicular current magnetoresistance in Co/Cu/NiFeCo/Cu multilayered microstructures.* Applied Physics Letters, 1995. **66**: p. 2579.

122. Bussman, K., et al., *CPP giant magnetoresistance of NiFeCo/Cu/CoFe/Cu multilayers.* IEEE Transactions on Magnetics, 1998. **34**: p. 924.

123. Chen, J., et al., *Analytical method for two dimensional current crowding effect in magnetic tunnel junctions.* Journal of Applied Physics, 2003. **91**: p. 8783.

124. Dieny, B., et al., *Effect of scattering at lateral edges on the current-perpendicular-to-plane giant magnetoresistance of submicronic pillars.* Journal of Applied Physics, 2001. **89**: p. 7302.

125. Leung, C.W., et al., *In situ fabrication of a cross-bridge Kelvin resistor structure by focused ion beam microscopy.* Nanotechnology, 2004. **15**: p. 786.

126. Han, G.C., et al., *Fabrication of sub-50 nm current-perpendicular-to-plane spin valve sensors.* Thin Solid Films, 2006. **505**: p. 41.

127. Whitney, T.M., et al., *Fabrication and magnetic properties of arrays of metallic nanowires.* Science, 1993. **261**: p. 1316.

128. Blondel, A., et al., *Giant magnetoresistance of nanowires of multilayers.* Physics Letters, 1994. **65**: p. 3019.

129. Piraux, L., et al., *Giant magnetoresistance in magnetic multilayered nanowires.* Applied Physics Letters, 1994. **65**: p. 2484.

130. Liu, K., et al., *Perpendicular giant magnetoresistance of multilayered Co/Cu nanowires. .* Physical Review B., 1995. **51**: p. 7381.

131. Blondel, A., et al., *Wire-shaped magnetic multilayers for "current perpendicular to plane" magnetoresistance measurements. .* Journal of Magnetism and Magnetic Materials, 1995. **148**: p. 317.

132. Evans, P.R., G. Yi, and W. Schwarzacher, *Current perpendicular to plane giant magnetoresistance of multilayered nanowires electrodeposited in anodic aluminum oxide membranes.* Applied Physics Letters, 2000. **76**: p. 481.

133. Blondel, A., B. Doudin, and J.-P. Ansermet, *Comparative study of the magnetoresistance of electrodeposited Co/Cu multilayered nanowires made by single and dual bath techniques.* Journal of Magnetism and Magnetic Materials, 1997. **165**: p. 34.

134.	Doudin, B., et al., *Magnetic and transport properties of electrodeposited nanostructured nanowires.* IEEE Transactions on Magnetics, 1998. **34**: p. 968.

135.	Doudin, B., A. Blondel, and J.-P. Ansermet, *Arrays of multilayered nanowires.* Journal of Applied Physics, 1996. **79**: p. 6090.

136.	Doudin, B. and J.-P. Ansermet, *A new method to construct nanostructured materials of controlled morphology.* Nanostructured Materials, 1995. **6**: p. 521.

137.	Chlebny, I., B. Doudin, and J.-P. Ansermet, *Pore size distributions of nanoporous track-etched membranes.* Nanostructured Materials, 1993. **2**: p. 637.

138.	Nasirpouri, F., et al., *GMR in multilayered nanowires electrodeposited in track-etched polyester and polycarbonate membranes.* Journal of Magnetism and Magnetic Materials, 2007. **308**: p. 35.

139.	Dubois, S., et al., *Perpendicular giant magnetoresistance of NiFe/Cu multilayered nanowires.* Applied Physics Letters, 1997. **70**: p. 396.

140.	Dubois, S., et al., *Perpendicular giant magnetoresistance of NiFe/Cu and Co/Cu multilayered nanowires.* Journal of Magnetism and Magnetic Materials, 1997. **165**: p. 30.

141.	Voegeli, B., et al., *Electron transport in multilayered Co/Cu nanowires.* Journal of Magnetism and Magnetic Materials, 1995. **151**: p. 388.

142.	Tang, X.-T., G.-C. Wang, and M. Shima, *Layer thickness dependence of CPP giant magnetoresistance in individual Co/Ni/Cu mulltilayer nanowires grown by electrodeposition.* Physical Review B., 2007. **75**: p. 134404.

143.	Wang, H., et al., *Highly ordered nanometric spin-valve arrays: fabrication and giant magnetoresistance effect.* Chemical Physics Letters 2006. **419**: p. 273.

144.	Wang, H., et al., *Fabrication and magnetotransport properties of ordered sub-100 nm pseudo-spin-valve element arrays.* Nanotechnology, 2006. **17**: p. 1651.

145.	Wang, H., et al., *Synthesis and characterization of FeMn-pinned spin-valve arrays.* Applied Physics Letters, 2006. **89**: p. 052107.

146.	Ono, T. and T. Shinjo, *Magnetoresistance of multilayers prepared on microstructured substrates.* Journal of the Physical Society of Japan, 1995. **64**: p. 363.

147.	Ono, T., et al., *Magnetoresistance study of Co/Cu/NiFe/Cu multilayers prepared on V-groove substrates. .* Physical Review B., 1997. **55**: p. 14457.

148.	Shinjo, T. and T. Ono, *Nanostructured magnetism, studies of multilayers prepared on microstructured substrates. .* Journal of Magnetism and Magnetic Materials, 1998. **177-181**: p. 31.

149.	Gijs, M.A.M., et al., *Perpendicular giant magnetoresistance of Co/Cu multilayers deposited under an angle on grooved substrates.* Applied Physics Letters, 1995. **66**: p. 1839.

150.	Gijs, M.A.M., et al., *Perpendicular giant magnetoresistance using microlithography and substrate patterning techniques.* Journal of Magnetism and Magnetic Materials, 1995. **151**: p. 333.

151.	Oepts, W., et al., *Perpendicular giant magnetoresistance of Co/Cu multilayers on grooved substrates: systematic analysis of the temperature dependence of spin-dependent scattering.* Physical Review B., 1996. **53**: p. 14024.

152.	Levy, P.M., et al., *Electrical transport in corrugated multilayered structures.* Physical Review B., 1995. **52**: p. 16049.

153.	Piraux, L., S. Dubois, and A. Fert, *Perpendicular giant magnetoresistance in magnetic multilayered nanowires.* Journal of Magnetism and Magnetic Materials, 1996. **159**: p. L287.

154.	Piraux, L., et al., *Perpendicular magnetoresistance in Co/Cu multilayered nanowires.* Journal of Magnetism and Magnetic Materials, 1996. **156**: p. 317.

155.	Chiang, W.-C., et al., *Variation of multilayer magnetoresistance with ferromagnetic layer sequence: spin-memory loss effects.* Journal of Applied Physics, 1997. **81**: p. 4570.

156. Hsu, S.Y., et al., *Towards a unified picture of spin dependent transport in perpendicular giant magnetoresisance and bulk alloys.* Physical Review Letters, 1997. **78**: p. 2652.

157. Caballero, J.A., et al., *Magnetoresistance of NiMnSb-based multilayers and spin-valves.* Journal of Vacuum Science and Technology A., 1998. **16**: p. 1801.

158. Bozec, D., et al., *Comparative study of the magnetoresistance of MBE-grown multilayers: [Fe/Cu/Co/Cu]$_N$ and [Fe/Cu]$_N$[Co/Cu]$_N$.* Physical Review B., 1999. **60**: p. 3037.

159. Baxter, D.V., et al., *Resistance and spin-direction memory loss at Nb/Cu interfaces.* Journal of Applied Physics, 1999. **85**: p. 4545.

160. Fujiwara, H. and G.J. Mankey, *CPP spin-valve device.* 2000: US 6560077 B2.

161. Nagasaka, K., et al., *Giant magnetoresistance properties of specular spin valve films in a current perpendicular to plane structure.* Journal of Applied Physics, 2001. **89**: p. 6943.

162. Manchon, A., et al., *Interpretation of relationship between current perpendicular to plane magnetoresistance and spin torque amplitude.* Physical Review B., 2006. **73**: p. 184418.

163. Dassonneville, B., et al., *A way to measure electron spin-flipping at F/N interfaces and application to Co/Cu.* Applied Physics Letters, 2010. **96**: p. 022509.

164. Nguyen, H.Y.T., W.P. Pratt Jr, and J. Bass, *Spin-flipping in Pt and at Co/Pt interfaces.* Journal of Magnetism and Magnetic Materials, 2014. **361**: p. 30.

165. Gijs, M.A.M., et al., *Perpendicular giant magnetoresistance of microstructured pillars in Fe-Cr and Co-Cu magnetic multilayers.* Materials Science and Engineering B, 1995. **31**: p. 85.

166. Lee, S.F., et al., *Field-dependent interface resistance of Ag/Co multilayers.* Physical Review B., 1992. **46**: p. 548.

167. Hansen, M. and K. Anderko, *Constitution of Binary Alloys, 2nd. Edition*. 1958: McGraw-Hill.

168. Pratt Jr, W.P., et al., *Giant magnetoresistance with current perpendicular to the layer planes of Ag/Co and AgSn/Co multilayers (invited).* Journal of Applied Physics, 1993. **73**: p. 5326.

169. Bass, J., et al., *Studying spin-dependent scattering in magnetic multilayers by means of perpendicular (CPP) magnetoresistance measurements.* Materials Science and Engineering B, 1995. **31**: p. 77.

170. Yang, Q., et al., *Giant CPP-magnetoresistance of Ni/Ag multilayers.* Physica B, 1994. **194-196**: p. 327.

171. Moreau, C.E., et al., *Measurement of spin diffusion length in sputtered Ni films using a special exchange-biased spin valve geometry.* Applied Physics Letters, 2007. **90**: p. 012101.

172. Dassonneville, B., et al., *Unusual current perpendicular-to-plane magnetoresistance (MR) for thick Co layers--difference in MR for fcc and hcp Co.* IEEE Transactions on Magnetics, 2010. **46**: p. 1405.

173. Portier, X., et al., *HRTEM studyof the 'orange-peel' effect in spin valves.* Journal of Magnetism and Magnetic Materials, 1999. **198-199**: p. 110.

174. Pratt Jr, W.P., et al., *How predictable is the current perpendicular to plane magnetoresistance ? (invited).* Journal of Applied Physics, 1996. **79**: p. 5811.

175. Fert, A., J.L. Duvail, and T. Valet, *Spin relaxation effects in the perpendicular magnetoresistance of magnetic multilayers.* Physical Review B., 1995. **52**: p. 6513.

176. Bass, J., et al., *How to isolate effects of finite spin diffusion length on giant magnetoresistance in magnetic multilayers.* Journal of Applied Physics, 1994. **75**: p. 6699.

177. Fowler, Q., et al., *Spin-diffusion lengths in dilute Cu(Ge) and Ag(Sn) alloys.* Journal of Magnetism and Magnetic Materials, 2009. **321**: p. 99.

178. Hsu, S.Y., et al., *Spin-diffusion lengths of Cu$_{(1-x)}$Ni$_x$ using current perpendicular to plane magnetoresistance measurements of magnetic multilayers.* Physical Review B., 1996. **54**: p. 9027.

179.    Loloee, R., B. Baker, and W.P. Pratt Jr, *Spin-diffusion length in Cu(22.7%Ni) at 4.2K.* 2006. p. Unpublished.

180.    Gu, J.Y., et al., *Direct measurement of quasipartical evanescent waves in a dirty superconductor.* Physical Review B., 2002. **66**: p. 140507(R).

181.    Kurt, H., et al., *Spin-memory loss and CPP-magnetoresistance in sputtered multilayers with Au.* Journal of Applied Physics, 2003. **93**: p. 7918.

182.    Kurt, H., et al., *Spin-memory loss at 4.2K in sputtered Pd, Pt, and at Pd/Cu and Pt/Cu interfaces.* Applied Physics Letters, 2002. **81**: p. 4787.

183.    Eid, K., et al., *Current-perpendicular-to-plane-magnetoresistance properties of Ru and Co/Ru interfaces.* Journal of Applied Physics, 2002. **91**: p. 8102.

184.    Reilly, A.C., et al., *Giant magnetoresistance of currrent-perpendicular exchange-biased spin-valves of Co/Cu.* IEEE Transactions on Magnetics, 1998. **34**: p. 939.

185.    Eid, K., et al., *Absence of mean-free-path effects in CPP magnetoresistance of magnetic multilayers.* Physical Review B., 2002. **65**: p. 054424.

186.    Galinon, C., et al., *Pd/Ag and Pd/Au interface specific resistances and interfacial spin-flipping.* Applied Physics Letters, 2005. **86**: p. 182502.

187.    Sharma, A., et al., *Conduction electron scattering and spin-flipping at sputtered Al/Cu interfaces.* Journal of Applied Physics, 2011. **109**: p. 053903.

188.    Bozec, D., et al., *The effect of non-local electron scattering on the current-perpendicular-to-the-plane magnetoresistance of magnetic multilayers.* Journal of Physics Condensed Matter, 2000. **12**: p. 4263.

189.    Eid, K., et al., *CPP magnetoresistance of magnetic multilayers: mean-free-path is not the culprit.* Journal of Magnetism and Magnetic Materials, 2001. **224**: p. L205.

190.    Eid, K., et al., *Further evidence agains mean-free-path effects in the CPP-MR.* Journal of Magnetism and Magnetic Materials, 2002. **240**: p. 171.

191.    Michez, L.A., et al., *Magnetoresistance of magnetic multilayers containing three types of magnetic layers.* Physical Review B., 2003. **67**: p. 092402.

192.    Michez, L.A., et al., *Direct evidence for mean-free-path effects in the magnetoresisance of magnetic multilayers with the current perpendicular to the planes.* Physical Review B., 2004. **70**: p. 052402.

193.    Michez, L.A., et al., *Magnetoresistance of magnetic multilayers: a phenomenological approach.* Journal of Physics Condensed Matter, 2006. **18**: p. 4641.

194.    Michez, L.A., et al., *Dependence of the magnetoresistance of magnetic multilayers on the number of magnetic layers.* Physical Review B., 2008. **77**: p. 012408.

195.    Michez, L.A., et al., *New results for the dependence of the magnetoresistance of magnetic multilayers on the layer thickness.* Europhysics Letters, 2008. **83**: p. 57007.

196.    Vouille, C., et al., *Inverse CPP-GMR in (A/Cu/Co/Cu) multilayers (A = NiCr, FeCr, FeV) and discussion of the spin asymmetry induced by impurities.* Journal of Applied Physics, 1997. **81**: p. 4573.

197.    Vouille, C., et al., *Microscopic mechanisms of giant magnetoresistance. .* Physical Review B., 1999. **60**: p. 6710.

198.    Park, W., et al., *Test of unified picture of spin dependent transport in perpendicular (CPP) giant magnetoresistance and bulk alloys.* Journal of Applied Physics, 1999. **85**: p. 4542.

199.    Reilly, A.C., et al., *Perpendicular giant magnetoresistance of $Co_{91}Fe_9$/Cu exchange-biased spin-valves: further evidence for a unified picture.* Journal of Magnetism and Magnetic Materials, 1999. **195**: p. L269.

200.    Johnson, M., *Spin coupled resistance observed in ferromagnet-superconductor-ferromagnet trilayers.* Applied Physics Letters, 1994. **65**: p. 1460.

201. Sharma, A., et al., *Specific resistance and scattering asymmetry of Py/Pd, Fe/V, Fe/Nb, and Co/Pt interfaces.* Journal of Applied Physics, 2007. **102**: p. 113916.

202. Ahn, C., K.-H. Shin, and W.P. Pratt Jr, *Magnetotransport properties of CoFeB and Co/Ru interfaces in the current-perpendicular-to-plane geometry.* Applied Physics Letters, 2008. **92**: p. 102509.

203. Theodoropoulou, N., et al., *Specific resistance, scattering asymmetry, and some thermal instability, of Co/Al, Fe/Al, and Co$_{91}$Fe$_9$/Al interfaces. .* IEEE Transactions on Magnetics, 2007. **43**: p. 2860.

204. Sharma, A., et al., *Current-perpendicular-to-plane (CPP) magnetoresistance of ferromagnetic (F)/Al interfaces (F = Py, Co, Fe, and Co$_{91}$Fe$_9$) and structural studies of Co/Al and Py/Al..* Physical Review B., 2008. **77**: p. 224438.

205. Kim, D.K., et al., *CPP transport properties of Ni/Ru and Co$_{90}$Fe$_{10}$/Cu interfaces.* IEEE Transactions on Magnetics, 2010. **46**: p. 1374.

206. Theodoropoulou, N., et al., *Interface specific resistance and scattering asymmetry of Permalloy/Al.* Journal of Applied Physics, 2006. **99**: p. 08G502.

207. Santamaria, J., et al., *Interfacially dominated giant magnetoresistance in Fe/Cr superlattices.* Physical Review B., 2001. **65**: p. 012412.

208. Xia, K., et al., *First-principles scattering matrices for spin transport.* Physical Review B., 2006. **73**: p. 064420.

209. Elliott, R.P., *Constitution of Binary Alloys, First Supplement.* 1965: McGraw-Hill.

210. Nguyen, H.Y.T., W.P. Pratt Jr, and J. Bass, *Spin relaxation at sputtered metallic interfaces.* Applied Physics A (Materials Science and Processing), 2013. **111**: p. 361.

211. Delille, F., et al., *Thermal variation of current perpendicular-to-plane giant magnetoresistance in laminated and nonlaminated spin valves.* Journal of Applied Physics, 2006. **100**: p. 013912.

212. Nguyen, H.Y.T., et al., *Spin-flipping at sputtered Co/Ag interfaces.* Physical Review B., 2012. **86**: p. 064413.

213. Nguyen, H.Y.T., et al., *Conduction electron spin-flipping at sputtered Co$_{90}$Fe$_{10}$/Cu interfaces.* Journal of Applied Physics, 2011. **109**: p. 07C903.

214. Khasawneh, M.A., et al., *Spin-memory loss at Co/Ru interfaces.* Physical Review B., 2011. **84**: p. 014425.

215. Rojas-Sanchez, J.-C., et al., *Spin pumping and inverse spin Hall effect in platinum: the essential role of spin-memory loss at metallic interfaces.* Physical Review Letters, 2014. **112**: p. 106602.

216. Liu, Y., et al., *Interface enhancement of Gilbert damping from first principles.* Physical Review Letters, 2014. **113**: p. 207202.

217. Yuasa, H., et al., *Output enhancement of spin-valve giant magnetoresistance in current-perpendicular-to-plane geometry.* Journal of Applied Physics, 2002. **92**: p. 2646.

218. Mathon, J., *Large enhancement of the perpendicular giant magnetoresistance in pseudorandom magnetic multilayers.* Physical Review B., 1996. **54**: p. 55.

219. Mathon, J., *Ab initio calculation of the perpendicular giant magnetoresistance of finite Co/Cu(001) and Fe/Cr(001) superlattices with fluctuating layer thicknesses.* Physical Review B., 1997. **55**: p. 960.

220. Chiang, W.-C., et al., *Perpendicular giant magnetoresistance of Co/Cu multilayers with fluctuating Co layer thicknesses. .* Proceedings of the Materials Research Society, 1997. **475**: p. 451.

221. Schep, K.M., P.J. Kelly, and G.E.W. Bauer, *Giant magnetoresistance without defect scattering.* Physical Review Letters, 1995. **74**: p. 586.

222. Tsoi, M.V., A.G.M. Jansen, and J. Bass, *Search for point-contact giant magnetoresistance in Co/Cu multilayers.* Journal of Applied Physics, 1997. **81**: p. 5530.

223.    Wellock, K., et al., *Giant magnetoresistance of magnetic multilayer point contacts.* Physical Review B, 1999. **60**: p. 10291.

224.    Theeuwen, S.J.C.H., et al., *Local probing of the giant magnetoresistance.* Applied Physics Letters, 2000. **77**: p. 2370.

225.    Gu, J.Y., et al., *Enhancing current-perpendicular magnetoresistance in permalloy-based exchange-biased spin valves by increasing spin-memory loss.* Journal of Applied Physics, 2000. **87**: p. 4831.

226.    Rottmayer, R. and J.-G. Zhu, *A new design for an ultra-high density magnetic recording head using a GMR sensor in the CPP mode.* IEEE Transactions on Magnetics, 1995. **31**: p. 2597.

227.    Spallas, J.P., et al., *Perpendicular current giant magnetoresistance in 0.4 micro-m diameter Co-Cu multilayer sensors.* IEEE Transactions on Magnetics, 1996. **32**: p. 4710.

228.    Spallas, J.P., et al., *Improved performance of Cu-Co CPP GMR sensors.* IEEE Transactions on Magnetics, 1997. **33**: p. 3391.

229.    Nakamoto, K., et al., *CPP-GMR read heads with a current screen layer for 300 Gb/in$^2$ recording.* IEEE Transactions on Magnetics, 2008. **44**: p. 95.

230.    Takagishi, M., et al., *Magnetoresistance ratio and resistance area design of CPP-MR film for 2-5 Tb/in$^2$ read sensors.* IEEE Transactions on Magnetics, 2010. **46**: p. 2086.

231.    Eid, K., W.P. Pratt Jr, and J. Bass, *Enhancing current-perpendicular-to-plane magnetoresistance (CPP-MR) by adding interfaces within ferromagnetic layers.* Journal of Applied Physics, 2003. **93**: p. 3445.

232.    Oshima, H., et al., *Spin filtering effect at inserted interfaces in perpendicular spin valves.* Physical Review B., 2002. **66**: p. 140404(R).

233.    Oshima, H., et al., *Perpendicular giant magnetoresistance of CoFeB single and dual spin-valve films.* Journal of Applied Physics, 2002. **91**: p. 8105.

234.    Yuasa, H., et al., *Effect of inserted Cu on current-perpendicular-to-the-plane magnetoresistance of Fe$_{50}$Co$_{50}$ spin valves.* Journal of Applied Physics, 2003. **93**: p. 7915.

235.    Yuasa, H., et al., *The number of Cu lamination effect on current-perpendicular-to-plane giant-magnetoresistance of spin valves with Fe$_{50}$Co$_{50}$ alloy.* Journal of Applied Physics, 2005. **97**: p. 113907.

236.    Saito, M., *CPP mode magnetic sensing element including a multilayer free layer biased by an antiferromagnetic layer.* 2003: US6947263.

237.    Heijden, P., R. Rottmayer, and M. Seigler, *High resistance CPP transducer in a read/write head*. 2001: US20020052330.

238.    Dieny, B., et al., *Multilayered structures comprising magnetic nano-oxide layers for current perpendicular to plane GMR heads.* 2001: US6888703.

239.    Tanaka, A., et al., *Spin-valve heads in the current-perpendicular-to-the-plane mode for ultrahigh-density recording. .* IEEE Transactions on Magnetics, 2002. **38**: p. 84-88.

240.    Takagishi, M., et al., *The applicability of CPP-GMR heads for magnetic recording.* IEEE Transactions on Magnetics, 2002. **38**: p. 2277.

241.    Eglehoff, W.F.J., et al., *Oxygen as a surfactant in growth of giant magnetoresistance spin valves.* Journal of Applied Physics, 1997. **82**: p. 6142.

242.    Hoshiya, H., K. Hoshino, and Y. Tsuchiya, *Spin-valve head containing partial current-screen-layer, product method of said head, and current-screen method.* 2003: US20040042127.

243.    Sugawara, T., *Current-perpendicular-to-the-plane structure magnetoresistive element having the free and/or pinned layers being made of a granular film which includes an electrically conductive magnetic material and a dielectric material.* 2003: US7002781.

244.    Fujiwara, H., et al., *CPP spin-valve element.* 2003: US7538987.

245. Oshima, H., et al., *Current-perpendicular spin valves with partially oxidized magnetic layers for ultrahigh-density magnetic recording.* IEEE Transactions on Magnetics, 2003. **39**: p. 2377.

246. Tanaka, A., et al., *Readout performance of confined-current-path current-perpendicular-to-plane heads.* IEEE Transactions on Magnetics, 2004. **40**: p. 203.

247. Funayama, T., et al., *Current-perpendicular-to-plane magnetoresistance effect device with double current control layers.* 2004: US7405906.

248. Horng, C. and R.-Y. Tong, *Novel process and structure to fabricate CPP spin valve heads for ultra-high recording density*. 2004: US20050201022.

249. Li, M., et al., *Seed/AFM combination for CPP GMR device.* 2004: US20060007605.

250. Fukuzawa, H., et al., *MR ratio enhancement by NOL current-confined-path structures in CPP spin valves.* IEEE Transactions on Magnetics, 2004. **40**: p. 2236.

251. Fukuzawa, H., et al., *Nanoconstricted structure for current-confined path in current-perpendicular-to-plane spin valves with high magnetoresistance.* . Journal of Applied Physics, 2005. **97**: p. 10C509.

252. Fukuzawa, H., et al., *Large magnetoresistance ratio of 10% by $Fe_{50}Co_{50}$ layers for current-confined-path current-perpendicular-to-plane giant magnetoresisance spin-valve films.* Applied Physics Letters, 2005. **87**: p. 082507.

253. Nakamoto, K., et al., *CPP-GMR reader and wraparound shield writer for perpendicular recording.* IEEE Transactions on Magnetics, 2005. **41**: p. 2914.

254. Hoshino, K., et al., *CPP-GMR with oxidized CoFe layer on various lower-electrode materials.* IEEE Transactions on Magnetics, 2005. **41**: p. 2926.

255. Carey, M., et al., *Current-perpendicular-to-the-plane spin-valve (CPP-SV) sensor with curent-confining apertures concentrated near the sensing edge.*, USA, Editor. 2005: US20070097558.

256. Fukusawa, H., H. Yuasa, and H. Iwasaki, *Current-perpendicular-to-plane magnetoresistive element in which the magnetization direction of an intermediate metallic magnetic layer is twisted.*, USA, Editor. 2005: US8213129.

257. Jogo, A., et al., *A Co-SiO$_2$ granular materials as a new current-confining layer for current perpendicular-to-plane spin valves.* IEEE Transactions on Magnetics, 2006. **42**: p. 2356.

258. Zhang, K., et al., *Method to form a current confining path of a CPP GMR device.* . 2006: US7610674.

259. Hoshiya, H., et al., *CPP-GMR magnetic head having GMR-screen layer.* 2006: US7697245.

260. Nowak, J.J., et al., *Current perpendicular to plane magnetoresistive sensor pre-product with current confining path precursor.* 2006: US8027129.

261. Sato, J., K. Matsushita, and H. Imamura, *Effective resistance mismatch and magnetoresistance of a CPP-GMR system with current-confined paths.* IEEE Transactions on Magnetics, 2007. **44**: p. 2608.

262. Fukuzawa, H., H. Yuasa, and H. Iwasaki, *CPP-GMR films with a current-confined-path nano-oxide layer (CPP-NOL).* Journal of Physics D: Applied Physics, 2007. **40**: p. 1213.

263. Yuasa, S. and A. Fukushima, *CPP type giant magneto-resistance element and magnetic sensor.* 2007: US20080026253.

264. Zhang, K., et al., *Novel CPP device with an enhanced dR/R ratio.* 2007: US20080278864.

265. Funayama, T., K. Tateyama, and M. Takagishi, *CCP-CPP-GRM head assembly, magnetic recording/reproducing apparatus, and specification method of appropriate sense current direction of CCP-CPP-GMR head.* 2007: US7514916.

266. Berthold, T.R., et al., *Method for making a current-perpendicular-to-the-plane giant magnetoresistance (CPP-GMR) sensor with a confined-current-path (CCP).* 2008: US8178158.

267. Dieny, B., B. Rodmacq, and F. Emult, *Spin valve magnetoresistive device with conductive-magnetic material bridges in a dielectric or semiconductor layer alternatively of magnetic material.*, USA, Editor. 2008: US7830640 B2.

268. Wang, C.C., et al., *Magnetoresistance calculation in current-perpendicular-to-plane giant magnetoresistance spin valves with current-confined paths.* Journal of Applied Physics, 2009. **105**: p. 013909.

269. He, Q., Y. Chen, and J. Ding, *CCP-CPP magnetoresistive reader with high GMR value.* 2009: US 8551626.

270. Yuasa, H., et al., *Enhancement of magnetoresistance by hydrogen ion treatment for current-perpendicular-to-plane giant magnetoresistive films with a current-confined-path nano-oxide layer.* Applied Physics Letters, 2010. **97**: p. 112501.

271. Zeng, D.G., et al., *Numerical failure analysis of current-confined-path current perpendicular-to-plane giant magnetoresistance spin-valve read sensors under high current density.* Journal of Applied Physics, 2011. **109**: p. 033901.

272. Zhang, K., M. Li, and Y.-H. Chen, *Multiple CCP layers in magnetic read head devices.* 2011: US7990660.

273. Zhang, K., et al., *CPP device with a plurality of metal oxide templates in a confining current path (CCP) spacer.* 2011: US20110265325.

274. Lederman, M.M. and D.J. Kroes, *Thin film giant magnetoresistive CPP transducer with flux guide yoke structure.* 1996: US5627704.

275. Dykes, J.W. and Y.K. Kim, *Current perpendicular-to-the-plane spin valve type magnetoresistive transducer.* 1996: US5668688.

276. Pohm, A.V., et al., *Two leg, side by side, 0.6 TO 1.0 micron wide, high output, vertical, GMR head sensors.* . IEEE Transactions on Magnetics, 1998. **34**: p. 1486.

277. Barr, R.A., B.W. Crue, and M. Zhao, *CPP magnetoresistive device with reduced edge effect and method for making same.* 1998: US6198609.

278. Barr, R.A., B.W. Crue, and M. Zhao, *Current perpendicular to plane magnetoresistive device with low resistance lead.* 1999: US6134089.

279. Knapp, K.E. and R.A. Barr, *CMP magnetoresistive device and method for making same.* 2000: US6233125.

280. Mao, S., *Current perpendicular-to-plane spin valve head.* 2000: US6466419.

281. Li, S. and S. Araki, *Dual spin valve CPP MR with flux guide between free layers thereof.* 2000: US6680827.

282. Smith, N. and B. Yang, *In-stack single-domain stabilization of free layers for CIP and CPP spin-valve or tunnel-valve read heads.* 2001: US6473279.

283. Fontana Jr., R.E., et al., *CPP magnetoresistive sensor with in-stack longitudinal biasing and overlapping magnetic shield.* 2001: US 6680832.

284. Khizroev, S., et al., *Perpendicular magnetic recording head with a magnetic shield to reduce side reading.* 2001: US6738233.

285. Furukawa, A., et al., *Maximization of the output of shielded current perpendicular to the plane spin valve heads by optimum sense current and hard bias.* Journal of Applied Physics, 2002. **91**: p. 7270.

286. Matsuzono, A., et al., *Study on requirements for shielded current perpendicular to the plane spin valve heads based on dynamic read tests.* Journal of Applied Physics, 2002. **91**: p. 7267.

287. Nishiyama, Y., *CPP magnetic sensing element and method for making the same.* 2002: US20030053269.

288. Heinonen, O.G., M. Seigler, and E.W. Singleton, *Current-perpendicular-to-the-plane spin valve reader with reduced scattering of majority spin electrons.* 2002: US6791805.

289. Gill, H.S., *CPP sensor with dual self-pinned AP pinned layer structures.* 2002: US 6781798.
290. Pinarbasi, M., *Current perpendicular to the planes (CPP) spin valve sensor with in-stack biased free layer and self-pinned antiparallel (AP) pinned layer structure.* 2002: US6741432.
291. Gill, H.S., *Current perpendicular to the planes (CPP) sensor with free layer stabilized by current field.* 2002: US6819530.
292. Mauri, D. and T. Lin, *High resistance sense current perpendicular-to-plane (CPP) giant magnetoresistance (GMR) head.* 2002: US 7133264.
293. Jiang, Y., et al., *Enhancement of current-perpendicular-to-plane giant magnetoresistance by synthetic antiferromagnet free layers in single spin-valve films.* Applied Physics Letters, 2003. **83**: p. 2874.
294. Mao, S., et al., *Vertical GMR recording heads for 100 Gb/in$^2$.* IEEE Transactions on Magnetics, 2003. **39**: p. 2396.
295. Zheng, Y., K. Ju, and O. Voegeli, *Process of making a GMR improvement in CPP spin valve head by inserting a current channeling layer (CCL).* 2003: US7040005.
296. Hasegawa, N., *CPP magnetic sensing element in which pinned magnetic layers of upper and lower multilayer films are magnetized antiparallel to each other, method for making the same, and magnetic sensing device including the same.* 2003: US20030143431.
297. Nishiyama, Y., *GMR magnetic sensing element having an antiferromagnetic layer extending beyond the track width and method for making same.* 2003: US20030231436.
298. Saito, M., et al., *Narrow track current-perpendicular-to-plane spin valve GMR heads.* IEEE Transactions on Magnetics, 2004. **40**: p. 207.
299. Zhu, J.-G., et al., *Current induced noise in CPP spin valves.* IEEE Transactions on Magnetics, 2004. **40**: p. 2323.
300. Kasiraj, P. and S. Maat, *Current-perpendicular-to-plane magnetoresistive sensor with free layer stabilized against vortex magnetic domains generated by the sense current.* 2004: US7057862.
301. Carey, M.J., et al., *Current-perpendicular-to-plane magnetoresistive sensor with free layer stabilized by in-stack orthogonal magnetic coupling to an antiparallel pinned biasing layer.* 2004: US7106561.
302. Saito, M., et al., *CPP giant magnetoresistive head having antiferromagnetic film disposed in rear of element.* 2004: US7220499.
303. Saito, M., et al., *Self-pinned CPP giant magnetoresistive head with antiferromagnetic film absent from current path.* 2004: US7599155.
304. Guo, Y. and L.Y. Zhu, *Method to achieve both narrow track width and effective longitudinal stabilization in a CPP GMR read head.* 2004: US7246427.
305. Li, M., et al., *Fabrication method for an in-stack stabilized synthetic stitched CPP GMR head.* 2004: US7180716.
306. Gill, H.S., *CPP differential GMR sensor having antiparallel stabilized free layers for perpendicular recording.* 2004: US7242556.
307. Gill, H., *CPP GMR with hard magnet in stack bias layer.* 2004: US20050238924.
308. Gill, H., *Method and apparatus for providing a dual current-perpendicular-to-plane (CPP) GMR sensor with improved top pinning.* 2004: US20050213258.
309. Maat, S., et al., *Current perpendicular to the plane spin valves utilizing ultrastrong Ir-coupled antiparallel pinned layers for thick reference layer stabilation and high magnetoresistance.* Journal of Applied Physics, 2005. **98**: p. 073905.
310. Maat, S., et al., *Ultrathin CoPt-pinned current perpendicular to the plane spin-valves.* Journal of Applied Physics, 2005. **98**: p. 113907.
311. Freitag, J.M., et al., *Current perpendicular to plane magnetoresistive sensor having a shape enhanced pinned layer and an in stack bias structure.* 2005: US7522391.

312. Zhang, K., et al., *FCC-like trilayer AP2 structure for CPP GMR EM improvement.* 2005: US20070070556.

313. Smith, N., et al., *Angular dependence of spin-torque critical currents for CPP-GMR read heads.* IEEE Transactions on Magnetics, 2005. **41**: p. 2935.

314. Smith, N., et al., *Thermal and spin-torque noise in CPP (TMR and/or GMR) read sensors.* IEEE Transactions on Magnetics, 2006. **42**: p. 114.

315. Smith, N., *Modeling and measurement of spin torques in current-perpendicular-to-plane giant magnetoresistive (invited).* Journal of Applied Physics, 2006. **99**: p. 08Q703.

316. Gill, H.S., *Current perpendicular to plane (CPP) magnetoresistive sensor with free layer biasing by exchange pinning at back edge.* 2006: US7405909.

317. Carey, M.J., J.R. Childress, and S. Maat, *Current-perpendicular-to-plane (CPP) magnetoresistive sensor with improved ferromagnetic free layer structure.* 2006: US7551409.

318. Lin, T., *Current-perpendicular-to-plane sensor epitaxially grown on a bottom shield.* 2006: US7796364.

319. Childress, J.R., et al., *Dual current-perpendicular-to-plane giant magnetoresistive sensors for magnetic recording heads with reduced sensitivity to spin-torque-induced noise.* Journal of Applied Physics, 2006. **99**: p. 085305.

320. Childress, J.R., et al., *Fabrication and recording study of all-metal dual-spin-valve CPP read heads.* IEEE Transactions on Magnetics, 2006. **42**: p. 2444.

321. Childress, J.R., et al., *All-metal current-perpendicular-to-plane giant magnetoresistance sensors for narrow-track magnetic recording.* IEEE Transactions on Magnetics, 2008. **44**: p. 90.

322. Maat, S., M.J. Carey, and J.R. Childress, *Magnetotransport properties and spin-torque effects in current perpendicular to the plane spin valves with Co-Fe-Al magnetic layers. .* Journal of Applied Physics, 2007. **101**: p. 093905.

323. Maat, S., M.J. Carey, and J.R. Childress, *Magnetotransport properties and spin-torque effects in current perpendicular to the plane spin valves with Co-Fe-Al magnetic layers (Erratum).* Journal of Applied Physics, 2007. **102**: p. 049902.

324. Gill, H.S. and J. Pemsiri, *Current perpendicular to plane magnetoresistive sensor with reduced read gap.* 2007: US7961440.

325. Carey, M.J., et al., *High-output current-perpendicular to the plane giant magnetoresistance sensor with synthetic-ferrimagnet free layer and enhanced spin-torque critical currents.* Applied Physics Letters, 2008. **93**: p. 102509.

326. Smith, N., et al., *Coresonant enhancement of spin-torque critical currents in spin valves with a synthetic ferrimagnet free layer.* Physical Review Letters, 2008. **101**: p. 247205.

327. Maat, S., et al., *Suppression of spin torque noise in current perpendicular to the plane spin-valves by addition of Dy cap layers.* Applied Physics Letters, 2008. **93**: p. 103506.

328. Nikolaev, K., et al., *Heusler alloy based current-perpendicular-to-the-plane giant magnetoresistance heads for high density magnetic recording.* Journal of Applied Physics, 2008. **103**: p. 07F533.

329. Liu, F., S. Li, and L.L. Chen, *Current perpendicular-to-plane read sensor with back shield.* 2008: US8077435.

330. Lin, C.C. and M. Li, *CPP with elongated pinned layer.* 2008: US7978441.

331. Min, T., et al., *CPP magnetic recording head with self-stabilizing vortex configuration.* 2008: US8018690.

332. Zeng, D.G., K.-W. Chung, and S. Bae, *Thermomigration-induced magnetic degradation of current perpendicular to the plane giant magnetoresistance spin-valve read sensors operating at high current density.* Journal of Applied Physics, 2009. **106**: p. 113908.

333. Zeng, D.G., et al., *Numerical simulation of current density induced magnetic failure for giant magnetoresistance spin valve read sensors.* Journal of Applied Physics, 2010. **108**: p. 023903.

334. Zeng, D.G., et al., *Effects of media stray field on electromigration characteristics in current-perpendicular-to-plane giant magnetoresistance spin-valve read sensors.* Journal of Applied Physics, 2012. **111**: p. 093921.

335. Carey, M.J., et al., *Co$_2$MnGe-based current-perpendicular-to-the-plane giant-magnetoresistance spin-valve sensors for recording head applications. .* Journal of Applied Physics, 2011. **109**: p. 093912.

336. Caballero, J.A., et al., *Effect of deposition parameters on the CPP GMR of NiMnSb-based spin-valve structures.* Journal of Magnetism and Magnetic Materials, 1999. **198-199**: p. 55.

337. Seyama, Y., A. Tanaka, and M. Oshiki, *Giant magnetoresistance properties of CoFe/Cu multilayer in the CPP (current perpendicular to the plane) geometry.* IEEE Transactions on Magnetics, 1999. **35**: p. 2838.

338. Hosomi, M., et al., *Film structure dependence of the magnetoresistive properties in current perpendicular to plane spin valves and its relation with current in plane magnetoresistive properties.* Journal of Applied Physics, 2002. **91**: p. 8099.

339. Hoshiya, H. and K. Hoshino, *Current-perpendicular-to-the-plane giant magnetoresistance with half-metal materials laminated between CoFe layers.* Journal of Applied Physics, 2004. **95**: p. 6774.

340. Hoshiya, H. and K. Hoshino, *Current-perpendicular-to-the-plane spin-valve films with iron-added magnetite layers.* Journal of Applied Physics, 2005. **97**: p. 10C504.

341. Aoshima, K.-I., et al., *Low resistance spin-valve-type current-perpendicular-to-plane giant magnetoresistance with Co$_{75}$Fe$_{25}$.* Journal of Applied Physics, 2005. **97**: p. 10C507.

342. Hasegawa, N., et al., *CPP magnetic detecting element including a CoFe pinned layer or free layer.* 2005: US20060044705.

343. Nagasaka, K., et al., *CPP-GMR technology for future high-density magnetic recording.* Fujitsu Science and technology journal, 2006. **42**: p. 149.

344. Oshima, H., et al., *Novel synthetic ferrimagnet pinned layers for metallic CPP spin valves requiring high pinning-field and output.* IEEE Transactions on Magnetics, 2005. **41**: p. 2929.

345. Jogo, A., et al., *Current-perpendicular spin valves with high-resistivity ferromagnetic metals for ultrahigh-density magnetic recording.* Journal of Magnetism and Magnetic Materials, 2007. **309**: p. 80.

346. Singh, L.J., et al., *Magnetoresistance of spin valve structures based on the full Heusler alloy Co$_2$MnSi..* Journal of Applied Physics, 2006. **100**: p. 013910.

347. Yakushiji, K., et al., *Current-perpendicular-to-plane magnetoresistance in epitaxial Co$_2$MnSi/Cr/Co$_2$MnSi trilayers.* Applied Physics Letters, 2006. **88**: p. 222504.

348. Kodama, K., et al., *Current-perpendicular-to-plane giant magnetoresistance of a spin valve using Co$_2$MnSi Heusler alloy electrodes. .* Journal of Applied Physics, 2009. **105**: p. 07E905.

349. Furubayashi, T., et al., *Current-perpendicular-to-plane giant magnetoresistance in spin-valve structures using epitaxial Co$_2$FeAl$_{0.5}$Si$_{0.5}$ trilayers.* Applied Physics Letters, 2008. **93**: p. 122507.

350. Furubayashi, T., et al., *Structure and transport properties of current-perpendicular-to-the-plane spin valves using Co$_2$Fe(Al$_{0.5}$Si$_{0.5}$) Heusler alloy electrodes.* Journal of Applied Physics, 2010. **107**: p. 113917.

351. Tripathy, D. and A.O. Adeyeye, *Current-perpendicular-to-plane giant magnetoresistance in half-metallic pseudo-spin-valve structures.* Journal of Applied Physics, 2008. **103**: p. 07D702.

352. Sakuraba, Y., et al., *Mechanism of large magnetoresisance in Co$_2$MnSi/Ag/Co$_2$MnSi devices with current perpendicular to the plane.* Physical Review B., 2010. **82**: p. 094444.

353. Maat, S., M.J. Carey, and J.R. Childress, *Current perpendicular to the plane spin valves with CoFeGe magnetic layers.* Applied Physics Letters, 2008. **93**: p. 143505.

354. Mizuno, T., et al., *Transport and magnetic properties of CPP-GMR sensor with CoMnSi Heusler alloy.* IEEE Transactions on Magnetics, 2008. **44**: p. 3584.

355. Nikolaev, K., et al., *'All-Heusler alloy" current-perpendicular-to-plane giant magnetoresistance.* Applied Physics Letters, 2009. **94**: p. 222501.

356. Sakuraba, Y., et al., *Enhancement of spin-asymmetry by L2₁-ordering in Co₂MnSi/Cr/Co₂MnSi current-perpendicular-to-plane magnetoresistance devices.* Applied Physics Letters, 2009. **94**: p. 012511.

357. Iwase, T., et al., *Large interface spin-asymmetry and magnetoresistance in fully epitaxial Co₂MnSi/Ag/Co₂MnSi current-perpendicular-to-plane magnetoresistive devices.* Applied Physics Express, 2009. **2**: p. 063003.

358. Nakatani, T.M., et al., *Bulk and interfacial scattering in current-perpendicular-to-plane giant magnetoresistance with Co₂Fe(Al₀.₅Si₀.₅) Heusler alloy layers with Ag spacer. .* Applied Physics Letters, 2010. **96**: p. 212501.

359. You, C.Y., et al., *Current-perpendicular-to-the-plane giant magnetoresistance of an all-metal spin valve structure with Co₄₀Fe₄₀B₂₀ magnetic layer.* Applied Physics Letters, 2010. **96**: p. 142503.

360. Shimazawa, K., et al., *CPP-GMR film with ZnO-based novel spacer for future high-density magnetic recording.* IEEE Transactions on Magnetics, 2010. **46**: p. 1487.

361. Takahashi, Y.K., et al., *Large magnetoresistance in current-perpendicular-to-plane pseudospin valve using a Co₂Fe(Ge₀₅Ga₀.₅) Heusler alloy.* Applied Physics Letters, 2011. **98**: p. 152501.

362. Takahashi, A., et al., *Erratum: Large Magnetoresistance in current-perpendicular-to-plane pseudo spin valve using a Co₂Fe(Ge₀.₅Ga₀.₅) Heusler alloy,APL 98, 152501 (2011).* Applied Physics Letters, 2011. **98**: p. 189901.

363. Hase, N., et al., *Enhancement of current-perpendicular-to-plane giant magnetoresistance by insertion of Co₅₀Fe₅₀ layers at the Co₂Mn(Ga₀.₅Sn₀.₅) interface.* Journal of Applied Physics, 2011. **109**: p. 07E112.

364. Sato, J., et al., *Large magnetoresistance effect in epitaxial Co(2)Fe(0.4)Mn(0.6)Si/Ag/Co(2)Fe(0.4)Mn(0.6)Si devices.* Applied Physics Express, 2011. **4**: p. 113005.

365. Sakuraba, Y., et al., *Extensive study of giant magnetoresistance properties in half-metallic Co(2)(Fe,Mn)Si-based devices.* Applied Physics Letters, 2012. **101**: p. 252408.

366. Hase, n., et al., *Effect of NiAl underlayer and spacer on magnetoresistance of current-perpendicular-to-plane spin valves using Co(2)Mn(Ga(0.5)Sn(0.5).* Journal of Magnetism and Magnetic Materials, 2012. **324**: p. 440-444.

367. Goripati, H.S., et al., *Current-perpendicular-to-plane giant magnetoresistance in Co₂Fe(Ga₁₋ₓGeₓ) Heusler alloy.* Journal of Applied Physics, 2013. **113**: p. 043901.

368. Li, S., et al., *Enhancement of giant magnetoresistance by L2₁ ordering in Co₂Fe(Ge₀.₅Ga₀.₅) Heusler alloy current-perpendicular-to-plane pseudo spin valves.* Applied Physics Letters, 2013. **103**: p. 042405.

369. Takahashi, Y.K., et al., *Structure and magnetoresistance of current-perpendicular-to-plane pseudo spin valves using Co₂Mn(Ga₀.₂₅Ge₀.₇₅) Heusler alloy.* Journal of Applied Physics, 2013. **113**: p. 223901.

370. Du, Y., et al., *Polycrystalline current-perpendicular-to-plane giant magnetoresistance pseudo spin-valves using Co₂Mn(Ga₀.₂₅Ge₀.₇₅) Heusler alloy.* Journal of Applied Physics, 2013. **114**: p. 053910.

371. Du, Y., et al., *001 textured polycrystalline current-perpendicular-to-plane pseudo spin-valves using Co₂Fe(Ga₀.₅Ge₀.₅) Heusler alloy.* Applied Physics Letters, 2013. **103**: p. 202401.

372.     Chen, J., et al., *Crystal orientation dependence of current-perpendicular-to-plane giant magnetoresistance of pseudo spin-valves with epitaxial Co$_2$Fe(Ge$_{0.5}$Ga$_{0.5}$) Heusler alloy layers.* Journal of Applied Physics, 2014. **115**: p. 233905.

373.     Chen, J., et al., *Crystal Orientation dependence of band matching in all-B2-trilayer current-perpendicular-to-plane giant magnetoresistance pseudo spin-valves using Co$_2$Fe(Ge$_{0.5}$Ga$_{0.5}$) Heusler alloy and NiAl spacer.* Journal of Applied Physics, 2015. **117**: p. 17C119.

374.     Du, Y., et al., *Large magnetoresistance in current-perpendicular-to-plane pseudo spin-valves using Co$_2$Fe(Ga$_{0.5}$Ge$_{0.5}$) Heusler alloy and AgZn spacer.* Applied Physics Letters, 2015. **107**: p. 112405.

375.     Furubayashi, T., et al., *Enhancement of current-perpendicular-to-plane giant magnetoresistance in Heusler-alloy based pseudo spin valves by using a CuZn spacer.* Journal of Applied Physics, 2015. **118**: p. 163901.

376.     Diao, Z., et al., *Half-metal CPP GMR sensor for magnetic recording.* Journal of Magnetism and Magnetic Materials, 2014. **356**: p. 73-81.

377.     Narisawa, H., T. Kubota, and K. Takanashi, *Current perpendicular to film plane type giant magnetoresistance effect using a Ag-Mg spacer and Co(2)Fe(0.4)Mn(0.6)Si Heusler alloy electrodes.* Applied Physics Express, 2015. **8**: p. 063008.

378.     Read, J.C., et al., *Current-perpendiculr-to-the-plane giant magnetoresistance in spin-valves with AgSn alloy spacers.* Journal of Applied Physics, 2015. **118**: p. 043907.

379.     Ziman, J.M., *The Physics of Metals: 1. Electrons*. 1969: Cambridge University Press.